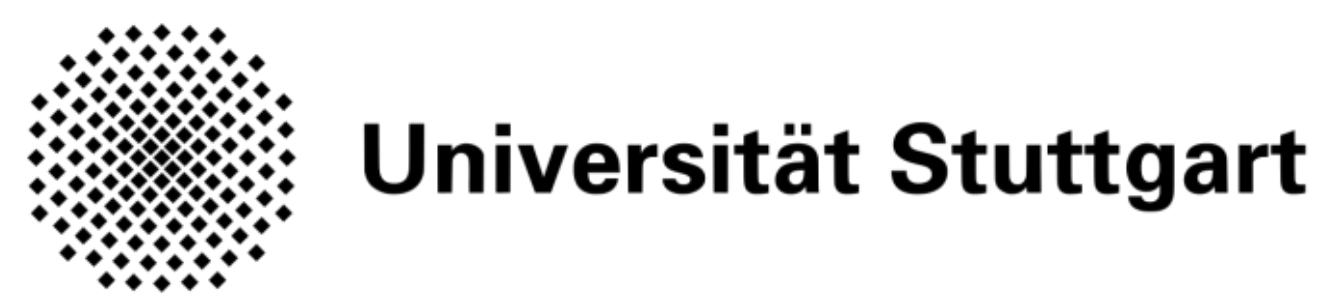

# Integrating Large Language Model Agents with Digital Twins for Industrial Autonomous Systems

Von der Graduate School of Excellence advanced Manufacturing Engineering
der Universität Stuttgart zur Erlangung der Würde eines
Doktor-Ingenieurs (Dr.-Ing.) genehmigte Abhandlung

Vorgelegt von
Yuchen Xia
aus China

| | |
|---|---|
| Hauptberichter: | Prof. Dr.-Ing. Dr. h. c. Michael Weyrich |
| Mitberichter: | Prof. Dr.-Ing. Steffen Becker |

Tag der mündlichen Prüfung: 2026.03.23

Institut für Automatisierungs- und Softwaresysteme (IAS)
der Universität Stuttgart

2026

This thesis was developed during my work as academic staff at the Institute of Industrial Automation and Software Engineering (IAS) at the University of Stuttgart.

My heartfelt gratitude goes to my family for their selfless support and unspoken sacrifices, which have always been my deepest motivation and have allowed me to fully dedicate myself to the successful completion of my doctoral studies.

I would like to sincerely thank my doctoral advisor, Prof. Dr.-Ing. Dr. h.c. Michael Weyrich, for his continuous support and for providing an excellent academic environment that enabled me to pursue my doctoral research and fully develop my potential.

I am also grateful to Dr.-Ing. Nasser Jazdi for his thoughtful support. His mentorship has been both academically inspiring and personally meaningful, guiding me in developing my ideas and offering valuable life insights.

I would also like to thank my colleagues at IAS for the collaborative and supportive working environment, as well as for the many valuable discussions.

Furthermore, I would like to thank the students who completed their master's theses with me. Teaching and working with them has been a wonderful social experience, providing a strong sense of connection and serving as a great source of motivation and inspiration throughout my doctoral journey.

Stuttgart, March 2026

Yuchen Xia

# Table of Contents

# Abstract

Industrial automation is being reshaped by advances in digitalization and the growing use of cyber-physical systems. Modern production environments demand higher adaptability, faster reconfiguration, and more intuitive system-supported human-machine interaction. These requirements exceed the capabilities of traditional rule-based systems, which are inherently rigid and rely heavily on manually engineered fixed logic. As a result, such systems are unable to autonomously adjust their behavior to accommodate the variability and dynamism of modern production environments.

The central problem addressed in this dissertation is that current industrial automation systems lack a systematic approach for integrating adaptive and generalizable reasoning capabilities. Such reasoning capabilities are required to enable the systems to autonomously interpret, plan, and execute variable user tasks under dynamically changing system conditions and across heterogeneous components.

To address this problem, this dissertation proposes a generalizable three-layer framework that integrates large language models (LLMs), digital twins, and automation systems into an autonomous system. Within this framework, autonomy is conceptualized as a design property that is assigned to system components and enabled by LLM-based reasoning to achieve adaptive and objective-oriented system behavior. Furthermore, the Task–Process–Service–Resource (TPSR) model is introduced as a unified mechanism for transforming user tasks into executable processes. Four functional roles through which LLMs contribute to task automation are identified: process orchestration, service matching, digital resource generation, and agent-as-a-service. Five peer-reviewed studies instantiate and refine these concepts through iterative design cycles based on the design science research methodology.

The developed concepts are demonstrated through case studies and prototype implementations. These demonstrations show that the proposed approach enables adaptive task planning, event-driven control, simulation-based parameterization, and digital model generation. Across these case studies and evaluations, the systems achieve high task executability, command correctness, and content-generation accuracy, automating a substantial portion of the required manual work.

The resulting framework and design knowledge enable the integration of flexible LLM-based reasoning capabilities into industrial automation systems and thereby improve their adaptability and usability, particularly in use cases such as process planning, system control, and information model generation. Limitations arise from the dependency on precise digital representations, the computational demands of LLMs, and the continued need for human intervention in safety-critical conditions. These limitations delineate the boundary conditions of the designed systems and indicate directions for future research.

# Zusammenfassung

Die industrielle Automatisierung wird zunehmend durch Digitalisierung und den wachsenden Einsatz cyberphysischer Systeme geprägt. Moderne Produktionsumgebungen erfordern eine höhere Anpassungsfähigkeit, schnellere Rekonfigurierbarkeit und eine intuitivere Interaktion zwischen Menschen und Automatisierungssystemen. Diese Anforderungen übersteigen die Leistungsfähigkeit traditioneller, regelbasierter Systeme, die inhärent starr sind und stark auf manuell entwickelte, festgelegte Logik angewiesen bleiben. Dadurch fehlt solchen Systemen die Fähigkeit, ihr Verhalten autonom anzupassen, sodass sie die Variabilität und Dynamik moderner Produktionsumgebungen nicht bewältigen können.

Das zentrale Problem dieser Dissertation besteht darin, dass aktuelle industrielle Automatisierungssysteme über keinen systematischen Ansatz zur Integration solcher adaptiven und generalisierbaren Reasoning-Fähigkeiten verfügen. Solche Reasoning-Fähigkeiten sind erforderlich, um die Systeme in die Lage zu versetzen, variierende Nutzeraufgaben unter dynamisch veränderlichen Systembedingungen und über heterogene Komponenten hinweg autonom zu interpretieren, zu planen und auszuführen.

Die Dissertation leistet hierzu einen Beitrag in Form eines generalisierbaren, dreischichtigen Rahmenwerks, das Large-Language-Modelle (LLMs), digitale Zwillinge und Automatisierungssysteme zu einem integrierten autonomen System zusammenführt. In diesem Rahmen wird Autonomie als Gestaltungseigenschaft konzipiert, die gezielt einzelnen Systemkomponenten zugewiesen wird und durch LLM-basiertes Reasoning zur Erreichung eines adaptiven und zielorientierten Systemverhaltens genutzt wird. Darüber hinaus wird das Task-Process-Service-Resource-Modell als einheitlicher Mechanismus zur Überführung von Benutzeraufgaben in ausführbare Prozesse eingeführt. Vier funktionale Rollen, durch die LLMs zur Aufgabenautomatisierung beitragen, werden identifiziert: Prozessorchestrierung, Service-Matching, Generierung digitaler Ressourcen sowie Agent-as-a-Service. Fünf begutachtete Studien instanziieren und verfeinern diese Konzepte in iterativen Designzyklen auf Grundlage der Design-Science-Research-Methodik.

Die entwickelten Konzepte werden anhand von Fallstudien und prototypischen Implementierungen demonstriert. Diese Demonstrationen zeigen, dass der entwickelte Ansatz adaptive Aufgabenplanung, ereignisgesteuerte Steuerung, simulationsbasierte Parametrierung und digitale Modellgenerierung ermöglicht. In den Fallstudien und Prototypenauswertungen erreichen die Systeme eine hohe Ausführbarkeit von geplanten Aufgaben, eine hohe Korrektheit der Steuerungsbefehle und eine hohe Genauigkeit der Inhaltsgenerierung, wodurch ein großer Teil der ansonsten erforderlichen manuellen Arbeit automatisiert wird.

Der resultierende Rahmen und das daraus gewonnene Designwissen ermöglichen die Integration flexibler, LLM-basierter Reasoning-Fähigkeiten in industrielle Automatisierungssysteme und verbessern dadurch deren Anpassungsfähigkeit und Nutzbarkeit, insbesondere in Anwendungsfällen wie der Prozessplanung, der Systemsteuerung und der Informationsmodellgenerierung. Einschränkungen ergeben sich aus der Abhängigkeit von präzisen digitalen Repräsentationen, den hohen Rechenanforderungen von LLMs sowie dem weiterhin erforderlichen menschlichen Eingreifen in sicherheitskritischen Situationen. Diese Einschränkungen umreißen die Randbedingungen der entworfenen Systeme und weisen auf zukünftige Forschungsrichtungen hin.

# 1 Introduction

Industrial automation is undergoing a profound transformation. Driven by advances in digitalization, connectivity, and artificial intelligence, production systems are evolving from rigidly structured control systems toward intelligent and adaptive cyber-physical systems [1–5]. The Industry 4.0 initiative exemplifies this shift, emphasizing digitalized and interoperable production systems that enable efficiency, flexibility, and end-to-end data integration [1,2,5–7]. However, this transformation also introduces new complexity to system design, operation, and management. Three persistent challenges emerge: adaptability, complexity management, and knowledge scarcity.

### Adaptability under Dynamic External Conditions

Traditional automation architectures are optimized for stability and operate under fixed workflows and predefined logic. When products, resources, or environmental conditions change, such systems require manual reconfiguration or redeployment [5]. Achieving continuous operation under variable conditions requires automation systems capable of runtime reconfiguration [5,8–14]—that is, dynamic adjustment of processes, services, and resources in response to changing objectives and conditions [3–5,11,14,15].

### Complexity Management of Internally Heterogeneous System Components

Modern industrial environments combine heterogeneous components including machines, robotic units, software modules, and human–machine interfaces [5,9,10,12,16–18]. These components often have distinct data formats and control semantics [2,3,7,15,19] and coordinating them demands semantic integration and context-aware information processing so that the system can align distributed functionalities toward a common objective [3,10,20,21]. Conventional automation logic, based on rule-driven control, lacks the capacity to interpret such heterogeneous representations and to autonomously coordinate operations at scale [5,9,10,22].

### Knowledge Scarcity and Usability Bottlenecks

Automation still relies heavily on expert knowledge for system design and operation [5,15,17,23]. Such expertise is fragmented and costly, limiting adaptability and hindering timely problem solving [8,9,15,22]. To enhance usability, automation systems should incorporate mechanisms for intelligent reasoning and knowledge-driven task solving [3,6,22,24], assisting users to achieve task objectives with reduced engineering effort [4,5].

Collectively, these challenges highlight the need for novel design approaches for autonomous systems that enable (1) runtime adaptability under dynamic conditions [5,12], (2) systematic management of heterogenous components and their complexity [5,9,12], and (3) intelligent reasoning that solve user tasks [5].

## 1.1 Research Conceptualization

To address the identified challenges, the following sections conceptualize the foundational definitions of autonomy in automation and describe how large language model (LLM) agents and digital twins form the core building blocks of such systems.

### 1.1.1 Automation System, Autonomous System, and Automaton

Automation in the industrial context is defined as the creation and application of technology to monitor and control the production process and delivery of products and services [25]. Similarly, the IEC vocabulary standard [26] defines "automatic" as "pertaining to a process or equipment that, under specified conditions, functions without human intervention". Synthesizing these standard definitions, an automation system can be defined as a coordinated hardware–software infrastructure that monitors, controls, and regulates processes through sensors, actuators, and the underlying control logic with minimal human input. Within this overarching concept of automation systems, two conceptual subclasses can be distinguished, as illustrated in Figure 1.

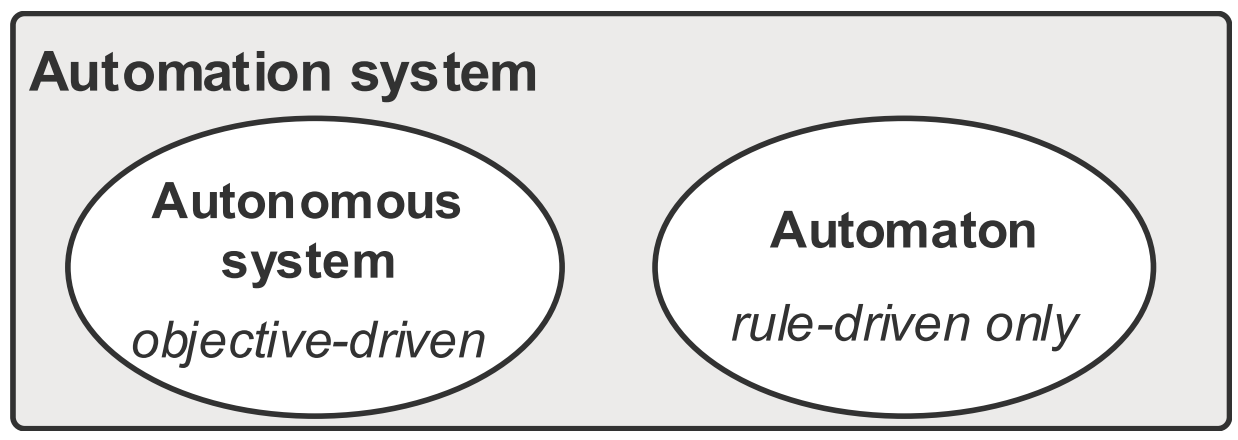

**Figure 1 Definitions: conceptual distinction between autonomous system and automaton as mutually exclusive subclasses of automation systems.**

An automaton is a traditional automation system that operates in closed and fully predictable environments. Its behavior, as defined in [26], is rule-driven, governed by predefined logic and fixed relationships [15,21]. While such systems ensure reliability under fixed conditions, their behavior cannot adapt to objectives or conditions that fall outside predefined scenarios [4,5,8,9,15,17,22].

An autonomous system, by contrast, is an automation system with the capability for context-aware and objective-oriented reasoning. It operates under open and partially unpredictable input conditions, dynamically interprets context and adjusts its behavior to achieve objectives [4,14]. In this sense, autonomy implies the freedom of runtime flexibility in reasoning and adaptive decision-making within a defined functional role [5,10].

In this dissertation, an autonomous system is defined as a specialized automation system that operates under open and partially unpredictable input conditions, capable of context-aware and flexible runtime reasoning to achieve its objectives [5,27,28]. Achieving this level of autonomy calls for two key innovations: a new paradigm of software design centered on the concept of autonomous agents [20,29], and the integration of general-purpose reasoning capabilities [4,30] enabled by LLMs.

## 1.1.2 Autonomous Agent

An autonomous agent (see Figure 2) is defined as a software component with a defined functional role that interprets contextual information and performs flexible reasoning to produce outputs aligned with its task objective [29,31]. In this work, context is defined as the digital representation that captures the system's structural information, its dynamic state, and the conditions of its surrounding environment at a given moment. The agent continuously acquires this context, reasons over it, and generates outputs that influence the system's behavior in pursuit of its objective [21,29].

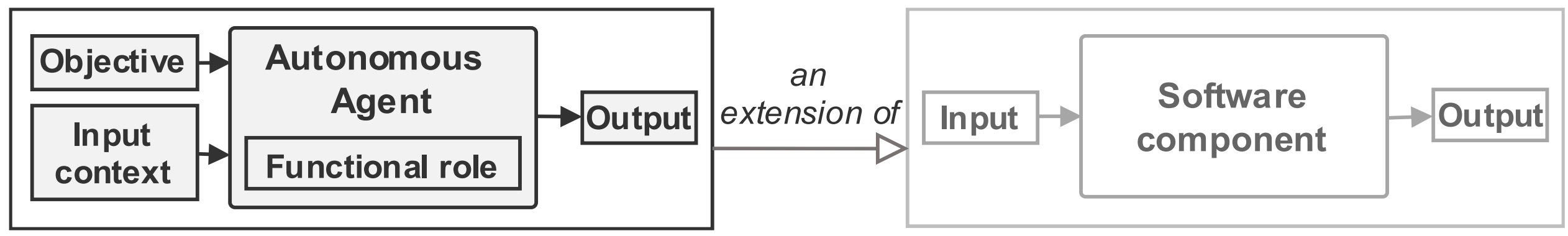

**Figure 2 An autonomous agent is conceptualized as a specialized software component that is extended with objectives and contextual input to produce output through reasoning.**

This conception of the autonomous agent forms the design foundation for the autonomous system framework developed in this dissertation: by embedding context-aware and objective-oriented reasoning into modular software components, autonomous agents enable flexible, purpose-driven information processing within software-supported automation systems in dynamic environments.

In this regard, reasoning is defined as the process of transforming inputs into objective-aligned outputs through intermediate steps or tokens [31,32], as illustrated in Figure 3.

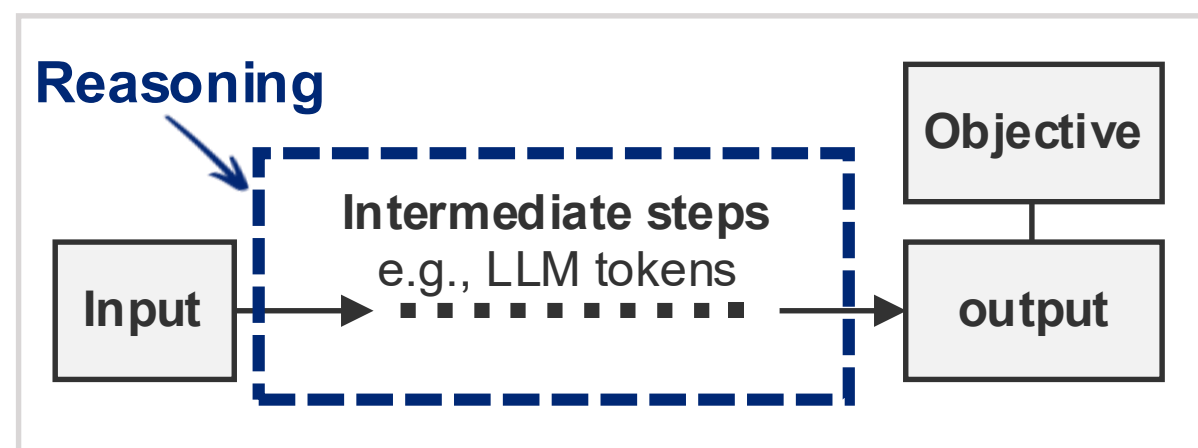

**Figure 3 Schematic illustration of reasoning: transforming input into objective-aligned output via intermediate steps or LLM tokens.** Adapted from [32], with modifications.

## 1.1.3 LLM Agent

An agent's autonomy is an intentionally designed property that allows the agent to flexibly achieve its objectives. The LLM agent integrates an LLM as its reasoning core [5,7,27,28,33], and the LLM provides the general-purpose reasoning capability that enables the autonomous agent to generate outputs aligned with its objectives [21,29,33–35]. In this paradigm, automation is redefined around flexible tasks rather than fixed processes [31,35]. A task represents an objective-oriented unit of work that the system interprets and solves at runtime [36].

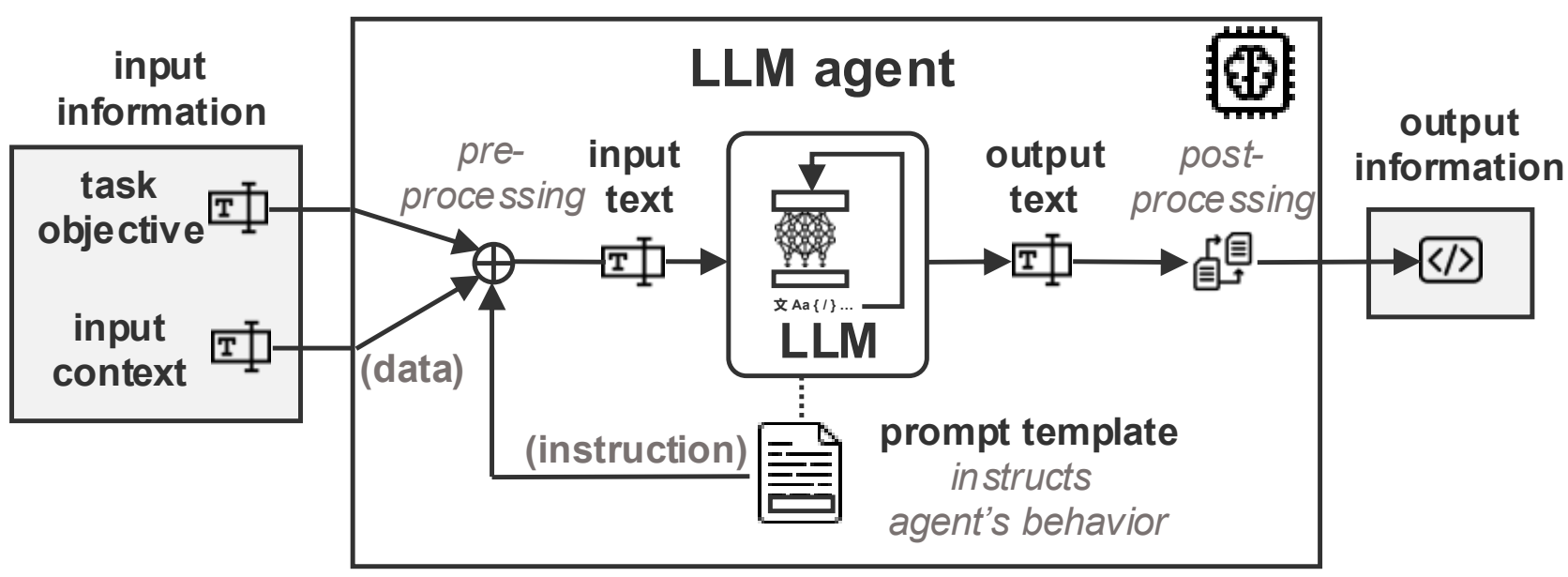

**Figure 4 LLM agent as an object-oriented information-processing software component**. The agent receives input information comprising a task objective and an input context, which are treated as data. The LLM is provided with a prompt template that defines the agent's behavior, which is treated as instruction. This information is combined through pre-processing into structured input text, upon which the LLM performs reasoning to generate output text. The output can then be post-processed into executable or machine-readable formats for downstream processing

As illustrated in Figure 4, an LLM agent receives text as input, combines it with a function-specific prompt, and performs reasoning using the LLM. The resulting output text serves as the agent's reasoning result that can be utilized by other components within the system [5,7,27,28]. This mechanism is characterized by several desirable properties:

- **Semantic reasoning.** Through LLMs' large-scale pre-training and reinforcement learning alignment processes, they develop semantic understanding and general-purpose reasoning capabilities that allow them to produce objective-aligned outputs beyond what rule-based systems can achieve [7,35,37,38].
- **Context awareness.** LLM reasoning is conditioned by arbitrary runtime input, ensuring that the agent output remains grounded in the current system context [33,35–37,39].
- **Prompt programmability.** The functional behavior of an LLM can be configured through a text-based prompt template that specifies instructions, while the dynamic input text provides runtime data [7,37,38]. This allows for the programmability of the LLM agent.
- **Interface operability.** Output from the LLM can be parsed into structured formats that meet downstream interface requirements, such as JSON objects, structured code, command line instructions or formatted data structures [7]. This enables the LLM agent to interoperate seamlessly with other system components [33,35,37,38].

Through these properties, LLM agents act as intelligent, context-aware information processors within a digitalized environment [29,31,35,36,40]. Their operation relies on digital inputs and outputs, which in turn require an underlying digital infrastructure that supports the overall information-processing workflow [36].

### 1.1.4 Digital Twin

Digital twins provide a digital environment [1,19,41,42] where LLM agents perceive inputs (perception) and produce outputs that trigger changes (actions). They synchronize data with the physical system [5,13,19,23,30,41–44] and can functionally provide two core components: **representation** and **service** (see Figure 5).

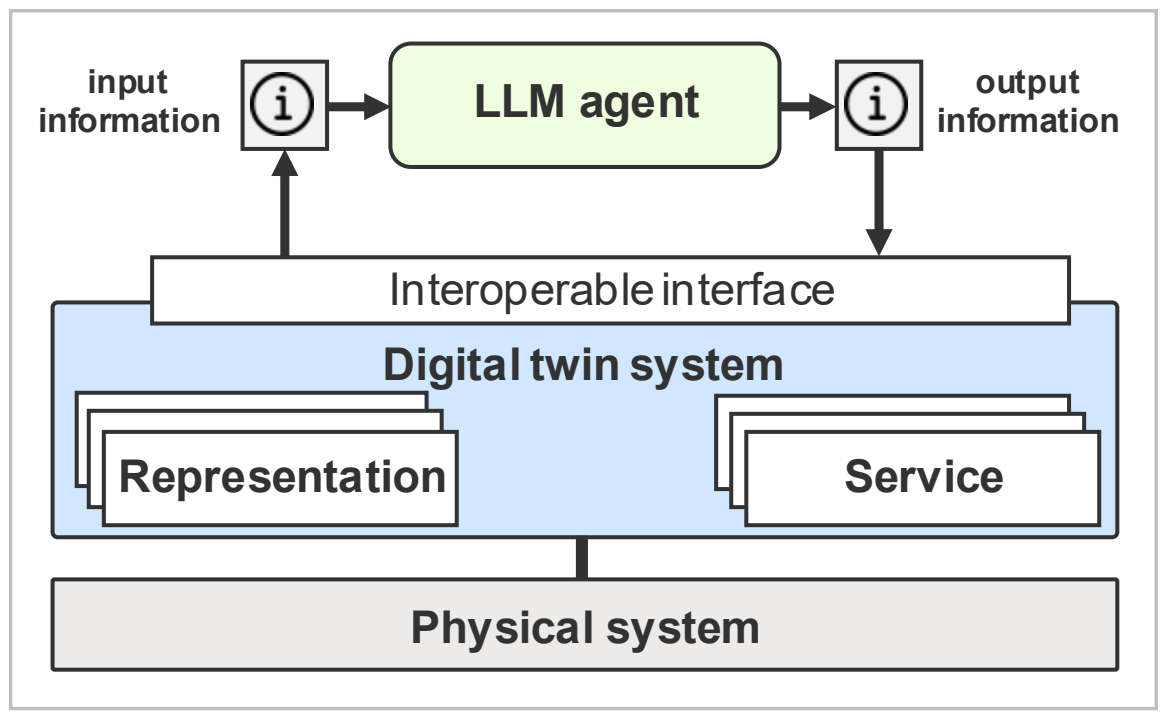

**Figure 5 Conceptual framework illustrating the interaction between the LLM agent, the digital twin system, and the physical system.** The digital twin mediates between the physical and software layers by providing representations and services through an interoperable interface, supplying contextual input to the LLM agent and receiving schema-conformant output information for further system operations.

**Representation** supplies the LLM agent with the necessary input about the current state of the system. These include real-time observations, such as sensor readings and system states, as well as model-based representations that describe system composition and functions [24,41,42,44].

**Service** exposes functions that can be invoked to interact with system components, enabling the translation of LLM-generated outputs into operations within the broader software-supported automation system [41,42,44].

Representation and service form the foundational pillars of the digital twin system. By mediating between perception and action, the digital twin system enables LLM agents to reason contextually and to transform their reasoning outcomes into system behavior.

## 1.2 Problem Awareness—Research Gaps

The conceptual foundations of autonomous systems have been established in the preceding section; however, existing research remains fragmented across three areas: automation technology in the industrial automation domain, systems modeling in the software domain, and large language models in the natural language processing domain. This disciplinary isolation has led to conceptual and architectural discontinuities that hinder the development of coherent intelligent automation solutions.

Three persistent limitations can be observed:

- **Industrial automation** remains largely rule-based and scenario-specific [2,9]. It performs reliably under fixed conditions but cannot easily adapt to unforeseen tasks or contexts [2,5,22].
- **Digital twins** provide representations and interfaces, enabling data exchange and system interoperability [3,24,30]. However, their roles remain primarily descriptive and reactive;

therefore, integrating intelligent capabilities is essential for enabling autonomous decision-making [41,42,44].

- **Large language models** offer powerful reasoning and generative capabilities. As an emerging disruptive innovation, they reshape the boundaries of intelligent automation; however, their concrete integration into industrial automation systems has not been systematically investigated [5,22].

These gaps reveal the absence of a unifying architectural framework, as well as the design knowledge required to construct it, that can systematically connect digital twin infrastructures with the reasoning capabilities of LLM agents. This dissertation is therefore motivated by the need to develop novel approaches and to establish a unified foundation for designing LLM-driven autonomous systems.

## 1.3 Research Questions

Building upon the identified challenges and gaps, this dissertation addresses the overarching question: *How can LLMs and digital twins be systematically integrated into industrial automation systems to enable adaptive and flexible task automation?*

Four specific research questions (RQs) guide the investigation, each targeting a key aspect of the problem:

- RQ1—System design: What architectural design enables the integration of LLM agents, digital twins, and automation systems into a coherent framework?
- RQ2—Task automation: What mechanisms can systematically operationalize tasks into executable task-solving processes by utilizing the reasoning capability of LLM agents?
- RQ3—Functional roles of LLMs: What functional roles can LLMs fulfill within this framework to enable task automation?
- RQ4—Case study and evaluation: How can the effectiveness of LLM-integrated automation be evaluated across typical industrial automation use cases?

## 1.4 Research Methodology—Design Science Research (DSR)

Building on the initial problem awareness, this dissertation adopts Design Science Research (DSR) [45–49] as its overarching methodology to structure the systematic investigation. Within this paradigm, research progresses through iterative design cycles in which problem analysis, artifact construction, evaluation, and reflection jointly contribute to the creation of validated design knowledge, which is consolidated across the studies of this cumulative dissertation (see Figure 6). To be scientifically meaningful, the **Design Cycle** must be embedded within two external cycles: the **Relevance Cycle**, which connects the research to practical requirements and field testing in the application environment, and the **Rigor Cycle**, which grounds the artifacts in established theories and methods and incorporates the results back into the knowledge base:

- **Relevance Cycle** links to **Practical Relevance** by addressing concrete problems and opportunities in the application environment and demonstrating their utility through empirical case testing.
- **Rigor Cycle** links to **Knowledge Base** through building on established theories, methods, and prior meta-artifacts, while integrating newly generated design knowledge back into the base.

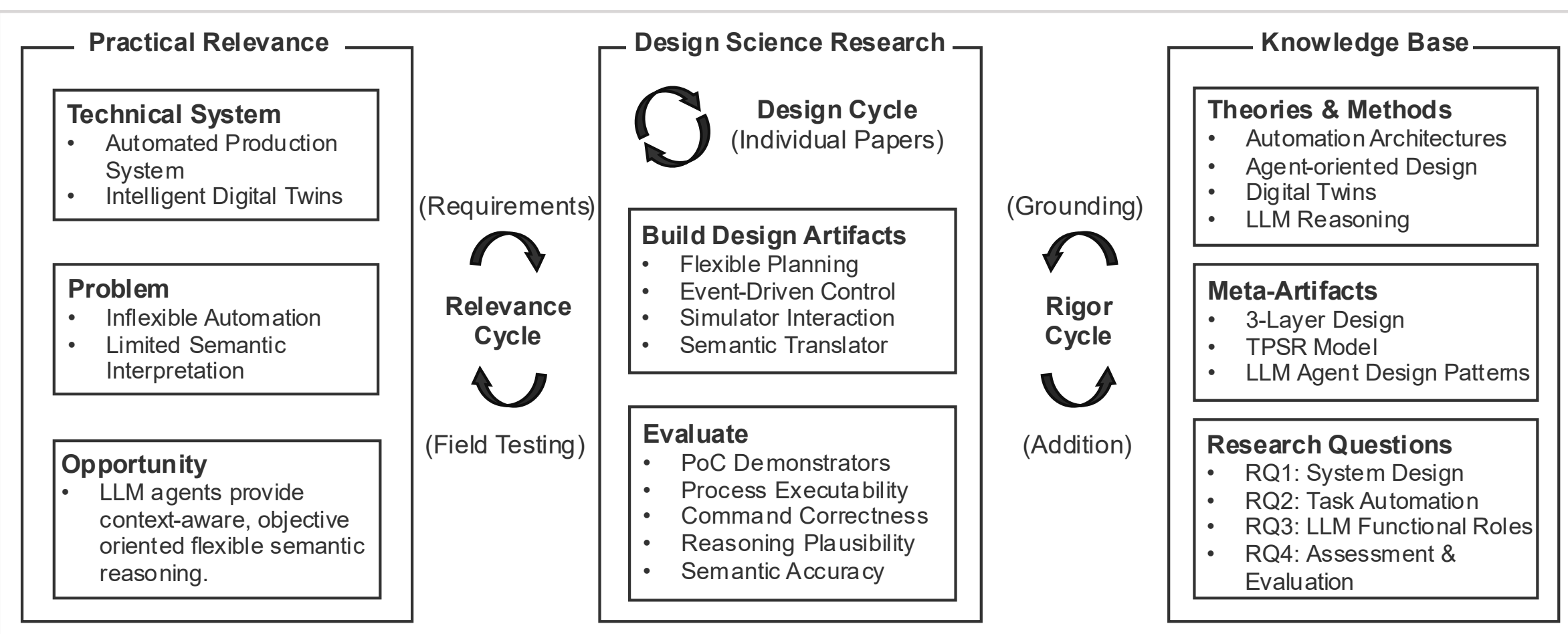

**Figure 6 Dissertation overview mapped onto the DSR framework, linking practical relevance, iterative artifact design, and the knowledge base.** Framework illustration adapted from [45,48]

In addition to the three-cycle framework, this dissertation follows the established DSR knowledge-creation process at the level of each individual study [49,50]. Figure 7 summarizes this process: each included paper starts from a concrete technical system, problem, and opportunity, goes through a design cycle of problem awareness, design, development, and evaluation, and thereby yields novel practical solutions as well as generalizable design science knowledge that contributes to the existing knowledge base.

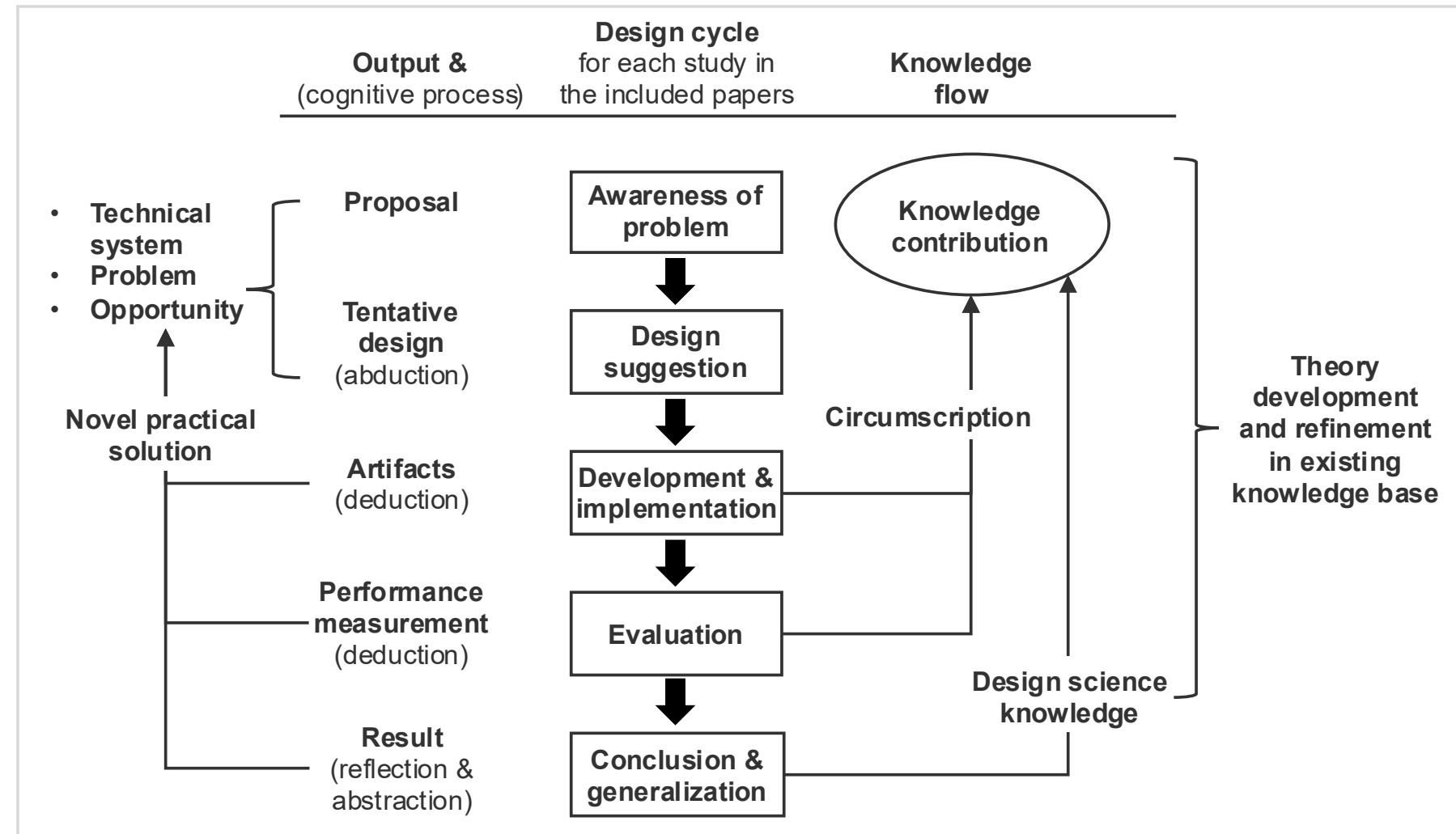

**Figure 7 Knowledge creation process of Design Science Research at the level of each included study, linking practical problems, design cycle activities, and resulting knowledge contributions.** Methodologies and illustration adapted from [49,50].

## 1.5 Critical Analysis of Existing Design Approaches

From a DSR perspective [48,49], this section examines existing design approaches in industrial automation and digital twin technologies to establish the background knowledge base. It evaluates how current standards, architectures, and modeling methods address practical needs and identifies the methodological and technological limitations that constrain their applicability. By contrasting these limitations with the capabilities afforded by LLMs, the section derives the unmet design requirements and reveals new opportunities for LLM-integrated approaches.

### 1.5.1 Automation System Architecture and Agent-oriented Design

Existing research on the design of automation systems has progressed from the rigid hierarchical structures of the traditional automation pyramid toward distributed, adaptive, and intelligent paradigms such as Holonic Manufacturing Systems (HMS), Cyber-Physical Systems (CPS), and Service-oriented Multi-agent Systems (SoMaS).

The ISA-95/IEC 62264 [51] framework represents a canonical hierarchical architecture that standardizes system layers and their interactions, thereby improving modularity, coordination, and overall system manageability. This hierarchy is often illustrated as an automation pyramid [52]. However, these hierarchical architectures are tightly coupled, making reconfiguration costly [53,54], and data exchange across layers is often constrained or even unavailable [54]. Consequently, these architectures provide only limited support for dynamically changing production scenarios [54–56].

Holonic Manufacturing Systems (HMS) architectures [57] introduce modularity and distributed autonomy to improve flexibility and reconfigurability, but their communication and coordination mechanisms remain complex, which has hindered their broader industrial adoption [56–59].

To improve data availability and connectivity, Cyber-Physical Systems (CPS) architectures bridge the physical and cyber worlds, enabling real-time data acquisition and dynamic reconfiguration to support dynamic system adaptation [16,53,55,60]. Despite these strengths, they still rely on cumbersome and effort-intensive integration processes to handle heterogeneous subsystems and diverse data representations at the software level, which prevents them from achieving semantic interoperability [16,55,60,61].

Software designs such as agent-oriented and service-oriented paradigms have informed the development of Service-oriented Multi-agent Systems (SoMaS) [53–55], which further promote decentralization, autonomy, and service reuse at the software design layer [16]. However, these agents still suffer from limited semantic reasoning capabilities [61]. Most agent systems rely on rule-driven mechanisms—such as IF–THEN rules or finite state machines—where intelligence is effectively constrained by a predefined rule base, making it difficult to scale or adapt to dynamically changing context [7,28,54,61]. Run-time responsiveness is also restricted, as service orchestration typically depends on predesigned logic rather than context-aware decision-making [28,62].

In light of these methodological developments and persistent limitations, these findings highlight the need for more flexible and semantically capable reasoning mechanisms than those supported by existing methods, thereby motivating the exploration of LLM-based agent systems as a promising approach.

### 1.5.2 Intelligent Semantic Digital Twin

Digital Twins (DT) are characterized by their ability to construct high-fidelity virtual representations of physical assets and processes. This capability is enabled by the systematic integration of heterogeneous data, including sensor measurements, operating states, and engineering models. Through this integration, the digital twin forms a coherent digital representation that bridges physical systems and high-level software functions, thereby ensuring data availability and supporting interoperable control, while also enabling advanced functions such as real-time monitoring, virtualized process evaluation, and predictive analysis [41,63–69]. Although digital twins can provide rich representation and synchronization, their designs lack built-in mechanisms for intelligent data utilization [41,64,66,67]. As a result, advanced data processing is delegated to dedicated AI modules rather than treated as inherent functions of the digital twin.

When viewed from an intelligence-integration perspective, a deeper limitation becomes evident: the absence of generalizable semantic interpretation [41,64,68,69]. Effective data utilization relies on deriving machine-interpretable meaning from underlying representations, yet the data accessible through digital twins often remain heterogeneous and context-dependent, often demanding locally tailored processing that may not scale across systems [63,64,66,68,69]. To address this limitation, recent work on semantic digital twins incorporates semantic modeling constructs that unify heterogeneous information representations [24,41,70]. Techniques including model transformation [71–74], ontologies [68,70,71,75,76], knowledge graphs [24,71,75,76], and standardized vocabularies [7,77,78] support interoperable automated data processing across heterogeneous systems. However, fixed mapping rules and semantic standardization often lead to lossy semantic translations and substantial engineering overhead in maintaining the rule base, thereby constraining scalable and adaptive processing of heterogeneous data [63,66,69–74,76]. As a result, integrating intelligence into digital twins relies on strong domain priors and predefined rule-based mappings, yielding narrow and task-specific intelligence with limited generalization [63,64,66,67,71,74]. The lack of semantic interpretability further prevents the development of adaptive decision-making, restricting autonomous adjustment to evolving operational contexts [63,66]. Consequently, even AI-augmented digital twins—often described as “intelligent” or “cognitive”—remain incapable of context-aware, flexible semantic reasoning [68–70,75].

Synthesizing these limitations reveals persistent capability gaps in semantic interpretation, adaptive intelligence, and cross-context generalization within current digital twin methodologies. Such gaps highlight the need for more flexible and context-aware forms of reasoning, providing the methodological motivation for integrating LLM-based intelligence.

### 1.5.3 Motivation for Integration of LLMs in Industrial Automation

LLMs represent a disruptive innovation in machine intelligence. Unlike algorithmic or data-driven techniques, their intelligence emerges from semantic interpretation and context-aware reasoning rather than explicit programming. This paradigm has not been anticipated by the established design methodologies in industrial automation, where intelligence has traditionally been engineered through deterministic logic, symbolic rules, or statistical models. Integrating LLM agents into industrial automation therefore raises a fundamental methodological question: how can LLM semantic reasoning be coherently embedded within structured automation systems? This question has remained open since the emergence of LLMs. The included paper [5] presents the first systematic and experimental approach to tackling this challenge in industrial automation. The originality of this work has been independently recognized in peer-reviewed surveys and structured literature reviews [79–86]. Follow-up studies [7,27,28,62] further advanced this line of research within this cumulative dissertation.

## 1.6 Design Cycles of the Included Papers within the DSR Framework

Each paper in this cumulative dissertation represents a DSR design cycle, as listed in Table 1. Collectively, the studies develop, implement, and evaluate artifacts for LLM-integrated industrial automation, generating design knowledge through iterative evaluation and reflection.

**Table 1 Overview of the included papers mapped to the Design Science Research (DSR) cycle.**

| Paper Cycle | Developed Design Artifacts | Addressed Problem with Case Study | Evaluation | Generalizable Design Principle |
|---|---|---|---|---|
| **A** [5] | Multi-agent process orchestration framework[1] | Dynamic task decomposition and process planning | Proof-of-concept with prototype | Three-layer LLM–DT–automation structure |
| **B** [28] | Event-driven control framework[2] | Dynamic context-aware control | Proof-of-concept with prototype + test cases + training | Perception–action loop for LLM agents |
| **C** [27] | Simulation-integrated multi-agent system[3] | Dynamic simulation interaction | Simulative prototype | Perception–action loop for LLM agents |
| **D** [7] | LLM-based semantic translator[4] | Semantic interoperability | Test cases + human annotation | LLM as semantic translator |
| **E** [62] | Meta design architecture | Methodology generalization across use cases | – | LLM multi-agent system design; Three paradigms of process modeling |

[1] https://github.com/YuchenXia/GPT4IndustrialAutomation
[2] https://github.com/YuchenXia/LLM4IAS
[3] https://github.com/YuchenXia/LLMDrivenSimulation
[4] https://github.com/YuchenXia/AASbyLLM

Papers A–D address the four research questions (RQ1—system design; RQ2—task automation; RQ3—functional roles of LLMs; RQ4—case studies and evaluation) across problem contexts and different technical systems. Each paper contributes to these questions from the perspective of its specific use case, provides system designs and insights that jointly extend the overall design knowledge. Paper E synthesizes the recurring regularities across these studies and derives the generalizable design knowledge.

## 1.7 Structure of the Dissertation

The dissertation is organized into five chapters, each corresponding to a core element of the Design Science Research methodology:

- **Chapter 1** introduces the research context and conceptual foundations, frames the research questions, and establishes the initial problem awareness. It discusses the practical relevance, reviews the existing knowledge base to identify methodological limitations, and derives the design requirements that motivate the use of large language models as a key enabling technology for integrating LLM agents into industrial autonomous systems.
- **Chapter 2** presents and summarizes the five peer-reviewed publications (Paper A–Paper E, as listed in Table 1), each constituting one iterative DSR design cycle and contributing artifacts toward answering RQ1–RQ4 at a case-specific level.
- **Chapter 3** synthesizes the design knowledge derived from these cycles, introducing the generalized three-layer architecture and the Task–Process–Service–Resource (TPSR) model, and explaining the functional roles of LLM agents in task automation, addressing RQ1–RQ3 at a meta level.
- **Chapter 4** synthesizes the assessment criteria and reports the empirical evaluation of the implemented artifacts across representative use cases, addressing the evaluation dimension of DSR and RQ4 at a meta level.
- **Chapter 5** concludes the dissertation, summarizes the practical and theoretical contributions, and provides a research outlook.

This structure reflects the cumulative nature of the dissertation, linking individual studies into a consistent research trajectory and a unified body of design knowledge.

# 2 Included Publications

This cumulative dissertation includes five peer-reviewed publications [5,7,27,28,62] (Papers A–E), which are embedded in the Design Science Research (DSR) framework [45–49]. The studies iteratively design, implement, and evaluate artifacts for LLM-integrated industrial automation, and together produce generalizable design knowledge.

## 2.1 Overview of Included Publications

The five included papers are listed in Table 2. The full texts are provided in Appendix—Reprinted Publications

**Table 2 Publications included in the dissertation**

| | **Included publications** |
|---|---|
| A | Towards autonomous system: flexible modular production system enhanced with large language model agents |
| | © 2023 IEEE. Reprinted, with permission, from<br>***Y. Xia**, M. Shenoy, N. Jazdi and M. Weyrich, "Towards autonomous system: flexible modular production system enhanced with large language model agents," 2023 IEEE 28th International Conference on Emerging Technologies and Factory Automation (ETFA), Sinaia, Romania, 2023, pp. 1-8, doi: 10.1109/ETFA54631.2023.10275362.* |
| B | Control Industrial Automation System with Large Language Model Agents |
| | © 2025 IEEE. Reprinted, with permission, from<br>***Y. Xia**, N. Jazdi, J. Zhang, C. Shah and M. Weyrich, "Control Industrial Automation System with Large Language Model Agents," 2025 IEEE 30th International Conference on Emerging Technologies and Factory Automation (ETFA), Porto, Portugal, 2025, pp. 1-8, doi: 10.1109/ETFA65518.2025.11205539.* |
| C | LLM experiments with simulation: Large Language Model Multi-Agent System for Simulation Model Parametrization in Digital Twins |
| | © 2024 IEEE. Reprinted, with permission, from<br>***Y. Xia**, D. Dittler, N. Jazdi, H. Chen and M. Weyrich, "LLM experiments with simulation: Large Language Model Multi-Agent System for Simulation Model Parametrization in Digital Twins," 2024 IEEE 29th International Conference on Emerging Technologies and Factory Automation (ETFA), Padova, Italy, 2024, pp. 1-4, doi: 10.1109/ETFA61755.2024.10710900.* |
| D | Generation of Asset Administration Shell with Large Language Model Agents: Toward Semantic Interoperability in Digital Twins in the Context of Industry 4.0 |
| | © 2024 IEEE. Reprinted, with permission, from<br>***Y. Xia**, Z. Xiao, N. Jazdi and M. Weyrich, "Generation of Asset Administration Shell With Large Language Model Agents: Toward Semantic Interoperability in Digital Twins in the Context of Industry 4.0," in IEEE Access, vol. 12, pp. 84863-84877, 2024, doi: 10.1109/ACCESS.2024.3415470.* |
| E | An Architecture for Integrating LLMs with Digital Twins and Automation Systems |
| | © 2025 IEEE. Reprinted, with permission, from<br>***Y. Xia**, N. Jazdi and M. Weyrich, "An Architecture for Integrating Large Language Models with Digital Twins and Automation Systems," 2025 IEEE 30th International Conference on Emerging Technologies and Factory Automation (ETFA), Porto, Portugal, 2025, pp. 1-8, doi: 10.1109/ETFA65518.2025.11205636.* |

Paper A [5] presents the first proof-of-concept study integrating LLM agents with industrial automation systems through a digital-twin-mediated architecture. The implemented prototype demonstrates autonomous and flexible production planning on a modular, automated production system. Paper B [28] extends this system design to dynamic, event-driven, closed-loop control and additionally introduces model training on a collected dataset. The quantitative evaluation of the training shows significant accuracy improvements in control command generation over baseline models. Paper C [27] investigates the integration approach of LLM agents for controlling a simulator within digital twin environments, without necessarily involving a physical system. Paper D [7] investigates the semantic translation capabilities of LLM agents, focusing on how they interpret and generate information models to enable semantic interoperability in digital twins. Paper E [62] synthesizes key design patterns for integrating LLMs into industrial automation systems, providing an architectural perspective that unifies insights from the preceding studies.

## 2.2 Paper A: Towards Autonomous System: Flexible Modular Production System Enhanced with LLM Agents

**Relevance:** Modern automation systems struggle to adapt to frequently changing production demands, as process planning is still largely predefined and manually reconfigured. Existing rule-based control frameworks built on fixed workflows lack the intelligence to autonomously interpret new task objectives and orchestrate the process operations needed to achieve them. To address this limitation, this study [5] explores whether LLM agents can function as general-purpose reasoning components that translate natural-language task input from users into machine-executable process plans, thereby improving the adaptability and useability of a modular production system.

**Design:** The designed system is an LLM multi-agent framework that integrates LLM agents that plan and control production processes within a modular automation system. As illustrated in Figure 8, the agents are mapped to automation modules. Their skills and interfaces are encapsulated according to the representations provided by the digital twin system. The multi-agent system comprises role-specific agents that interpret user tasks and decompose them into executable module-level skills and component-level functionalities, which can be executed via the interfaces exposed by the digital twin. Each agent is realized as a software component configured by a devised prompt template that defines its functional role, objectives, operational context, and output schema.

**Implementation:** The design is instantiated and deployed on a real-world modular production facility equipped with Asset Administration Shell (AAS) middleware that exposes standardized service interfaces for each automation module. In this setup, a user provides a task objective in natural language, which is interpreted by LLM agents powered by the OpenAI *text-davinci-003* model to orchestrate operation processes to accomplish the task. A demonstration of the system with further implementation details is available in a public GitHub repository.[5]

[5] https://github.com/YuchenXia/GPT4IndustrialAutomation

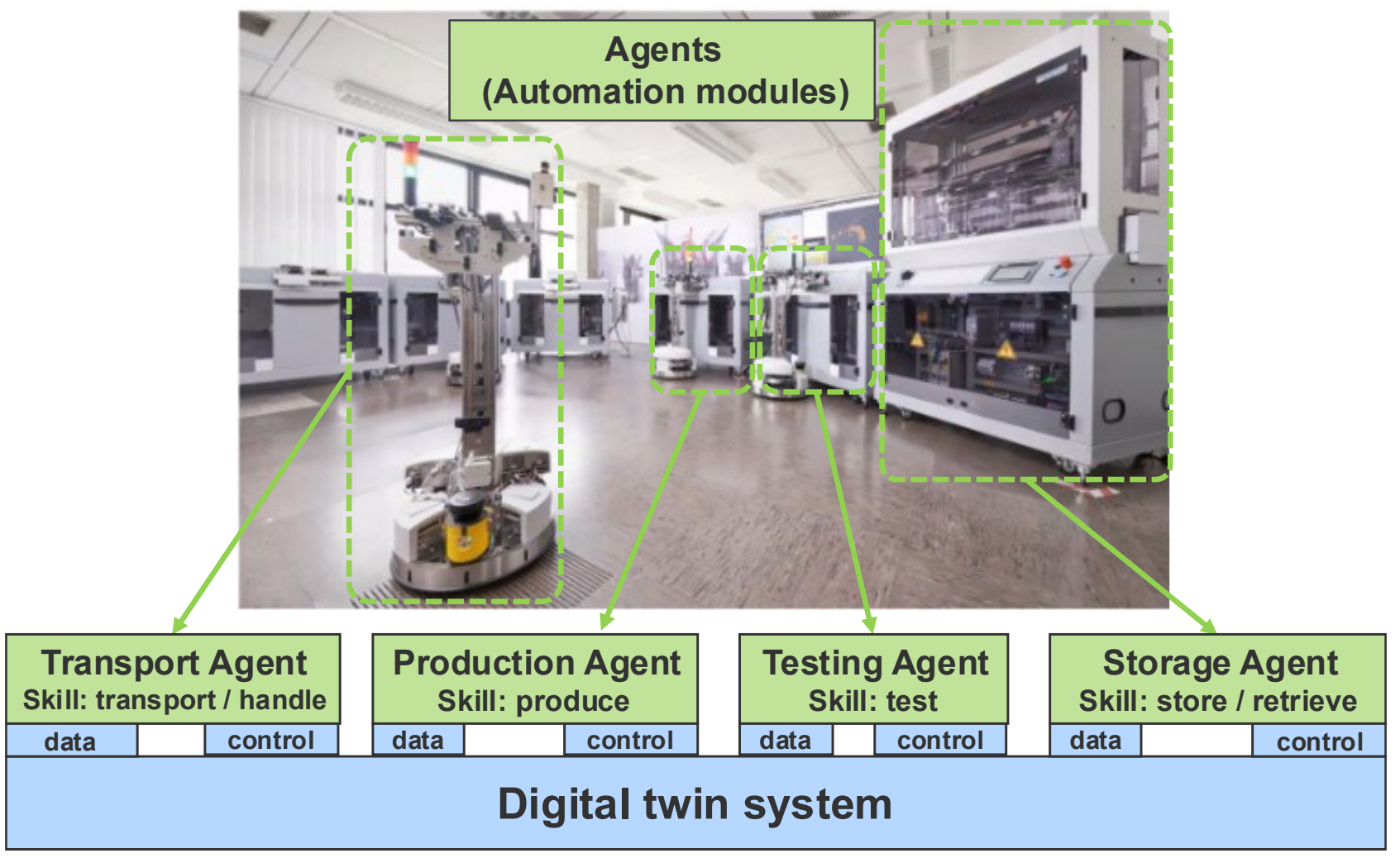

**Figure 8 Modular production system with LLM-integrated agents mediated by the digital twin.** Each automation module (transport, production, testing, and storage) is represented as an agent with defined skills and bidirectional data/control interfaces connected through the digital twin system.

**Evaluation:** The developed system is evaluated as the first proof-of-concept in a real modular automation system to verify feasibility. The prototype system successfully handles tasks that are not predefined and dynamically creates production processes by orchestrating higher-level skills and atomic functionalities exposed through the digital twin service interfaces. The testing results demonstrate two key outcomes:

- The integration mechanism successfully incorporates LLM agents into the automation system, enabling the system to respond to user tasks and execute automated operations that meet the task objective.
- The process executability is evaluated with an 88% success rate in generating error-free, executable process plans that fulfill user task objectives across 50 test cases.

**Incremental reflection and generalization:** Several insights have emerged that are generalizable beyond the specific experimental context and contribute to broader design knowledge:

- First, enabling LLM agents to flexibly plan and execute production processes within an automation system requires a three-layer integration (LLM system ↔ digital twin system ↔ physical system), in which the digital twin system provides both system representations and service interfaces, thereby linking text-based reasoning to executable operations.
- Second, enabling flexible automation within complex systems requires a modular system design. This modularity is also reflected in the multi-agent system, in which agents operate at different responsibility scopes and flexibly control the automation modules.
- Third, the task-solving workflow follows a structured task decomposition procedure. A user-defined task is first interpreted by a manager agent, which generates a process plan (see Figure 21 and Figure 24). This plan is subsequently refined and executed by operator agents through the functionality services provided by the digital twin system.

- Fourth, the LLM agents function as information processors guided by structured prompts, enabling context-aware and instruction-compliant reasoning in a programmable way.

**Boundary conditions and limitations:** The availability of interoperable, semantically described service interfaces and high-fidelity digital twin representations is a prerequisite for LLM agents to interpret system context, perform semantic reasoning, and interact effectively with the automation system. The system achieved an 88% executability rate, demonstrating that the approach enables adaptive task automation but still falls short of guaranteeing consistently reliable behavior, thereby necessitating human oversight.

## 2.3 Paper B: Control Industrial Automation System with LLM Agents

**Relevance:** Paper B [28] deepens the investigation in Paper A [5], addressing the same practical problem of enabling flexible autonomy in industrial automation systems. While Paper A demonstrates that LLM agents can generate adaptive process plans, it also reveals an inherent limitation of the underlying technical approach: the method supports flexible planning but lacks the dynamic, event-driven control required to react to runtime changes in real-world production environments. Building on the insights from Paper A, Paper B examines whether LLM agents can autonomously generate correct control actions in a continuous decision cycle under dynamically changing conditions. Furthermore, because state-of-the-art LLMs are pretrained on general-purpose datasets rather than industrial control data, this study also examines whether model training can improve control-command generation accuracy

**Design:** The designed system of this study is an event-driven control framework that integrates LLM agents into a perception–action loop mediated by the digital twin system, as illustrated in Figure 9. The system's perception and action are mediated by two components:

- **Event Log Memory** maintains time-ordered event histories and exposes them through a publish–subscribe mechanism, enabling role-specific LLM agents to subscribe to context-relevant topics and receive filtered event streams for decision-making. Based on these events, the agents generate schema-conformant control commands.
- **Command Handler** maps the parsed commands to the digital twin service endpoints and invokes the required module-level and component-level actions to close the perception–action loop.

In contrast to Paper A's process-planning design pattern [5], this design emphasizes online control decision cycles: operations are dynamically executed at runtime as a result of actionable decisions derived from the evolving event context.

Additionally, the study investigates Supervised Fine-Tuning (SFT), in which a base LLM is trained on collected event–command pairs to improve command generation accuracy for the control task.

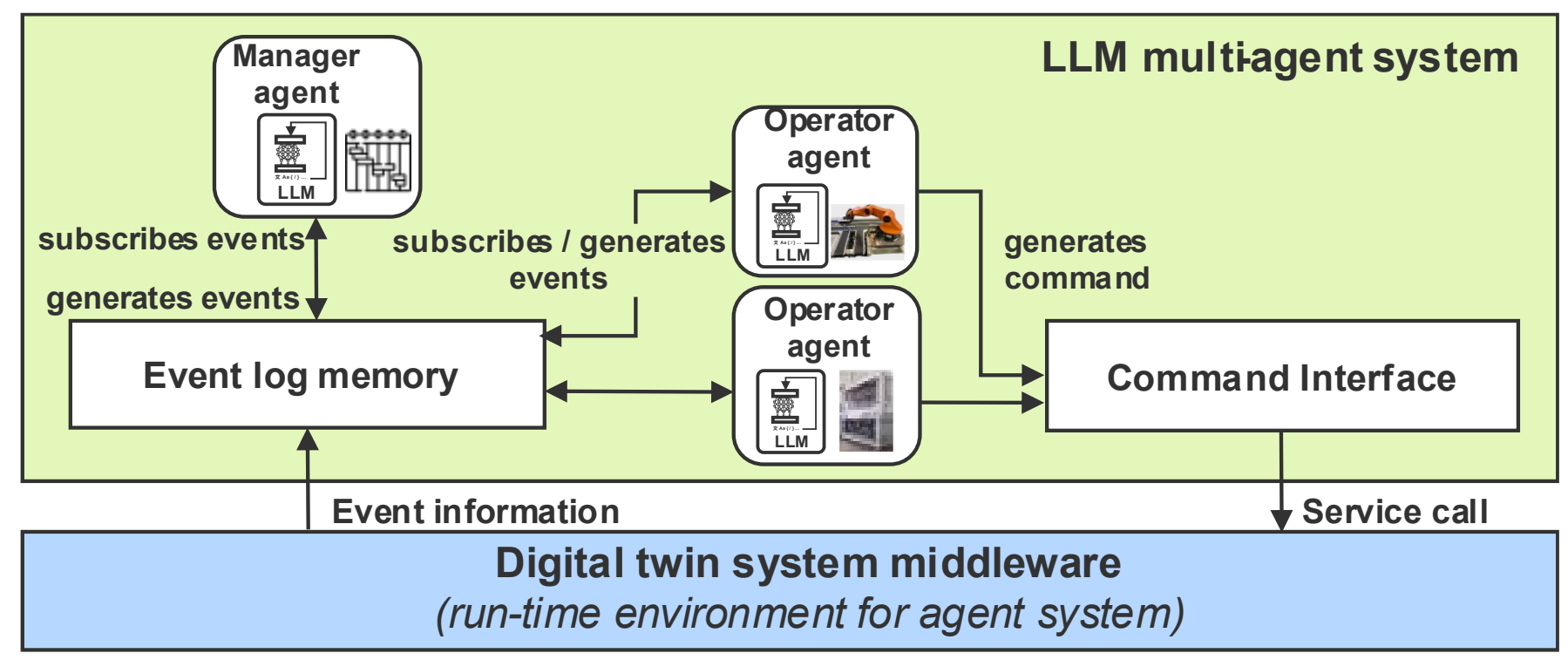

**Figure 9 Information flow and communication in the LLM multi-agent system.**

**Implementation:** The framework is realized on the same modular production system coupled with an AAS-based digital twin middleware. As illustrated in Figure 10, raw PLC signals, ROS signals and system control states are continuously monitored by a *Data Observer* component, which translates them into semantically annotated events for the role-specific agents. These agents subscribe to context-relevant event topics and generate control commands accordingly. A *Command Handler* component parses the agents' outputs (expressed in JSON schemas for function calls) and binds them to OPC UA and ROS service endpoints to invoke the corresponding operations.

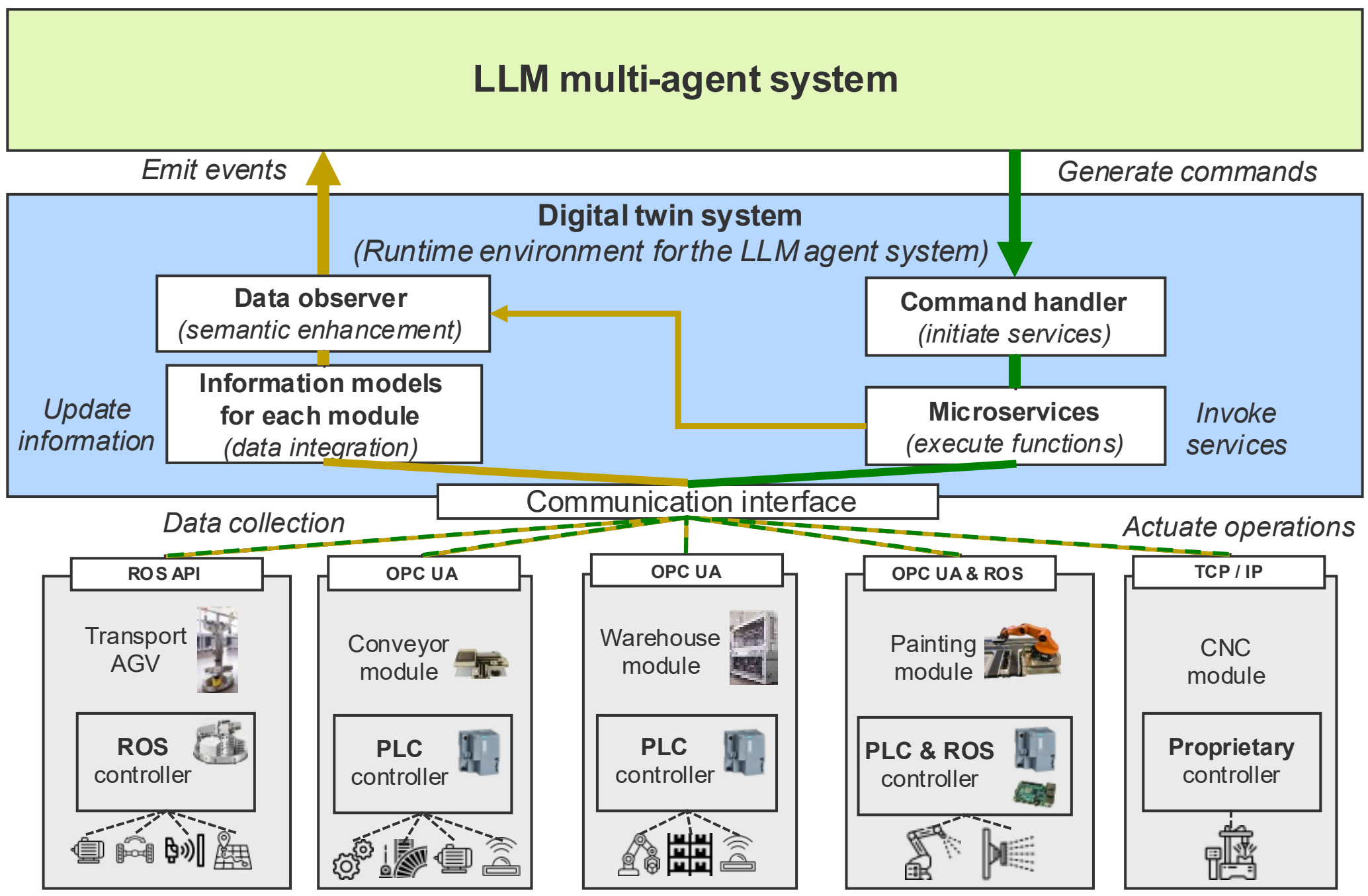

**Figure 10 Event-driven control architecture integrating LLM agents with the digital twin middleware.** The LLM agent system interacts with the digital twin through a perception–action loop: observed data are translated into semantically annotated events and emitted to the agents, while generated commands are executed via microservices and communication interfaces to control automation modules [28].

To improve command-generation accuracy, the study applies supervised fine-tuning to several state-of-the-art LLMs (GPT-4o, Llama-3-8B, Llama-3-70B, Qwen2-7B, Qwen2-72B, and Mistral-7B) using a task-specific dataset that consists of event–command pairs. Their performance is evaluated in terms of reasoning plausibility and command correctness. The fine-tuned models are then compared with their respective base models to quantify performance improvements. The demonstration of the implemented system as well as the dataset are released on a GitHub repository. [6]

**Evaluation:** The artifact is tested to assess its feasibility for enabling LLM-driven control in a modular production environment. The assessment focuses on two aspects: (1) the functional validation of the event-driven control loop, and (2) the testing of the effect of fine-tuning on command generation accuracy for industrial control tasks. A quantitative evaluation is performed using a dataset that consists of 100 test cases containing structured event–command pairs, using two key metrics:

- Command correctness rate: the percentage of generated commands that matched the reference command in syntax and semantics.
- Reason plausibility score: a rating from human evaluators (1–5 Likert scale) assessing the soundness of the logical justification for each command generated by LLMs.

Baseline evaluation across pre-trained models revealed notable performance variance. GPT-4o achieved the highest accuracy (81%) and the highest reason plausibility score (4.7/5), followed by Llama-3-70B (75%), while smaller models such as Llama-3-8B exhibited substantially lower reliability (37%). When supervised fine-tuning was applied to the collected dataset, accuracy increased markedly, reaching 95–100% for the larger models such as the fine-tuned GPT-4o and Llama-3 models, which demonstrated particularly strong performance.

Overall, the evaluation shows that (1) LLM agents can perform context-aware, event-driven control through the system's digital twin, and (2) supervised fine-tuning effectively trains base LLMs for control command generation, improving command correctness and reasoning plausibility. These findings validate the feasibility of the system design and its empirical performance.

**Incremental reflection and generalization:** This study extends the process-planning approach of Paper A [5] by showing that an event-driven, closed-loop control framework enables adaptive and context-aware automation with finer-grained responses to dynamic events. The incremental and generalized contributions to design knowledge can be summarized in two aspects:

- First, the results show that LLM agents can operate as continuous decision-making components through the iterative cycle of "perceive → reason → act → perceive", the agents generate control decisions that are handled by the digital twin and executed on the physical system. This demonstrates a reusable mechanism pattern for LLM-agent-driven automation.
- Second, supervised fine-tuning of pre-trained LLMs on event-command datasets effectively adapts general-purpose models to domain-specific control tasks, improving command accuracy rates. Combined with structured prompting, fine-tuning provides a methodological approach for developing specialized LLM agents with higher task accuracy.

**Boundary conditions and limitations:** The effectiveness of this continuous control critically relies on semantically enriched event information in the digital twin layer, which ensures that each agent receives context relevant to its functional role. Such contextual precision is essential for correct LLM reasoning and, consequently, for the stable operation of the overall LLM-driven automation system. Moreover, although training can improve task performance, it simultaneously induces persistent, global shifts in model behavior that affect how the LLM responds across all possible inputs. These shifts are desired only when the operational input distribution is sufficiently known.

[6] https://github.com/YuchenXia/LLM4IAS

## 2.4 Paper C: LLM Experiments with Simulation: Large Language Model Multi-Agent System for Simulation Model Parametrization in Digital Twins

**Relevance:** While [5] and [28] demonstrate that LLM agents can plan and control automation processes via digital twins, experimentation on physical hardware remains slow and setup-intensive. A simulation environment that replaces the physical system removes this dependency and enables rapid, low-cost digitalized experimentation. Paper C [27] addresses this need by investigating how LLM agents can operate in such a virtual setting, interacting with a simulator to interpret simulation state and generate control parameters to achieve a given task objective.

**Design:** The artifact developed in this study is a simulation-integrated LLM multi-agent system that realizes a closed perception–reasoning–action loop with a simulation environment, as illustrated in Figure 11. The LLM multi-agent system consists of specialized agents assigned to the roles of observation, reasoning, decision, and summarization, each instantiated through a structured prompt template defining its functional role, objective, and output schema. The simulator provides snapshot representations of the system state. The agents iteratively process these snapshots to reason over system conditions, generate hypotheses, and adjust control parameters through the dynamic interaction with the simulator.

**Implementation:** The system is implemented and demonstrated through a case in which a simulator models a container mixing process, where the task is to achieve a homogeneous mixture by dynamically adjusting input sequencing and mixing intensity. The simulator continuously outputs structured state snapshots that reflect the container's current state and key indicators of mixture uniformity. An LLM multi-agent system interacts with the simulator through an iterative perception–reasoning–action loop. In each iteration, the observation agent extracts relevant insights about the state information, the reasoning agent interprets mixture uniformity and hypothesizes parameter adjustments, and the decision agent generates control commands to control the simulation parameters. All intermediate reasoning steps and actions are recorded and subsequently summarized by a summarization agent to produce a readable explanation of the overall parameterization process. The system is implemented as an open-source software demonstrator.[7]

[7] https://github.com/YuchenXia/LLMDrivenSimulation

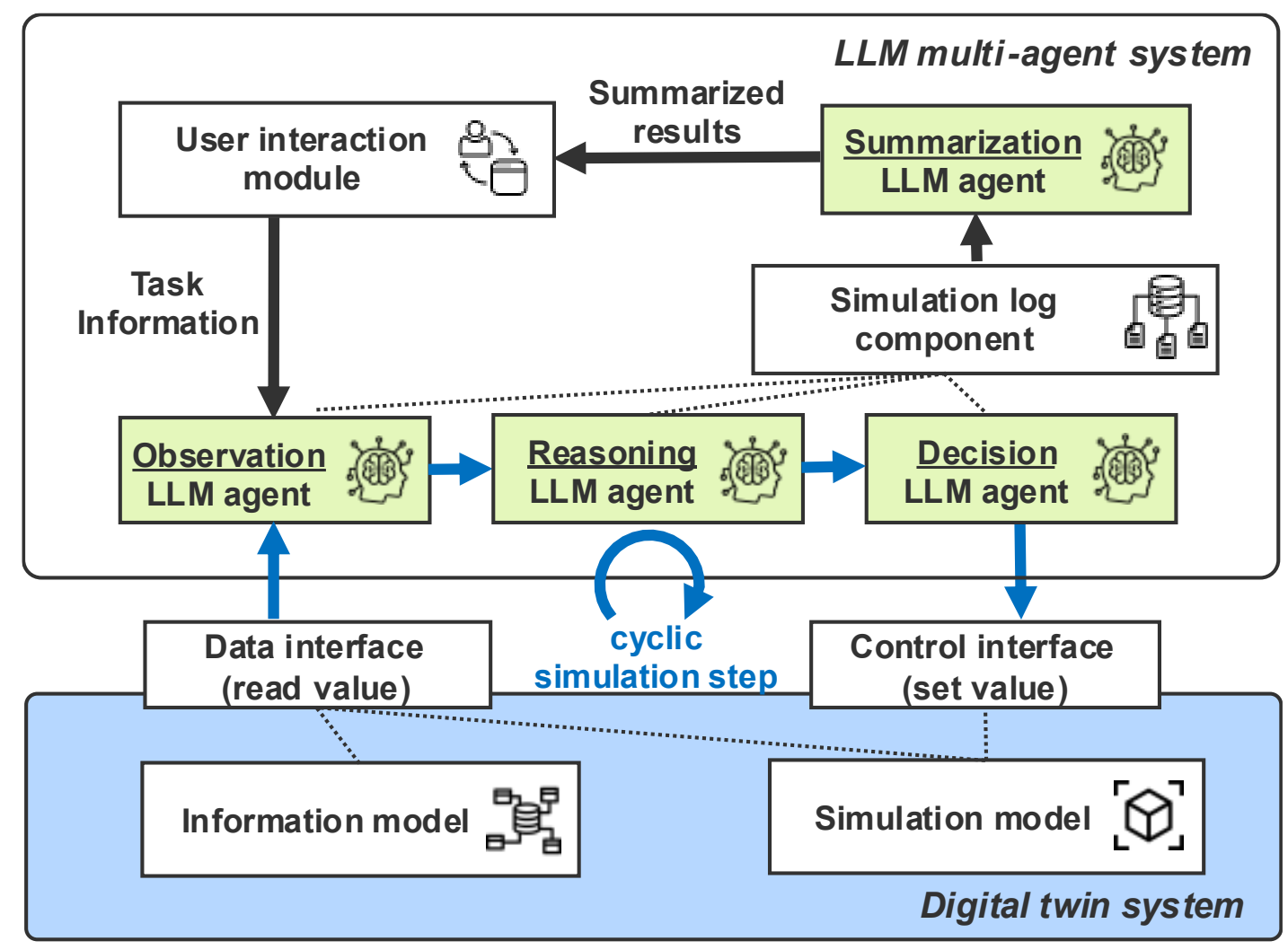

**Figure 11 Multi-agent system design for simulation-integrated reasoning.** Specialized LLM agents are organized into observation, reasoning, and decision roles to process simulation states, generate parameter adjustments, and control the simulation model. A summarization agent further consolidates outcomes into structured reports, enabling complex reasoning and analysis services within digital twin environments.

**Evaluation:** The evaluation is conducted qualitatively to verify the feasibility of the closed-loop reasoning and its effective interaction with the simulation model. The results confirm that LLM agents can autonomously perform process parameterization, thereby providing a proof-of-concept for LLM-driven autonomous experimentation with simulation model.

**Incremental reflection and generalization:** This study extends earlier work by demonstrating how LLM agents can interface with a simulated environment without relying on physical hardware. It demonstrates the use of LLM agents for simulation-based experimentation, where agents observe, evaluate, hypothesize, and adjust parameters through repeated perception–reasoning–action cycles. The study also demonstrates how state-snapshot representations discretize continuous simulator dynamics into structured text inputs that facilitate LLM reasoning.

**Boundary conditions and limitations:** The effectiveness of this approach is contingent upon the simulation providing state information that faithfully represents the underlying physical process. Since LLM agents operate exclusively on discrete state and text-based reasoning rather than mathematical optimization, missing, distorted, or oversimplified representations can result in parameter adjustments that appear plausible but are physically invalid, limiting their suitability for tasks requiring precise quantitative tuning. Accordingly, the method is appropriate for heuristic parameter exploration and hypothesis testing in simulation, with the resulting parameter choices requiring validation through real-world experimentation.

## 2.5 Paper D: Generation of Asset Administration Shell with LLM Agents: Toward Semantic Interoperability in Digital Twins

**Relevance:** While Papers A–C [5,27,28] demonstrated that LLM agents can operate effectively when supported by a digital twin, such operation requires well-structured and semantically coherent digital representations. In practice, however, industrial data are frequently fragmented across incompatible, vendor-specific formats, limiting semantic interoperability and hindering automated information exchange [9]. The Asset Administration Shell (AAS) provides a standardized framework [87] to unify these representations, but its adoption is constrained by the manual effort required to extract and formalize information from unstructured data sources. This creates a persistent gap between the intended goal of achieving interoperability in digital twin systems and the effort needed to create the underlying semantic models. To bridge this gap, Paper D [7] investigates whether LLM agents can serve as semantic translators that automatically extract, interpret, and formalize technical data from heterogeneous textual sources into standardized, machine-interpretable AAS models, thereby operationalizing semantic interoperability by leveraging the generative capabilities of LLMs.

**Design:** This study develops an LLM-based retrieval-augmented framework that automatically generates Asset Administration Shell (AAS) model instances from unstructured technical documentation, such as datasheets and technical manuals. The core design employs LLM agents to interpret raw textual descriptions of technical components and generate structured data models containing the key data elements required for AAS modeling, including property names, values, concept definitions, affordances, value types, units, and source descriptions. Through this process, standardized information models are produced in the form of AAS model instances, as illustrated in Figure 12. The designed system also provides configuration options, including the ability to select and compare different LLM backends and to activate or deactivate the retrieval function within the processing pipeline.

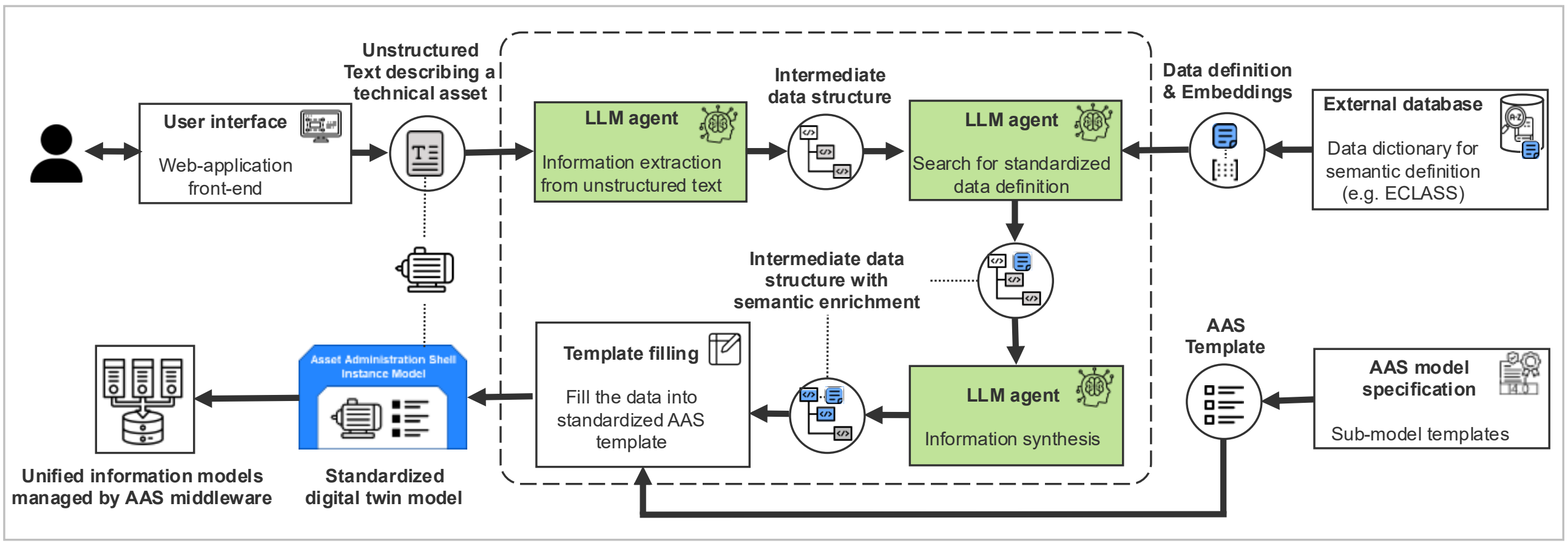

**Figure 12 Retrieval-augmented LLM agent framework for automated generation of Asset Administration Shell (AAS) instance models.** Unstructured textual descriptions of technical assets are processed through a multi-agent pipeline for information extraction, semantic enrichment, and template filling, producing standardized AAS models that enable semantic interoperability in digital twin systems.

**Implementation:** The designed system is implemented as the "AASbyLLM" web application, which processes unstructured technical documentation and automatically generates AAS model instances. The application follows a retrieval-augmented information processing pipeline in which one agent extracts candidate properties, another retrieves similar domain concept definitions from the ECLASS standard dictionary [88], and a synthesis agent consolidates these intermediate results into a structured data model. This data model is then mapped into the AAS JSON schema to produce complete, machine-readable AAS model instances. The system supports multiple LLM backends and configurable retrieval settings, and it provides an annotation user interface (see Figure 29) for evaluating model output quality across different configurations. Users can review and download the generated AAS model instance directly from the web application. The framework has been tested using documentation from a variety of industrial automation components. All source code and evaluation data are publicly available to ensure transparency and reproducibility.[8]

**Evaluation:** The system was evaluated by using several LLM models to process technical documentation from 20 industrial automation components and generate corresponding AAS model instances. Performance was compared between configurations with and without retrieval augmentation. The generated AAS data elements were reviewed by human evaluators across 2,259 individual data fields for correctness and semantic accuracy. Two aggregate metrics were used: the effective generation rate, which reflects how many instance samples are correct and immediately usable without further manual modification, and a helpfulness score, which measures whether the generated descriptions improve user understanding.

Across models, effective generation rates ranged from 62% to 79%, with larger LLMs consistently performing better. Retrieval augmented generation (RAG) notably improved smaller models but offered limited gains for larger ones. Overall, the evaluation shows that the system can automate a substantial portion of AAS model creation and reduce manual effort by producing standardized, semantically coherent information model instances suitable for usage in interoperable digital twins.

**Incremental reflection and generalization:** This study shows that LLM agents can generate standardized and semantically coherent information models from unstructured technical text, substantially reducing the manual effort required for model creation. By producing semantically aligned models, the approach enables interoperability across digital twin systems. The proposed agent-driven workflow, supported by retrieval augmentation, provides a reusable mechanism for automatically transforming diverse documentation into standardized machine-interpretable models, thereby improving the scalability and consistency of digital twin integration.

**Boundary conditions and limitations:** The system's performance depends on the availability of sufficiently detailed textual documentation and on the domain knowledge learned by the LLMs during their training. Incomplete documentation or insufficient model knowledge directly degrades the quality of the generated results. The benefit of retrieval augmentation is also conditional: it can compensate for weaker models but offers limited gains for stronger ones. The retrieval process may introduce noise if the retrieved material from the database is not aligned with the task. Finally, although the approach automates a major portion of the information model instance creation process, human review remains necessary to ensure semantic correctness and consistency.

[8] https://github.com/YuchenXia/AASbyLLM

## 2.6 Paper E: An Architecture for Integrating LLMs with Digital Twins and Automation Systems

**Relevance:** While earlier studies [5,7,27,28] have demonstrated feasibility across planning, control, simulation, and semantic translation, their architectural designs remained tailored to specific cases. Paper E [62] addresses this gap by abstracting these case-specific designs into a generalized architecture that articulates how LLM agents can interact with digital twins to operationalize tasks.

**Incremental reflection and generalization:** Paper E distills three recurring interaction patterns that characterize how LLM agents orchestrate and execute processes through digital twins to solve tasks:

- **Planning-based modeling** structures a process as a predefined plan sequence of actions, allowing LLMs to reason about procedural dependencies, generate the plan sequence, and subsequently execute it.
- **Event-driven modeling** represents a process as a stream of contextual events, allowing agents to make discrete, iterative decisions in response to changing system conditions, forming a perception–action loop.
- **State-based modeling** models the task solving as a state–transition cycle, where system states are discretely represented at each step to support iterative LLM reasoning to trigger transitions.

Together, these patterns form unified reference models that abstract the design approaches used in the preceding studies, as illustrated in Figure 13.

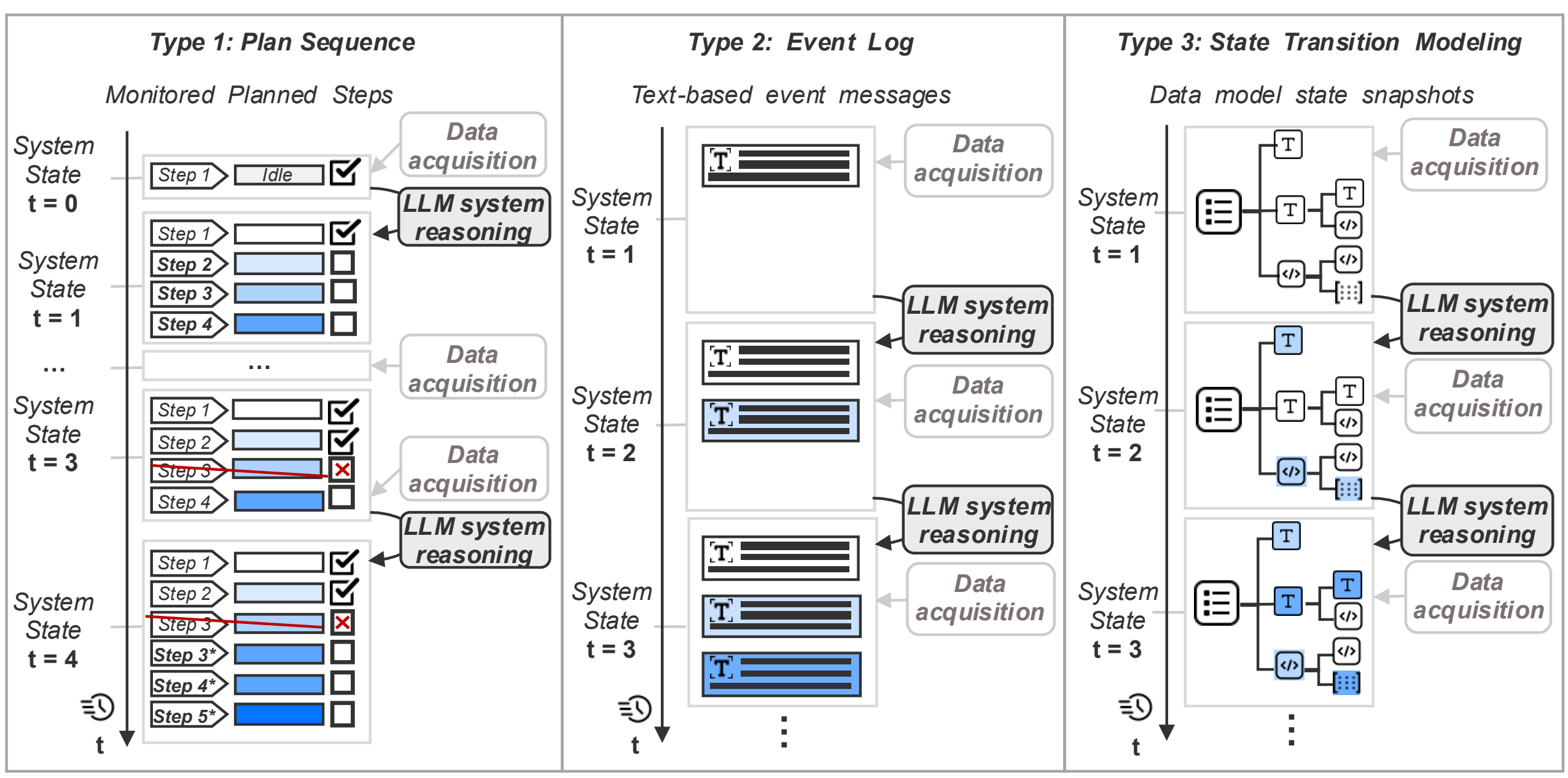

**Figure 13 Three paradigms of process modeling for LLM integration.** (1) Plan sequence modeling represents a predefined process plan whose steps are monitored and dynamically replanned when deviations occur. (2) Event-log modeling represents the process as a chronological stream of event messages. (3) State transition modeling represents the process as a sequence of temporally ordered system state snapshots for reasoning. All three paradigms discretize a overall process into process segments.

# 3 Synthesis of Design Knowledge from the Included Publications

This chapter synthesizes the key insights across the five studies presented in Chapter 2, covering production planning, event-driven control, simulation-based parameterization, and semantic translation, and abstracts these case-specific results into generalizable design knowledge. By distilling the recurring design patterns across these studies, the chapter derives a unified framework for LLM-enhanced automation systems consisting of three interrelated aspects: (1) architectural integration of **physical layer**, **digital twins layer**, and **LLM system layer**; (2) the **Task–Process–Service–Resource (TPSR) model** as a mechanism for operationalizing task automation; and (3) the **functional contributions of LLMs**, expressed as four distinct functional roles within the TPSR framework: process orchestration, service matching, digital resource generation, and agent-as-a-service.

## 3.1 Three-Layer Integration of LLMs, Digital Twins, and Physical Systems

LLMs provide context-aware, objective-oriented semantic reasoning capabilities that enable the realization of intelligent automation systems. Developing such systems requires a structured organization of components and their functions. The architecture is organized into three layers: the Physical Layer, the Digital Twin Layer, and the LLM System Layer, as illustrated in Figure 14. The Digital Twin Layer virtualizes the physical environment into digital representations that LLM agents can process and interact with. It enables bidirectional interoperability: perceptual information flows upward through the digital twin, while decisions and control commands generated by the LLM system are transmitted downward for execution.

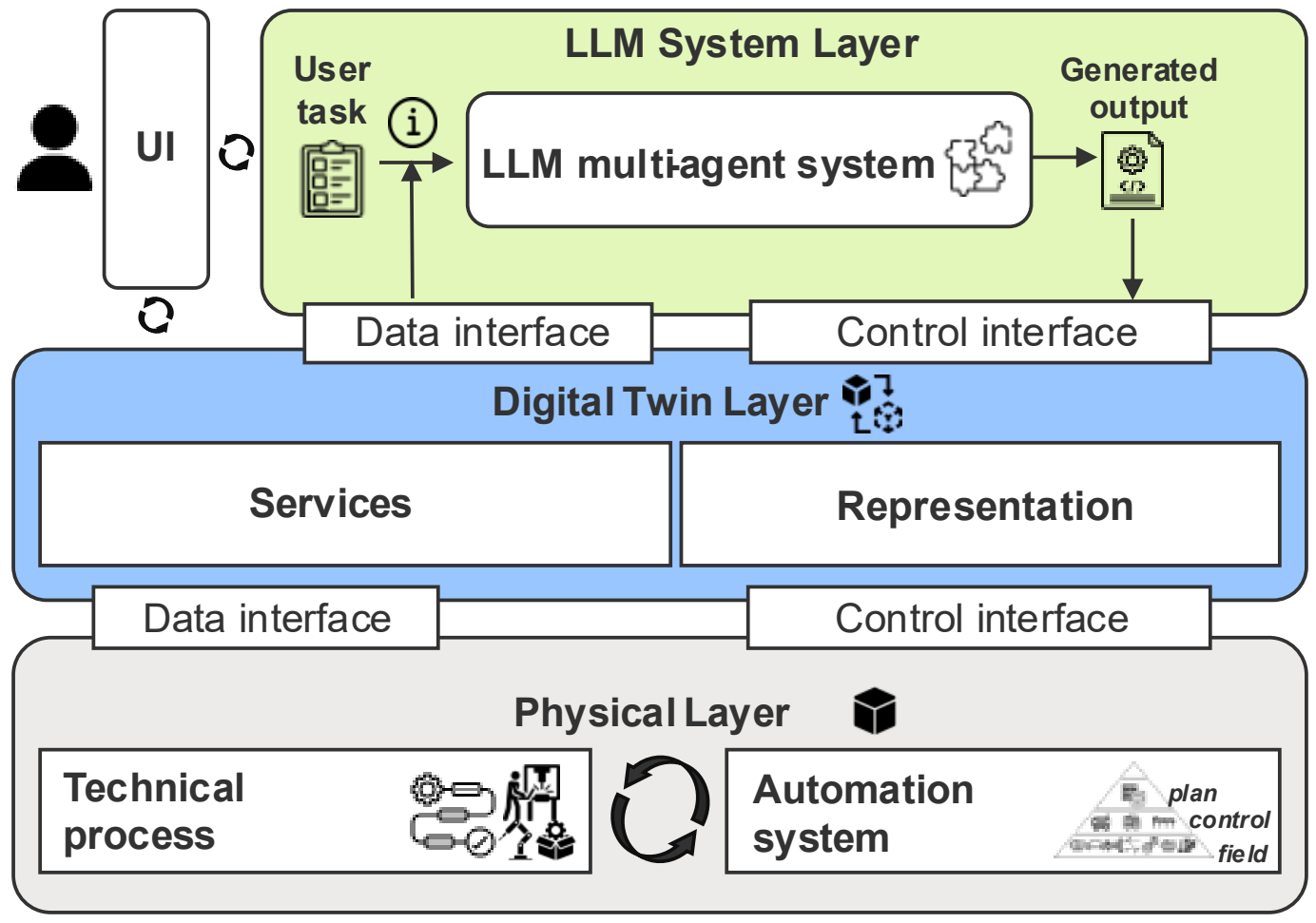

**Figure 14 Three-layer integration of LLM system, digital twin, and physical layers.** The architecture organizes system functionality into three layers: the LLM system layer provides reasoning and task-solving through multi-agent information processing, the digital twin layer mediates with services and representations, and the physical layer anchors execution in technical processes and automation systems.

### 3.1.1 Physical Layer

The physical layer forms the physical foundation of the architecture, comprising the **technical process** and the **automation system**, as illustrated in Figure 14.

In general, a **technical process** is the sequence of physical and informational transformations through which material, energy, or data are converted, transported, or stored to achieve a specific industrial objective. The **automation system** executes and controls these technical processes by coordinating sensors, actuators, and controllers. The organization of automation systems follows the hierarchical layered structure of the automation pyramid, as illustrated in Figure 15: devices in field level perform direct sensing and actuation, control level handles real-time logic, and higher levels manage manufacturing planning and enterprise coordination.

Within this conceptualized architecture in Figure 15, each entity within the automation pyramid has a defined modularity that enables flexible composition and orchestration of components to execute technical processes. A physical module, its digital twin, and its LLM agent share the same modularity. This modular alignment determines how digital twins represent physical assets and encapsulate their operational functions in software, thereby shaping how LLM agents are designed and integrated.

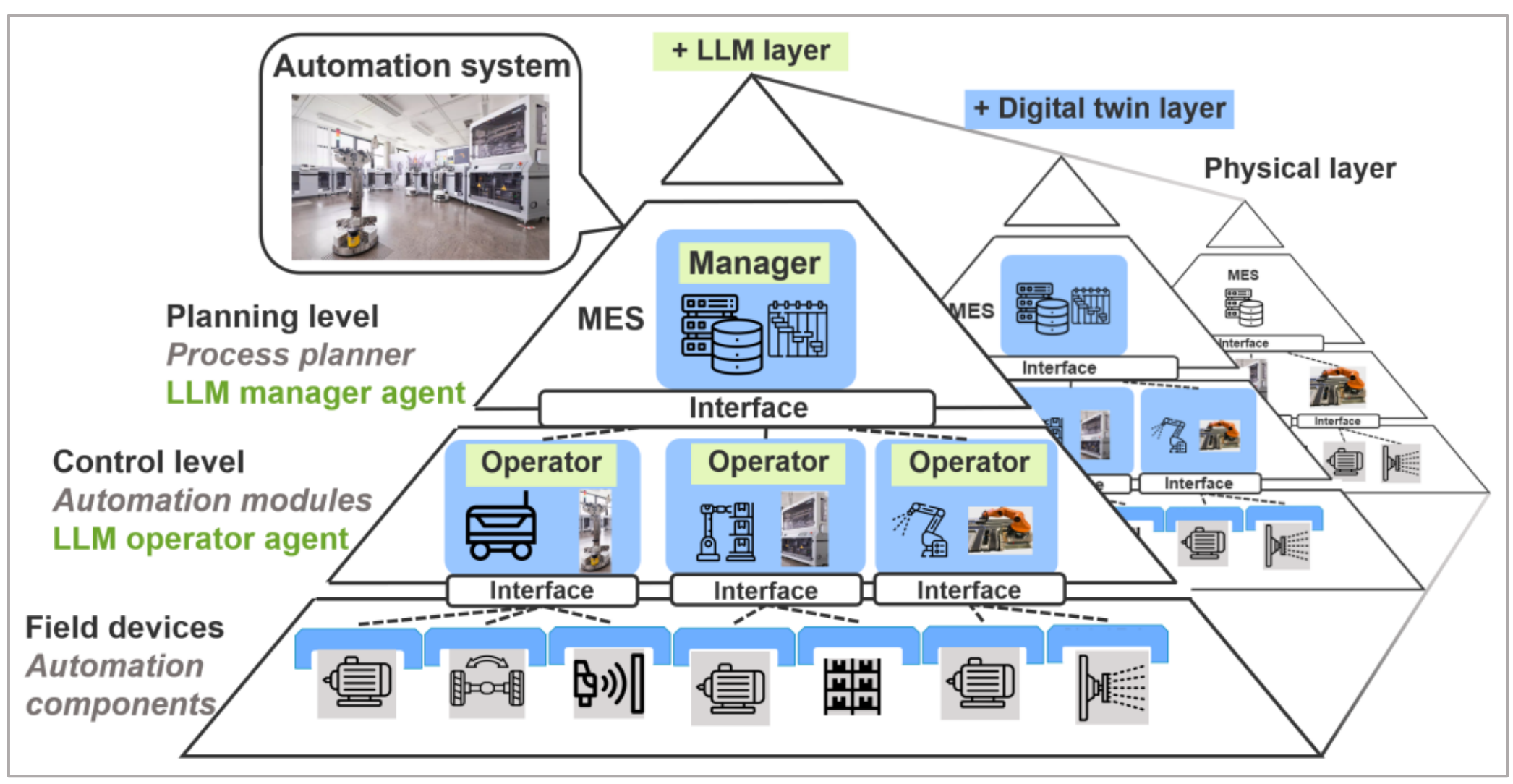

**Figure 15 Orthogonal relationship between the layered automation pyramid and the proposed three-layer architecture.** It illustrates how the modularity of the physical automation system extends through the digital twin and LLM-agent layers. The agent, digital twin, and automation module share a consistent modular structure across all layers.

### 3.1.2 Digital Twin Layer

The digital twin layer constitutes the essential mediation that provides the LLM system with a virtual runtime environment. It bridges the physical system and the LLM system by maintaining digital **representations**, ensuring **semantic interoperability** and exposing **service** interfaces.

**Representations** maintain synchronized digital models of physical entities and system states. They serve as the fundamental data resource for monitoring and reasoning.

As each subsystem operates within a distinct functional context, its data representation methods and data structures can differ accordingly. Data must therefore be translated across these heterogeneous contexts from the source to the target to ensure information usability. This gives rise to a semantic translation problem, which is also referred to as the problem of semantic interoperability.

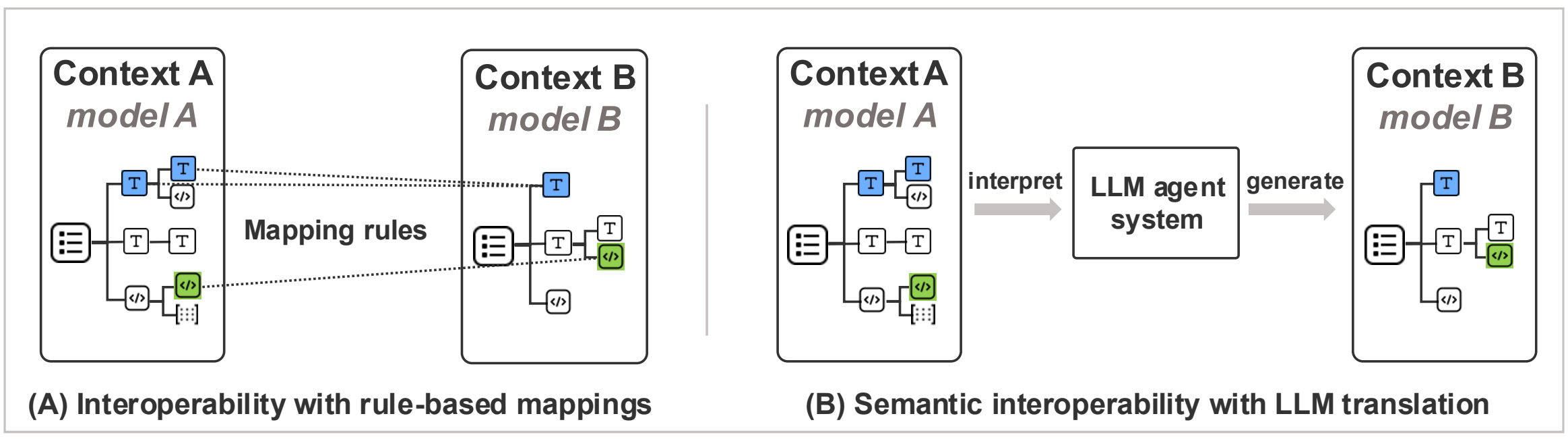

**Figure 16 Two approaches to achieving interoperability: (A) rule-based mappings and (B) semantic interoperability through LLM interpretation and translation.**

Semantic interoperability can generally be achieved through two alternative mechanisms, as illustrated in Figure 16.

- **Rule-based mappings** rely on predefined transformation rules that convert data from a source representation to a target representation.
- **LLM-based translation** uses LLMs to interpret source data and generate semantically aligned target representations.

The first rule-based mapping approach is applied in Paper B [28], where the data observer translates identical changes in system state into different event descriptions in different contexts using predefined rules. For example, a change in a low-level sensor signal, such as a flip in a raw binary value, is mapped to different component-level state changes with annotated event descriptions for higher-level processing, as illustrated in Figure 17.

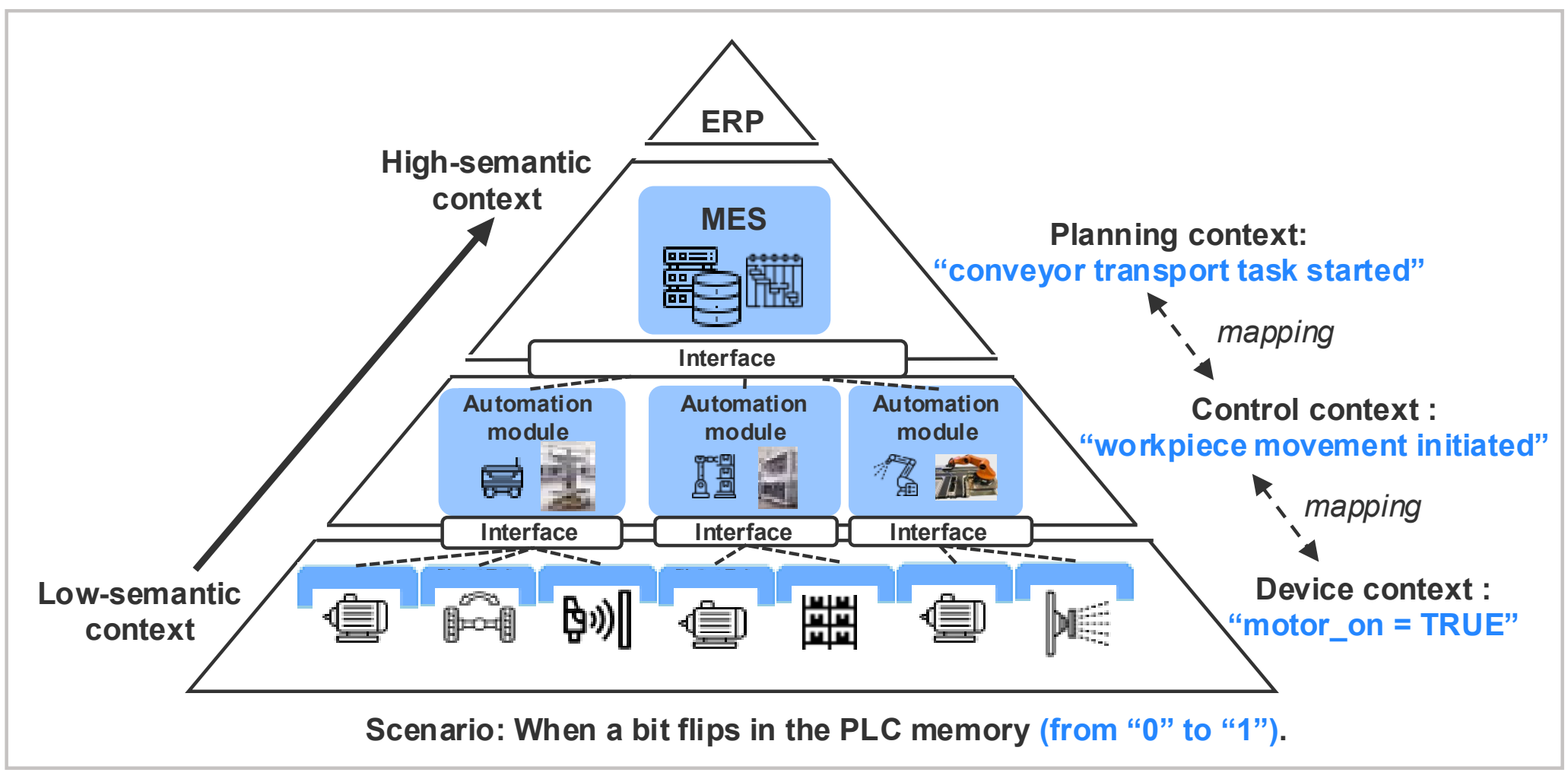

**Figure 17 Example of semantic context mapping across the automation pyramid.** A bit-level change in a field controller is reinterpreted at the control and planning contexts through mappings, illustrating semantic interoperability across abstraction layers. This results in semantic enrichment across system contexts for the same underlying change.

The second approach, LLM-based semantic translation, is explored in [7]. Instead of relying on predefined mapping rules, LLM agents interpret source model information and generate semantically consistent target models, thereby automating data translation across heterogeneous modeling contexts.

**Services** provide standardized interfaces for data exchange and control. Through these interfaces, the software-supported automation system's functionalities can be invoked to query states, configure operational parameters, and issue control commands.

Entities at different levels within the automation pyramid have different levels of functional abstraction (see Figure 15). Consequently, services are defined with varying granularity [5]. This is illustrated in Figure 18 on two levels:

- **Skills of automation modules** correspond to the control level of the automation pyramid and represent composite service operations that can be performed by an automation module to accomplish higher-level tasks, such as transporting or assembling workpieces.

- **Functionalities of automation components** correspond to the field level and represent atomic service operations executed directly by individual automation components, such as actuating a motor, loading or unloading parts, or moving a robot arm.

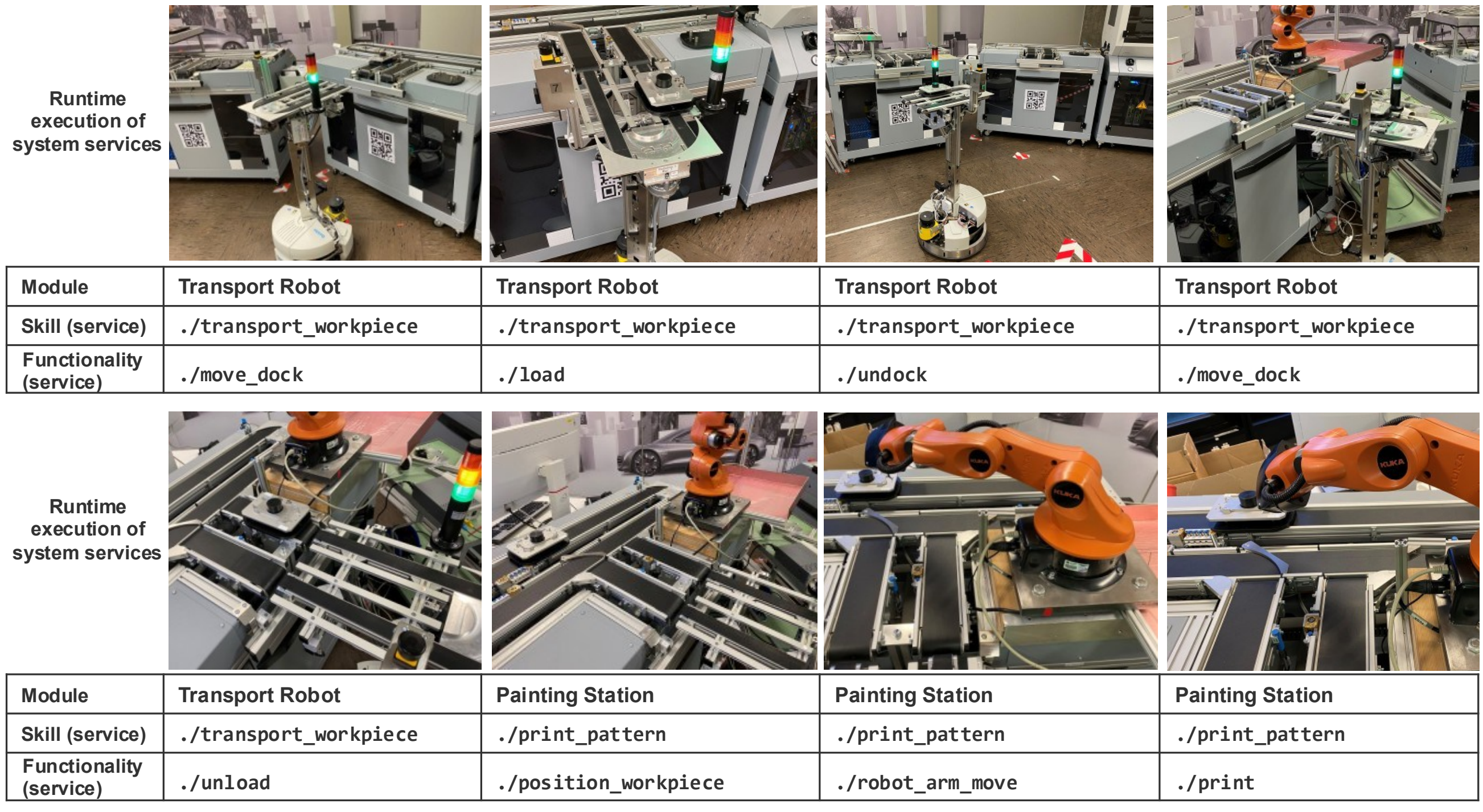

| Module | Transport Robot | Transport Robot | Transport Robot | Transport Robot |
| --- | --- | --- | --- | --- |
| Skill (service) | ./transport_workpiece | ./transport_workpiece | ./transport_workpiece | ./transport_workpiece |
| Functionality (service) | ./move_dock | ./load | ./undock | ./move_dock |

| Module | Transport Robot | Painting Station | Painting Station | Painting Station |
| --- | --- | --- | --- | --- |
| Skill (service) | ./transport_workpiece | ./print_pattern | ./print_pattern | ./print_pattern |
| Functionality (service) | ./unload | ./position_workpiece | ./robot_arm_move | ./print |

**Figure 18 Module-level skills and component-level functionalities in a modular automation system.** Skills represent composite operations within a module, while functionalities represent atomic actions executed by individual devices.

When both representations and services are made accessible to an agent, they jointly establish a perception–action loop that enables agents' interaction with the environment. Through representations, the agent perceives and interprets the system state, while through services, it executes operations. Within this architecture, the digital twin middleware serves as the runtime environment mediating this bidirectional information flow. In [28], this information-processing mechanism is systematically realized, as detailed in Figure 19.

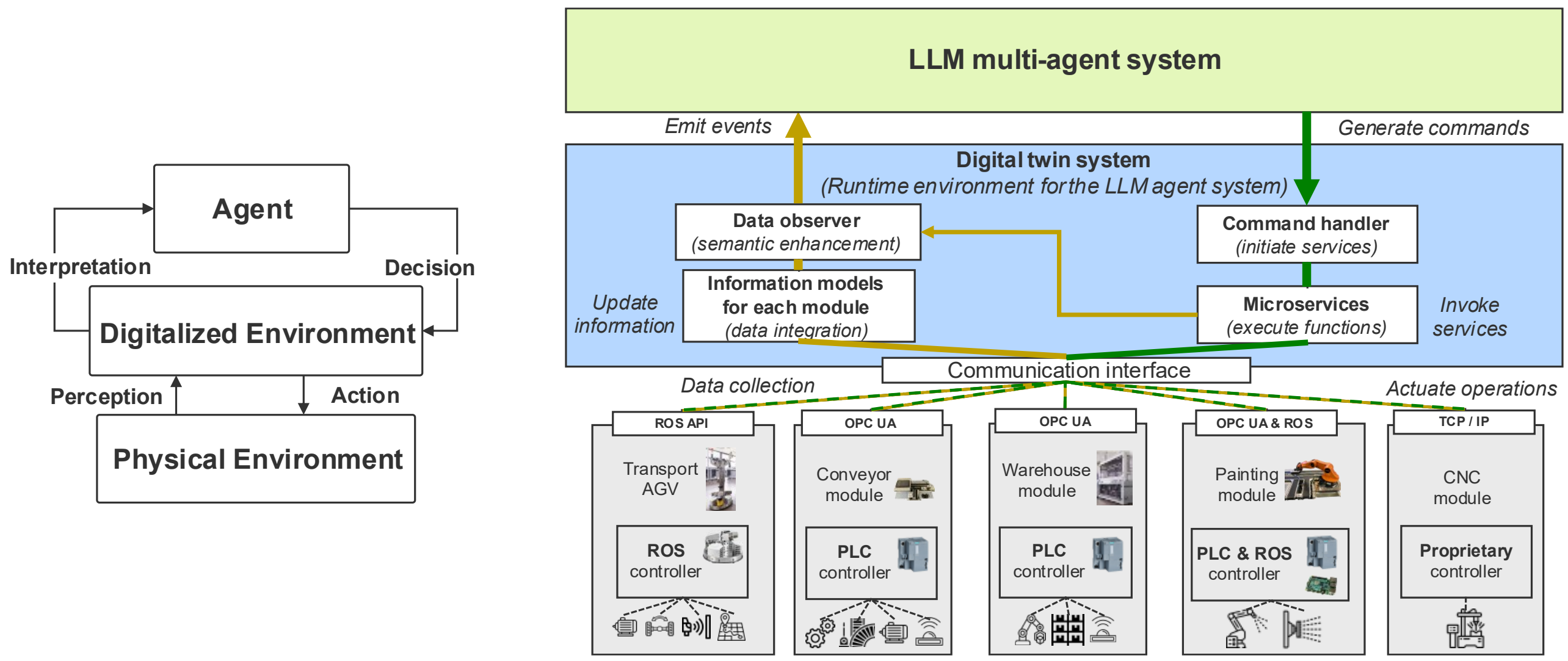

**Figure 19 Information-processing pipeline illustrating the perception–action loop of LLM agents integrated with the digital twin middleware.** The left side shows the general interaction flow, while the right side depicts the detailed implementation in [28].

### 3.1.3 LLM System Layer

As a fundamental building block, an LLM agent is a software component that embeds an LLM as its reasoning core. It receives the task objective and the current context as structured-text inputs. The agent processes this input information through a prompt template to perform semantic reasoning and generate outputs that can be parsed into executable commands, structured data, or code. This modular component design, together with its input/output interface, provides the architectural compatibility required for integrating agents into software-supported automation systems.

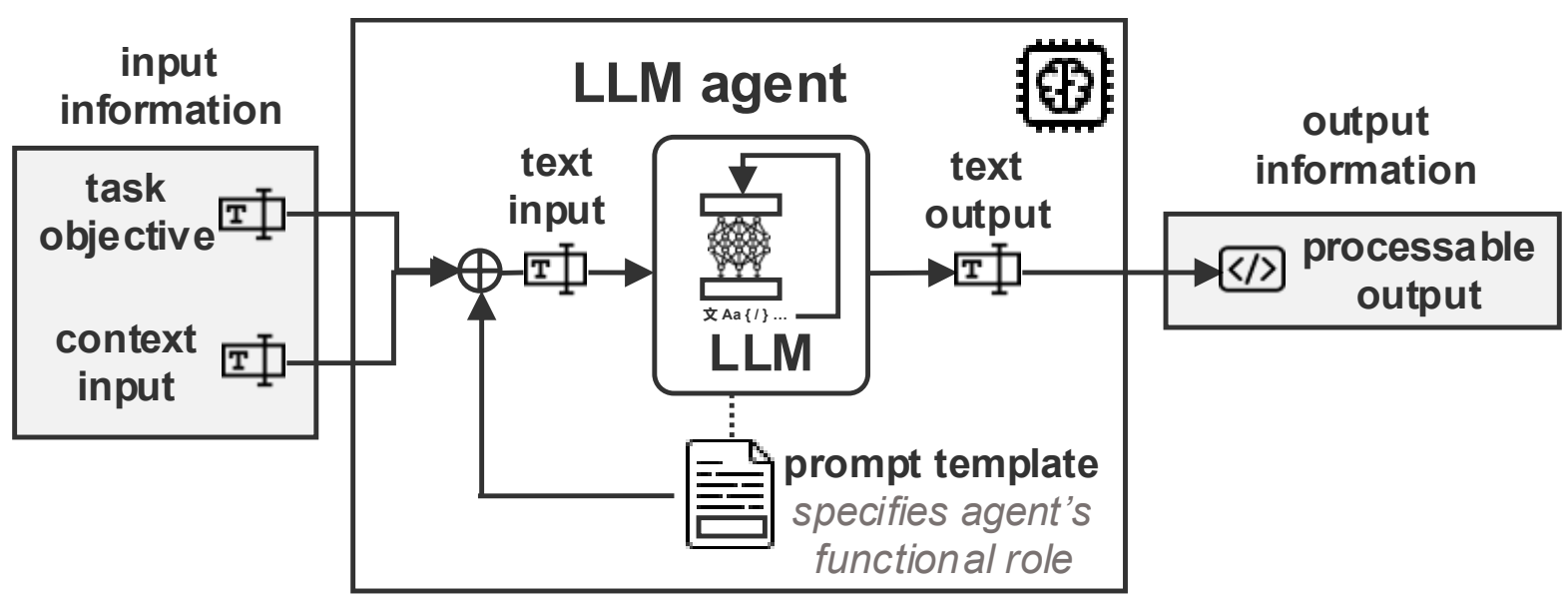

**Figure 20 Information-processing mechanism of an LLM agent.** The agent receives input information comprising a task objective and an input context. Guided by a prompt template that instructs the agent's behavior, the LLM performs reasoning to generate output text. The output can then be post-processed into executable or machine-readable formats for downstream processing.

Prompt design is the basic mechanism for programming the behavior of an LLM agent. The prompt template typically follows a schema consisting of (1) role and objective definition, which specifies the agent's responsibility and objectives; (2) context information that provides the run-time data and environmental state for reasoning; (3) task-specific instructions that define the reasoning to be performed; and (4) a schema definition that specifies the desired format of the output.

Substantial flexibility exists in prompt design, allowing the behavior of each agent to be tailored to its functional role. Detailed prompt implementations for each study are provided in the respective papers [5,7,27,28] and their accompanying open-source repositories.

When the complexity of a task exceeds the reasoning capacity of a single agent, the task can be decomposed into more manageable subtasks within an information processing pipeline. Each subtask is handled by a specialized software component or an LLM agent. Two complementary approaches can therefore be applied:

- **Integration of other software components:** Some subtasks can be better solved by conventional software tools—such as search functions, text manipulation utilities, or mathematical algorithms—rather than by an LLM. These components can be incorporated into the information processing pipeline to handle deterministic rule-based operations, while the LLM agents can be particularly designed to handle subtasks that require flexible reasoning.
- **Design of a multi-agent system:** When tasks involve complex reasoning across multiple problem dimensions, a multi-agent system can be employed. In this approach, multiple LLM agents are instantiated, each responsible for a specific aspect of information processing. The agents exchange structured messages to coordinate their reasoning and achieve a shared system objective. As demonstrated in [5] (see Figure 21), a manager agent decomposes high-level tasks into executable plans, while operator agents reason over individual plan segments and invoke atomic interface operations. In [28] (see Figure 22), specialized agents for observation, reasoning, decision, and summarization collaborate within a simulation-integrated digital system environment, iteratively interpreting system states and issuing control commands to the simulator through a perception–reasoning–action loop.

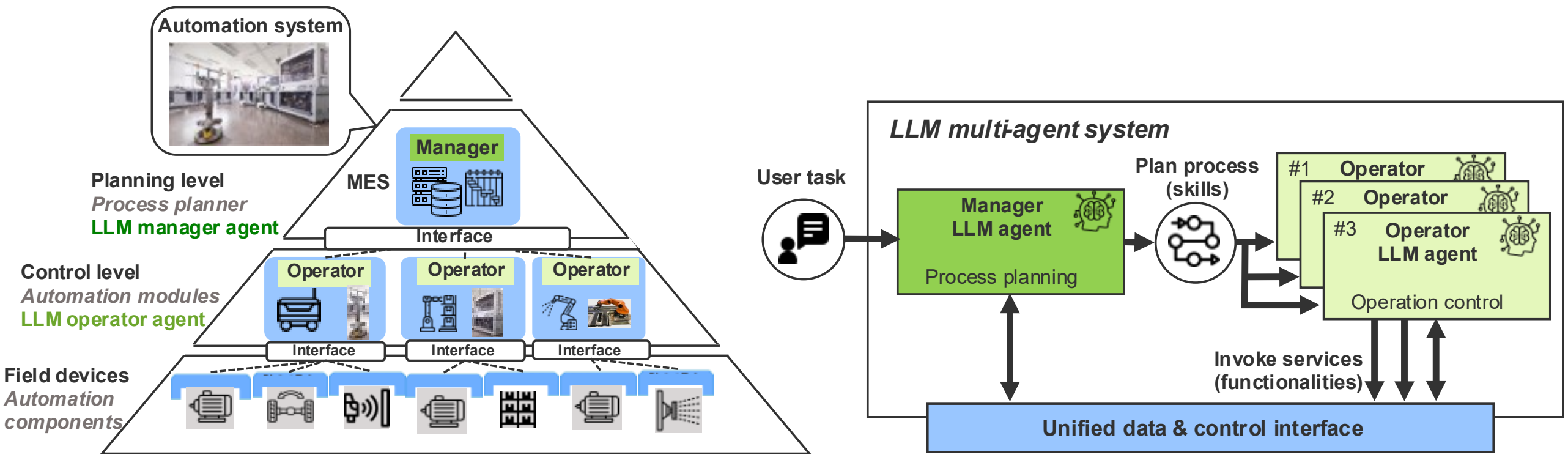

**Figure 21 Hierarchical manager–operator multi-agent paradigm aligned with the automation pyramid.** LLM manager agents operate at the planning level to decompose tasks, while LLM operator agents at the control level translate these tasks into service calls for field devices, thereby coordinating automation resources through digital twin interfaces.

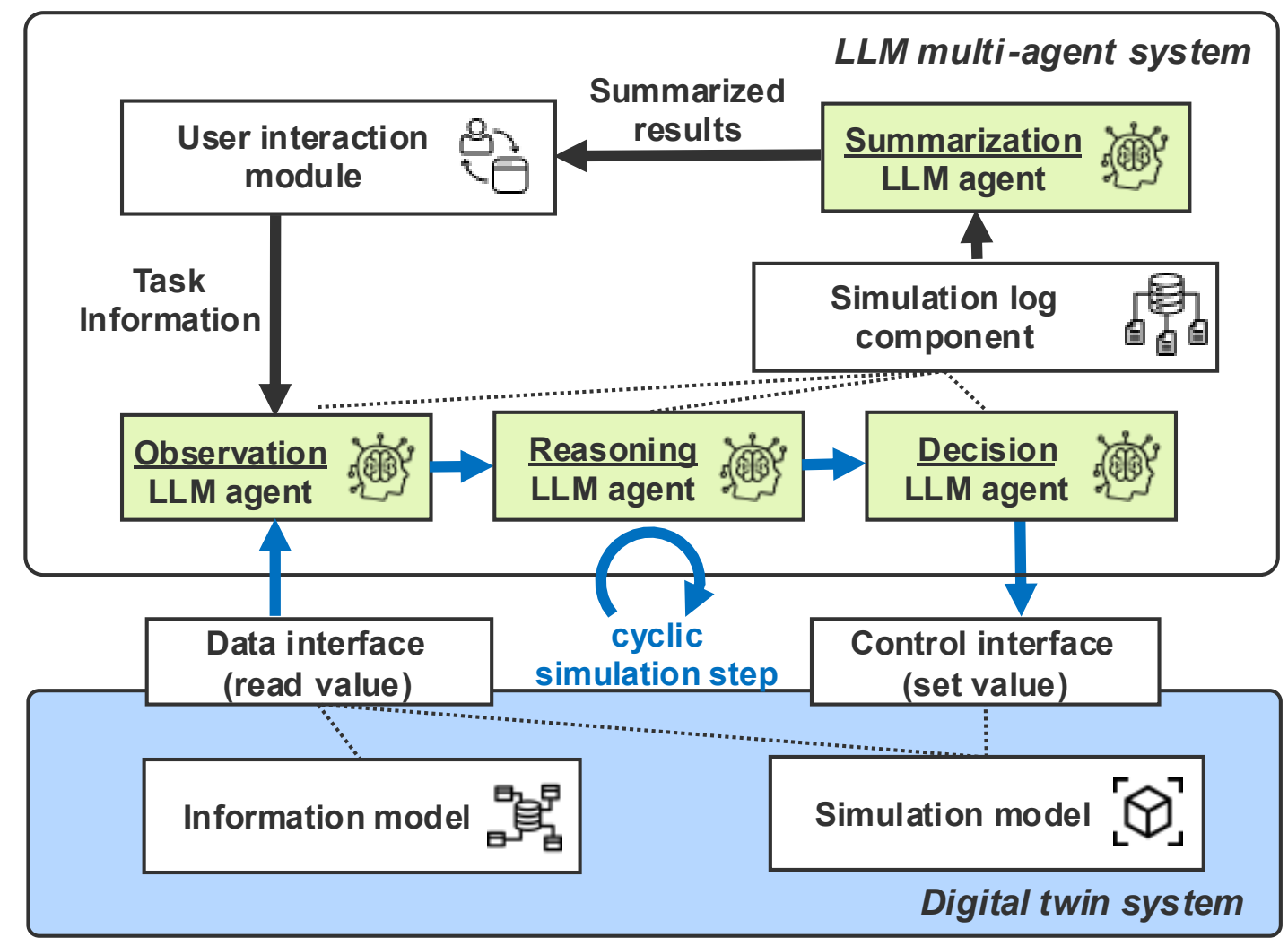

**Figure 22 Multi-agent LLM architecture integrated with a digital twin system.** The architecture implements a closed perception–reasoning–decision loop in which specialized LLM agents observe system states, perform reasoning, generate control actions, and summarize simulation outcomes.

## 3.2 The Task–Process–Service–Resource Model for Task Automation

While the three-layer architecture defines the structural organization of the system, an operational model is required to model how user tasks can be translated into executable system behavior. To formalize this translation process, the **Task–Process–Service–Resource (TPSR)** model is proposed as a unified conceptual framework. Figure 23 presents an overview of this framework.

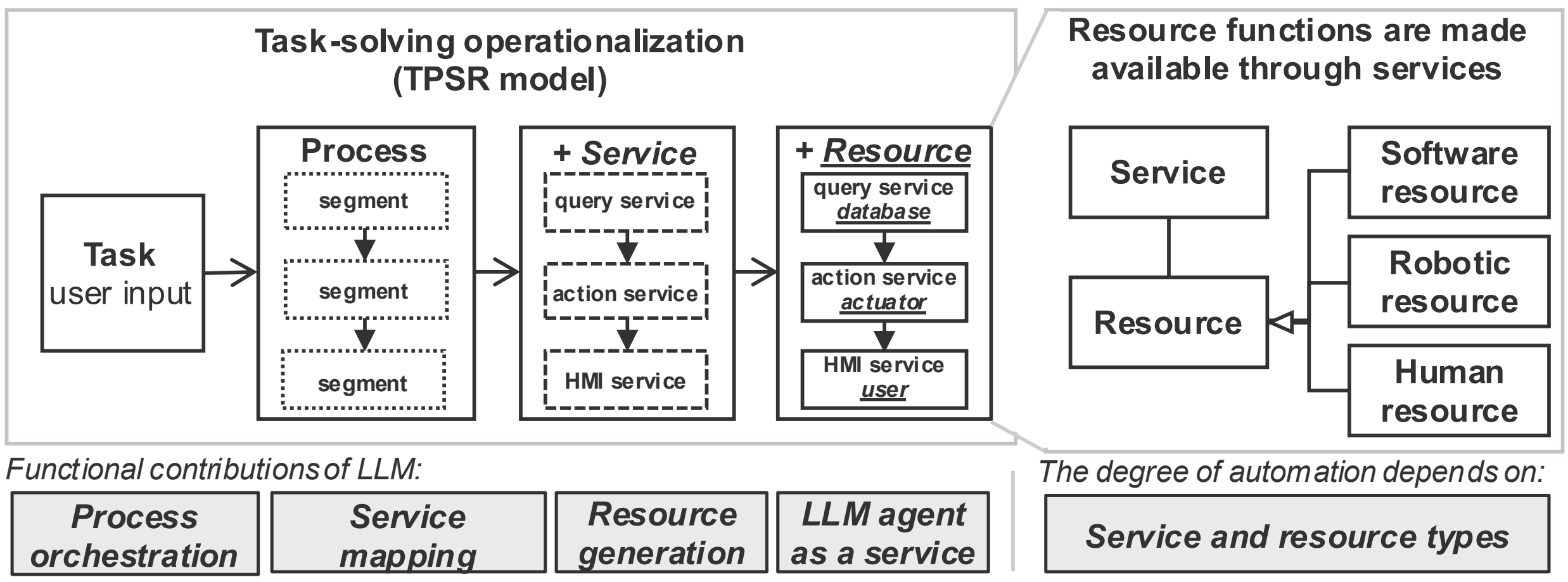

**Figure 23 Task–Process–Service–Resource (TPSR) model with functional contributions of LLM agents.** A user task is decomposed into process segments, which are realized through services bound to software, robotic, or human resources. The degree of automation depends on the types of services and resources. LLM agents can contribute to the task-solving through process orchestration, service matching, digital resource generation, and agent-as-a-service.

### 3.2.1 TPSR Model Definition

The TPSR model structures a workflow from the task input to the operation execution into four parts:

- **Task**: A task represents the user's need and determines the objective of the system.
- **Process**: A process is a logical decomposition of a task into an ordered sequence of segments, each representing a discrete operational step required to achieve the task objective.
- **Service**: A service is a modular and callable software function that can realize a defined process segment and is provided through the digital twin software system.
- **Resource**: A resource is a concrete entity that executes the operations defined by services, thereby realizing their functional effects in the system.

  Resources are classified into three types, each corresponding to distinct modalities of operation execution:

  - **Robotic resource:** A robotic resource is a physical entity capable of interacting with the real world through actuation or sensing. It performs tangible operations under software control. Examples include actuators, sensors, industrial robots, and other programmable automation modules.
  - **Software resource:** A software resource is a computational entity that performs non-physical operations such as data processing, reasoning, simulation, or control logic computation. Examples include databases, algorithms, digital models, and simulators.
  - **Human resource:** A human resource is a human actor who performs manual tasks outside the scope of full automation. It can interact with the automation system through Human–Machine Interface (HMI), contributing to activities such as physical actions, decision-making, validation, and exception handling.

### 3.2.2 Task Solving Mechanism Based on the TPSR Model

To operationalize the Task–Process–Service–Resource model, a systematic mechanism is required. The essential problem can therefore be formulated as transforming a user task into an executable process, in which each process segment is realized as a service and provisioned with the appropriate resources.

**Task identification:** Upon receiving the user input in natural language, the system infers the user's need, defines the corresponding system objective, and then generates a task description that aims to meet the identified user need.

From **Task** to **Process:** The second stage focuses on translating the described task into a formally structured process model by decomposing it into a sequence of discrete process segments required to achieve the defined objective.

From **Process Segment** to **Service:** In the third stage, each process segment is mapped to a service that defines a required function. Depending on implementation status, a service may be concrete or abstract. A concrete service is pre-implemented and exposed through the system's service interface, whereas an abstract service is specified by its required inputs and outputs and remains to be implemented.

From **Service** to **Resource:** A service is realized by allocating the necessary resources and by executing the control logic that coordinates their operation. These resources collectively deliver the intended operational outcome.

## 3.2.3 Process Modeling

A user task must be translated into an executable process that organizes both information processing and physical operations into a defined sequence. This requires a systematic process modeling approach. In conventional systems engineering, this work requires a designer specifying process models using notations such as BPMN [89], Petri nets [90], or SysML activity and state diagrams [91] to define an automated system's behavior. Analogously, a task-solving LLM agent must also orchestrate a process that organizes the required operations in an automation system. In [62], three paradigms are synthesized from the included papers: (1) planning and replanning mechanisms [5], (2) event-driven dynamic decision-making [28], and (3) state-based transition invocation [27]. Each paradigm employs a distinct process representation and is elaborated in the following subsections:

### Type 1—Planning

In the modular production system described in [5], the process is modeled as a predefined plan sequence of operational steps, and plans are generated at two different abstraction levels. At the automation-module level, a general plan specifies how modules should be orchestrated to perform coarse-grained skills. Within each module, a more detailed plan is generated to execute fine-grained functionalities executed by individual automation components. LLM agents are used to generate these plans at their respective abstraction levels, as illustrated in Figure 24.

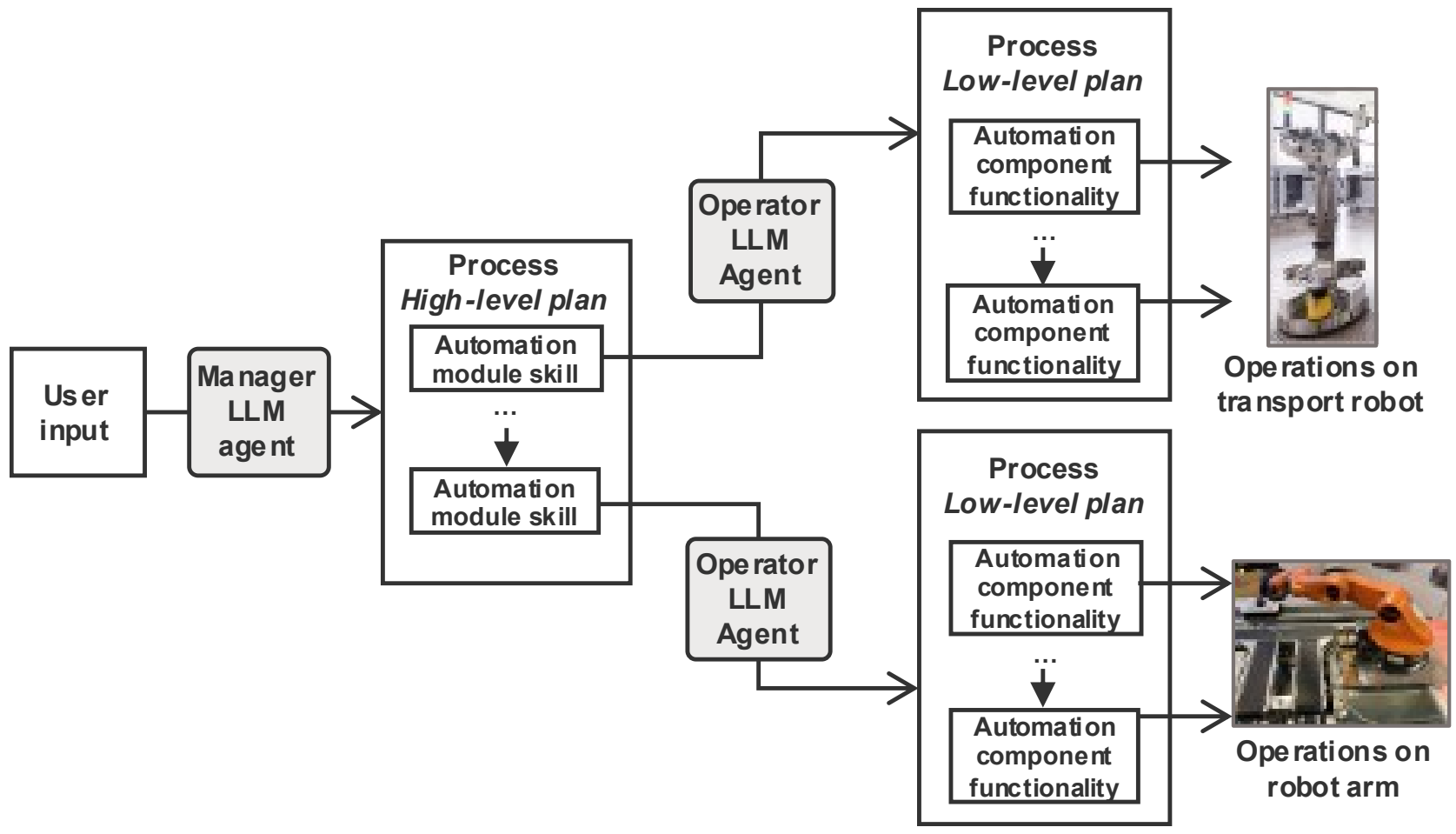

**Figure 24 Two-phase task decomposition with LLM agents.** The manager agent interprets user input into a high-level process composed of machine skills, while operator agents refine each skill into low-level service calls that control robotic resources such as transport robots and robot arms.

### Type 2—Ordered event messages

In [28], an event-driven process modeling method is developed to implement the TPSR model within an event-driven control framework. In this paradigm, the digital twin middleware emits a continuous stream of time-stamped textual event messages that represent temporally ordered changes in the system. These events convey semantically annotated information about the current system and are published to a shared event log memory. Each LLM agent subscribes to event log topics relevant to its functional role, interprets the contextual information, and generates executable service calls as outputs to invoke the corresponding robotic, software, or HMI services. The outcomes of these service executions are subsequently recorded as new events, forming a continuously evolving event stream that records the temporal progression of the process, as illustrated in Figure 25.

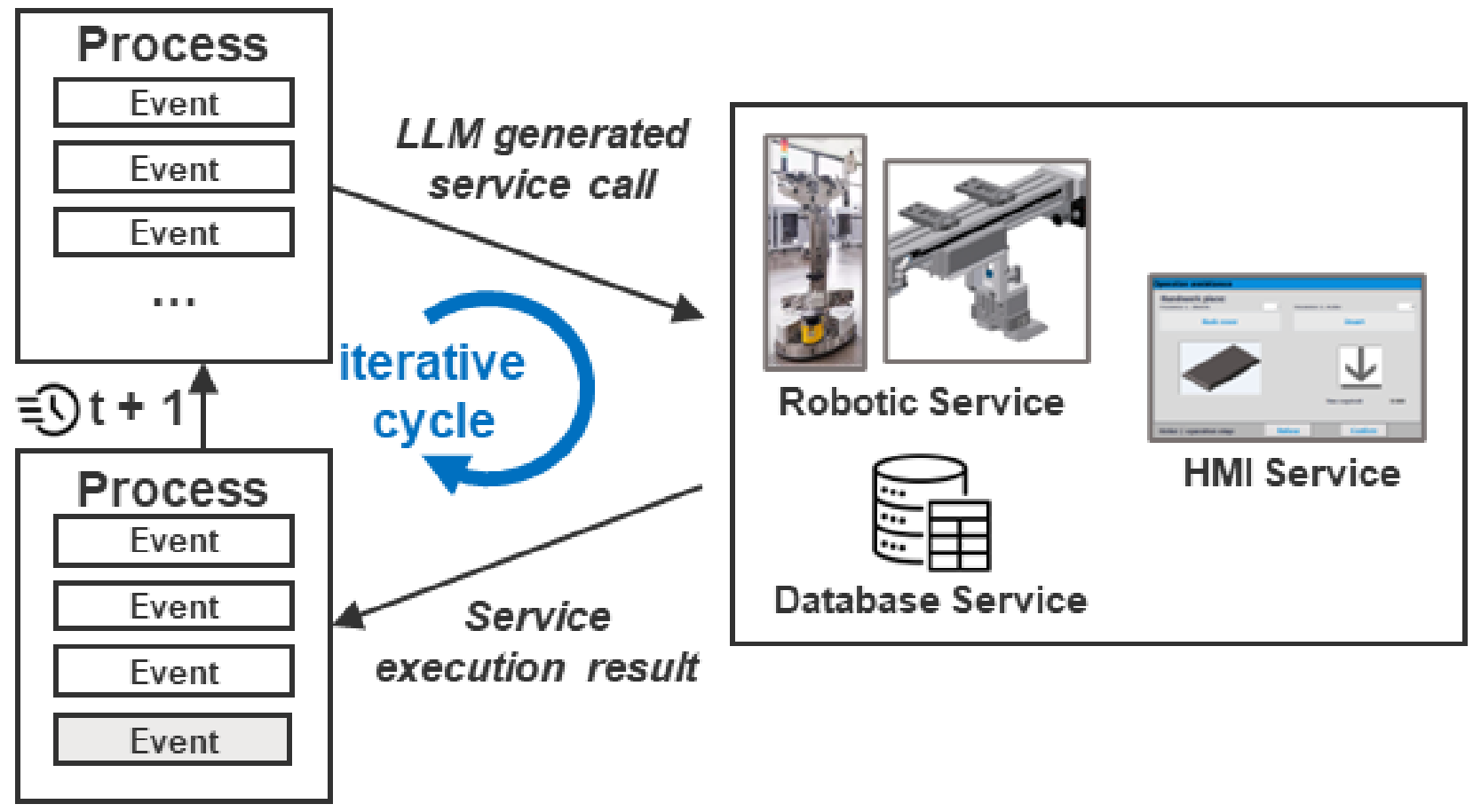

**Figure 25 Event-driven task execution with iterative perception–action cycle.** LLM agents interpret temporally ordered events and generate service calls to robotic, database, or HMI services. The results of service execution are recorded as new events, forming an iterative loop that enables adaptive task execution.

This mechanism establishes an iterative perception–reasoning–action cycle, enabling adaptive task execution in response to dynamically changing system conditions. In this process modeling method, process segments are not predefined within a plan but are dynamically generated.

### Type 3—System state and transition

In [27], a state–transition process modeling approach was developed to integrate LLM agents with a simulator. In this paradigm, the information in the simulation system is represented through a data structure that captures discrete snapshots of the system state at each time point. These state snapshot representations are iteratively provided to the LLM agents, allowing them to reason over the current system conditions and generate parameterized control commands. The execution of these commands updates the system's behavior (i.e., simulation steps), resulting in a state transition and a new system state that is again fed back to the LLM agent as observation. Through this iterative state–reason–transition cycle, the LLM system observes and controls the process, as illustrated in Figure 26. This approach enables the LLM to interact with simulation models to analyze system behavior and reason about control strategies to achieve a defined task objective.

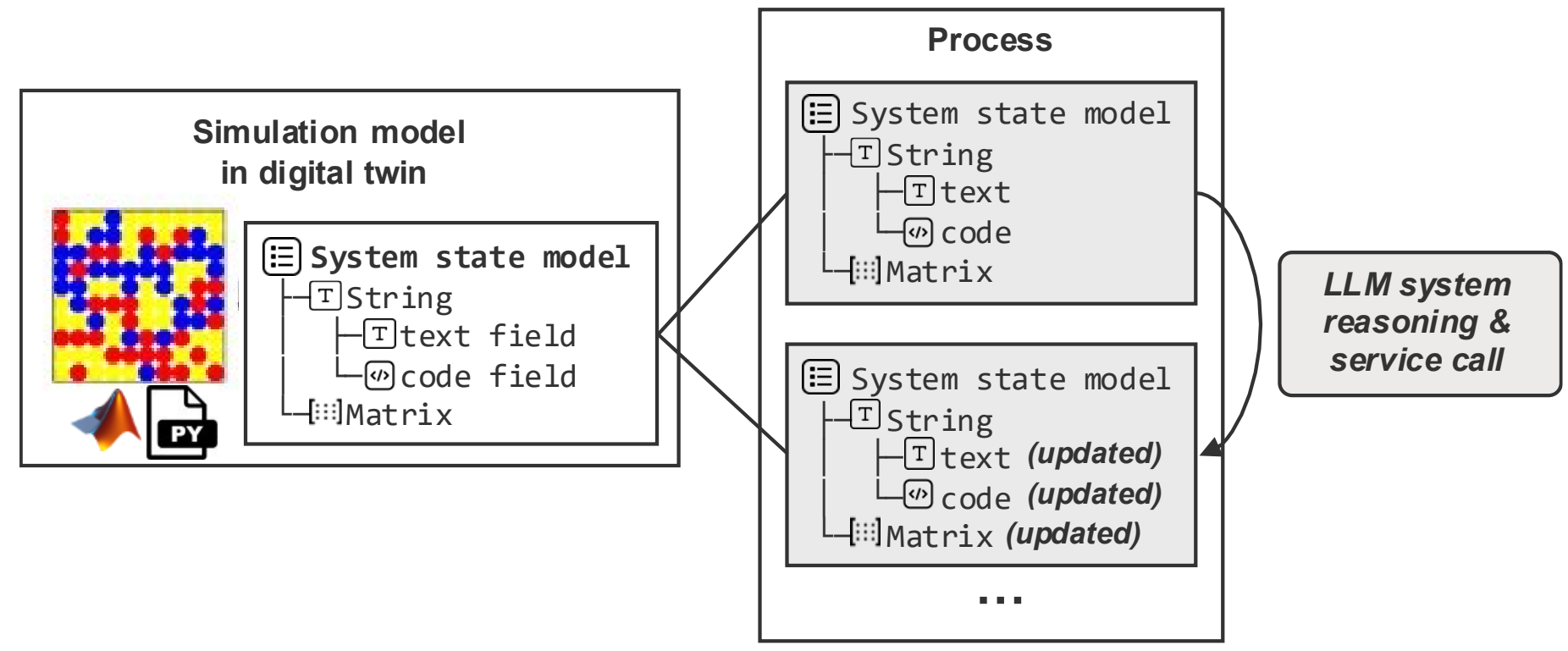

**Figure 26 Iterative interaction between LLM system and simulation models.** Discrete system states from the digital twin simulation are represented in structured information models and provided to the LLM system for reasoning. Generated service calls update the simulation, producing new states that are iteratively fed back to the LLM for continuous control and interpretation.

## 3.3 Functional Roles of LLMs in Task Automation

Building on the system architecture and the TPSR model, this section addresses Research Question RQ3 by explaining how LLMs contribute to task automation in four distinct functional roles: **process orchestration**, **service matching**, **digital resource generation**, and **agent-as-a-service**, as summarized in Figure 27.

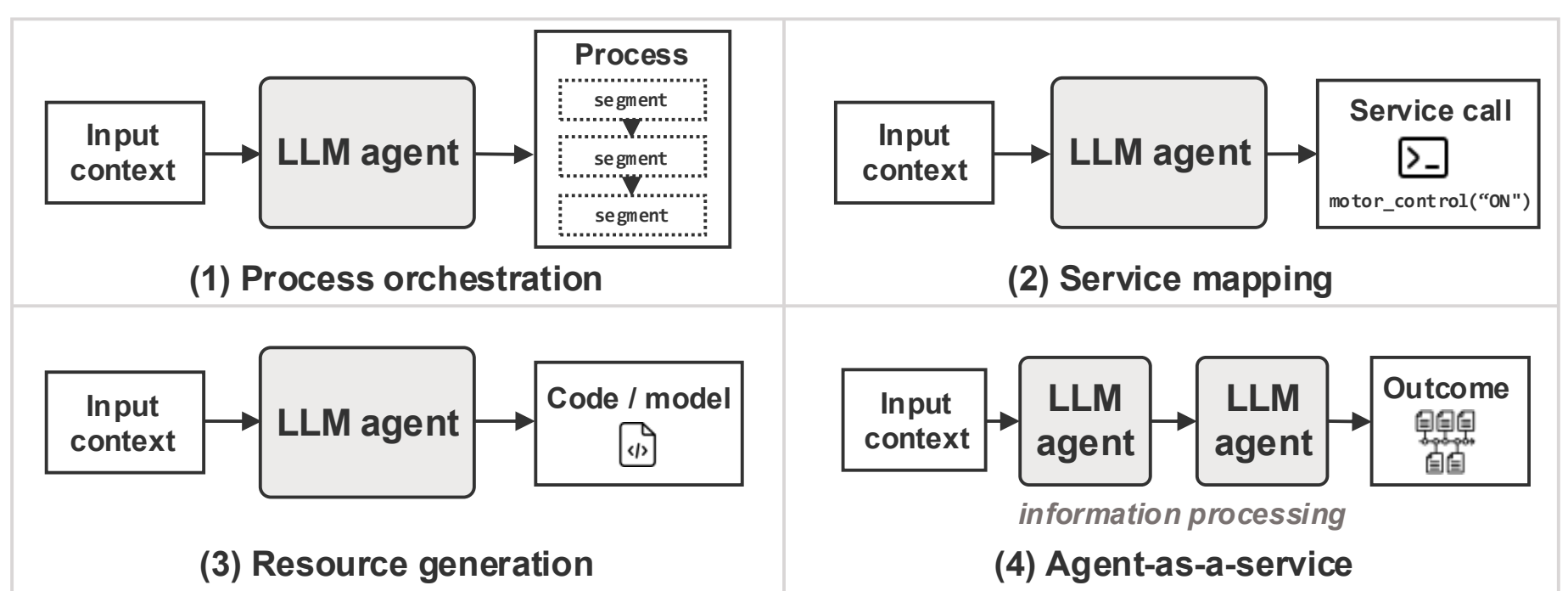

**Figure 27 Four functional roles of LLM agents in task automation.** (1) Process orchestration: interpreting tasks and generating structured process segments. (2) Service matching: binding sub-tasks or process steps to executable services. (3) Resource generation: creating new artifacts or code when services are missing. (4) Agent-as-a-Service: providing reasoning and information processing as a callable service.

### 3.3.1 Process Orchestration

Process orchestration leverages the semantic reasoning capabilities of LLMs to interpret user tasks and translate them into structured and executable processes. Within the TPSR framework, this usage primarily supports the “Task → Process” transformation, bridging the gap between task objectives and operational workflows. As demonstrated in [5], a plan can be generated as an orchestrated production process that decomposes user tasks into coordinated operational steps.

### 3.3.2 Service Matching

Service matching refers to the reasoning and selection process by which LLMs identify and invoke executable services aligned with the current task objective or the process requirements. Within the TPSR framework, this role supports both the "Task → Service" and "Process → Service" transformation. It ensures that tasks or process segments are dynamically mapped to appropriate system services for execution. Similarly, [28] and [27] illustrate how LLMs interpret task objectives and system states to determine suitable control services for invocation. This capability enables adaptive binding between abstract tasks and concrete operational functions, thereby supporting flexible task execution in dynamic environments.

### 3.3.3 Digital Resource Generation

Digital resource generation refers to the capability of LLMs to create new digital artifacts, such as information models, service specifications, and executable code, that enable new or enhanced functionalities within a software-supported automation system. As demonstrated in [7], an LLM multi-agent system is developed to automatically create standardized information models in the form of Asset Administration Shell (AAS) instances from unstructured technical documentation. The generated artifacts can be integrated into digital twin environments to enhance semantic interoperability and to broaden the system's digital resource foundation for supporting automated functions.

Resource generation further includes the automatic creation of executable code, enabling new functions such as algorithms and scripts to be generated on demand at runtime, thereby extending system capabilities beyond existing software modules and libraries [92–95].

### 3.3.4 LLM as Information Processor and Agent-as-a-Service

The LLM itself can be encapsulated as an information-processing resource and exposed as a callable service within the TPSR framework. In this configuration, the LLM serves as a callable service endpoint that performs on-demand flexible reasoning for other components within an information processing pipeline. In [27], LLM agents serve as reasoning components within a perception–action loop. They interpret system states to derive insights, produce intermediate reasoning results, and generate decision command code to control the simulator. Similarly, in [7], LLM agents act as information processors that extract, transform, and synthesize content from unstructured data, enabling the system to produce standardized information model instances.

# 4 Assessment and Evaluation

This chapter synthesizes the assessment framework and presents the evaluation results of the systems developed in the included studies. Section 4.1 introduces the conceptual and methodological criteria used to assess the validity of the system design from both **qualitative** and **quantitative** perspectives. Section 4.2 applies these criteria empirically through controlled experiments that test the system's performance in **process planning**, **event-driven control**, and **semantic data translation**. Section 4.3 synthesizes the **limitations** and **threats to validity** identified across the included studies, forming the generalizable boundary conditions for design knowledge.

## 4.1 Assessment Criteria and Framework

The proposed LLM-enhanced task automation system is assessed using a combined qualitative and quantitative framework. Qualitatively, the assessment focuses on **task executability** and **degree of automation**, evaluating whether user-defined tasks can be operationalized and to what extent their execution can be automated. Quantitatively, the system's operational performance is evaluated through controlled experiments, using metrics such as **correctness**, **reasoning plausibility**, and **semantic accuracy** across planning, control, and data translation tasks. Together, these criteria assess both the soundness of the system design and the effectiveness of its execution.

### 4.1.1 Qualitative Criteria: Task Executability and Degree of Automation

#### Task Executability

The executability of a task determines whether a user-defined goal can be transformed into an executable process within the TPSR framework. This assessment verifies whether the complete "Task → Process → Service → Resource" chain can be instantiated.

The logical foundation of task executability rests on two principles: **temporal ordering** and **process discretization**:

- Temporal ordering: the underlying proposition is that real-world processes unfold over time. For a task to be executable, it must be convertible into an operation that can take place in reality and follows an ordered sequence. This temporal progression shall be formally represented as a process model.
- Process discretization: To represent a real phenomenon, it needs to be expressed as a finite sequence of symbolic elements that together form a representation. When this representation is constructed from an appropriate abstraction perspective, the resulting symbolic sequence defines a bounded subspace within the conceptual space. This boundedness establishes a logically tractable scope within which a modular entity can be described, reasoned about, or controlled. This allows for the definition of a discretized modular entity within a process as a process segment, enabling systematic connection with other discretized process segments.

On the basis of these philosophical and logical principles, a task is executable in the TPSR framework if the following four conditions are satisfied:

(1) Process initialization: The task can be translated into an ordered process that consists of discretized process segments.

(2) Service representation: Each process segment can be mapped to a defined service whose description specifies its function, inputs, outputs, and execution interface.

(3) Service concretization: Each represented service can be further concretized into an implemented form—either as an automated service or as a manual service through HMI.

(4) Resource availability: The necessary computational, physical, or human-operated resources are available and capable of realizing the functional effects defined by the service within the system.

Representative examples of task executability failures resulting from violations of these conditions are illustrated in Table 3.

**Table 3 Representative scenarios illustrating task executability failures within the TPSR framework.**

| Unmet condition | Example scenario | Failure example explanation |
| --- | --- | --- |
| (1) Process initialization | *"[Task]: Optimize factory efficiency"* | The task is too abstract and cannot be decomposed into a defined process sequence. |
| (2) Service representation | *"[Process segment]: Inspect workpiece quality"* | The system lacks a defined service interface for the required process segment. |
| (3) Service concretization | *"[Service call]: Send alarm email to user"* | The defined service has no executable implementation at runtime. |
| (4) Resource availability | *"[Service call]: Start conveyor"* | The physical device is offline or occupied, blocking the service execution. |

Task executability was applied in [5] as a measure to evaluate whether the generated production plans can be executed within an automated production system, where 88% of the generated operations were carried out successfully. In [28], executability was further examined in an event-driven control setting. The detailed evaluation results are summarized in Section 4.2.1.

### Degree of Automation

The IEC vocabulary standard [26] defines the degree of automation as "the ratio of automatically executed functions to the total number of system functions", thereby conceptualizing the degree of automation as a proportion. In the context of the included studies, the concept is used as a means to analyze whether the activities involved in task solving can be executed autonomously without human involvement. In this respect, the degree of automation can be assessed at two levels:

**Automation of process orchestration:** This level assesses the system's ability to automatically translate a user task into a structured process model composed of ordered process segments that satisfy the task objective. Here, LLM agents function as reasoning components that generate executable process model from textual task descriptions, as demonstrated in [5].

**Automation of process execution:** Once a process model is available, this level evaluates the extent to which its process segments can be executed automatically via available services and resources. Services are categorized into physically automated services, digitally automated services, and Human–Machine Interface (HMI) services, according to the degree of required human involvement as listed in Table 4. The degree of automation can then be analyzed by examining how process segments are distributed across these categories.

**Table 4 Classification of service modalities within the TPSR framework for task automation.**

<table>
<tr><th colspan="2">Service Type</th><th>Description</th></tr>
<tr><td rowspan="2">Automated Service</td><td>Physically Automated Services</td><td>These services interface directly with physical hardware components, such as robotic components or sensors, and execute tangible physical actions or acquire real-world data.</td></tr>
<tr><td>Digitally Automated Services</td><td>These services are purely computational operations that perform tasks such as algorithmic processing, database querying, and other software-based functions, without direct interaction with physical entities.</td></tr>
<tr><td>Manual Service</td><td>HMI Services</td><td>The presence of an HMI implies that a process contains at least one process segment that falls outside the automation boundary and requires human participation. From a system design perspective, a manual operation can be encapsulated as an HMI service.</td></tr>
</table>

Drawing on insights from the included studies, LLM agents increase the degree of automation beyond conventional systems by enabling autonomous task interpretation, process orchestration and dynamic service coordination. The resulting operations are executed either automatically or via HMI, depending on infrastructure availability.

### 4.1.2 Quantitative Criteria: Correctness, Reasoning Plausibility, Semantic Accuracy

The quantitative assessment defines measurable criteria for evaluating the performance of the LLM-agent-driven automation system across three functional use cases: process planning, event-driven control, and semantic data translation, as summarized in Figure 28.

#### Correctness

Correctness quantifies whether the command outputs generated by the system match valid and executable actions. It reflects the extent to which an LLM agent can produce process plans or control commands that are both syntactically correct and semantically aligned with the intended task objective. In practice, correctness is evaluated by comparing each system-generated command with a predefined reference command in a test case, as illustrated by the examples in Table 5. A case is counted as correct if the generated output matches the reference; otherwise, it is classified as incorrect. An overall correctness rate is calculated as the proportion of correctly generated commands to the total number of evaluated test cases (see Table 6).

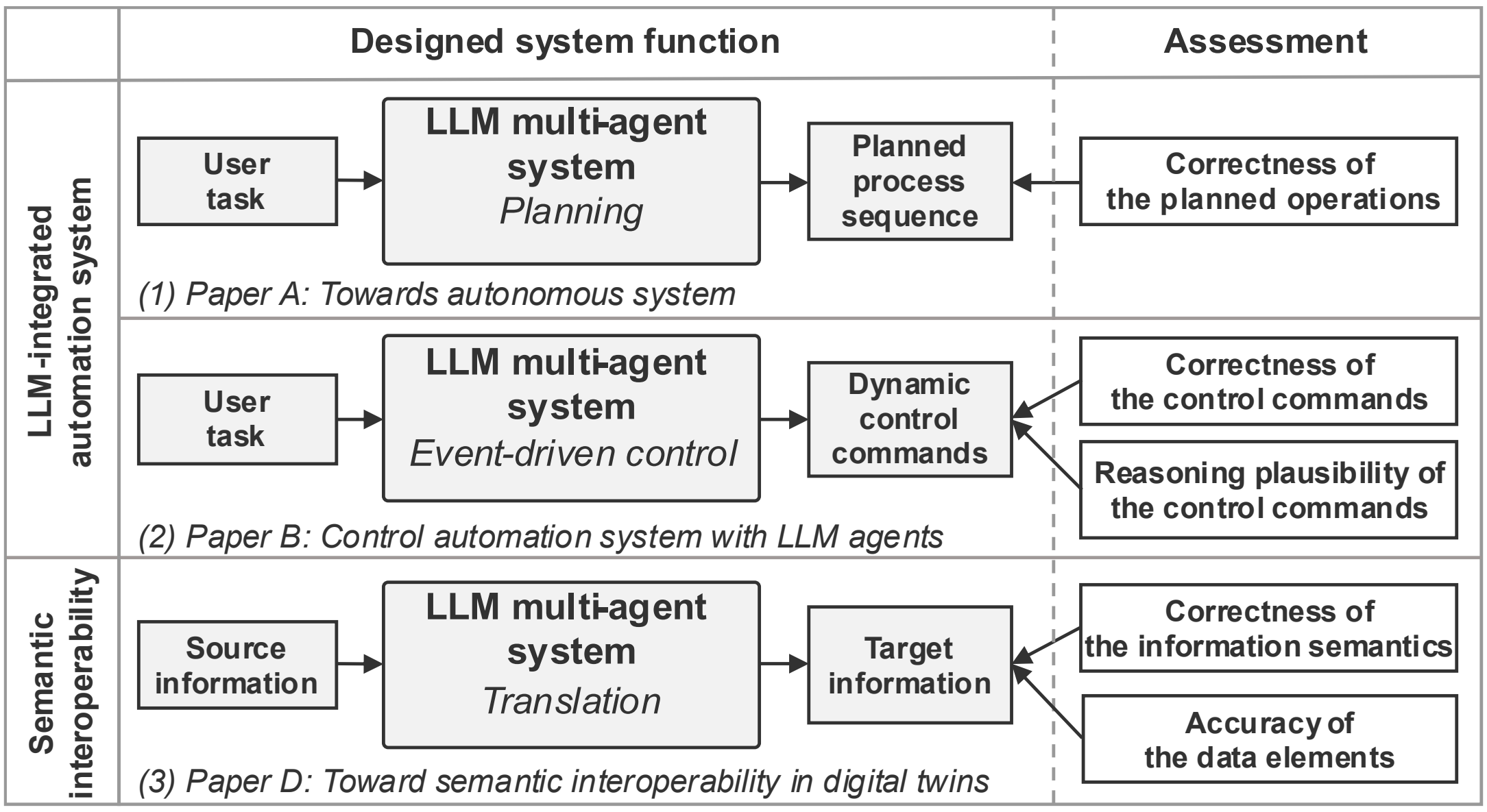

**Figure 28 Mapping of designed system functions to their corresponding assessment criteria across the included studies.** Each functional use case—process planning [5], event-driven control [28], and semantic data translation [7]—is evaluated through specific quantitative metrics of correctness, reasoning plausibility, and semantic accuracy.

## Reasoning Plausibility

Reasoning plausibility assesses the logical soundness and contextual adequacy of the LLM agent's justification (see the "generated reason" column in Table 5) accompanying each generated command in a test case, as described in Paper B. For each test case, evaluators review the generated reason and assign a score on a 5-point scale. Higher plausibility reflects both correct control command generation and the presence of contextually justified reasoning.

**Table 5 Examples of test cases used to evaluate command correctness and reasoning plausibility of LLM-generated control commands.**

| Test case | Reference correct control command | Generated control command | Correct-ness | Reference correct reason | Generated reason (as justification) | Reason plausibility |
|---|---|---|---|---|---|---|
| #1 | `RFID_read_workpiece_info()` | `RFID_read_workpiece_info()` | **TRUE** | "Workpiece is detected at the holder H2 on conveyor C1 and information of the workpiece needs to be read." | "The workpiece has been detected by BG51, need to read the workpiece information." | **5** |
| #21 | `alert_to_supervisor('conveyor C1 failure')` | `conveyor_1_run('reverse', 2)` | **FALSE** | "Conveyor C1 fails to move forward and an alert to supervisor is needed." | "Conveyor C1 failed to move forward, attempting reverse to clear potential jam." | **3** |

### Semantic Accuracy

Semantic accuracy evaluates whether the system correctly interprets source data and translates it into target information models that faithfully reflects the original meaning, as described in Paper D. It is assessed through manual inspection of generated data elements to identify inaccuracies and through ratings of the overall quality and clarity of the textual information generated.

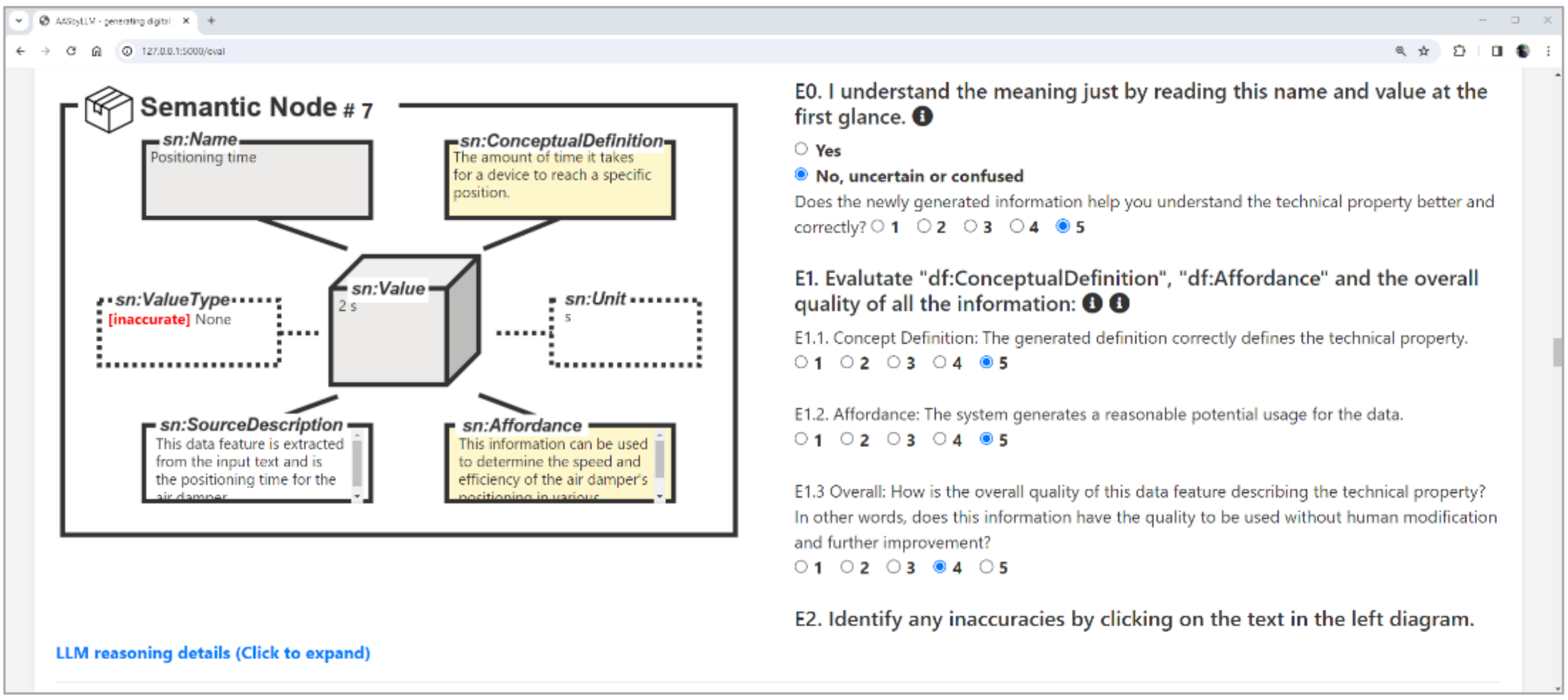

**Figure 29 The interactive evaluation interface used for assessing semantic accuracy.** Each generated semantic node is reviewed across multiple dimensions (including name, conceptual definition, unit, and affordance) to determine its semantic accuracy and overall interpretability.

As shown in Figure 27, human evaluators review each generated data structure via a graphical interface, identifying incorrect fields and rating conceptual clarity, affordance, and interpretability. This evaluation captures both field-level accuracy and the overall semantic fidelity of the generated model instances.

## 4.2 Evaluation Results and Discussion

This section presents the empirical evaluation of the LLM-integrated automation framework across representative use cases, including process planning and control[5,27,28], as well as semantic translation for interoperability [7]. The results address RQ3 and RQ4 by evaluating system performance with respect to the functional roles of LLM agents.

### 4.2.1 Process Planning and Execution Control (in Papers A and B)

As summarized in Figure 28, the systems developed in Paper A (Towards Autonomous System) [5] and Paper B (Control Industrial Automation System with LLM Agents) [28] investigate LLM-integrated automation in modular production environments. In this case, both studies operationalize user-defined tasks within an automated production environment. They focus respectively on process planning and event-driven control, providing complementary evaluations of LLM-integrated automation across different levels of task execution.

## Evaluation Setup and Key Results

**Evaluation of the generated planned operations:** In [5], the evaluation involves 50 distinct user tasks in natural language. Under this setup, 88% of the process sequences generated by LLM agents were successfully executed on the automation system and achieved task objectives.

**Evaluation of the generated control commands:** In [28], the LLM-agent-based decision-making mechanism is implemented in an event-driven framework and evaluated using 100 test cases, each representing a distinct operational scenario. For each case, the LLM multi-agent system observes a sequence of events and generates a corresponding control command.

The evaluation assesses both command execution correctness and reasoning plausibility across different LLMs, as summarized in Table 6. The test cases are grouped into two categories:

- **Standard operation test cases** represent routine production operational scenarios that follow predefined Standard Operating Procedures (SOPs). These cases are used to verify the agent's ability to execute typical control tasks correctly and consistently under normal system conditions.
- **Exception test cases**, by contrast, simulate unexpected or non-routine events such as missing parts, sensor failures, or abnormal task requests. These cases evaluate the agent's capacity for adaptive reasoning and its robustness in generating valid control commands under unforeseen conditions.

Each generated command is evaluated along two dimensions:

- **Execution correctness** measures whether the command generated by the LLM agent matches the reference command in the test dataset, indicating the system's ability to produce executable and correct control actions. Results are reported as percentages in Table 6.
- **Reasoning plausibility** assesses the logical soundness of the justification provided for each generated command. It is evaluated by human raters and reported on a five-point scale in Table 6.

To examine the effect of task-specific training achieved through supervised fine-tuning on a dataset, the evaluation has compared five state-of-the-art LLMs and their trained versions.

- **General-purpose models:** General-purpose LLMs were first tested without additional training to establish baseline performance across both standard and exception test cases.

- **Fine-tuned models:** Supervised fine-tuning was applied using a collected dataset of paired event–command examples collected from system operations. The fine-tuned models were then re-evaluated to quantify the improvement in command accuracy and reasoning plausibility.

**Table 6 Evaluation results for command generation using different LLMs.** The table reports the command correctness rate (%, value left) and reason plausibility on a five-point scale (value right) for different models. Results are shown for all test cases in total as well as separately for standard operation and exception test cases, both before and after supervised fine-tuning. [28]

| | Evaluation results for command generation using different LLMs based on 100 test cases | | | | |
|---|---|---|---|---|---|
| | ***GPT-4o*** | **Llama-3-70B-Instruct** | **Llama-3-8B-Instruct** | **Qwen2-7B-Instruct** | ***Mistral-7Bx8-Instruct-v0.2*** |
| Pre-trained (All test cases) | **81% \| 4.7** | **75% \| 4.3** | **37% \| 2.8** | **65% \| 3.7** | **29% \| 2.4** |
| Pre-trained (Standard operation test cases) | 100% \| 5.0 | 87% \| 4.5 | 53% \| 3.1 | 63% \| 3.6 | 34% \| 2.4 |
| Pre-trained (Exception test cases) | 41% \| 4.0 | 50% \| 3.8 | 3% \| 2.2 | 69% \| 4.0 | 19% \| 2.3 |
| Fine-tuned (All test cases) | **100% \| 5.0** | **95% \| 4.8** | **96% \| 4.9** | **97% \| 4.9** | **45% \| 3.1** |
| Fine-tuned (Standard operation test cases) | 100% \| 5.0 | 94% \| 4.8 | 99% \| 4.9 | 97% \| 4.9 | 61% \| 3.6 |
| Fine-tuned (Exception test cases) | 100% \| 5.0 | 97% \| 4.9 | 91% \| 4.7 | 97% \| 5.0 | 9% \| 2.3 |

### Model Adaptation through Fine-Tuning

Before fine-tuning, large general-purpose models achieved correctness rates of 65%–81%. After supervised fine-tuning on a task-specific input–output dataset, correctness increased to 95%–97%, indicating a substantial performance improvement, as shown in Table 6. When fine-tuning the GPT-4o model using OpenAI’s fine-tuning API, 100% correctness was observed. These results demonstrate that task-specific fine-tuning has enabled more accurate and contextually appropriate control decisions.

### Discussion of the Evaluation Results

The evaluation results provide empirical evidence for the feasibility of the proposed LLM-integrated automation framework. The 88% executability achieved in process planning in Paper A [5] demonstrates that natural-language user tasks can be transformed into executable operations, validating the integration of LLMs, digital twins, and automation systems. The event-driven control experiments in Paper B [28] further show that LLM agents can perform dynamic, context-aware decision-making within a perception–action loop. Model training substantially improves control performance, increasing correctness from 65–81% to 95–100% across different models. This demonstrates the effectiveness of adapting LLMs to domain-specific control tasks on the condition that a representative training dataset can be provided. Overall, these results indicate that the proposed framework enables adaptive, context-aware, and objective-oriented task solving beyond fixed rule-based logic, supporting its use as an assistant system for autonomous production planning and control.

## 4.2.2 Semantic Translation and Interoperability (in Paper D)

This section evaluates the semantic translation capability of LLM agents in generating digital twin information models. The designed system transforms unstructured textual descriptions of automation components into standardized Asset Administration Shell (AAS) model instances, thereby supporting interoperability across heterogeneous models.

### Evaluation Setup and Key Results

The evaluation uses data extracted from the documentation of 20 industrial automation components, including sensors, actuators, controllers, and network devices. Each component's documentation was processed by the LLM-agent-based system to generate corresponding AAS data elements. In total, 2,259 data elements were produced and manually evaluated by seven human evaluators, resulting in over 22,590 individual ratings.

Overall performance is evaluated using two quantitative metrics that capture the quality and usability of the generated AAS information model instances:

1. **Effective generation rate (pass rate):** This metric measures the proportion of correctly generated model elements in the AAS model that are semantically correct and ready for direct use without further modification. A generated model element is considered effective when all its data fields are evaluated as accurate, including name, definition, affordance, unit, and description, as shown in Figure 29.
2. **Helpfulness score:** This metric assesses the extent to which the generated content improves the clarity and interpretability of technical data. Human evaluators rate each generated information field on a five-point scale based on how much the translation enhances their understanding relative to the raw input.

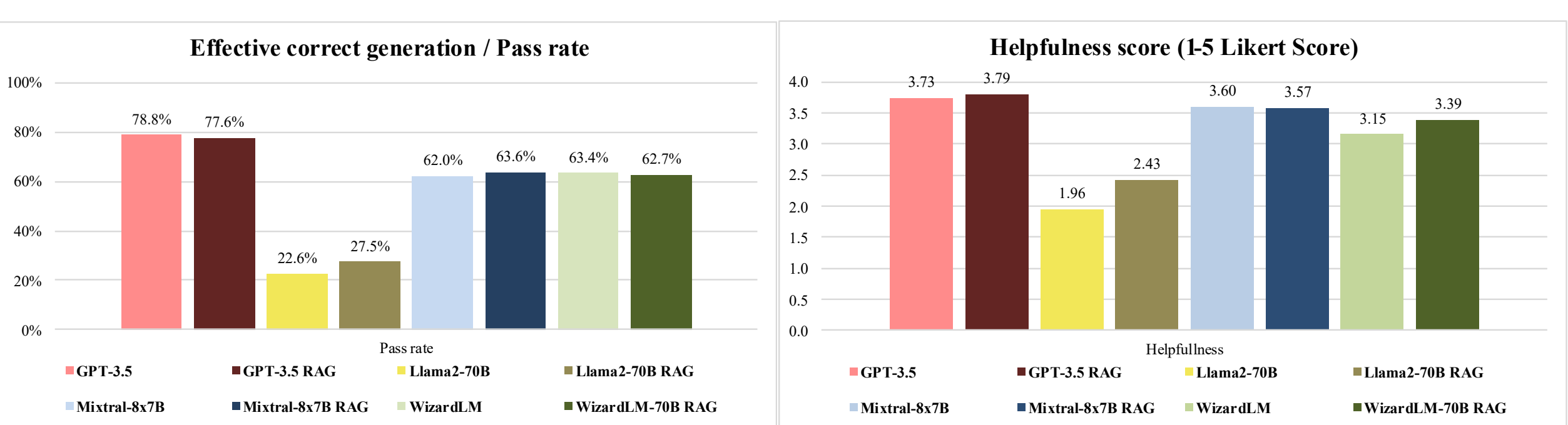

**Figure 30 Comparison of the effective generation rate (pass rate) and helpfulness score of generated AAS data elements across different LLMs with and without retrieval-augmented generation.** The left chart shows the proportion of correctly generated information fields that are complete and ready for use, while the right chart presents the average helpfulness scores (1–5 Likert scale) rated by human evaluators, indicating how effectively the generated content improves understanding of technical concepts [7].

In more detail, the generated model elements were manually evaluated at the data-field level to assess accuracy and completeness across key semantic dimensions, including name, concept definition, data usage (affordance), unit, and description. The distribution of inaccuracy rates across these dimensions for different LLMs and RAG configurations is shown in Figure 31.

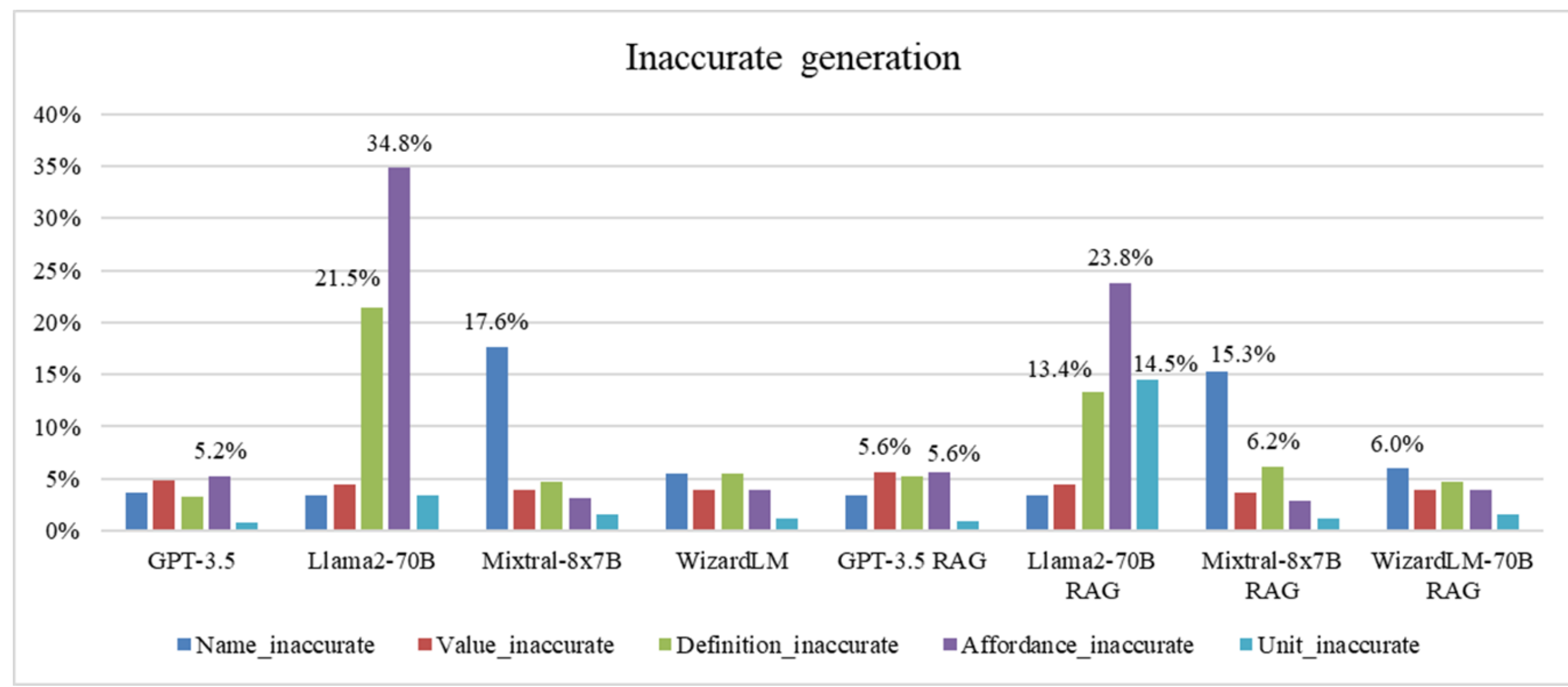

**Figure 31 Inaccurate generation ratios across different state-of-the-art LLMs with and without RAG**. The bars represent the percentage of incorrect outputs for each field in the generated AAS data elements, including name, value, definition, affordance, and unit [7].

For most data fields generated by the evaluated LLMs, accuracy exceeds 95%, with inaccuracy remaining below 5%, indicating generally reliable semantic translation performance. The observed outliers (i.e., values exceeding 10%) for LLaMA 2-70B and Mixtral-8x7B can largely be attributed to misinterpretations of prompt instructions by the models, indicating that these models struggle with the more demanding interpretation and translation task. For stronger models, retrieval-augmented generation does not lead to a clear additional performance gain.

## Discussion of the Evaluation Results

Based on the evaluation results in Figure 30 and Figure 31, the system generated semantically consistent and usable AAS model instances, achieving an effective generation rate of up to 79% with largely accurate data-field values. Helpfulness scores followed the same trend, indicating that LLMs that translate data more accurately also can help users better understand complex technical concepts.

The comparison between configurations with and without retrieval-augmented generation shows that RAG improves performance only conditionally. It provides marginal benefits for weaker models by providing supplementary contextual information, whereas stronger models such as GPT-3.5 and WizardLM-70B already achieve consistently high semantic accuracy without retrieval support. This indicates that well-trained LLMs possess sufficient intrinsic knowledge for semantic interpretation and that the marginal benefit of RAG diminishes as model capability increases.

The overall results confirm that LLM agents can accurately translate unstructured technical descriptions into standardized AAS information model instances. This demonstrates the system's ability to perform semantic translation and to enable interoperability across heterogeneous digital twin representations. These findings validate LLM agents as semantic translators and demonstrate their suitability for digital resource generation in industrial digitalization and automation applications.

## 4.3 Limitations, Threats to Validity and Boundary Conditions for Design Knowledge

The LLM-integrated automation system developed in this dissertation extends beyond conventional rule-based systems by enabling context-aware, objective-oriented task solving. This section outlines these limitations and clarifies the boundary conditions under which the design knowledge applies.

### 4.3.1 Limitations

Since the proposed automation systems were evaluated (as in Sections 4.1 and 4.2) through a finite number of prototypical implementations, controlled laboratory studies, and a limited set of test cases, the resulting evidence is necessarily constrained by these settings. The following thematic paragraphs outline the empirical and artefact-related limitations that shape the applicability and generalizability of the results.

**Prototype character and laboratory settings.** All implemented systems have been evaluated in controlled laboratory or prototypical environments rather than in full-scale industrial deployments. The modular production system used in Papers A [5] and B [28] is a laboratory facility; the container-mixing scenario in Paper C [27] is a simulation with a restricted decision space; and the AASbyLLM framework in Paper D [7] is validated by translating digital documentation of automation components into structured models that represent these components. These settings provide realistic but still constrained conditions and therefore do not capture the full diversity of industrial automation scenarios, domain-specific requirements, and edge cases that may occur in practice.

**Human oversight and safety constraints.** While the LLM agents can autonomously make decisions that automate task solving and generate recommendations, inherent informational uncertainty arises when they must reason over unknown information or unconstrained open input spaces [96], which leads to stochastic behaviors. Although issues of repeatability can be resolved in theory [97], unpredictable input disturbances can still induce non-deterministic outcomes, and human oversight therefore remains necessary. Across all studies, human operators remain in the loop to review and intervene when required. In the modular production environment in [5] and [28], experiments are continuously monitored; in the AASbyLLM system [7], human annotators validate the generated information models. The evaluation results do not imply 100% reliability, and the artifacts have not been assessed in unsupervised safety-critical settings in which LLM-generated outputs would be executed without human review. While the autonomy of LLM agents contributes to higher levels of automation beyond the limit of the rule-based systems, their decisions still shall be overseen by human users.

**Dependence on digital twin quality and textual representations.** All case studies assume the existence of a digital twin infrastructure with sufficient representational fidelity and accessible service interfaces. Papers A [5] and B [28] leverage an AAS-based middleware that exposes module skills and component functionalities as text descriptions; Paper C [27] assumes a simulation model whose states are available as structured textual data; and Paper D [7] relies on the AAS meta-model as a standardized target schema for translation. Overall, the LLM-based reasoning presupposes that relevant system knowledge is available in textual form, including natural-language task descriptions, structured model information, and essential system specification knowledge. In environments where digital representations are incomplete, inconsistent, or poorly documented, these preconditions may not be satisfied.

**Prompt design and model capabilities.** Prompts specify LLM-agent behavior, and thus system performance depends on the prompt templates and model configurations employed. The quantitative performance reported in Section 4.2 reflects a particular choice of LLMs, prompt templates, and configuration parameters. Model training also adapts LLM-agent behavior, and the command correctness and reasoning plausibility reported in Paper B [28] depend on the training data. While training yields performance gains, it induces persistent, global shifts in model behavior, altering how the LLM-agent responds across the entire input distribution. This makes such training desirable only when the task input distribution can be characterized. The semantic accuracy in Paper D [7] depends both on the underlying LLM and on whether a retrieval process is enabled through data injection in the prompt. Weaker models, especially smaller models deployed, yield worse performance profiles. Consequently, system behavior reflects dependencies on both model choices and the specific prompts adopted in the experiments, as documented in the respective papers and their accompanying repositories.

Taken together, these empirical and artefact-related limitations frame the conditions under which the studies are conducted and thereby inform the subsequent discussion of threats to validity.

### 4.3.2 Threats to Validity

**Construct validity.** In this dissertation, the central concepts underlying the proposed autonomous system design are defined with clear intent and meaning in Section 1.1. In the evaluation, metrics including task executability, command correctness, semantic accuracy, and reasoning plausibility are systematically operationalized and measured through observable system behavior following the criteria defined in Section 4.1. These clear operational definitions reduce the risk of inadequate construct specifications.

The residual threats are minimal and largely inherent to studies involving open-ended task inputs and natural-language flexibility. The task inputs used for testing task automation in Papers A [5], B [28] and C [27] cover representative scenarios but cannot exhaustively reflect the full variability of real-world conditions. In Paper D [7], minor ambiguities may also arise from variability in human annotators' subjective ratings when assessing semantic accuracy and effective generation rates (pass rate), although these risks are mitigated by involving multiple annotators and a large number of evaluation cases.

**Internal validity.** Internal validity concerns whether the effects identified in the evaluation can be attributed to the integration of LLM agents with digital twins rather than to alternative explanations or uncontrolled sources of variance. Residual threat factors to internal validity remain but are bounded: the stochasticity of LLM reasoning and the openness of natural-language inputs create run-to-run variation that introduces limited incidental variance; the representational approaches used for system events and states may implicitly encode prior domain knowledge that contributes to the evaluated decision-making performance; the particular LLM used to power the agents can be replaced by different models, making model choice a potential source of variance; and human annotators introduce some subjectivity when rating semantic accuracy or reasoning plausibility. These risks are mitigated by fixing configurations within controlled experiments in Paper B [28] and Paper D [7], using a large number of test cases, comparing relative performance across different test cases and model variants, involving multiple human evaluators, and applying consistent evaluation metrics that are grounded in executable, observable, and decidable system behavior.

**External validity.** External validity concerns the extent to which the findings generalize beyond the specific case studies and prototype settings investigated in this dissertation. First, the evaluated systems do not reflect the strict safety requirements of many industrial domains, where autonomous decisions produced by LLM-based systems cannot be accepted without certified safeguards and human oversight. As a result, transferability to such settings remains limited. Second, the proposed architecture presupposes a certain degree of component modularity and modular digital modelling (e.g., service-oriented interfaces, digital twin representations), so its applicability is reduced in tightly coupled legacy systems, which lack the structural modularity necessary to support flexible orchestration. Third, the benefits of LLM-based reasoning are most pronounced for flexible, knowledge-intensive tasks, while simple, repetitive, and deterministic tasks can typically be handled more efficiently and reliably by conventional rule-based control, limiting the practical relevance of LLM integration in those settings.

Overall, these considerations limit the strength and scope of the empirical claims and, from a design science perspective, provide the basis for articulating the identified limitations as explicit boundary conditions for the applicability of the generalizable design knowledge.

### 4.3.3 Boundary Conditions for Design Knowledge

The identified limitations and threats to validity also inform the resulting design knowledge by delineating the boundary conditions under which the proposed artifacts operate effectively.

First, the approach is particularly well-suited in **long-tail, open-ended application scenarios** where input conditions, operating states, and task cases cannot be exhaustively enumerated before deployment. In such contexts, rule-based logic becomes intractable, because the number of possible combinations of states and task variants grows combinatorially and cannot be fully anticipated at design time. Here, the semantic reasoning capabilities of LLM agents are well-suited to interpreting open-ended inputs and to composing context-dependent solutions at runtime. Conversely, for simple, repetitive tasks with a limited number of frequent operating scenarios, conventional rule-based PLC-control remains more efficient and reliable. The designed artifacts should therefore be instantiated preferentially in problem domains where long-tail distributions of input scenarios, large reasoning spaces, and knowledge-intensive information processing are prominent.

Second, the designed systems are **model-sensitive but should be designed to be model-agnostic**. In practice, system behavior depends on the capability of the chosen LLM and the prompt configuration used in each agent. As LLM technology evolves, different model families and variants can exhibit distinct performance characteristics. This leads to a design implication that LLM agents should be architected so that their interfaces and information-processing pipelines are independent of any specific model, meaning that the LLM should function as an interchangeable module within the agent system, allowing model substitution and prompt adjustment without modifying the surrounding system.

Third, the applicability of the approach **depends on the availability of digital representation infrastructures** for physical systems. The proposed architecture presupposes that system components are readily represented in modular digital twins or comparable structured models that expose machine-readable interfaces and semantic descriptions of capabilities and operational states. Digitalization is therefore a prerequisite for enabling LLM-based reasoning over physical systems. Moreover, this digitalization is essential for providing the representative datasets required for model training. The effectiveness of such training depends on having a comprehensive dataset and a well-characterized data distribution.

Finally, the **role of the user** must be explicitly considered. Users can supply task inputs, provide task objectives, and retain responsibility for oversight and final decisions. Accordingly, the proposed systems are best understood as intelligent assistant or recommendation systems that augment user productivity rather than replace human judgment. This aligns with safety requirements and established industrial practice: in many safety-critical or heavily regulated domains, control actions and system modifications must be validated by human operators or engineers. A boundary condition is therefore that LLM agents should operate within application settings that provide supervision, review, and override mechanisms.

Taken together, these boundary conditions specify where the proposed design artifacts can be deployed with justified confidence and where additional safeguards or complementary methods are required.

# 5 Conclusion and Outlook

This cumulative dissertation presents a multi-cycle Design Science Research project that investigates how LLMs can be integrated with digital-twin-supported automation systems to enable adaptive and flexible task automation. Across five peer-reviewed studies, the research develops and evaluates a series of artifacts that are progressively refined into generalizable design knowledge. Together, the outcomes demonstrate how LLM-based semantic reasoning can be operationalized within industrial automation systems and establish the basis for the overall contributions discussed in the following section.

## 5.1 Overall Contributions and Answers to Research Questions

This cumulative dissertation makes contributions to the conception, design, and implementation of autonomous systems for industrial automation. It defines autonomous systems as an interdisciplinary integration of automation systems, digital twins, and LLM agents, enabling context-aware and objective-oriented autonomy through LLM reasoning. The contributions can be summarized into four interrelated aspects:

### Conceptualization of Autonomy in Industrial Automation

Drawing on the design knowledge generalized across the included studies, this dissertation frames autonomy as a property emerging from the reasoning freedom structurally enabled by the system design in conjunction with the semantic reasoning capabilities of LLMs. It defines autonomous systems as automation systems capable of context-aware, objective-oriented decision-making at runtime. This conception extends traditional automation theory with a flexible reasoning dimension and provides the conceptual foundation reused throughout the dissertation. The key concepts are defined in Section 1.1.

### Three-Layer Integration Architecture

The dissertation introduces and validates a generalizable three-layer architecture that integrates the LLM multi-agent system, the digital twin mediation layer, and the physical automation system. The digital twins provide the semantic and operational grounding required for LLMs to interact with the physical environment. Building on this, the architecture establishes a reusable blueprint for integrating LLM agents into cyber-physical environments while remaining compatible with the traditional automation pyramid. This contribution directly answers RQ1 (System Design).

### Generalizable Task–Process–Service–Resource Model

The TPSR model defines an abstraction for transforming user-defined tasks into executable automation processes. It systematizes the relationships between tasks, processes, services, and resources, providing a conceptual bridge between operational task automation and objective-oriented reasoning. Within this model, task automation is framed as an operationalization chain ("Task → Process → Service → Resource"), and LLM agents fulfill four functional roles: they orchestrate processes by translating user tasks into process segments (process orchestration); match process segments to executable services (service matching); generate digital resources such as information models and code (digital resource generation); and provide on-demand reasoning within information-processing pipelines (agent-as-a-service). Together, these roles articulate how LLM reasoning can be utilized, and this contribution answers RQ2 (Task Automation) and RQ3 (Functional Roles of LLMs).

### System Design Artifacts, Reference Implementations and Evaluation Insights

The studies included in this cumulative dissertation deliver reusable system design knowledge and open-source software frameworks that serve as reference implementations for LLM-driven automation. Two major systems were developed to address two representative use cases: (1) the LLM multi-agent system for flexible modular production, validated in a laboratory facility with hardware implementation;[9] and (2) the AASbyLLM system for standardized information model generation, released as an open-source software implementation.[10] Each system is further accompanied by design blueprints, prompt templates, demonstrators, and evaluation datasets. These materials form a reference stack that supports replication and further development within the research community.

The empirical results across the case studies answer RQ4 (Evaluation). These results provide demonstrations as well as qualitative and quantitative evaluations, and the developed systems show high performance in the case studies: 88% task executability in flexible production planning, 95–100% control accuracy after fine-tuning in adaptive event-driven control, and approximately 79% valid information model generation for semantic interoperability. These results collectively substantiate the proof-of-concept and confirm the effectiveness of the proposed systems.

## 5.2 Outlook

Building on the results of this cumulative dissertation, several directions emerge for future research to further advance the technical limitations.

The first substantive research direction concerns safety assurance and verification. While this dissertation has shown that LLM agents can generate accurate plans, control commands, and semantic translations, their stochastic behavior under information uncertainty and the open-ended input space require systematic safeguards. Future work could therefore explore automated verification pipelines that combine simulation-based testing with formal checks on generated commands and process plans. Runtime monitoring, anomaly detection over dynamic system states, and reward modeling for model training may further help ensure compliance with safety requirements and operational constraints, enabling LLM-driven automation to be deployed with greater confidence in industrial environments.

[9] https://github.com/YuchenXia/GPT4IndustrialAutomation
[10] https://github.com/YuchenXia/AASbyLLM

A second direction relates to computational efficiency, model management, and the model-agnostic design of agents. The artifacts developed in this work assume the availability of sufficiently capable LLMs, and the evaluation has highlighted that different model families and sizes exhibit distinct performance profiles. Future research can therefore explore systematic strategies for model selection, examine performance trade-offs, and investigate hybrid setups that combine high-capacity and computationally lightweight models with dynamic task routing based on task difficulty.

A third direction focuses on advancing the digital twin layer that grounds LLM reasoning in physical reality. The proposed approach presupposes modular, semantically rich digital representations; however, many industrial environments still lack such infrastructures. Future work could therefore address methods for the automated creation of digital representations of physical systems, automated extraction of models from existing engineering artifacts, and multi-modal extensions of digital twins that include time-series data, CAD models, or visual information alongside text for reasoning. Such advances would also enhance data availability, enabling more comprehensive model training and adaptation for industrial automation settings.

Finally, the human role in LLM-driven autonomous systems warrants deeper investigation. This dissertation has treated LLM agents primarily as intelligent assistants whose outputs remain under user's control; future work can elaborate design patterns for human–AI collaboration, including interface concepts, explanation mechanisms, and responsibility allocation between operators, engineers, and autonomous agents. Empirical studies on exception recovery, trust calibration, and organizational adoption readiness could complement the technical contributions and help position LLM-integrated automation not merely as a technological innovation, but as a sustainable socio-technical system.

# References

[1] Derigent, W., Cardin, O., and Trentesaux, D., 2021, “Industry 4.0: Contributions of Holonic Manufacturing Control Architectures and Future Challenges,” J Intell Manuf, **32**(7), pp. 1797–1818. https://doi.org/10.1007/s10845-020-01532-x.

[2] Folgado, F. J., Calderón, D., González, I., and Calderón, A. J., 2024, “Review of Industry 4.0 from the Perspective of Automation and Supervision Systems: Definitions, Architectures and Recent Trends,” Electronics (Basel), **13**(4). https://doi.org/10.3390/electronics13040782.

[3] Sahlab, N., Jazdi, N., and Weyrich, M., 2021, “An Approach for Context-Aware Cyber-Physical Automation Systems,” IFAC-PapersOnLine, **54**(4), pp. 171–176. https://doi.org/https://doi.org/10.1016/j.ifacol.2021.10.029.

[4] Müller, M., Müller, T., Ashtari Talkhestani, B., Marks, P., Jazdi, N., and Weyrich, M., 2021, “Industrial Autonomous Systems: A Survey on Definitions, Characteristics and Abilities,” **69**(1), pp. 3–13. https://doi.org/doi:10.1515/auto-2020-0131.

[5] Xia, Y., Shenoy, M., Jazdi, N., and Weyrich, M., 2023, “Towards Autonomous System: Flexible Modular Production System Enhanced with Large Language Model Agents,” *2023 IEEE 28th International Conference on Emerging Technologies and Factory Automation (ETFA)*, pp. 1–8. https://doi.org/10.1109/ETFA54631.2023.10275362.

[6] Zheng, P., wang, H., Sang, Z., Zhong, R. Y., Liu, Y., Liu, C., Mubarok, K., Yu, S., and Xu, X., 2018, “Smart Manufacturing Systems for Industry 4.0: Conceptual Framework, Scenarios, and Future Perspectives,” Frontiers of Mechanical Engineering, **13**(2), pp. 137–150. https://doi.org/10.1007/s11465-018-0499-5.

[7] Xia, Y., Xiao, Z., Jazdi, N., and Weyrich, M., 2024, “Generation of Asset Administration Shell With Large Language Model Agents: Toward Semantic Interoperability in Digital Twins in the Context of Industry 4.0,” IEEE Access, **12**, pp. 84863–84877. https://doi.org/10.1109/ACCESS.2024.3415470.

[8] Marks, P., Hoang, X. L., Weyrich, M., and Fay, A., 2018, “A Systematic Approach for Supporting the Adaptation Process of Discrete Manufacturing Machines,” Res Eng Des, **29**(4), pp. 621–641. https://doi.org/10.1007/s00163-018-0296-5.

[9] Ribeiro, L., and Björkman, M., 2018, “Transitioning From Standard Automation Solutions to Cyber-Physical Production Systems: An Assessment of Critical Conceptual and Technical Challenges,” IEEE Syst J, **12**(4), pp. 3816–3827. https://doi.org/10.1109/JSYST.2017.2771139.

[10] Antons, O., and Arlinghaus, J. C., 2024, “Designing Distributed Decision-Making Authorities for Smart Factories – Understanding the Role of Manufacturing Network Architecture,” Int J Prod Res, **62**(1–2), pp. 204–222. https://doi.org/10.1080/00207543.2023.2217285.

[11] Mabkhot, M. M., Ferreira, P., Eaton, W., and Lohse, N., 2024, “Estimating Adaptation Effort in Industry 4.0-Enabled Systems: Introducing Two Complexity Indices with an Evolvable Network Graph Approach,” J Ind Inf Integr, **40**, p. 100616. https://doi.org/https://doi.org/10.1016/j.jii.2024.100616.

[12] Zhou, L., Jiang, Z., Geng, N., Niu, Y., Cui, F., Liu, K., and Qi, N., 2022, “Production and Operations Management for Intelligent Manufacturing: A Systematic Literature Review,” Int J Prod Res, **60**(2), pp. 808–846. https://doi.org/10.1080/00207543.2021.2017055.

[13] Zhu, Q., Huang, S., Wang, G., Moghaddam, S. K., Lu, Y., and Yan, Y., 2022, “Dynamic Reconfiguration Optimization of Intelligent Manufacturing System with Human-Robot Collaboration Based on Digital Twin,” J Manuf Syst, **65**, pp. 330–338. https://doi.org/https://doi.org/10.1016/j.jmsy.2022.09.021.

[14] Müller, T., Jazdi, N., Schmidt, J.-P., and Weyrich, M., 2021, “Cyber-Physical Production Systems: Enhancement with a Self-Organized Reconfiguration Management,” Procedia CIRP, **99**, pp. 549–554. https://doi.org/https://doi.org/10.1016/j.procir.2021.03.075.

[15] Banerjee, A., and Choppella, V., 2025, “A Knowledge-Driven Approach for Dynamic Reconfiguration of Control Design in Internet of Things and Cyber–Physical Systems,” IEEE Internet Things J, **12**(5), pp. 5615–5641. https://doi.org/10.1109/JIOT.2024.3487578.

[16] Cruz Salazar, L. A., Ryashentseva, D., Lüder, A., and Vogel-Heuser, B., 2019, “Cyber-Physical Production Systems Architecture Based on Multi-Agent’s Design Pattern—Comparison of Selected Approaches Mapping Four Agent Patterns,” The International Journal of Advanced Manufacturing Technology, **105**(9), pp. 4005–4034. https://doi.org/10.1007/s00170-019-03800-4.

[17] Suvarna, M., Yap, K. S., Yang, W., Li, J., Ng, Y. T., and Wang, X., 2021, “Cyber–Physical Production Systems for Data-Driven, Decentralized, and Secure Manufacturing—A Perspective,” Engineering, **7**(9), pp. 1212–1223. https://doi.org/https://doi.org/10.1016/j.eng.2021.04.021.

[18] Cañas, H., Mula, J., Campuzano-Bolarín, F., and Poler, R., 2022, “A Conceptual Framework for Smart Production Planning and Control in Industry 4.0,” Comput Ind Eng, **173**, p. 108659. https://doi.org/https://doi.org/10.1016/j.cie.2022.108659.

[19] Kumar, S., Savur, C., and Sahin, F., 2021, “Survey of Human–Robot Collaboration in Industrial Settings: Awareness, Intelligence, and Compliance,” IEEE Trans Syst Man Cybern Syst, **51**(1), pp. 280–297. https://doi.org/10.1109/TSMC.2020.3041231.

[20] Gaiardelli, S., Spellini, S., Panato, M., Lora, M., and Fummi, F., 2022, “A Software Architecture to Control Service-Oriented Manufacturing Systems,” *2022 Design, Automation & Test in Europe Conference & Exhibition (DATE)*, pp. 40–43. https://doi.org/10.23919/DATE54114.2022.9774522.

[21] Yousaf, S., Haque, H. M. U., Atif, M., Hashmi, M. A., Khalid, A., and Vinh, P. C., 2023, “A Context-Aware Multi-Agent Reasoning Based Intelligent Assistive Formalism,” Internet of Things, **23**, p. 100857. https://doi.org/https://doi.org/10.1016/j.iot.2023.100857.

[22] Yang, X., and Zhu, C., 2024, “Industrial Expert Systems Review: A Comprehensive Analysis of Typical Applications,” IEEE Access, **12**, pp. 88558–88584. https://doi.org/10.1109/ACCESS.2024.3419047.

[23] Bolender, T., Bürvenich, G., Dalibor, M., Rumpe, B., and Wortmann, A., 2021, “Self-Adaptive Manufacturing with Digital Twins,” *2021 International Symposium on Software Engineering for Adaptive and Self-Managing Systems (SEAMS)*, pp. 156–166. https://doi.org/10.1109/SEAMS51251.2021.00029.

[24] Sahlab, N., Kamm, S., Müller, T., Jazdi, N., and Weyrich, M., 2021, “Knowledge Graphs as Enhancers of Intelligent Digital Twins,” *2021 4th IEEE International Conference on Industrial Cyber-Physical Systems (ICPS)*, pp. 19–24. https://doi.org/10.1109/ICPS49255.2021.9468219.

[25] International Society of Automation (ISA), 2025, “What Is Automation?” [Online]. Available: https://www.isa.org/about-isa/what-is-automation.

[26] Deutsches Institut für Normung e.V. (DIN), 2014, “Internationales Elektrotechnisches Wörterbuch – Teil 351: Leittechnik (IEC 60050-351:2013),” (DIN IEC 60050-351:2014-09), p. 381. https://doi.org/10.31030/2159569.

[27] Xia, Y., Dittler, D., Jazdi, N., Chen, H., and Weyrich, M., 2024, “LLM Experiments with Simulation: Large Language Model Multi-Agent System for Simulation Model Parametrization in Digital Twins,” *2024 IEEE 29th International Conference on Emerging Technologies and Factory Automation (ETFA)*, pp. 1–4. https://doi.org/10.1109/ETFA61755.2024.10710900.

[28] Xia, Y., Jazdi, N., Zhang, J., Shah, C., and Weyrich, M., 2025, “Control Industrial Automation System with Large Language Model Agents,” *2025 IEEE 30th International*

*Conference on Emerging Technologies and Factory Automation (ETFA)*, pp. 1–8. https://doi.org/10.1109/ETFA65518.2025.11205539.

[29] He, J., Treude, C., and Lo, D., 2025, "LLM-Based Multi-Agent Systems for Software Engineering: Literature Review, Vision, and the Road Ahead," ACM Trans. Softw. Eng. Methodol., **34**(5). https://doi.org/10.1145/3712003.

[30] Jazdi, N., Ashtari Talkhestani, B., Maschler, B., and Weyrich, M., 2021, "Realization of AI-Enhanced Industrial Automation Systems Using Intelligent Digital Twins," Procedia CIRP, **97**, pp. 396–400. https://doi.org/https://doi.org/10.1016/j.procir.2020.05.257.

[31] Wang, L., Ma, C., Feng, X., Zhang, Z., Yang, H., Zhang, J., Chen, Z., Tang, J., Chen, X., Lin, Y., Zhao, W. X., Wei, Z., and Wen, J., 2024, "A Survey on Large Language Model Based Autonomous Agents," Front Comput Sci, **18**(6). https://doi.org/10.1007/s11704-024-40231-1.

[32] Zhou, D., 2025, "LLM Reasoning." [Online]. Available: https://dennyzhou.github.io/LLM-Reasoning-Stanford-CS-25.pdf.

[33] Xi, Z., Chen, W., Guo, X., He, W., Ding, Y., Hong, B., Zhang, M., Wang, J., Jin, S., Zhou, E., Zheng, R., Fan, X., Wang, X., Xiong, L., Zhou, Y., Wang, W., Jiang, C., Zou, Y., Liu, X., Yin, Z., Dou, S., Weng, R., Qin, W., Zheng, Y., Qiu, X., Huang, X., Zhang, Q., and Gui, T., 2025, "The Rise and Potential of Large Language Model Based Agents: A Survey," Science China Information Sciences, **68**(2), p. 121101. https://doi.org/10.1007/s11432-024-4222-0.

[34] Fan, H., Liu, X., Fuh, J. Y. H., Lu, W. F., and Li, B., 2025, "Embodied Intelligence in Manufacturing: Leveraging Large Language Models for Autonomous Industrial Robotics," J Intell Manuf, **36**(2), pp. 1141–1157. https://doi.org/10.1007/s10845-023-02294-y.

[35] Luo, J., Zhang, W., Yuan, Y., Zhao, Y., Yang, J., Gu, Y., Wu, B., Chen, B., Qiao, Z., Long, Q., Tu, R., Luo, X., Ju, W., Xiao, Z., Wang, Y., Xiao, M., Liu, C., Yuan, J., Zhang, S., Jin, Y., Zhang, F., Wu, X., Zhao, H., Tao, D., Yu, P. S., and Zhang, M., 2025, "Large Language Model Agent: A Survey on Methodology, Applications and Challenges," CoRR, **abs/2503.21460**. [Online]. Available: https://doi.org/10.48550/arXiv.2503.21460.

[36] Zhang, Z., Dai, Q., Bo, X., Ma, C., Li, R., Chen, X., Zhu, J., Dong, Z., and Wen, J.-R., 2025, "A Survey on the Memory Mechanism of Large Language Model-Based Agents," ACM Trans. Inf. Syst., **43**(6). https://doi.org/10.1145/3748302.

[37] Gao, C., Lan, X., Li, N., Yuan, Y., Ding, J., Zhou, Z., Xu, F., and Li, Y., 2024, "Large Language Models Empowered Agent-Based Modeling and Simulation: A Survey and Perspectives," Humanit Soc Sci Commun, **11**(1), p. 1259. https://doi.org/10.1057/s41599-024-03611-3.

[38] Kong, Y., Ruan, J., Chen, Y., Zhang, B., Bao, T., Shiwei, S., Qing, du G., Hu, X., Mao, H., Li, Z., Zeng, X., Zhao, R., and Wang, X., 2024, "TPTU-v2: Boosting Task Planning and Tool Usage of Large Language Model-Based Agents in Real-World Industry Systems," *Proceedings of the 2024 Conference on Empirical Methods in Natural Language Processing: Industry Track*, F. Dernoncourt, D. Preo\c{t}iuc-Pietro, and A. Shimorina, eds., Association for Computational Linguistics, Miami, Florida, US, pp. 371–385. https://doi.org/10.18653/v1/2024.emnlp-industry.27.

[39] Guo, T., Chen, X., Wang, Y., Chang, R., Pei, S., Chawla, N. V, Wiest, O., and Zhang, X., 2024, "Large Language Model Based Multi-Agents: A Survey of Progress and Challenges," *Proceedings of the Thirty-Third International Joint Conference on Artificial Intelligence*. https://doi.org/10.24963/ijcai.2024/890.

[40] Li, Y., Sun, L., and Zhang, Y., 2025, "MetaAgents: Large Language Model Based Agents for Decision-Making on Teaming," Proc. ACM Hum.-Comput. Interact., **9**(2). https://doi.org/10.1145/3711032.

[41] Ashtari Talkhestani, B., Jung, T., Lindemann, B., Sahlab, N., Jazdi, N., Schloegl, W., and Weyrich, M., 2019, "An Architecture of an Intelligent Digital Twin in a Cyber-Physical Production System," **67**(9), pp. 762–782. https://doi.org/doi:10.1515/auto-2019-0039.

[42] Löcklin, A., Müller, M., Jung, T., Jazdi, N., White, D., and Weyrich, M., 2020, "Digital Twin for Verification and Validation of Industrial Automation Systems – a Survey," *2020 25th IEEE International Conference on Emerging Technologies and Factory Automation (ETFA)*, pp. 851–858. https://doi.org/10.1109/ETFA46521.2020.9212051.

[43] Franceschi, P., Mutti, S., Ottogalli, K., Rosquete, D., Borro, D., and Pedrocchi, N., 2022, "A Framework for Cyber-Physical Production System Management and Digital Twin Feedback Monitoring for Fast Failure Recovery," Int J Comput Integr Manuf, **35**(6), pp. 619–632. https://doi.org/10.1080/0951192X.2021.1992666.

[44] Maschler, B., Braun, D., Jazdi, N., and Weyrich, M., 2021, "Transfer Learning as an Enabler of the Intelligent Digital Twin," Procedia CIRP, **100**, pp. 127–132. https://doi.org/https://doi.org/10.1016/j.procir.2021.05.020.

[45] Hevner, A. R., March, S. T., Park, J., and Ram, S., 2004, "Design Science in Information Systems Research," MIS Q., **28**(1), pp. 75–105.

[46] Wieringa, R., 2014, *Design Science Methodology for Information Systems and Software Engineering*, Springer.

[47] Peffers, K., Tuunanen, T., Rothenberger, M. A., and Chatterjee, S., 2007, "A Design Science Research Methodology for Information Systems Research," Journal of Management Information Systems, **24**(3), pp. 45–77. https://doi.org/10.2753/MIS0742-1222240302.

[48] Hevner, A. R., 2007, "A Three Cycle View of Design Science Research," Scandinavian journal of information systems, **19**(2), p. 4.

[49] Vaishnavi, V. K., and Kuechler, W., 2015, *Design Science Research Methods and Patterns: Innovating Information and Communication Technology, 2nd Edition*, CRC Press, Inc., USA.

[50] Vom Brocke, J., Hevner, A., and Maedche, A., 2020, *Design Science Research. Cases*, Springer.

[51] 2013, "Enterprise–Control System Integration — Part 1: Models and Terminology," (IEC 62264-1:2013).

[52] Martinez, E. M., Ponce, P., Macias, I., and Molina, A., 2021, "Automation Pyramid as Constructor for a Complete Digital Twin, Case Study: A Didactic Manufacturing System," Sensors, **21**(14). https://doi.org/10.3390/s21144656.

[53] Lyu, G., and Brennan, R. W., 2025, "Multi-Agent Modelling of Cyber-Physical Systems for IEC 61499-Based Distributed Intelligent Automation," Int J Comput Integr Manuf, **38**(5), pp. 596–622. https://doi.org/10.1080/0951192X.2023.2294442.

[54] Nagorny, K., Colombo, A. W., and Schmidtmann, U., 2012, "A Service- and Multi-Agent-Oriented Manufacturing Automation Architecture: An IEC 62264 Level 2 Compliant Implementation," Comput Ind, **63**(8), pp. 813–823. https://doi.org/https://doi.org/10.1016/j.compind.2012.08.003.

[55] Seitz, M., Gehlhoff, F., Cruz Salazar, L. A., Fay, A., and Vogel-Heuser, B., 2021, "Automation Platform Independent Multi-Agent System for Robust Networks of Production Resources in Industry 4.0," J Intell Manuf, **32**(7), pp. 2023–2041. https://doi.org/10.1007/s10845-021-01759-2.

[56] Thomas, A., Borangiu, T., and Trentesaux, D., 2017, "Holonic and Multi-Agent Technologies for Service and Computing Oriented Manufacturing," J Intell Manuf, **28**(7), pp. 1501–1502. https://doi.org/10.1007/s10845-015-1188-4.

[57] Van Brussel, H., Wyns, J., Valckenaers, P., Bongaerts, L., and Peeters, P., 1998, "Reference Architecture for Holonic Manufacturing Systems: PROSA," Comput Ind, **37**(3), pp. 255–274. https://doi.org/https://doi.org/10.1016/S0166-3615(98)00102-X.

[58] Derigent, W., Cardin, O., and Trentesaux, D., 2021, "Industry 4.0: Contributions of Holonic Manufacturing Control Architectures and Future Challenges," J Intell Manuf, **32**(7), pp. 1797–1818. https://doi.org/10.1007/s10845-020-01532-x.

[59] Colombo, A. W., Schoop, R., and Neubert, R., 2006, “An Agent-Based Intelligent Control Platform for Industrial Holonic Manufacturing Systems,” IEEE Transactions on Industrial Electronics, **53**(1), pp. 322–337. https://doi.org/10.1109/TIE.2005.862210.
[60] Ma, J., Wang, Q., and Zhao, Z., 2017, “SLAE–CPS: Smart Lean Automation Engine Enabled by Cyber-Physical Systems Technologies,” Sensors, **17**(7). https://doi.org/10.3390/s17071500.
[61] Karaduman, B., Tezel, B. T., and Challenger, M., 2023, “Rational Software Agents with the BDI Reasoning Model for Cyber–Physical Systems,” Eng Appl Artif Intell, **123**, p. 106478. https://doi.org/https://doi.org/10.1016/j.engappai.2023.106478.
[62] Xia, Y., Jazdi, N., and Weyrich, M., 2025, “An Architecture for Integrating Large Language Models with Digital Twins and Automation Systems,” *2025 IEEE 30th International Conference on Emerging Technologies and Factory Automation (ETFA)*, pp. 1–8. https://doi.org/10.1109/ETFA65518.2025.11205636.
[63] Zheng, X., Lu, J., and Kiritsis, D., 2022, “The Emergence of Cognitive Digital Twin: Vision, Challenges and Opportunities,” Int J Prod Res, **60**(24), pp. 7610–7632. https://doi.org/10.1080/00207543.2021.2014591.
[64] Xu, H., Wu, J., Pan, Q., Guan, X., and Guizani, M., 2023, “A Survey on Digital Twin for Industrial Internet of Things: Applications, Technologies and Tools,” IEEE Communications Surveys & Tutorials, **25**(4), pp. 2569–2598. https://doi.org/10.1109/COMST.2023.3297395.
[65] He, B., and Bai, K.-J., 2021, “Digital Twin-Based Sustainable Intelligent Manufacturing: A Review,” Adv Manuf, **9**(1), pp. 1–21. https://doi.org/10.1007/s40436-020-00302-5.
[66] Tao, F., Xiao, B., Qi, Q., Cheng, J., and Ji, P., 2022, “Digital Twin Modeling,” J Manuf Syst, **64**, pp. 372–389. https://doi.org/https://doi.org/10.1016/j.jmsy.2022.06.015.
[67] Kreuzer, T., Papapetrou, P., and Zdravkovic, J., 2024, “Artificial Intelligence in Digital Twins—A Systematic Literature Review,” Data Knowl Eng, **151**, p. 102304. https://doi.org/https://doi.org/10.1016/j.datak.2024.102304.
[68] D’Amico, R. D., Erkoyuncu, J. A., Addepalli, S., and Penver, S., 2022, “Cognitive Digital Twin: An Approach to Improve the Maintenance Management,” CIRP J Manuf Sci Technol, **38**, pp. 613–630. https://doi.org/https://doi.org/10.1016/j.cirpj.2022.06.004.
[69] Mihai, S., Yaqoob, M., Hung, D. V, Davis, W., Towakel, P., Raza, M., Karamanoglu, M., Barn, B., Shetve, D., Prasad, R. V, Venkataraman, H., Trestian, R., and Nguyen, H. X., 2022, “Digital Twins: A Survey on Enabling Technologies, Challenges, Trends and Future Prospects,” IEEE Communications Surveys & Tutorials, **24**(4), pp. 2255–2291. https://doi.org/10.1109/COMST.2022.3208773.
[70] Jinzhi, L., Zhaorui, Y., Xiaochen, Z., Jian, W., and Dimitris, K., 2022, “Exploring the Concept of Cognitive Digital Twin from Model-Based Systems Engineering Perspective,” The International Journal of Advanced Manufacturing Technology, **121**(9), pp. 5835–5854. https://doi.org/10.1007/s00170-022-09610-5.
[71] Listl, F. G., Dittler, D., Hildebrandt, G., Stegmaier, V., Jazdi, N., and Weyrich, M., 2024, “Knowledge Graphs in the Digital Twin: A Systematic Literature Review About the Combination of Semantic Technologies and Simulation in Industrial Automation,” IEEE Access, **12**, pp. 187828–187843. https://doi.org/10.1109/ACCESS.2024.3514923.
[72] Platenius-Mohr, M., Malakuti, S., Grüner, S., Schmitt, J., and Goldschmidt, T., 2020, “File- and API-Based Interoperability of Digital Twins by Model Transformation: An IIoT Case Study Using Asset Administration Shell,” Future Generation Computer Systems, **113**, pp. 94–105. https://doi.org/https://doi.org/10.1016/j.future.2020.07.004.
[73] Ferko, E., Berardinelli, L., Bucaioni, A., Behnam, M., and Wimmer, M., 2024, “Towards Interoperable Digital Twins: Integrating SysML into AAS with Higher-Order Transformations,” *2024 IEEE 21st International Conference on Software Architecture Companion (ICSA-C)*, pp. 342–349. https://doi.org/10.1109/ICSA-C63560.2024.00063.

[74] Schmidt, C., Volz, F., Stojanovic, L., and Sutschet, G., 2023, "Increasing Interoperability between Digital Twin Standards and Specifications: Transformation of DTDL to AAS," Sensors, **23**(18). https://doi.org/10.3390/s23187742.

[75] Feng, J., Tang, H., Zhou, S., Cai, Y., and Zhang, J., 2025, "Cognitive Digital Twins of the Natural Environment: Framework and Application," Eng Appl Artif Intell, **139**, p. 109587. https://doi.org/https://doi.org/10.1016/j.engappai.2024.109587.

[76] Karabulut, E., Pileggi, S. F., Groth, P., and Degeler, V., 2024, "Ontologies in Digital Twins: A Systematic Literature Review," Future Generation Computer Systems, **153**, pp. 442–456. https://doi.org/https://doi.org/10.1016/j.future.2023.12.013.

[77] Xia, Y., Jazdi, N., and Weyrich, M., 2022, "Automated Generation of Asset Administration Shell: A Transfer Learning Approach with Neural Language Model and Semantic Fingerprints," *2022 IEEE 27th International Conference on Emerging Technologies and Factory Automation (ETFA)*, pp. 1–4. https://doi.org/10.1109/ETFA52439.2022.9921637.

[78] Shi, D., Meyer, O., Oberle, M., and Bauernhansl, T., 2025, "Dual Data Mapping with Fine-Tuned Large Language Models and Asset Administration Shells toward Interoperable Knowledge Representation," Robot Comput Integr Manuf, **91**, p. 102837. https://doi.org/https://doi.org/10.1016/j.rcim.2024.102837.

[79] Liu, Y., Ji, T., Guo, X., Xu, X., and Polzer, J., 2025, "A Survey of Cognitive Digital Twin and the Potential Use of LLMs," Manuf Lett, **44**, pp. 1242–1253. https://doi.org/https://doi.org/10.1016/j.mfglet.2025.06.144.

[80] Wang, L., Ma, C., Feng, X., Zhang, Z., Yang, H., Zhang, J., Chen, Z., Tang, J., Chen, X., Lin, Y., Zhao, W. X., Wei, Z., and Wen, J., 2024, "A Survey on Large Language Model Based Autonomous Agents," Front Comput Sci, **18**(6), p. 186345. https://doi.org/10.1007/s11704-024-40231-1.

[81] Li, X., Wang, S., Zeng, S., Wu, Y., and Yang, Y., 2024, "A Survey on LLM-Based Multi-Agent Systems: Workflow, Infrastructure, and Challenges," Vicinagearth, **1**(1), p. 9. https://doi.org/10.1007/s44336-024-00009-2.

[82] Javaid, S., Fahim, H., He, B., and Saeed, N., 2024, "Large Language Models for UAVs: Current State and Pathways to the Future," IEEE Open Journal of Vehicular Technology, **5**, pp. 1166–1192. https://doi.org/10.1109/OJVT.2024.3446799.

[83] Qin, B., Pan, H., Dai, Y., Si, X., Huang, X., Yuen, C., and Zhang, Y., 2024, "Machine and Deep Learning for Digital Twin Networks: A Survey," IEEE Internet Things J, **11**(21), pp. 34694–34716. https://doi.org/10.1109/JIOT.2024.3416733.

[84] Ma, Y., Zheng, S., Yang, Z., Zheng, P., Leng, J., and Hong, J., 2025, "Leveraging Large Language Models in next Generation Intelligent Manufacturing: Retrospect and Prospect," J Manuf Syst, **82**, pp. 809–840. https://doi.org/https://doi.org/10.1016/j.jmsy.2025.07.019.

[85] Xu, W., Liu, M., Sokolsky, O., Lee, I., and Kong, F., 2024, "LLM-Enabled Cyber-Physical Systems: Survey, Research Opportunities, and Challenges," *2024 IEEE International Workshop on Foundation Models for Cyber-Physical Systems & Internet of Things (FMSys)*, pp. 50–55. https://doi.org/10.1109/FMSys62467.2024.00013.

[86] Ouerghemmi, C., and Ertz, M., 2025, "Integrating Large Language Models into Digital Manufacturing: A Systematic Review and Research Agenda," Computers, **14**(8). https://doi.org/10.3390/computers14080318.

[87] Industrial Digital Twin Association, 2024, *Specification of the Asset Administration Shell – Part 1: Metamodel*, Frankfurt am Main, Germany. [Online]. Available: https://industrialdigitaltwin.org/wp-content/uploads/2024/06/IDTA-01001-3-0-1_SpecificationAssetAdministrationShell_Part1_Metamodel.pdf.

[88] ECLASS e.V., 2022, "ECLASS Standard Release 13.0." [Online]. Available: https://eclass.eu/support/content-creation/release-process/information-on-releases/release-130.

[89] Object Management Group (OMG), 2014, "Business Process Model and Notation (BPMN) Version 2.0.2."

[90] Murata, T., 1989, "Petri Nets: Properties, Analysis and Applications," Proceedings of the IEEE, **77**(4), pp. 541–580. https://doi.org/10.1109/5.24143.

[91] Object Management Group (OMG), 2019, "OMG Systems Modeling Language (SysML) Specification, Version 1.6."

[92] Hou, X., Zhao, Y., Liu, Y., Yang, Z., Wang, K., Li, L., Luo, X., Lo, D., Grundy, J., and Wang, H., 2024, "Large Language Models for Software Engineering: A Systematic Literature Review," ACM Trans. Softw. Eng. Methodol., **33**(8). https://doi.org/10.1145/3695988.

[93] Shi, L., Tang, Z., Zhang, N., Zhang, X., and Yang, Z., 2025, "A Survey on Employing Large Language Models for Text-to-SQL Tasks," ACM Comput. Surv., **58**(2). https://doi.org/10.1145/3737873.

[94] Yang, W., Some, L., Bain, M., and Kang, B., 2025, "A Comprehensive Survey on Integrating Large Language Models with Knowledge-Based Methods," Knowl Based Syst, **318**, p. 113503. https://doi.org/https://doi.org/10.1016/j.knosys.2025.113503.

[95] Mustapha, K. B., 2025, "A Survey of Emerging Applications of Large Language Models for Problems in Mechanics, Product Design, and Manufacturing," Advanced Engineering Informatics, **64**, p. 103066. https://doi.org/https://doi.org/10.1016/j.aei.2024.103066.

# Appendix—Reprinted Publications

IEEE copyright and credit notice for reuse of published material in a dissertation:

**Publication A: © 2023 IEEE. Reprinted, with permission, from:**

Y. Xia, M. Shenoy, N. Jazdi and M. Weyrich, "Towards autonomous system: flexible modular production system enhanced with large language model agents," 2023 IEEE 28th International Conference on Emerging Technologies and Factory Automation (ETFA), Sinaia, Romania, 2023, pp. 1–8, doi: 10.1109/ETFA54631.2023.10275362.

**Publication B: © 2025 IEEE. Reprinted, with permission, from:**

Y. Xia, N. Jazdi, J. Zhang, C. Shah and M. Weyrich, "Control Industrial Automation System with Large Language Model Agents," 2025 IEEE 30th International Conference on Emerging Technologies and Factory Automation (ETFA), Porto, Portugal, 2025, pp. 1–8, doi: 10.1109/ETFA65518.2025.11205539.

**Publication C: © 2024 IEEE. Reprinted, with permission, from:**

Y. Xia, D. Dittler, N. Jazdi, H. Chen and M. Weyrich, "LLM experiments with simulation: Large Language Model Multi-Agent System for Simulation Model Parametrization in Digital Twins," 2024 IEEE 29th International Conference on Emerging Technologies and Factory Automation (ETFA), Padova, Italy, 2024, pp. 1–4, doi: 10.1109/ETFA61755.2024.10710900.

**Publication D: © 2024 IEEE. Reprinted, with permission, from:**

Y. Xia, Z. Xiao, N. Jazdi and M. Weyrich, "Generation of Asset Administration Shell With Large Language Model Agents: Toward Semantic Interoperability in Digital Twins in the Context of Industry 4.0," IEEE Access, vol. 12, pp. 84863–84877, 2024, doi: 10.1109/ACCESS.2024.3415470.

**Publication E: © 2025 IEEE. Reprinted, with permission, from:**

Y. Xia, N. Jazdi and M. Weyrich, "An Architecture for Integrating Large Language Models with Digital Twins and Automation Systems," 2025 IEEE 30th International Conference on Emerging Technologies and Factory Automation (ETFA), Porto, Portugal, 2025, pp. 1–8, doi: 10.1109/ETFA65518.2025.11205636.

# Towards autonomous system: flexible modular production system enhanced with large language model agents

Yuchen Xia
*Institute of Industrial Automation and Software Engineering*
*University of Stuttgart*
Stuttgart, Germany
yuchen.xia@ias.uni-stuttgart.de

Manthan Shenoy
*Institute of Industrial Automation and Software Engineering*
*University of Stuttgart*
Stuttgart, Germany
st175289@stud.uni-stuttgart.de

Nasser Jazdi
*Institute of Industrial Automation and Software Engineering*
*University of Stuttgart*
Stuttgart, Germany
nasser.jazdi@ias.uni-stuttgart.de

Michael Weyrich
*Institute of Industrial Automation and Software Engineering*
*University of Stuttgart*
Stuttgart, Germany
michael.weyrich@ias.uni-stuttgart.de

*Abstract* — **In this paper, we present a novel framework that combines large language models (LLMs), digital twins and industrial automation system to enable intelligent planning and control of production processes. We retrofit the automation system for a modular production facility and create executable control interfaces of fine-granular functionalities and coarse-granular skills. Low-level functionalities are executed by automation components, and high-level skills are performed by automation modules. Subsequently, a digital twin system is developed, registering these interfaces and containing additional descriptive information about the production system. Based on the retrofitted automation system and the created digital twins, LLM-agents are designed to interpret descriptive information in the digital twins and control the physical system through service interfaces. These LLM-agents serve as intelligent agents on different levels within an automation system, enabling autonomous planning and control of flexible production. Given a task instruction as input, the LLM-agents orchestrate a sequence of atomic functionalities and skills to accomplish the task. We demonstrate how our implemented prototype can handle un-predefined tasks, plan a production process, and execute the operations. This research highlights the potential of integrating LLMs into industrial automation systems in the context of smart factory for more agile, flexible, and adaptive production processes, while it also underscores the critical insights and limitations for future work. Demos at: https://github.com/YuchenXia/GPT4IndustrialAutomation**

***Keywords— autonomous system, intelligent agent, GPT, digital twin, Asset Administration Shell, smart factory***

## I. INTRODUCTION

Flexible production has emerged as a significant aspect of modern manufacturing environments in response to changing market demands and product customization requirements. Manufacturers need to adapt quickly to market changes and to stay competitive. This leads the manufacturer to consider diversifying their products and providing customized manufacturing services, which requires an agile production system and efficient management of the complexity of the production.

However, there are several technical challenges for deployment of agile and flexible production in reality: First of all, flexible production requires **seamless integration** of diverse technologies solution, e.g., robotics, automation, planning algorithms etc. Secondly, the production equipment and manufacturing processes need to be **reconfigurable** [1][2], which requires modular processes and systems as well as reconfigurable machines. Furthermore, automated flexible production also requires **quick changeover** [3] after decision-making to adapt the production against the changing requirements. Eventually, a highly **knowledgeable workforce** in every complicated technology with high availability to manage and supervise the complex system is too luxurious to be true. Traditional production systems frequently face difficulties in fulfilling these requirements due to their inflexible [1], dedicated workflows and restricted adaptability, as well as the absence of domain-specific knowledge in reconfiguring the production facility.

To tackle these challenges and requirements, we propose a novel solution: a large language model (LLM) enhanced automated modular production system for flexible manufacturing.

Our messages and contributions from this paper are summarized as follows:

(1) We demonstrate with a representative use case explaining why and how large language models can be used to achieve a higher level of intelligence and adaptability of industrial automation systems by planning and controlling the production, especially in the context of flexible production scenarios.

(2) We structure the system design according to the automation pyramid, illustrating a feasible technical approach to integrate LLMs into automation system.

(3) We prefer the more scalable in-context-learning approach over the fine-tuning approach, and the task-specific knowledge is injected into a LLM in prompt. As prompt engineering is an emerging field with little standardization, we devise a structured prompt template for this use case, drawing on insights from existing research in Natural Language Processing.

## II. BACKGROUND

In this section, we start by discussing why and how modular production systems can meet the requirements seamless integration and reconfigurability for flexible production. Then we emphasize the importance of modular query and control interfaces to allow the LLM to access information about the physical production processes and to adapt the production to changing requirements. Last but not least, we provide a brief overview of LLMs and the fundamental reasons why they have the potential to handle domain-specific tasks in industrial automation.

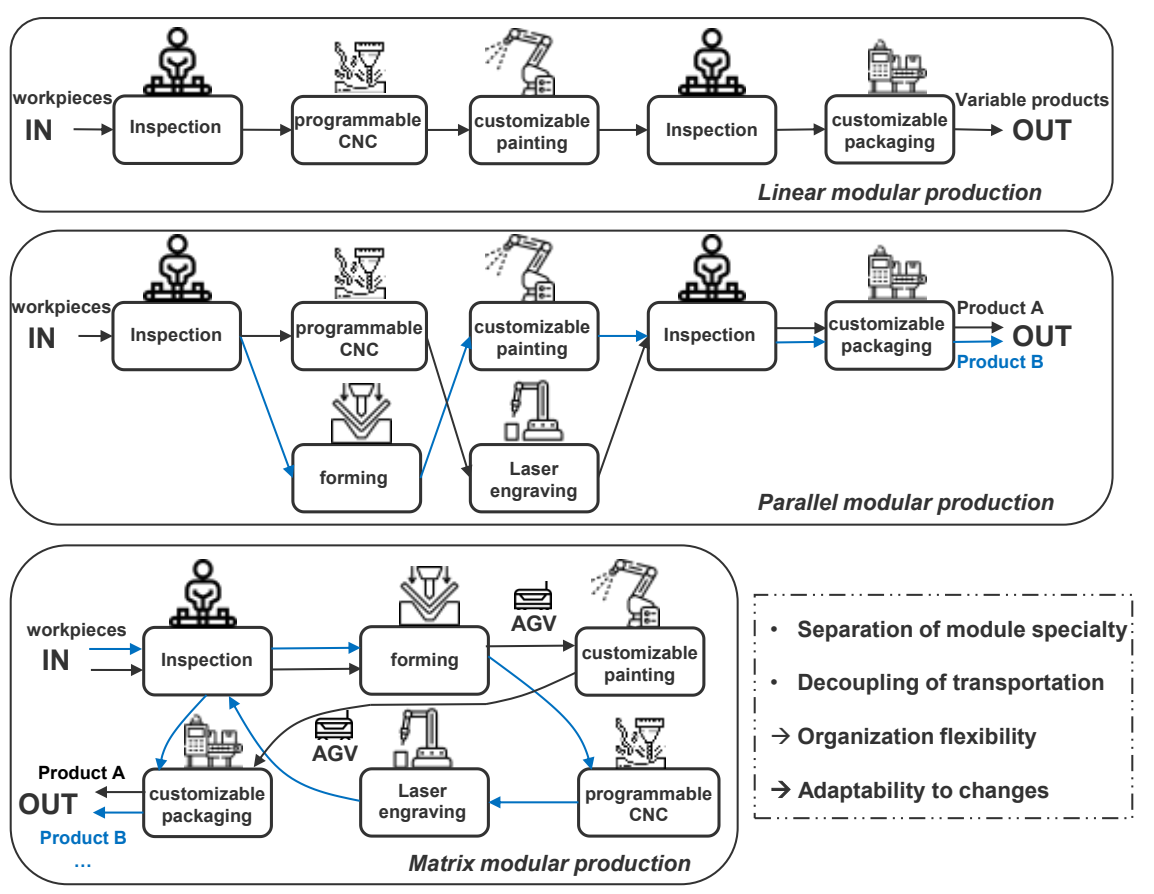

*Figure 1 Three structure types of modular production system*

### *A. Modular Production Systems*

Modular production systems are developed to address the challenges of flexibility, scalability, and adaptability in manufacturing. These systems consist of a series of modular modules, which can be easily reconfigured, replaced, or updated to accommodate varying production requirements. In the following introduction, we categorize them into three types: linear, parallel and matrix modular production.

#### *1) Linear modular production*

The production process follows a step-by-step sequence, with each module performing its designated task before passing the workpiece to the next module. In discrete production like automotive assembly [4], the material can be processed differently with variable process module. In continuous production in the process industry, the process plants can be designed modular to decrease efforts and cost in system planning, integration, and configuration [5].

#### *2) Parallel modular production*

The production system enables multiple modules and lines to operate concurrently on a variety of tasks. In comparison, parallel modular production supports simultaneous production operations, allowing a workpiece to be processed by multiple modules [4], which further increases flexibility. To effectively combine various production modules, additional transportation systems are necessary for seamless process automation.

#### *3) Matrix modular production*

The matrix modular production [6] decouples the logistics tasks from production and changes the rigid line structure into matrix structure, which consist of modular production cells and automated transportation systems (often by applying **A**utomated **G**uided **V**ehicle). These systems comprise independent modules that can be reconfigured and combined to execute a wider range of production tasks. As various production modules with different specialties can be rearranged, added, or removed with minimal impact on the overall system, the matrix production has the potential to quickly adapt to diverse requirements, customer preferences, and market demands. Some literature also refers to this production type as Matrix Manufacturing Systems (MMS) [7].

Despite the structural superiority of matrix modular production for flexible reconfiguration, planning and process orchestration for customized production tasks still rely on the accumulated expertise within a company. Identifying a feasible solution to a problem can be time-consuming if any part of the required knowledge is unavailable or if there is a lack of effective communication among experts.

*Table 1 Comparison of different types of modular production against the changing customer demand examples*

| Case examples | Requirements | Linear MP | Parallel MP | Matrix MP |
|---|---|---|---|---|
| Customer wants the packaging material to be paper instead of plastics. | Variation of machine functionalities | + | + | + |
| Customer wants an engraved logo on the product instead of painted logo. | Variation of certain process steps | - | + | + |
| Customer wants a special HiL-quality test on the product in the middle of production process. | Variation of ochestration of processes | - | - | + |
| Customer returns the product due to a quality fault and demands reprocessing | Variation of problem-solving process | - | - | ○ |
| + : Requirements fulfilled without change of production system | | | | |
| – : Requirements hardly fulfilled due to the inflexible material flows | | | | |
| ○ : Requirements can be fulfilled with experts intervene and effort | | | | |

Large Language Models (LLMs) possess the capability to interpret information, generate reasoning insights, and assist in decision-making processes. Trained on vast amounts of data, LLMs can understand and process complex information across various domains. By harnessing the interpretation and reasoning abilities of LLMs, the planning and process orchestration can be streamlined. This can lead to faster problem-solving and better adaptation to customer demands.

### *B. Digital Twins*

Despite the vast knowledge and reasoning capabilities of large language models (LLMs), a critical question remains: How can the LLMs access real-world information and effectively address tasks in practical settings?

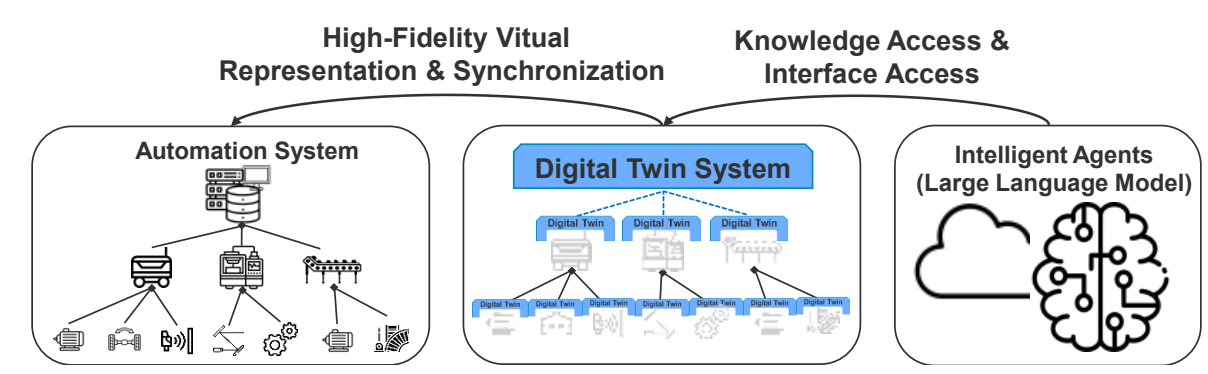

*Figure 2 The overall framework of the general concepts and their relationships*

The current state-of-the-art automation systems are not fully equipped to offer comprehensive descriptive information about production and unified accessible interfaces for querying and controlling physical processes. We developed a digital twin system to bridge the gap between LLMs and the physical world, as shown in Figure 2. Digital twins are synchronized virtual representations of physical assets or processes [8]. The digital twin system contains descriptive information about the production and exposes unified interfaces to LLM for manipulating the physical system. We lay special stress on the synchronization characteristics because it is fundamental to allow the reactive intelligent behavior of an autonomous system.

On one hand, automation systems are enhanced with digital twins and LLMs to unlock the potential of data- and AI-driven smart factories. On the other hand, LLMs interact with physical environments by having an embodiment in reality through the established infrastructure that combines automation systems and digital twins. This approach equips an artificial "brain" with mechatronic "hands" and "eyes" for more intelligent interaction.

### C. *LLM in automated Production Systems*

LLMs can be utilized to interpret complicated information, generate insights, and support decision-making processes in industrial automation systems.

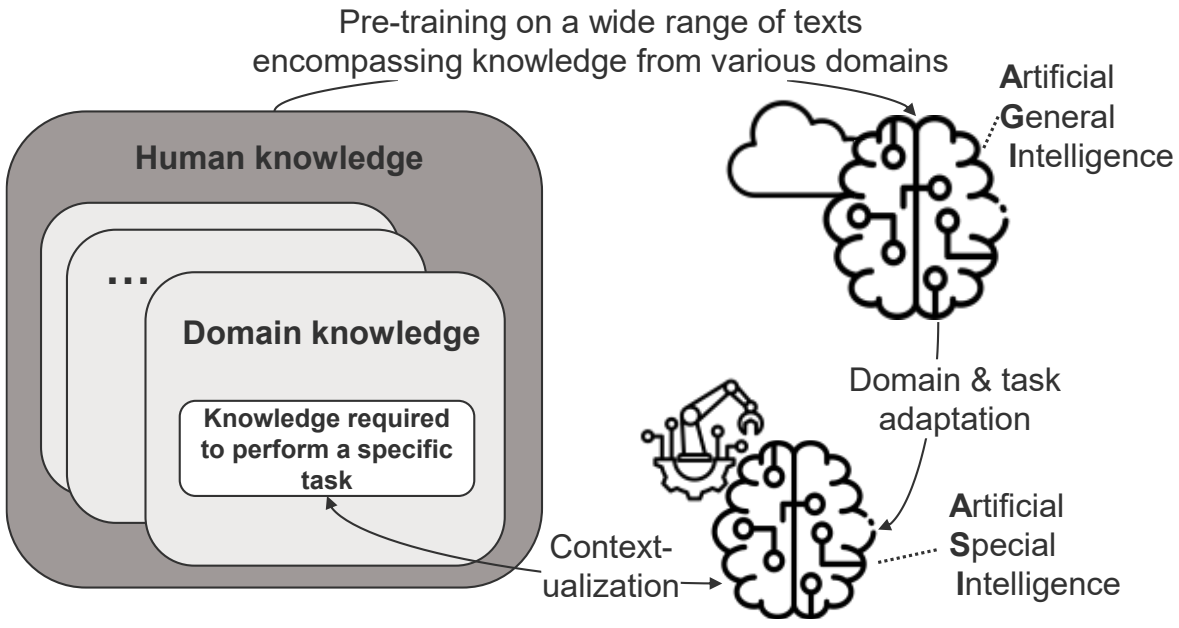

*Figure 4 Underlying general mechanism enabling LLMs to address domain-specific task.*

As illustrated in Figure 3, LLMs are deep learning models trained on vast amounts of text data, enabling them to generate human-like responses and handle complex language patterns across various NLP tasks. Recent advancements in NLP research have uncovered a promising finding: as the neuron size of LLMs increases, well-trained models gain the ability to interpret the meaning conveyed through language and demonstrate a capability of **approximating** human knowledge behind the language representation—a capability beyond the languages processing and not observed in smaller neural networks [9][10]. This development allows LLMs to perform general reasoning tasks effectively. Furthermore, as the training data for LLMs includes scientific papers, books, Q&A forums, and software code, LLMs are also informed with diverse domain-specific knowledge, which can be utilized for executing engineering related tasks.

By employing prompt engineering techniques [11], we develop multiple intelligent agents at both the MES level and the automation module level within the automation pyramid. These agents are specifically designed to manage production tasks within their respective scopes.

## III. Methods

In this section, we explain how we connect the LLM to the digital twin infrastructure with prompt engineering, allowing intelligent agents to manage and control the production operations to solve an unforeseen problem.

### A. *Integrate the information and expose the service interfaces of the digital twin*

First and foremost, the large language model agent requires high-fidelity information to accurately comprehend the production system. Thus, a data infrastructure that houses comprehensive information about the production system is fundamental. We model the production system in a digital twin system in a modular and cascaded manner. These modular digital twins contain detailed information about their represented assets.

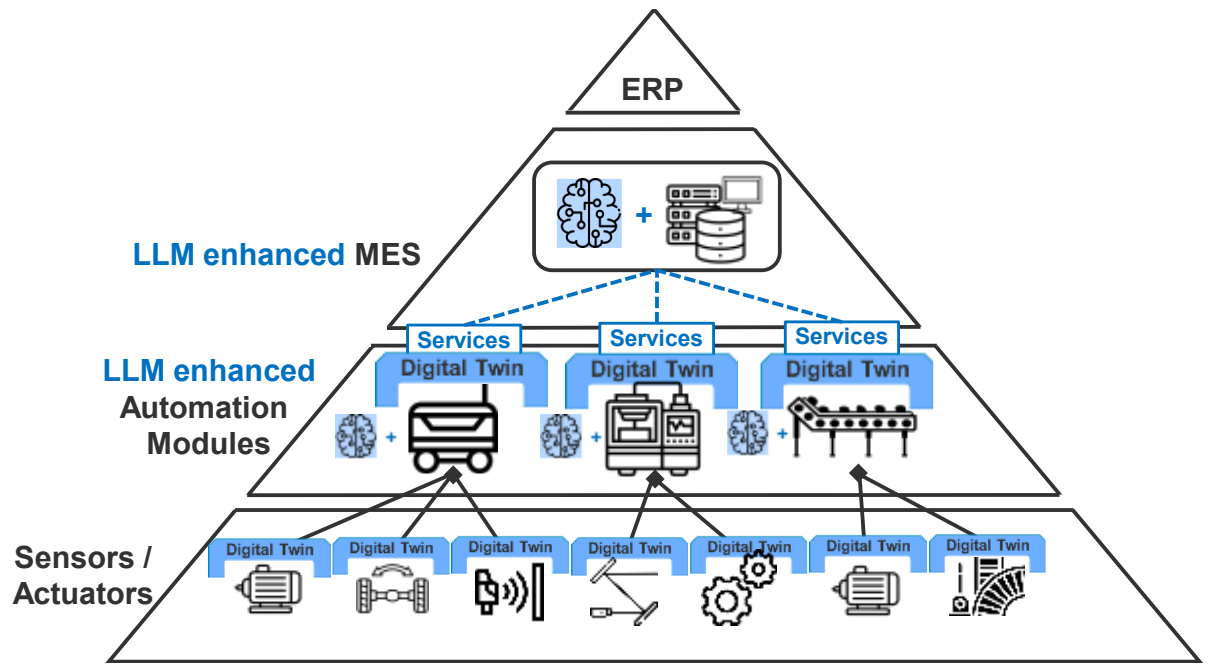

*Figure 3 Integration of LLM-agents and digital twins in automation systems for enhanced intelligence*

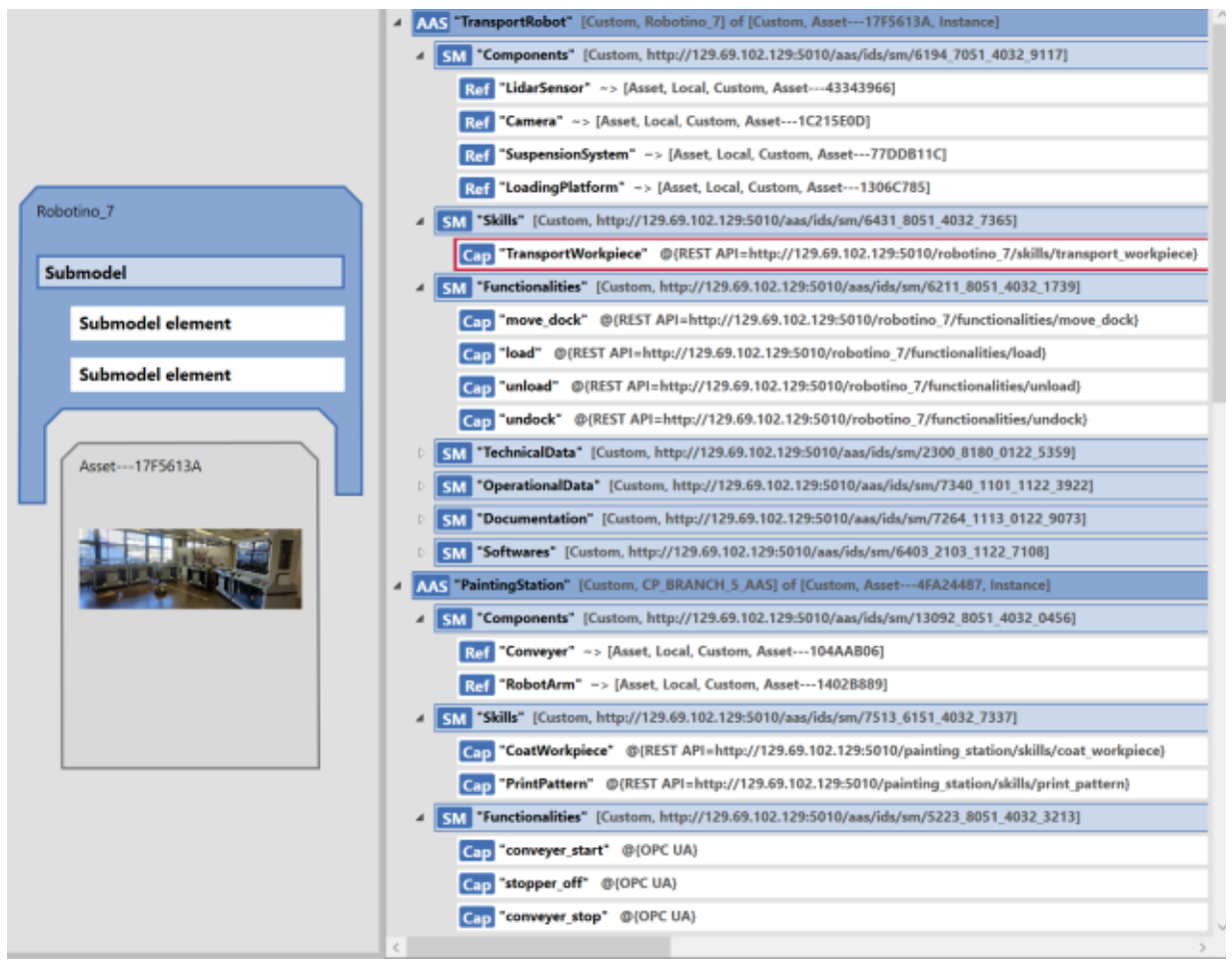

*Figure 5 Digital twin representation of the production system modeled with asset administration shell.*

The descriptive information in the digital twin system is modeled in the form of Asset Administration Shells (AAS) [1] and managed with an AAS-middleware [2]. Within the AAS, query and command services are referenced as URLs, which are semantically annotated with interface description in the skill sub-model. These interfaces enable the querying of asset states and control over automation system functionalities through RESTful service calls.

As shown in Figure 5, the digital twin of an automation module *"Transport Robot Robotino_7"* contains the cascaded information about its *"skills"*, the references to its *"components"*, callable *"functionalities"* interface, and other comprehensive information in sub-models *"technical data" "operational data" "documentation"* and *"software"* for further information.

Based on the descriptive knowledge about the assets and the callable interfaces provided by the digital twin system, it is possible to build two types of intelligent agents: A **manager**

[1] The Asset Administration Shell comprises extensive information pertaining to an asset, and it organizes this data into sub-models based on various aspects.

[2] We used Basyx AAS-middelware.

**agent** that works **on the top of** the automation modules, orchestrating diverse **skills** of the automation modules to plan the production; and several **operator agents** work **within** a particular automation module, orchestrating diverse **functionalities** to execute a given skill, as shown in Figure 5 and 6. Designing more than one agent is necessary to break down the challenging task into several manageable sub-tasks.

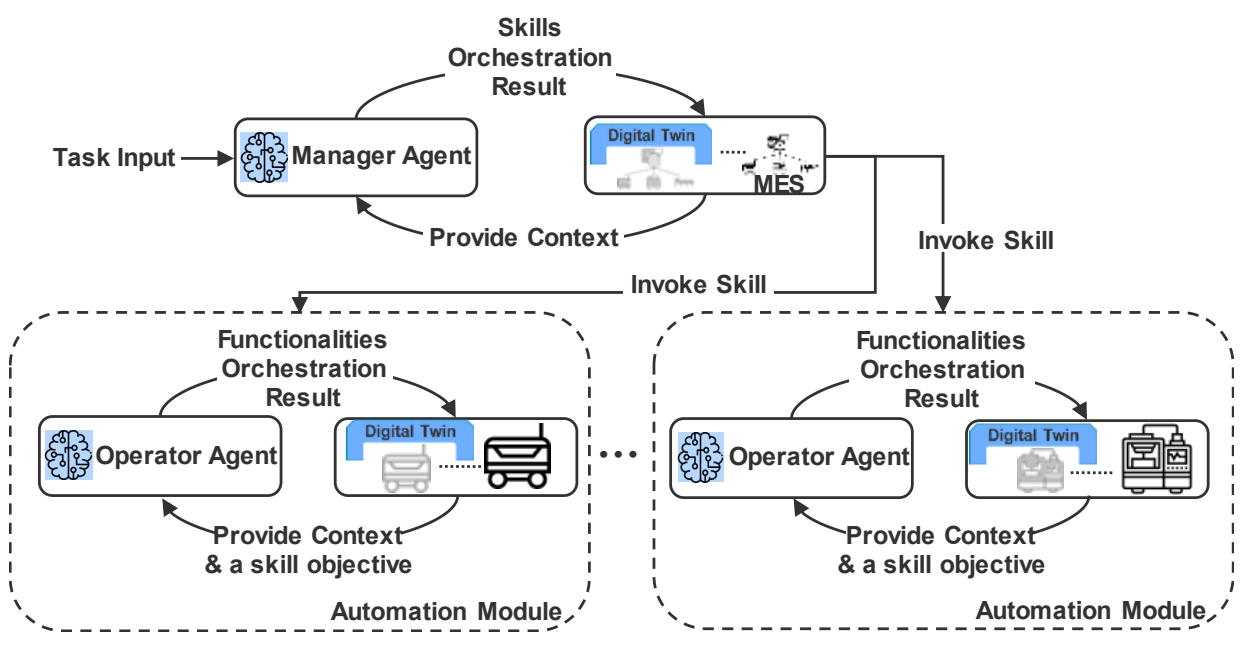

*Figure 6 Interactions between LLM-agents and digital twins of automation modules and components*

## B. Creating a LLM agent to adapt to a specific task with prompt engineering

We create these agents by contextualizing a GPT-model with prompts. The structure of the designed prompt and the interactions between digital twin and GPT-agent through the prompt are shown in Figure 7. More detailed examples are shown in the next *implementation section*.

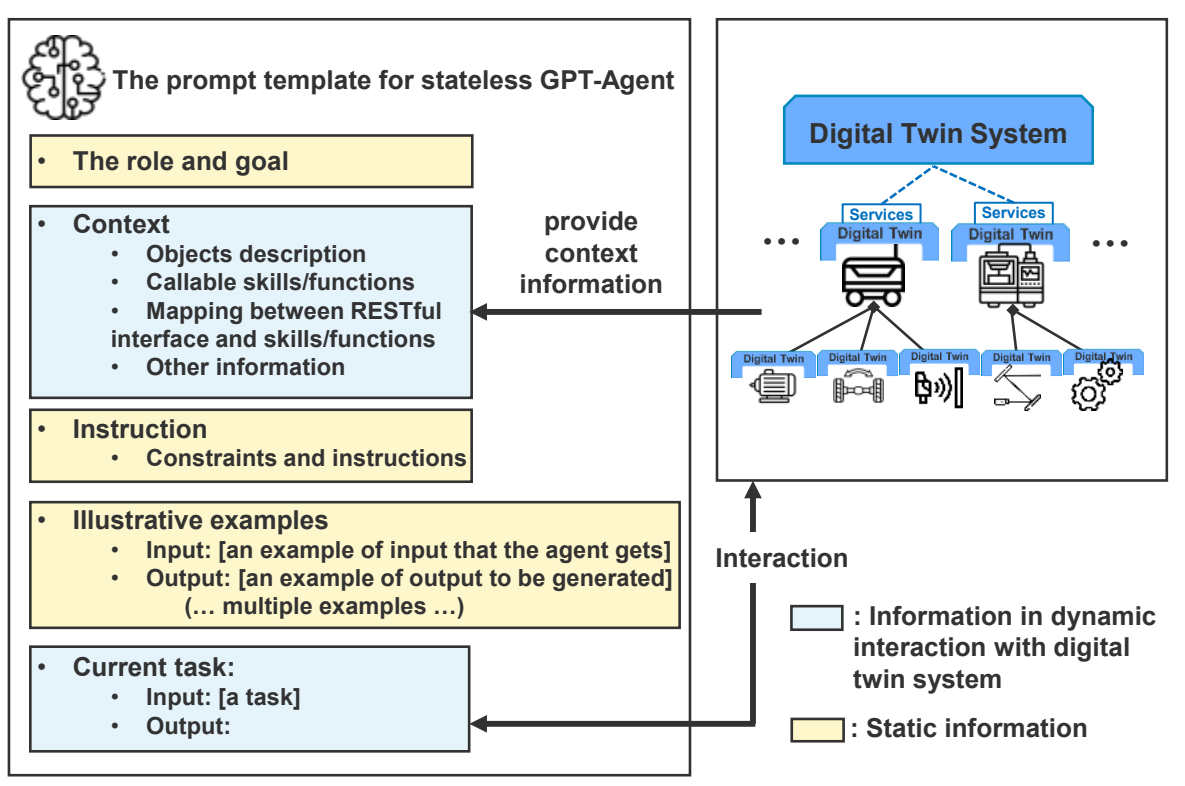

*Figure 7 structure and content elements of the prompt and the interaction between the digital twin and GPT-agent through the prompt*

The digital twin and the GPT-agents are connected via a prompt, which is sent to a LLM to initiate a response from the model. The prompt also serves as a trigger for the LLM to generate contextually appropriate text based on the given input. As shown in Figure 7, our designed prompts are composed of five distinct sections that target specific core working mechanisms of the GPT models. In the following texts, we explain them along with the design reasons. To increase the readability, we put further in-depth explanations related to NLP in footnotes.

### 1) *The role and the goal*

This section of the prompt outlines the **role** and **objective** of an agent in natural language[3], providing the model with clarity regarding the particular task that is expected to carry out.

By defining the role and objective concisely, the text can effectively convey intentions and expectations with fewer tokens[4]. This offers two benefits: Firstly, it enables the model to better align with the general requirements, producing outputs that adhere to the desired role (e.g., operator, manager, advisor) and goal (e.g., performing operations, finding solutions, providing suggestions). Secondly, concise text concentrates meaning within fewer tokens, allowing the model to infer stronger related connections between text elements with higher attention weight [5] while preventing the dilution of the model's attention.

### 2) *Context*

This section presents information derived from digital twins, aiming to supply descriptive information about the particular production system that the model needs for effective reasoning. As shown in table 2 and 3 in the *Implementation Section*, the knowledge should at least contain the objects description, the skills and functions of the objects and the mappings to the service interfaces of the executable operations. However, as the knowledge **representation** in digital twins' software system and texts in natural language are different, the information from the digital twin model shall be converted in text form in natural language, e.g., with fill-in-the-template mechanism and concatenation of text strings.

The converted information from digital twins offered in this context section serves two purposes: first, it enables the model to comprehend the production system's operations, incorporating additional information about the particular system. Secondly, as GPT has been trained across a wide range of subjects, it possesses extensive **general knowledge** that is implicitly stored within its model weights [12]. The descriptive information provided in the prompt guide GPT to "concentrate" [6] on the related knowledge embedded within the model when generating text. In this sense, this process actualizes the combination of the **general knowledge** of GPT with the **special knowledge** specified in prompt to execute reasoning for specific task.

### 3) *Instructions*

This section aims to guide the GPT-agent's behavior by specifying the desired output formats and establishing boundaries for the generated content. We also encourage the model to "think step-by-step"[7], a widely adopted strategy among researchers, to facilitate logically structured reasoning

[3] This portion of the prompt targets the zero-shot learning capability [16], which enable large language models (LLMs) to perform previously unencountered tasks without explicit examples, particularly when the model has undergone instruct-fine-tuning optimization [17].

[4] A token can be defined as a basic meaningful unit of the text in input and output.

[5] Attention mechanism is analyzed and visualized in [18].

[6] This concentration is based on the mechanism of auto-regression process in text generation [13], where predicted text is continuously generated based on previous seen tokens.

[7] Also referred to as "*chain-of-thought-prompting*" [19] which significantly improves accuracy in performing complex reasoning [20][21]. Models that trained on code generation typically exhibit superior performance in step-by-step reasoning [19].

and break down complex problems into a series of smaller intermediate tasks.

*4) Illustrative examples*

This section provides verified concrete illustrative instances that demonstrate the desired input-output pattern. This can be beneficial in several ways: first of all, it constrains the structure of the text to be generated. Furthermore, the model's performance can be improved even with a limited number of examples[8]. Last but not least, the examples can help further specify the context information and disambiguate the abstract information provided so far, and during our experimenting, we observed an increase in misunderstood or irrelevant response from the model without representative instance examples.

*5) Input and Output interaction pattern*

While the previous prompt sections focus on configuring the LLM to comprehend the task, this prompt section focuses on instructing the LLM to generate texts based on specific input. The input can be a user request to the *"manager agent"* to perform production task by orchestrating the skills, or it could be a skill demand to an *"operator agent"* to perform a skill by orchestrating the functionalities within an automation module.

We intentionally leave the prompt incomplete, ending with "*Output:*". By this means, the GPT's fundamental mechanism of next-token prediction [12][13] is addressed, upon which the model has been trained and optimized. Essentially, the agent carries out its designated task by completing the entire prompt and continuing writing the texts after the cue-word "*Output:*".

## IV. Implementation and Experiments

In this section, we first illustrate our methods with two examples of the prompt, then explain the system components and their interaction with diagram and show the implemented demonstrator of a matrix modular production facility in our laboratory.

Table 2 and 3 contain minimum prompts we specified to prompt the GPT-model *"text-davinci-003"* to solve the production planning and execution problem. Readers can reproduce the results of agents output by sending the prompt text to the GPT-model for text generation[9].

*Table 2 Prompt input example of the stateless manager-agent.*

| **The Prompt for the manager agent:** |
|---|
| Role and goal:<br>You are a manager of a production system. Your goal is to design an efficient production process based on a given task. You should take into account the provided context, instructions, and examples. Following these, you generate an output of a production process. |
| Context:<br>(1) A production process consists of one or more process steps.<br>(2) There are two types of process steps, one type is transportation process step, another type is production process step.<br>(3) If the next production process is executed in a different production module, transportation process between two production processes is necessary.<br>(4) The transportation step can be executed with a transport robot.<br>(5) Transportation step is not considered as production process step.<br>(6) A production process always begins with a skill of the storage module and ends with a skill of the storage module.<br>(7) This production system that you manage consists of several production modules. Each of these production modules has one or more skills to execute a production process step.<br>(8) Each process step can be executed with one skill of a module.<br>(9) The production process should only contain the necessary steps that are necessary to satisfy a task specified in the input.<br>The production modules are described as following:<br>(10) An inspection module. It has the following skills: (I1) check the raw material, (I2) check the faulty material, (I3) test the quality of the material.<br>(11) A storage module. It has the following skills: (S1) retrieve a workpiece, (S2) store a workpiece.<br>(12) A transport robot. It has the following skills: (T1) transport workpiece between different modules. (T2) leave the production area.<br>(13) A CNC machine module. It has the following capabilities: (M1) drilling, (M2) milling, (M3) polishing.<br>(14) A painting module. It has the following skills: (P1) coat the material, (P2) print a pattern on the surface with paint.<br>(15) A laser machine module. It has the following skills: (L1) print a pattern on the surface with laser. |
| Instructions:<br>As a manager of this production system, please arrange a production process based on the input.<br>Only use the skills that are given in the context section.<br>Only use the skills that are necessary to carry out the task.<br>Give an explanation with short reason in list form.<br>You should think step-by-step. |
| Examples:<br>Input:<br>{produce a steel sheet with a hole}<br>Output:<br>{(S1) – (T1) – (I1) – (T1) – (M1) – (T1) – (I3) – (T1) – (S2)}<br>Explanation:<br>(S1) retrieve a workpiece from storage module.<br>(T1) transport the workpiece from storage module to inspection module.<br>(I1) check the raw material to make sure it's suitable for production.<br>(T1) transport the material to the CNC machine module.<br>(M1) drill a hole on the steel sheet.<br>(T1) transport the material back to the inspection module.<br>(I3) test the quality of the material to ensure the final product meets the desired specifications.<br>(CT1) transport the material back to storage module to store the final product.<br>(…)<br>Input: {produce a steel nameplate and there should be a painted logo.}<br>Output:<br>{(S1) – (T1) – (I1) – (T1) – (M2) – (T1) – (L1) – (T1) – (I3) – (T1) – (S2)}<br>Explanation:<br>(S1) retrieve a workpiece from storage module.<br>(T1) transport the workpiece from storage module to inspection module.<br>(I1) check the raw material to make sure it's suitable for production.<br>(T1) transport the material to the CNC machine module.<br>(M2) mill the steel sheet to create the nameplate.<br>(T1) transport the material to the laser machine module.<br>(L1) print the logo on the nameplate using laser.<br>(T1) transport the material back to the inspection module.<br>(I3) test the quality of the material to ensure the final product meets the desired specifications.<br>(T1) transport the material back to storage module to store the final product. |
| Input: {the customer returned a wood nameplate and said there should be a painted customer logo on the backside. The wood nameplate is now in the storage module.}<br>Output: |
| **The generated output by the agent:** |
| {(S1) – (T1) – (P2) – (T1) – (I3) – (T1) – (S2)}<br>Explanation:<br>(S1) retrieve the wood nameplate from storage module.<br>(T1) transport the workpiece from storage module to painting module.<br>(P2) print the customer logo on the backside of the nameplate with paint.<br>(T1) transport the material back to the inspection module.<br>(I3) test the quality of the material to ensure the final product meets the desired specifications.<br>(T1) transport the material back to storage module to store the final product. |

*Table 3 Prompt input example of the stateless operator-agent for the transport robot*

| **The Prompt for an operator agent (a transport robot):** |
|---|
| **Role and goal:**<br>You are an operator of a transport robot of a production system that performs skills to fulfill a transportation task. Your goal is to orchestrate the functionalities of this robot to perform a skill. You should take into account the provided context, instructions, and examples. Following these, you generate an output of a series of functionalities and provide the correct URLs to these functionalities. |
| **Context:**<br>There are several production modules in this production system, and they are storage module, inspection module, painting module, CNC machine module and laser machine module.<br>The transport robot only transports workpiece between two of the following modules: the inspection module, the painting module, the CNC machine module and the laser machine module.<br>The transport robot can perform the following skill(s): (T1) Transport workpiece.<br>A component named "functionality handler" is a logical component, and it is identified as "functionality_handler_001". It can control the actions of the transport robot.<br>This component "functionality_handler_001" can execute the following functionalities of Robotino_7:<br>(1) Functionality "move_dock" will move the transport robot to a module and dock it to the module. This functionality can be called using the URL "http://129.69.102.129:5010/robotino_7/functionalities/move_dock".<br>(2) Functionality "load" will take a workpiece from a module and load it to the transport robot. This functionality can be called using the URL "http://129.69.102.129:5010/robotino_7/functionalities/load".<br>(3) Functionality "unload" will unload a workpiece from the transport robot and give it to a module. This functionality can be called using the URL "http://129.69.102.129:5010/robotino_7/functionalities/unload".<br>(4) Functionality "undock" will detach the transport robot from a module, if the transport robot is docked to a module. This functionality can be called using the URL "http://129.69.102.129:5010/robotino_7/functionalities/undock". |
| **Instructions:**<br>As an operator of this transport robot, please arrange a series of functionalities to solve the task specified in the input.<br>Only use the functionalities that are given.<br>Only use the functionalities that are necessary to carry out the task.<br>Specify the start module and target module in the output steps when necessary.<br>Give an explanation with short reason in list form.<br>You shall think step-by-step.<br>Follow the text structure and syntax in the examples. |
| **Examples:**<br>Input:<br>(T1) Transport the workpiece from laser machine module to the CNC module.<br>Output:<br>To transport the workpiece from the laser machine module to the CNC module, the following steps shall be executed: |

[8] This ability is also termed as termed as "few-shot learning" [22].

[9] These prompts are simplified from our hardware-dependent use case, and in order to improve the readability of the paper while retain the transferability of the method, too detailed information is cut short, such as the long identifiers and the amount of examples. The most convenient way to reproduce the generated results could be using the ChatGPT web-application and paste the prompt input into the conversation. Demos at: https://github.com/YuchenXia/GPT4IndustrialAutomation

(1) Move the transport robot to the laser machine module and dock it. Call the functionality "move_dock" using the URL "http://129.69.102.129:5010/robotino_7/functionalities/move_dock" to move the robot to the laser machine module and dock it.
(2) Load the workpiece from the laser machine module onto the transport robot. Call the functionality "load" using the URL "http://129.69.102.129:5010/robotino_7/functionalities/load" to load the workpiece onto the robot.
(3) Undock the transport robot from the laser machine module. Call the functionality "undock" using the URL "http://129.69.102.129:5010/robotino_7/functionalities/undock" to detach the robot from the laser machine module.
(4) Move the transport robot to the CNC module and dock it. Call the functionality "move_dock" using the URL "http://129.69.102.129:5010/robotino_7/functionalities/move_dock" to move the robot to the CNC module and dock it.
(5) Unload the workpiece from the transport robot onto the CNC module. Call the functionality "unload" using the URL "http://129.69.102.129:5010/robotino_7/functionalities/unload" to unload the workpiece from the robot onto the CNC module.

Input: {(T1) Transport the workpiece from the storage module to the painting module.}
Output:

**The generated output by the agent:**

To transport the workpiece from the storage module to the painting module, the following steps shall be executed:
(1) Move the transport robot to the storage module and dock it. Call the functionality "move_dock" using the URL "http://129.69.102.129:5010/robotino_7/functionalities/move_dock" to move the robot to the storage module and dock it.
(2) Load the workpiece from the storage module onto the transport robot. Call the functionality "load" using the URL "http://129.69.102.129:5010/robotino_7/functionalities/load" to load the workpiece onto the robot.
(3) Undock the transport robot from the storage module. Call the functionality "undock" using the URL "http://129.69.102.129:5010/robotino_7/functionalities/undock" to detach the robot from the storage module.
(4) Move the transport robot to the painting module and dock it. Call the functionality "move_dock" using the URL "http://129.69.102.129:5010/robotino_7/functionalities/move_dock" to move the robot to the painting module and dock it.
(5) Unload the workpiece from the transport robot onto the painting module. Call the functionality "unload" using the URL "http://129.69.102.129:5010/robotino_7/functionalities/unload" to unload the workpiece from the robot onto the painting module.

**The parsed result in JSON-file for control services invocation:**
**(by using regular expression to convert the texts into JSON in python)[10]**

[{"step": 1, "description": "Move the transport robot to the storage module and dock it", "action": "move_dock", "url": "http://129.69.102.129:5010/robotino_7/functionalities/move_dock"},{"step": 2, "description": "Load the workpiece from the storage module onto the transport robot", "action": "load", "url": "http://129.69.102.129:5010/robotino_7/functionalities/load"},{"step": 3, "description": "Undock the transport robot from the storage module", "action": "undock", "url": "http://129.69.102.129:5010/robotino_7/functionalities/undock"},{"step": 4, "description": "Move the transport robot to the painting module and dock it", "action": "move_dock", "url": "http://129.69.102.129:5010/robotino_7/functionalities/move_dock"},{"step": 5, "description": "Unload the workpiece from the transport robot onto the painting module", "action": "unload", "url": "http://129.69.102.129:5010/robotino_7/functionalities/unload"}]

In the prompts, we allow updates to the information in the context section in accordance with the descriptive information and semantic annotations in the digital twin system. This is implemented in a python program with query-based template-filling and text string concatenation (cf. Figure 8). The program sends the prompt to GPT-model and parses the returned text into structured production steps with identifiers.

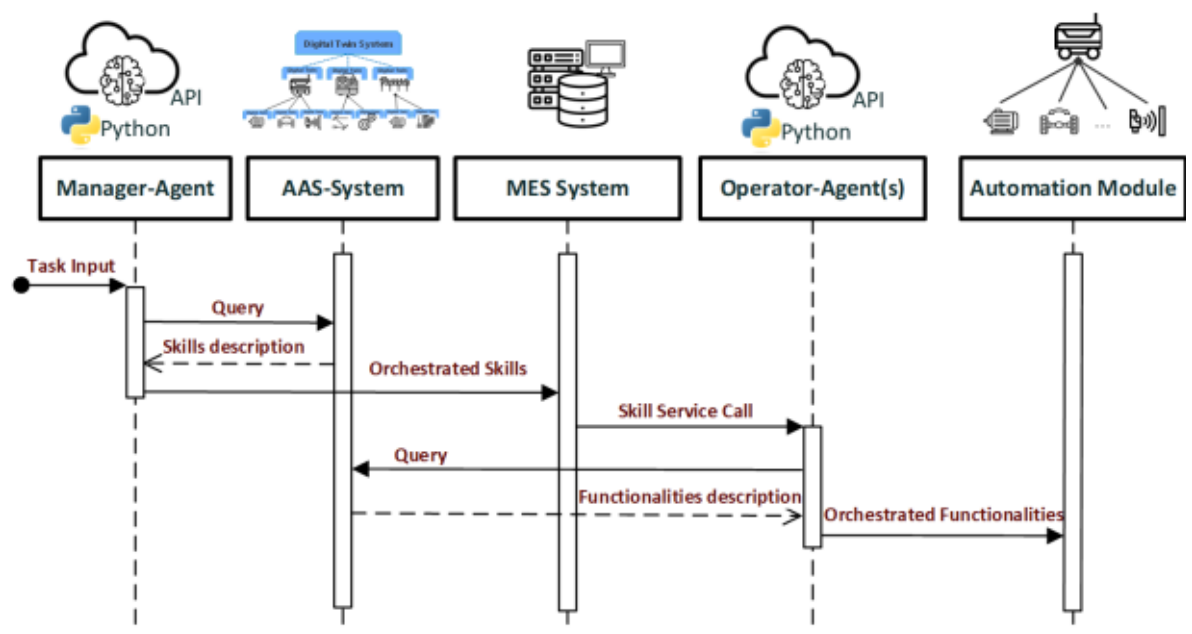

*Figure 9 Sequence diagram illustrating the interaction between different components of the prototype, from task input to operations execution.*

Several software components are implemented to realize our system design. The sequence diagram in Figure 8 shows the essential software components in our prototypical demonstrator and their interactions.

At the beginning of the process, a task input is handled by the manager agent. It queries a digital twin system (AAS-System) to retrieve the skills description of the automated production system and uses this data to update the context information in the specified prompt. The specified prompt is transmitted to the service API of a GPT-Model[11], and the GPT-model returns with the generated output texts. By parsing the generated texts, the manager agent obtains a sequence of skills to fulfill the task. The orchestrated skills are passed to the MES-System, based on which the MES-system invokes the skill service calls on one or more operator agent(s). On receiving the skill service calls, the operator agent retrieves the necessary information about the functionalities in an automation module from the digital twin system (AAS-System), and then orchestrates the functionalities to execute the requested skill. The orchestrated process is finally executed by the automation modules, as shown in Figure 9.

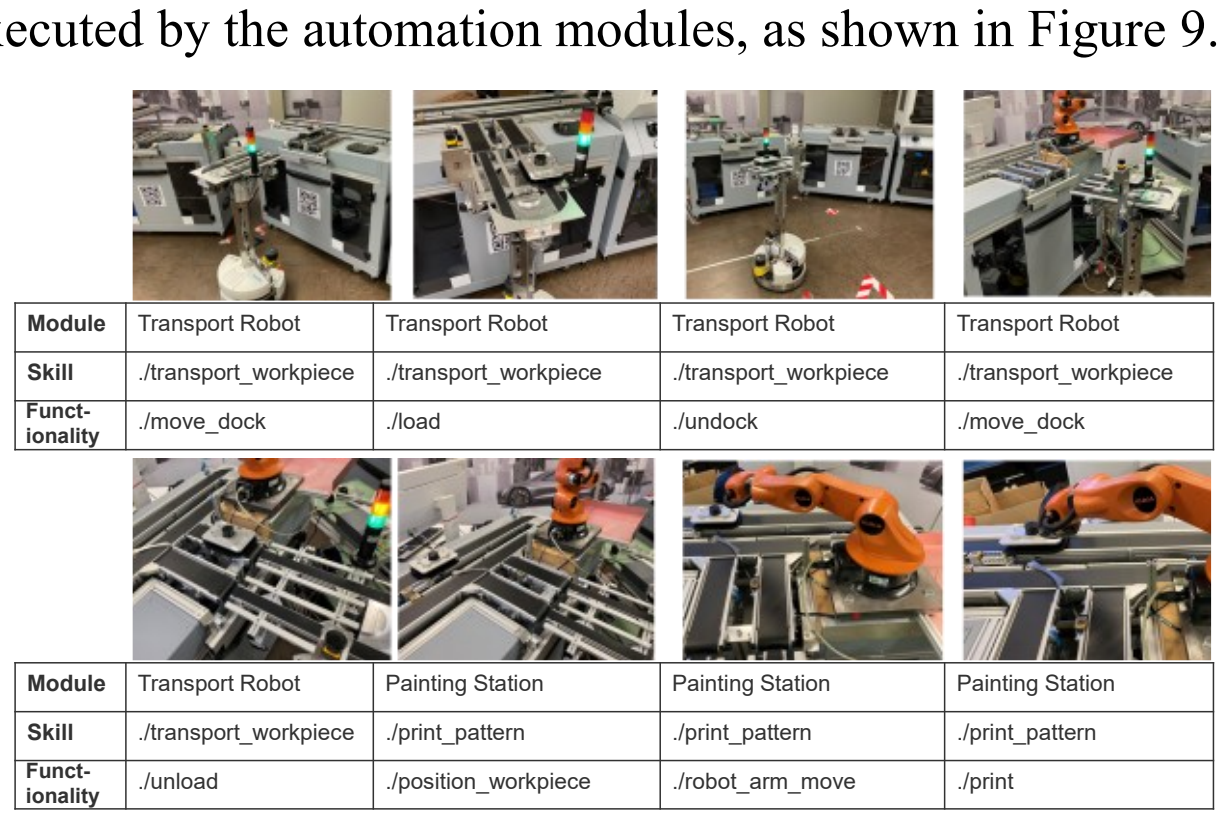

| Module | Transport Robot | Transport Robot | Transport Robot | Transport Robot |
|---|---|---|---|---|
| Skill | ./transport_workpiece | ./transport_workpiece | ./transport_workpiece | ./transport_workpiece |
| Funct-ionality | ./move_dock | ./load | ./undock | ./move_dock |

| Module | Transport Robot | Painting Station | Painting Station | Painting Station |
|---|---|---|---|---|
| Skill | ./transport_workpiece | ./print_pattern | ./print_pattern | ./print_pattern |
| Funct-ionality | ./unload | ./position_workpiece | ./robot_arm_move | ./print |

*Figure 8 Execution of production following the orchestrated processes directed by LLM-agents.*

## V. DISCUSSION

In the previous sections, we introduced a novel framework that integrates large language models (LLMs) with digital twin systems to enable intelligent management and control in industrial automation. Our method leverages prompt engineering to create LLM agents capable of adapting to specific tasks based on the information provided by digital twins. Through a case study involving a matrix modular production facility demonstrator, we showcased how our approach can address unforeseen problem tasks, autonomously orchestrate a production plan, and execute them to provide customizable production services.

Several key insights emerged from our study, and we explain them in three parts: positive insights, difficulties and lessons learned, as well as limitations and future works.

### *A.* Positive Insights

#### *1) Enhanced reasoning and decision-making capabilities for flexible/agile production*

By contextualizing the LLM agents using prompt engineering, we observed the human-like problem-solving capabilities in domain-specific tasks of production management and control.

Based on an instruct-finetuned model "*text-davinci-003*", we iteratively refined the proposed prompt template, which consists of five distinct sections. Using this template and the information provided by the digital twins, the defined LLM-agent effectively generates contextually appropriate responses to plan and control the production system autonomously. This reduces the need for human effort and can lead to increased productivity, reduced operational costs, and minimized delay time in production processes.

[10] Used regular expression in Python code:
*\((\d+)\) (.+)\. Call the functionality "(\w+)" using the URL "(.+)"*

[11] We used the API of OpenAI's GPT-Model "text-davinci-003" for text generation in our prototypical implementation.

*2) Digital twin system bridges the information gaps for LLM-agent.*

As the traditional automation systems are not designed to host comprehensive information and knowledge about the production and the automation system itself, the development of a digital twin system (implemented with AAS) is necessary to provide the missing knowledge for the intelligent agents. The proposed architecture demonstrates the feasibility of establishing a bridge between LLMs and the physical production infrastructure with digital twin system. The modular and cascaded digital twins enable scalable communication, allowing the LLM-agents to access comprehensive information about the production system and interact with the physical world through service interfaces.

*3) Scalability, reuse of modular functionalities/skills to achieve adaptability*

Our approach demonstrates the necessities of developing scalable services interface of atomic functionalities/skills of automation components/modules. These interfaces are indispensable to enable intelligent LLM-agents to interact with the physical world. By dynamically orchestrating those atomic functionalities and skills, a higher level of adaptability of the production system can be achieved to meet the customized production demand.

### *B. Difficulteis and Lessons Learned*

*1) Retrofitting the system to acquire modular interfaces.*

Retrofitting existing production facilities to accommodate modular interfaces can be a challenging process, necessitating time-consuming engineering efforts. While creating query interfaces is typically straightforward, developing control interfaces demands a higher level of caution due to the complexity involved. In comparison to coarse-granular skill, fine-granular functionalities exhibit more dependencies and require a closer examination of hardware-related aspects. Some required interfaces for our use case have application-level dependencies and cannot be executed independently without using the delivered vendor-specific software application. The modularity of the functionalities might not have been considered during the design phase when the system was developed. These dependencies across different components and on different system levels are preventing us from easily creating scalable interaction interfaces for our intelligent agents.

In order to solve this problem, we are looking into the code modularization on the device level to create the functionality interfaces from the bottom. By building from the bottom up, these "atomic" code-level functionalities can be orchestrated to create the modular skills of an automation module, allowing for greater reusability, flexibility and adaptability in system integration.

*2) Knowledge representation and conversion in prompts*

The conversion between knowledge representation in digital twins' software systems and natural language is a challenging task. Inaccurate or lossy conversion might result in inaccurate interpretation of the LLM agents for the production system and negatively affect their performance. Moreover, it is difficult to assess whether the LLM agents have accurately interpreted the context information provided in the prompt. How to determine the point at which refining the prompt would no longer yield significant performance improvement remains an open question. In our case study, we refined our prompts until the agents could generate effective results repeatedly (confer *section C.3.*).

When working on the iterative refinement of the prompts, we believe that the essential task is to "**translate the languages**" (e.g., translate the code and information models into text in natural language). In the translation, the knowledge conveyed by both representation forms shall pertain. Notably, we also use our own domain knowledge in automation and production engineering to design and additionally guide the LLM-agents. To be accurate in detail, LLM-agents don't understand the knowledge, but rather they **approximate the knowledge** conveyed by the representation.

*3) High-quality data and high-fidelity digital twins*

Supplying high-quality data and accurate knowledge representation in digital twin is essential to allow the LLM-agents to perform correct decision. However, creating high-fidelity digital twins can be labor-intensive and time-consuming, which also requires a comprehensive understanding of the system's architecture, behavior, and dependencies.

In our implementation, the digital twin system provides the asset information and exposes the updated interfaces description. The dynamic operational data are not replicated in digital twins to avoid data-inconsistency issues. The operational data should be provided by the MES system and the reverse-engineered RESTful-interfaces. In this sense, the digital twin system supplies descriptive information and annotations about the production system and components to help the LLMs to interpret how the production works and how to control it. However, we have not included the operational data into the prompt so far, because the GPT-agent has difficulty interpreting the dynamic numeric data and it would require extra memory for the model to cache the time series data. It is also due to the fact that we designed the LLM-agent to perform stateless interactions.

### *C. Limitation and Future Work*

*1) Stateless interaction*

One notable limitation of our approach is the stateless interaction of the LLM agents because we only give all the context information at once through the API-call. The agent itself does not know how its output affects the production system. In order to keep track of the dynamic effects on environment and to perform more informed decision, the agent need to maintain a memory of the data about the production, which could require an extra software component (e.g., a database) or new mechanism to merge the LLM agents into the digital twin system.

*2) Non-deterministic results*

Another limitation of utilizing large language models (LLMs) is the inherent non-deterministic nature of the generated results. LLMs, such as GPT, are designed to predict the most probable next token in a sequence based on the input prompt and their extensive training data. Consequently, the outputs produced by LLMs can vary each time, even when the input remains the same. Although it is possible to set the model temperature to 0 to let the model generate more invariable output when given the same input, a decision-making process that only based on unexplained probability estimation would still be unreliable. As the task for customized production planning inherently involves non-deterministic input, additional mechanisms (extra constraints or guidance) are required to ensure a deterministic output and the predictability of the results.

*3) Data dilema for comprehensive testing and evaluation*
In contrast to general NLP tasks, where standard benchmarks and datasets are openly available, industrial automation tasks often involve heterogeneous data from specific components, complex hardware-level dependencies, a wide range of use cases, limited data collections, diversity of knowledge representation, and compliance with safety and reliability standards. Consequently, developing benchmarks for performance evaluation of LLM agents in these tasks necessitates future collaborative efforts. At this stage, we have not yet established an evaluation benchmark to comprehensively quantify the performance of LLMs in industrial automation tasks and can only provide a use-case-level proof-of-concept. For the given manager agent example in the *implementation section*, we evaluated 50 generated samples by the model "text-davinci-003": 96% of all the skill sequences are executable without error, 88% can solve the particular task and are able to produce the right product, however, only 6% of all the generated skills sequences use the minimal required steps to solve the task efficiently without unnecessary steps.

*4) The computational complexity of LLM models*
The most capable LLMs often come with considerable size and computational requirements, which can pose challenges for their local deployment in real-world industrial settings. Fine-tuning a smaller model with dedicated data for a specific domain could be a potential solution. However, we hypothesize that well-trained larger models with more neurons and trained on data from diverse knowledge domains possess a stronger general intelligence, which benefits the interpretation and reasoning in specialized domains and tasks. In other words, we assume that the larger models that perform better in general tasks may also yield better outcomes in domain-specific tasks. For simpler tasks, such as identifying semantic similarity in search queries or providing auto-completion recommendations [14][15], smaller embedding models are sufficient for functions with lower complexity.

## VI. Conclusion and Outlook

In conclusion, we have presented a novel framework that integrates large language models (LLMs) with digital twin systems for intelligent management and control in industrial automation. Our approach demonstrates the potential of LLM agents in making informed decisions based on the information provided by digital twins, leading to improved productivity, reduced operational costs, and minimized delay time in production planning processes.

Our study highlighted several positive insights, such as using LLMs to enhance the intelligence of the automation system, integrating of the LLMs into production system via digital twin systems, and the necessity of scalable and modular interfaces to increase the adaptability of production system.

We also encountered challenges, including retrofitting existing systems, translating knowledge representation in prompts, and accommodating high-quality data and high-fidelity models with digital twins.

Last but not least, we identified limitations and future work directions, such as stateless interaction, non-deterministic results, data dilemma for benchmarking, and the computational requirements of LLMs. As the field of AI and industrial automation technology advances, we believe that integrating LLMs into industrial automation systems will lead to more efficient, flexible, and adaptive production systems. To further uncover and realize the potential of application of large language models in autonomous systems for industrial applications, it is crucial to engage in collaborative research efforts spanning across interdisciplinary fields.

## References

[1] T. Müller, B. Lindemann, T. Jung, N. Jazdi, and M. Weyrich, "Enhancing an Intelligent Digital Twin with a Self-organized Reconfiguration Management based on Adaptive Process Models," *Procedia CIRP*, vol. 104, pp. 786–791, Jan. 2021, doi: 10.1016/J.PROCIR.2021.11.132.

[2] T. Müller, N. Jazdi, J. P. Schmidt, and M. Weyrich, "Cyber-physical production systems: enhancement with a self-organized reconfiguration management," *Procedia CIRP*, vol. 99, pp. 549–554, Jan. 2021, doi: 10.1016/J.PROCIR.2021.03.075.

[3] M. Müller, T. Müller, B. Ashtari Talkhestani, P. Marks, N. Jazdi, and M. Weyrich, "Industrial autonomous systems: A survey on definitions, characteristics and abilities," *At-Automatisierungstechnik*, vol. 69, no. 1, pp. 3–13, Jan. 2021.

[4] P. Foith-Förster and I. Thomas Bauernhansl, "Changeable and reconfigurable assembly systems – A structure planning approach in automotive manufacturing," pp. 1173–1192, 2015, doi: 10.1007/978-3-658-08844-6_81.

[5] A. Markaj, A. Fay, N. Schoch, K. Stark, and M. Hoernicke, "Intention-based engineering for the early design phases and the automation of modular process plants," *IEEE International Conference on Emerging Technologies and Factory Automation, ETFA*, vol. 2022-September, 2022, doi: 10.1109/ETFA52439.2022.9921599.

[6] P. Greschke, "Matrix-Produktion als Konzept einer taktunabhängigen Fließfertigung," 2016.

[7] M. Trierweiler and T. Bauernhansl, "Reconfiguration of Production Equipment of Matrix Manufacturing Systems," pp. 20–27, 2021, doi: 10.1007/978-3-662-62962-8_3.

[8] D. Dittler, P. Lierhammer, D. Braun, T. Müller, N. Jazdi, and M. Weyrich, "An Agent-based Realisation for a continuous Model Adaption Approach in intelligent Digital Twins," 2022, Accessed: Mar. 21, 2023. [Online]. Available: https://arxiv.org/abs/2212.03681v1

[9] G. Jawahar, B. Sagot, and D. Seddah, "What Does BERT Learn about the Structure of Language?," *ACL 2019 - 57th Annual Meeting of the Association for Computational Linguistics, Proceedings of the Conference*, pp. 3651–3657, 2019, doi: 10.18653/V1/P19-1356.

[10] J. Wei *et al.*, "Emergent Abilities of Large Language Models," Jun. 2022, Accessed: Mar. 21, 2023. [Online]. Available: https://arxiv.org/abs/2206.07682v2

[11] T. Gao, A. Fisch, and D. Chen, "Making Pre-trained Language Models Better Few-shot Learners," *ACL-IJCNLP 2021 - 59th Ann. Meet. Assoc. Comput. Linguist. & 11th Int. Joint Conf. Nat. Lang. Process., Proceedings of the Conference*, pp. 3816–3830, 2021, doi: 10.18653/V1/2021.ACL-LONG.295.

[12] A. Radford and K. Narasimhan, "Improving Language Understanding by Generative Pre-Training," 2018.

[13] A. Vaswani *et al.*, "Attention Is All You Need," *Adv Neural Inf Process Syst*, vol. 2017-December, pp. 5999–6009, Jun. 2017, Accessed: Mar. 21, 2023. [Online]. Available: https://arxiv.org/abs/1706.03762v5

[14] M. Both, J. Muller, and C. Diedrich, "Reducing configuration efforts in energy management systems based on natural language processing methods and asset administration shells," *IEEE International Conference on Emerging Technologies and Factory Automation, ETFA*, vol. 2022-September, 2022, doi: 10.1109/ETFA52439.2022.9921479.

[15] Y. Xia, N. Jazdi, and M. Weyrich, "Automated generation of Asset Administration Shell: a transfer learning approach with neural language model and semantic fingerprints," *IEEE International Conference on Emerging Technologies and Factory Automation, ETFA*, vol. 2022-September, 2022, doi: 10.1109/ETFA52439.2022.9921637.

[16] V. Sanh *et al.*, "Multitask Prompted Training Enables Zero-Shot Task Generalization," Oct. 2021, Accessed: Mar. 21, 2023. [Online]. Available: https://arxiv.org/abs/2110.08207v3

[17] J. Wei *et al.*, "Finetuned Language Models Are Zero-Shot Learners," Sep. 2021, Accessed: Mar. 21, 2023. [Online]. Available: https://arxiv.org/abs/2109.01652v5

[18] J. Vig, Y. Belinkov, H. John, and A. Paulson, "Analyzing the Structure of Attention in a Transformer Language Model," pp. 63–76, Jun. 2019, doi: 10.18653/v1/w19-4808.

[19] J. Wei *et al.*, "Chain-of-Thought Prompting Elicits Reasoning in Large Language Models," Jan. 2022, Accessed: Mar. 21, 2023. [Online]. Available: https://arxiv.org/abs/2201.11903v6

[20] T. Kojima, S. Shane Gu, M. Reid Google Research, Y. Matsuo, and Y. Iwasawa, "Large Language Models are Zero-Shot Reasoners," May 2022, Accessed: Apr. 11, 2023. [Online]. Available: https://arxiv.org/abs/2205.11916v4

[21] B. Prystawski and N. D. Goodman, "Why think step-by-step? Reasoning emerges from the locality of experience," Apr. 2023, Accessed: Apr. 11, 2023. [Online]. Available: https://arxiv.org/abs/2304.03843v1

[22] T. B. Brown *et al.*, "Language Models are Few-Shot Learners," *Adv Neural Inf Process Syst*, vol. 2020-December, May 2020, Accessed: Mar. 21, 2023. [Online]. Available: https://arxiv.org/abs/2005.14165v4

# Control Industrial Automation System with Large Language Model Agents

Yuchen Xia, Nasser Jazdi, Jize Zhang, Chaitanya Shah, Michael Weyrich
Institute of Industrial Automation and Software Engineering
University of Stuttgart
Stuttgart, Germany
yuchen.xia@ias.uni-stuttgart.de, nasser.jazdi@ias.uni-stuttgart.de, st171260@stud.uni-stuttgart.de, st181599@stud.uni-stuttgart.de, michael.weyrich@ias.uni-stuttgart.de

***Abstract*— Traditional industrial automation systems require specialized expertise to operate and complex reprogramming to adapt to new processes. Large language models offer the intelligence to make them more flexible and easier to use. However, LLMs' application in industrial automation settings is underexplored. This paper introduces a framework for integrating LLMs to achieve end-to-end control of industrial automation systems. At the core of the framework is an agent system designed for industrial automation tasks. A structured prompting method and an event-driven modeling mechanism provide the information for LLMs to perform reasoning on different context levels, allowing them to semantically interpret the information, generate production plans, and control operations on the automation system. Furthermore, this framework facilitates the creation of structured datasets for fine-tuning LLMs on this specific downstream application. Our contribution includes a formal system design, proof-of-concept implementation, and a method for generating task-specific datasets for LLM fine-tuning and testing. This approach enables a more adaptive automation system that can respond to spontaneous events, allowing intuitive system operation and configuration through natural language. Demo videos and detailed evaluation data are accessible on GitHub: https://github.com/YuchenXia/LLM4IAS.**

***Keywords*— *Large Language Model, Industrial Automation System, Digital Twin, Intelligent Robotics, Multi-Agent System***

## I. Introduction

Traditional industrial automation systems are rigid, requiring specialized expertise for any modification or reconfiguration. For instance, when the system needs to be adapted to produce new product variants or execute different operations, significant effort is required to design and implement the necessary changes. This process is often hampered by several challenges, including the need for an in-depth understanding of the complicated equipment and the time-consuming effort in translating user requirements into executable programs. These factors contribute to delays and increased costs, with reconfiguration often constrained by knowledge barriers, the intricate nature of reprogramming tasks, and possibly inefficient communication between user and programmer. As a result, traditional industrial automation systems are not only inflexible but also costly and time-inefficient when adapting to new demands [1].

Large language models offer transformative potential in industrial automation. They can perform reasoning based on the knowledge internalized during pre-training, interpret both general and domain-specific language, and generate on-demand responses to varied inputs. While LLMs have demonstrated their utility in general chatbot applications [2], their tailored application in industrial contexts remains underexplored. The challenge lies in effectively adapting these capabilities to deliver tangible value in the industrial domain. A structured approach is required to link the digital functionality of LLMs with the physical realm of industrial automation.

In this paper, a novel framework for controlling and configuring industrial automation equipment using large language models is presented, enabling more flexible and reasoning-driven automation systems. The paper's contribution includes:

- An integral system design for applying large language models in industrial automation.
- A proof-of-concept implementation on a physical production system with quantitative evaluation.
- A systematic approach for creating datasets for fine-tuning LLMs to adapt a general pre-trained model for this specific industrial application.

As a result, the LLM-controlled automation system can interpret user tasks specified in natural language, generate control commands, and execute operations on the physical shop floor. The configuration of control logic is enabled by prompt design, and further application-specific adaptation of the LLMs is enabled by supervised fine-tuning on data collected during system operation.

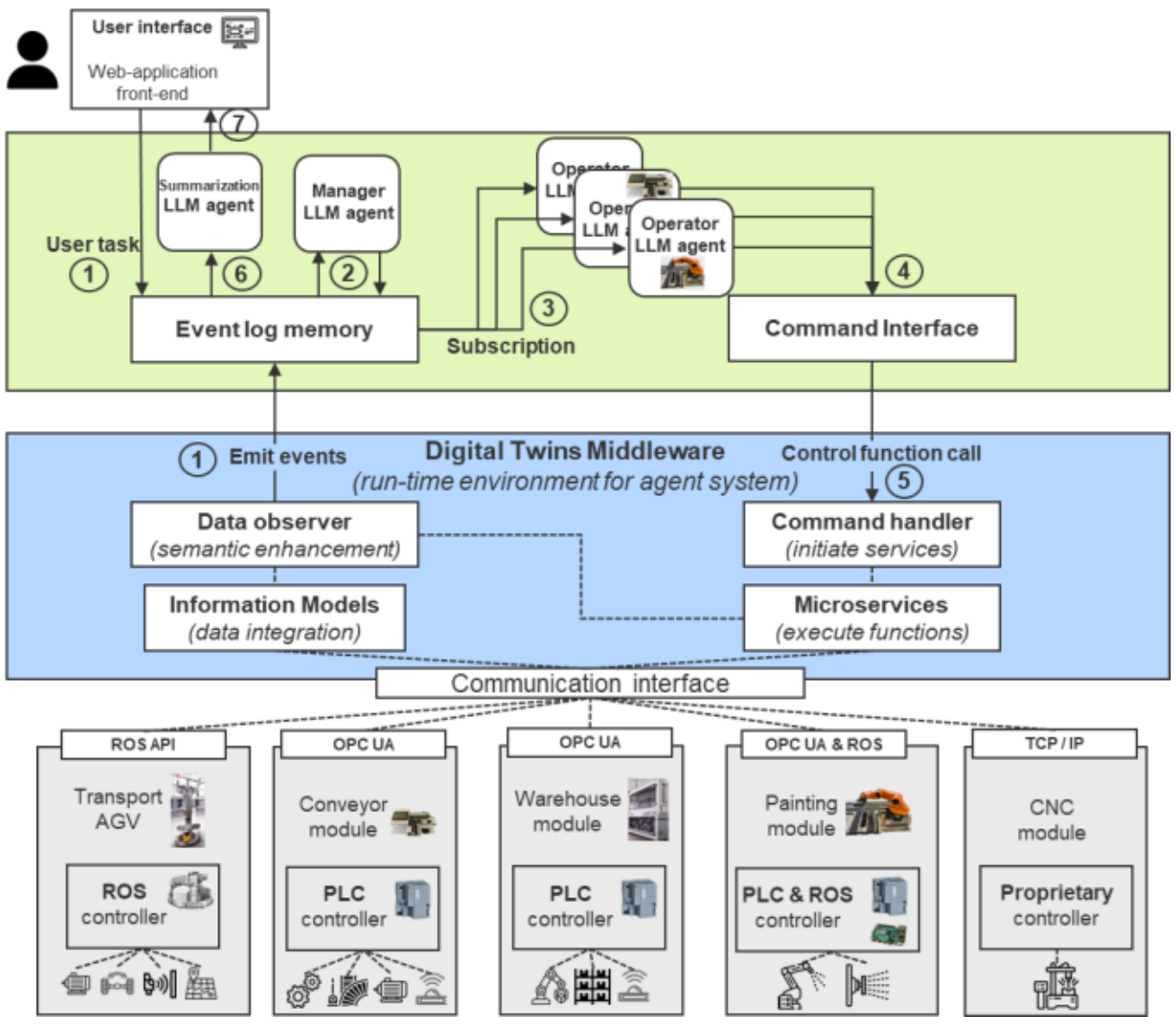

Figure 1 The overall system setup

## II. Requirements and System Setup

This section introduces the required technological foundation in typical industrial automation domain that enables the downstream application of LLM. Based on this foundation, we establish a system depicted in Figure 1.

### A. *Interoperability*

Interoperability is a fundamental prerequisite for implementing intelligent systems. This concept involves two key aspects: synchronized data acquisition and unified control interface.

PLCs are commonly central to industrial automation, serving as nodes for field data collection and control point exposure. Building on this, OPC UA [3] enables seamless interoperability by providing a unified interface to connect PLCs with higher-level systems across various devices and platforms. In robotics and automation systems that utilize ROS, unified data and control interfaces can be established via ROS communication mechanisms. For proprietary automation modules, communication can be achieved through industrial Ethernet TCP/IP.

Utilizing these technology stacks, various automation modules can be digitally integrated through a unified data and control interface. This integration facilitates the creation of a cyber environment, providing access to the physical system, also referred to as a Cyber-Physical System [4].

### B. *Digital twins and semantics*

A digital twin is software that provides a synchronized digital representation of physical assets [5], [6]. It maintains information models that integrate field data and offers services to other software components. Overall, it serves as the foundation for a runtime environment supporting high-level applications, especially indispensable for LLM system.

Another critical aspect for integrating LLM in industrial applications is data semantics. Systems operating at different levels often interpret the same data differently depending on the context [7]. For instance, "a bit flip from 0 to 1" in a PLC program indicates "motor on" at the field device level, "workpiece transport initiation" at a higher control level, and "logistics task starts" at the production planning level. These variations of semantics become more apparent when viewed through the lens of the automation pyramid, as illustrated in Figure 2.

This necessitates a semantic enhancement process across various abstraction levels. For this purpose, a data observer software component (refer to Figure 1) is developed. The data observer monitors data in information models and converts them into textual expressions on different semantic abstraction levels. As different LLM-agents have distinct requirements for how data and changes should be interpreted in their specific task contexts, we pre-define the rules determining which events should be emitted on data changes and customize the text content for each agent, as shown in Figure 3.

```
[Task Planner][Manager][12:04:23] task assigned: retrieve a 'white plastic cylinder' from the storage station.
[Storage Station][System][12:04:23] task received: retrieve a 'white plastic cylinder' from the storage station.
[Inspection Station][System][12:04:44] BG27 detects a workpiece at the outlet of conveyor C2.
[Storage Station][System][12:04:45] BG56 detects a carrier at the infeed of conveyor C1.
[Storage Station][Operator][12:04:45] Storage Station calls function: conveyor_1_run('forward', 13).
[Storage Station][Operator][12:04:45] Conveyor C1 starts running for 13 seconds.
[Storage Station][System][12:04:47] A carrier passes BG56.
[Storage Station][System][12:04:52] BG51 detects a carrier at the holder H2 on conveyor C1.
[Storage Station][Operator][12:04:53] Storage Station calls function: query_inventory_workpiece_position('white
 plastic cylinder').
[Storage Station][System][12:04:53] The 'white plastic cylinder' is located on shelf 'A_13'.
[Storage Station][Operator][12:04:54] Storage Station calls function: robot_arm_pick('A_13').
[Storage Station][System][12:04:54] Robot arm has started picking the workpiece from position A_13.
```

Figure 3 A snippet of the events subscribed by an LLM agent.

### C. *LLM agent*

Recent studies on LLM applications have focused on the concept of LLM agents [8]. LLMs have also been applied to enabling simulated task planning and execution in robotics [9], [10]. In the scope of this paper, an agent is defined as a software component that 1) is responsible for solving specific task requirements, 2) is associated with a physical asset and can be embodied in the form of an automation module. Building on our previous work [11], [12], which introduced a hierarchical manager/operator LLM agent structure for automation system planning and control, this paper presents a refined and more scalable system design. A new fundamental component of this system design is an event log memory, combined with subscription and broadcasting mechanism that provides time-ordered information to agents, thereby keeping them informed about ongoing activities.

Based on the tasks, the system features three distinct agent roles:

- **Manager Agent:** This agent responds to user commands or events and generates an operational plan. It assigns subtasks to operator agents through the event log and monitors the plan execution. This agent type leverages the reasoning capability of LLMs in problem-solving and planning.
- **Operator Agent:** This agent executes tasks assigned by the manager agent or reacts to events by generating function call commands to control the ongoing production process. The operator agents are embodied with diverse automation modules to execute operations. This agent type leverages the reasoning and code understanding capabilities to control the operations within the automation system.
- **Summarization Agent:** This agent subscribes to the event log and provides a summary of system

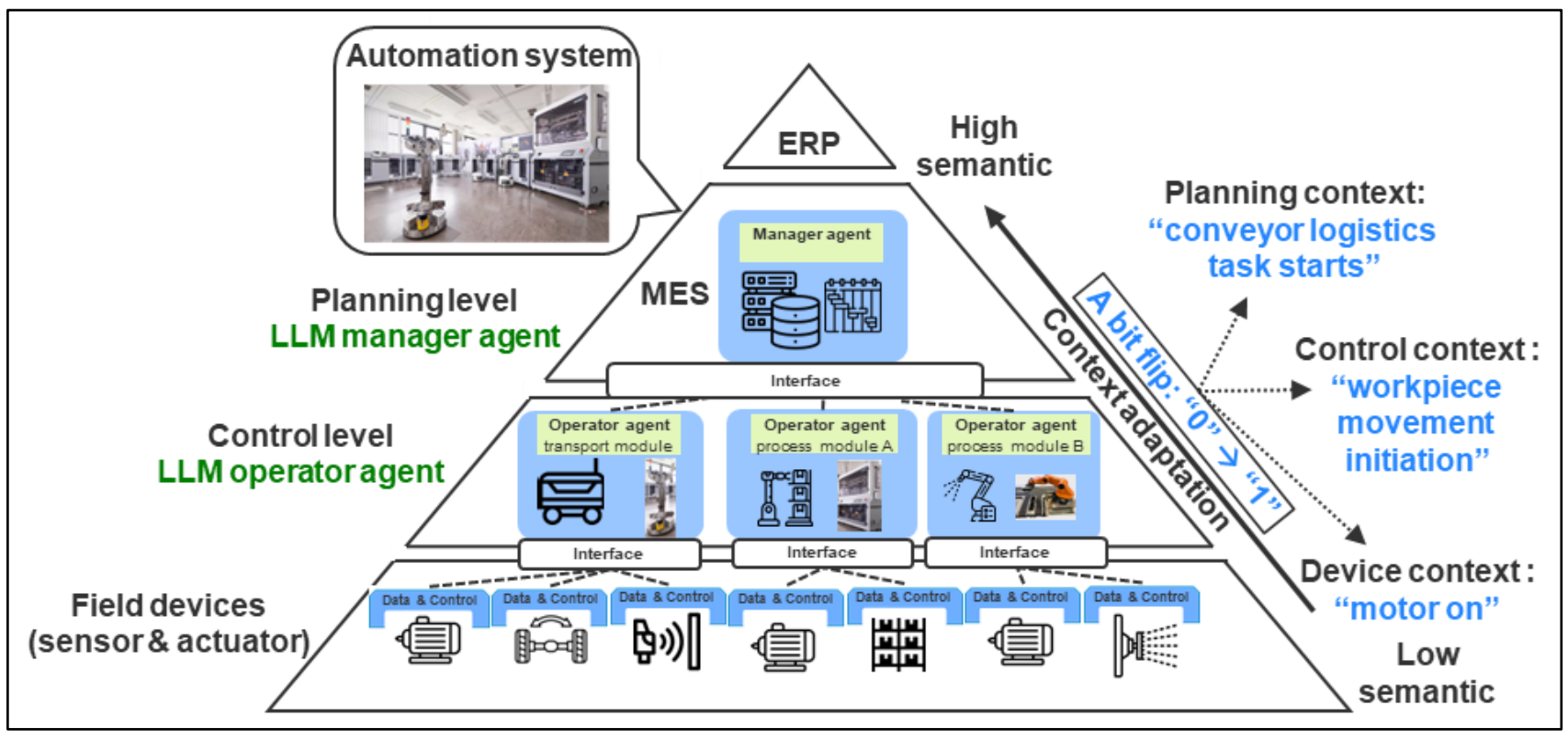

Figure 2 View of automation system modules and components from a hierarchy perspective: Automation Pyramid.

operations for the user. LLMs' long context understanding and alignment with human preferences are highly relevant for this task role. (refer to Figure 1)

Each agent type has clearly defined responsibilities using prompts and interacts with other system components.

*D. Data-driven LLM application and testing*

To assess the performance of the agent system performing automation tasks, the designed system (see Figure 1) is set up to generate test datasets (see Section IV). These datasets comprise pairs of system-generated events and agent prompts alongside the expected outputs. This enables systematic testing of the LLMs' responses under various operational conditions.

Furthermore, the established framework enables the runtime environment for LLM agents through digital twins of the industrial automation system. If historical records of changes in the information model in the digital twin system are available, these can be utilized to develop datasets tailored for downstream automation and robotics tasks. This not only allows for testing of the LLMs but also facilitates the post-training of general LLMs to adapt them for the use case of controlling automation system. This agent system primarily processes textual data, where the effectiveness of the system hinges on the LLMs' ability to accurately interpret and reason with the digital twin system generated textual events.

The following section further elaborates on the details of the proposed system for this application and provides a formal description of the framework's conceptual and structural composition.

## III. THE AGENT SYSTEM FRAMEWORK DESIGN

The organization of agent collaboration is crucial for effective application. For manufacturing process planning and control, we adopt a manager-operator model, applied across different abstraction levels following the automation pyramid (see Figure 2). Additionally, we introduce a summarization agent to generate reports based on the event log for users, as shown in Figure 1, though it is omitted in this section for brevity.

*A. Manager agent and operator agent for planning and control*

Our approach follows a manager-operator paradigm for task orchestration, as visualized in Figure 4, building on our previous works [11], [12]. The manager agent is a planning module responsible for processing user input tasks and decomposing them into sub-tasks to form a production plan. Operator agents are designed to control specific automation modules, receiving tasks from the manager and executing them accordingly.

Figure 4 illustrates how the agent system operates, describing its main conceptual components and their relationships. The system is designed to enable modular, flexible control over automation components by leveraging LLM-powered agents. The key components (in **bold**) are detailed as follows:

**The Agent System:** The overall system is controlled by several software agents, with each agent managing a dedicated automation module. This ensures responsibilities are cleanly separated and easy to scale.

**Automation Module:** Each **automation module** represents a specific piece of equipment or subsystem within the overall automation environment. An automation module is structured as follows:

- **Components**: The sensors, actuators, or other physical devices the agent can interact with.
- **Functions**: Operations that can be triggered on the module, such as starting a conveyor, picking a part, or processing a workpiece.

**Event Log and Subscriptions:** The entire system shares a global **event log**—a time-ordered list of all events emitted by the automation modules or agents. Every time the system perceives a change, or the system state updates, the corresponding event is added to this log with a timestamp.

**A subscription mechanism:** To remain focused and efficient, each agent only needs to be aware of the subset of events relevant to its role. This is achieved using a **subscription mechanism** and each agent subscribes to particular event types, filtering the global event log so it only observes changes within its specific area of responsibility.

**Prompt** Construction and LLM Decision-making: Whenever an agent needs to make a decision, it constructs a prompt that includes:

- Its current context (role and capabilities)
- The latest relevant events
- Applicable Standard Operating Procedures (**SOPs**)
- A record of the current system state as perceived by that agent

This prompt is then submitted to an LLM, which processes the information and generates the output that contains:

- A **reasoning statement** explaining the rationale behind its intended action
- A **function call**—the concrete command that will be executed on the automation module

Table 1 provides more details on the prompt design.

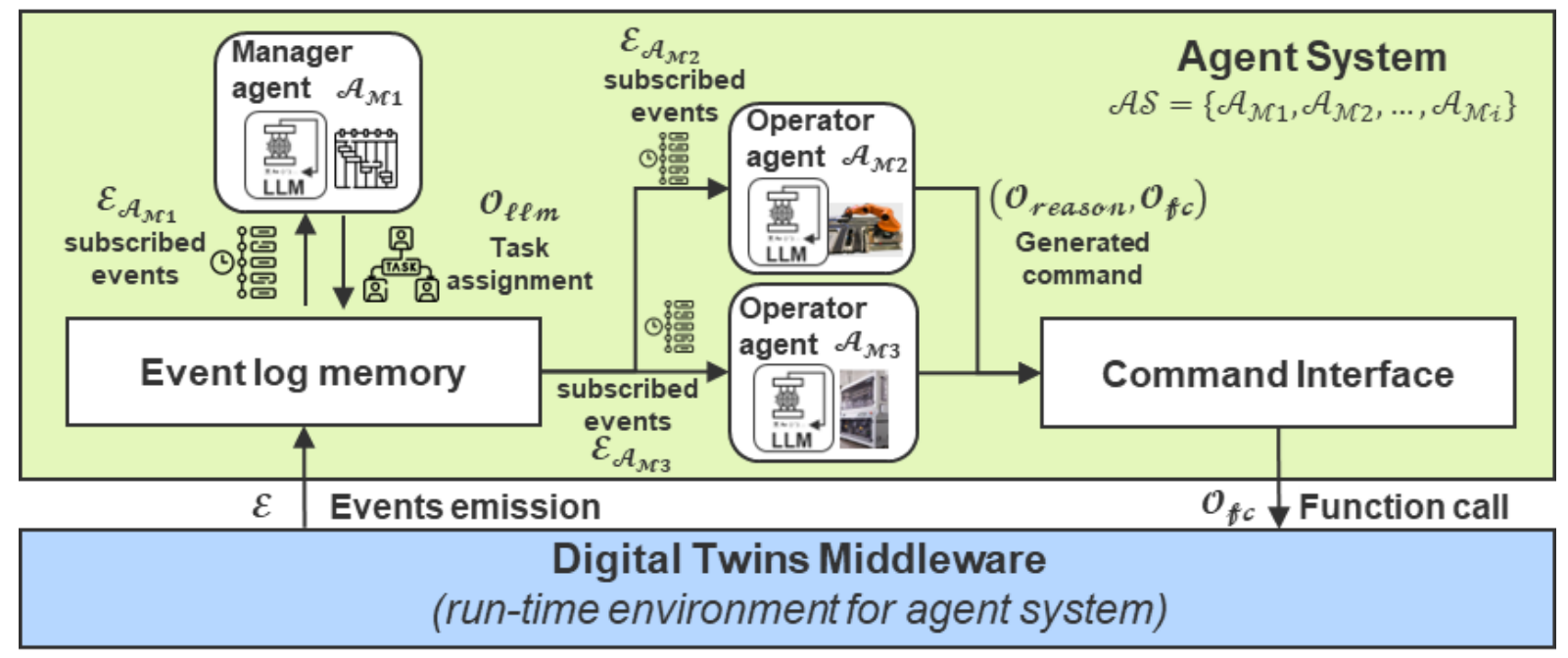

Figure 4 The agent system consisting of manager agent for planning and operator agents for controlling

## B. *Formal description of the LLM agent system*

This section introduces a formal specification of the agent framework's conceptual and structural composition and outlines the relationships between its components, which underpin its software-technical implementation.

In the context of this paper, an LLM agent is defined as a software component that processes textual data to control an automation module, with the LLM agent construct shown in Figure 5.

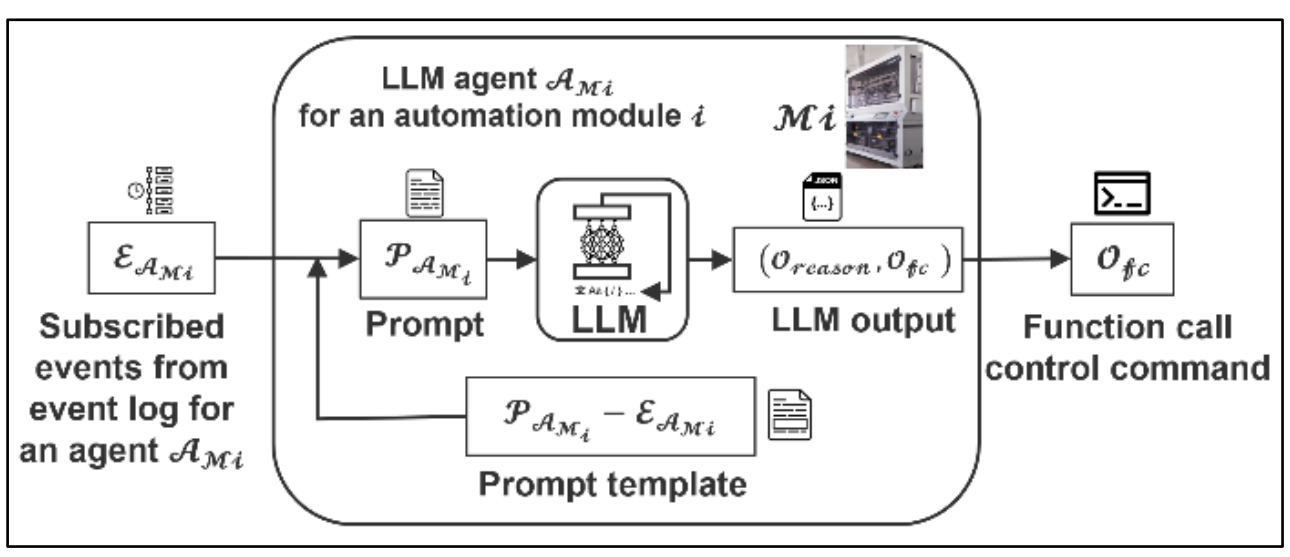

Figure 5 LLM agent processing cycle for command generation

- *Agent System* $\boldsymbol{\mathcal{AS}}$:

$$\mathcal{AS} = \{\mathcal{A}_{\mathcal{M}1}, \mathcal{A}_{\mathcal{M}2}, \dots, \mathcal{A}_{\mathcal{M}i}\}$$

An agent system $\mathcal{AS}$ consists of several agents, and each agent $\mathcal{A}_{\mathcal{M}i}$ (as illustrated in Figure 5) is responsible for controlling a specific automation module $\mathcal{M}i$.

- *Automation Module* $\boldsymbol{\mathcal{M}_i}$ :

$$\mathcal{M}_i = (\mathcal{C}_{\mathcal{M}_i}, \mathcal{F}_{\mathcal{M}_i}, \mathcal{E}_{\mathcal{M}_i})$$

Note that an overall automation system consists of multiple automation modules $\{\mathcal{M}_1, \mathcal{M}_2, \dots, \mathcal{M}_i\}$.

Each individual automation module $\mathcal{M}_i$ comprises a tuple, where:

- $\mathcal{C}_{\mathcal{M}_i}$ is the set of components of the $\mathcal{M}_i$.
- $\mathcal{F}_{\mathcal{M}_i}$ is the set of functions exposed by $\mathcal{M}_i$.
- $\mathcal{E}_{\mathcal{M}_i}$ is the set of events that are in the scope of $\mathcal{M}_i$.

The individual events in set $\mathcal{E}_{\mathcal{M}_i}$ are pre-defined, automatically emitted upon state changes in the module. This events generation mechanism maps low-semantic data to high-semantic information for process planning and control.

- *Event Log Memory* $\boldsymbol{\mathcal{E}}$ *that collects all the events:*

$$\mathcal{E} = \{(e_1, t_1), (e_2, t_2), \dots, (e_t, t_t) \mid t_1 < t_2 < \dots < t_t\}$$

The event log memory records a sequence of **all** events $(e_t, t_t)$ ordered chronologically, representing the history of state changes in the automation system.

- *Subscription Mechanism* $\boldsymbol{\mathcal{S}}$:

$$\mathcal{E}_{\mathcal{A}_{\mathcal{M}i}} = \mathcal{S}(\mathcal{A}_{\mathcal{M}i}) \subseteq \mathcal{E}$$

An agent $\mathcal{A}_{\mathcal{M}i}$ subscribes to the event log memory $\mathcal{E}$ to retrieve relevant events into its own event log memory $\mathcal{E}_{\mathcal{A}_{\mathcal{M}i}}$. $\mathcal{S}(\mathcal{A}_{\mathcal{M}i}) \subseteq \mathcal{E}$ denotes a selective function $\mathcal{S}$ that allows the agent $\mathcal{A}_{\mathcal{M}i}$ to subscribe to events from $\mathcal{E}$, thereby limiting the agent's scope of observation to relevant events.

- *Agent Prompt* $\boldsymbol{\mathcal{P}_{\mathcal{A}_{\mathcal{M}i}}}$:

$$\mathcal{P}_{\mathcal{A}_{\mathcal{M}i}} = \text{Textual}(\mathcal{R}_{\mathcal{A}_{\mathcal{M}i}}, \mathcal{C}_{\mathcal{M}_i}, \mathcal{F}_{\mathcal{M}_i}, \mathcal{SOP}_{\mathcal{A}_{\mathcal{M}i}}, \mathcal{E}_{\mathcal{A}_{\mathcal{M}i}})$$

A textual represented prompt $\mathcal{P}_{\mathcal{A}_{\mathcal{M}_i}}$ for an agent $\mathcal{A}_{\mathcal{M}i}$ that integrates the elements listed in Table 1.

- *LLM Generated Output* $\boldsymbol{\mathcal{O}_{\ell\ell m}}$:

$$\mathcal{O}_{\ell\ell m} = (\mathcal{O}_{reason}, \mathcal{O}_{fc})$$

$\mathcal{O}_{\ell\ell m}$ consists of LLM's generated reasoning and a function call.

## C. *The prompting method*

The design of the prompt is pivotal in this framework. The prompt serves several purposes except as merely instruction for LLM agents, but also to 1) incorporate knowledge about the automation system, 2) serve as an integration point to connect the LLM with real-time events, and 3) later serves as prefix when training a LLM in a supervised fine-tuning manner, and, notably, tokens in the prompt are not included in the loss calculation. The construct of the prompt is described in Table 1.

Table 1 Structure of the Prompt

| Prompt Section | *Definition* | *Typical Examples* [a] |
|---|---|---|
| **R**ole Definition $\mathcal{R}_{\mathcal{A}_{\mathcal{M}i}}$ | Describe the role and the task responsibility of the agent. | You are the operator of an automation module called "Storage Station", responsible for handling of workpieces and directing material transport on conveyors. |
| **C**omponent Description $\mathcal{C}_{\mathcal{M}_i}$ | Entries of component description about the sensors and actuators. | - BG56 is a proximity sensor located at one end of the Conveyor C1… - H1 is a holder located in the middle of the Conveyor C2 at the export verification point, Holder H1 can hold the workpiece in position. - TF81 is an RFID sensor located in the middle of the Conveyor C1 and at Holder H2 of the pick and place point; it can read the workpiece information. |
| Callable **F**unction $\mathcal{F}_{\mathcal{M}_i}$ | The parametrized functions provided by digital twins middleware that can be invoked by the agent. | - conveyor_1_run(direction, time) - This function is used to start Conveyor C1 and run it in a specified direction for a specified duration. |
| **S**tandard **O**peration **P**rocedure $\mathcal{SOP}_{\mathcal{A}_{\mathcal{M}i}}$ | Specify the behavior of the agent under normal operation. | - After detecting a carrier at the entrance, transport the carrier to the pick and place point. - When the carrier arrives at the pick and place point, query the workpiece position in the storage. |
| Auxiliary Instruction | Other instructions for guiding LLM to generate output with desiered format. | - A series of events will provide you the information about the current state of the system. - You should follow the following input and output pattern to generate your response in JSON format. First provide a short reason and then generate a function call. |
| **E**vent Log Input $\mathcal{E}_{\mathcal{A}_{\mathcal{M}i}}$ | The dynamic information that require an agent to react. | [Manager][12:04:23] task assigned: retrieve a 'white plastic cylinder' from the storage station. [System][12:04:23] task received: retrieve a 'white plastic cylinder' from the storage station. [System][12:04:23] BG56 detects a carrier at the infeed of conveyor C1. |
| **O**utput (to be generated) $\mathcal{O}_{\ell\ell m}$ | The generated command by the agent to control the system. | { "reason": "Carrier detected at entrance, initiate transport to pick and place point", "command": "conveyor_1_run('forward', 13)"} |

[a] *a.* more comprehensive examples can be accessed from ***GitHub***

The following sections offer a detailed explanation of several specialized parts in this prompt design.

### 1) *Event log*

This design of the event log is driven by the rationale that **time** and **information** are indispensable in control and planning tasks. The event log provides the LLM with dynamic information about production operations in a time-sequential order, organized as **events**. Based on experimentation and evaluation, this is a succinct way to represent the required information in text form. The first label indicates the scope of the message subscription, the second label indicates the source of the message, and the third label is a timestamp, accompanied by a textual description of the event occurring at that moment, as illustrated in Figure 6.

### 2) *Event emission from digital twin middleware*

An event emission is automatically triggered upon state changes in the information models, messages from other agents, or the initiation or completion of an operation. These event descriptions are pre-defined in the data observer and use natural language to describe high-level semantic information about system activities. The emitted events are recorded in the event log and are then consumed by the LLM agents according to their subscription scopes, allowing the agents to generate appropriate commands in response.

```
[Storage Station][System][12:05:00] BG26 detects a workpiece at the infeed of the conveyor C2.
[Storage Station][Operator][12:05:01] Storage Station calls function: conveyor_2_run('forward', 13).
[Storage Station][Operator][12:05:01] Conveyor C2 starts running for 13 seconds.
[Storage Station][System][12:05:03] A workpiece passes BG26.
[Storage Station][System][12:05:07] BG21 detects a workpiece at the Holder H1 on conveyor C2.
[Storage Station][Operator][12:05:07] Storage Station calls function: export_verify().
[Storage Station][System][12:05:07] This workpiece is verified to export from the storage station.
[Storage Station][Operator][12:05:08] Storage Station calls function: conveyor_2_run('forward', 8).
[Storage Station][Operator][12:05:08] Conveyor C2 starts running for 8 seconds.
[Storage Station][Operator][12:05:09] Storage Station calls function: H1_release().
[Storage Station][System][12:05:09] Holder H1 is released.
[Storage Station][System][12:05:10] A workpiece passes BG21.
```

Figure 6 A snippet of the subscribed events from the digital twin middleware for “Storage Station”.

#### 3) Standard Operation Procedure (SOP)

In manufacturing, Standard Operating Procedure (SOP) provides guidelines for consistent and safe operation. In our approach, we repurpose this concept of “SOP” to refer to instructions for the LLM on how to respond to specific conditions during machine operation.

Similar to how traditional SOPs guide human operators, the SOP in the prompt informs the LLM about standard protocols for operating machines in specified situations.

In contrast to “n-shot” prompting methods , this SOP-based approach enables the configuration of LLMs at a higher level of abstraction and allows for integration of behavioral knowledge. From the perspective of the user who needs to modify system behaviors, this approach facilitates programming the system using natural language by specifying SOP for LLM agents.

For situations that are not considered in SOP, the LLM agents usually, according to our experiment observation, generate commands based on common sense, enabling them to handle scenarios flexibly based on the information given in the prompt, though the responses are not always accurate.

#### 4) Reason in LLM output

Before generating a command, an LLM is instructed to first generate a reason. This serves three main purposes: First, this design mirrors the Reflection-Act (ReAct) process [13] in a minimalist manner. Second, it enhances transparency and explainability, allowing for deeper evaluation by comparing the generated reason with the reference reason. Third, reference reasons in the dataset can provide additional knowledge during fine-tuning and enable calculating loss and weight update on more data, helping the LLM to learn how and why specific planning and control decisions are made.

## IV. Dataset Creation

This section explains how the dataset is created and organized for testing and training the LLM, as illustrated in Figure 7.

### A. Dataset creation based on the agents system and prompting method

The LLM agents generate function calls based on the information provided in the prompt. During operation, the underlying digital twin system automatically emits new events in the event log. The agent is continuously updated with real-time textual information to generate a response, which includes a function call to control the equipment and a brief explanation for the reason.

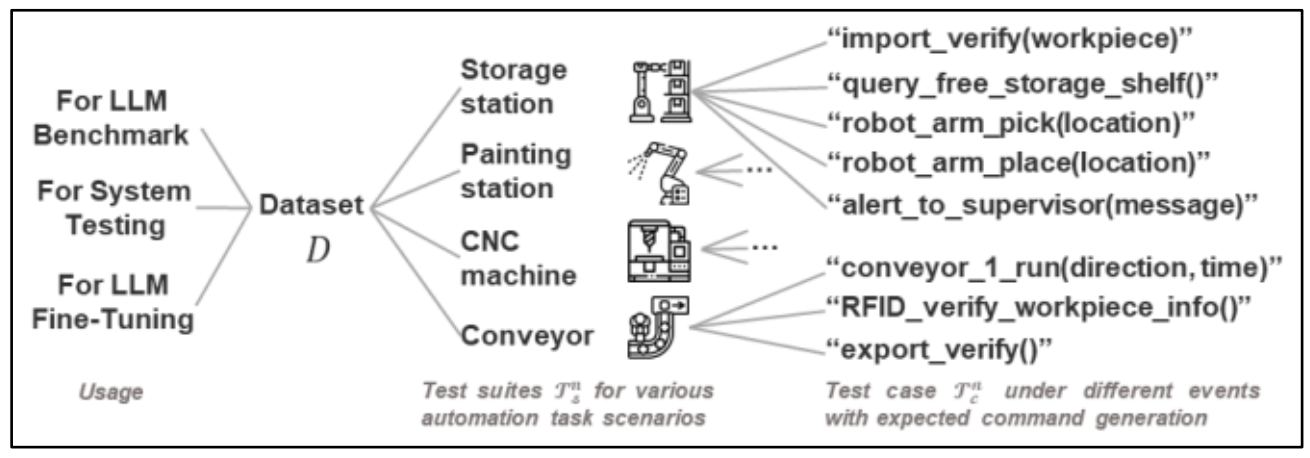

Figure 7 Structured organization of the dataset for testing and training

### B. Structured dataset

To systematically evaluate and further improve our LLM-based automation system, we construct a structured dataset that supports both **testing** and **fine-tuning**. The hierarchy and components of this dataset are as follows:

- **Test Point**: A *test point* represents a single step in agent interaction. It consists of the agent’s prompt (which includes its current role, available components, callable functions, standard operating procedures, and event history) together with a set of newly observed events. Each test point essentially captures the context in which the agent must make a decision.
- **Test Case**: A *test case* pairs a test point with the corresponding reference output from a human or documented ground truth. This output includes: (1) the function call that should be executed in response, and (2) a brief reasoning explaining the choice. In other words, each test case presents the LLM with a situation and the correct response.
- **Test Suite**: Several test cases that focus on a particular operational scenario are grouped into a *test suite*. Each test suite covers all the typical and unexpected events relevant to a specific workflow, such as inventory management or anomaly handling.
- **Dataset**: The *complete dataset* is made up of multiple test suites, thus covering a wide range of task scenarios and variations. This dataset is used not only to quantify the system’s accuracy, but also to serve as a resource for further LLM training.

Importantly, this dataset has dual purposes. First, it enables systematic testing, providing a basis to benchmark model performance across a variety of operational situations. Second, as highlighted in Figure 8, it can be directly used as training data for supervised fine-tuning of the LLM. By learning from these labeled cases, the LLM can better adapt to the specific operation sequence of the automated process.

### C. Formal description of the dataset

A formal specification of the conceptual and structural composition that underpins the software technical implementation is provided as follows:

- *Test Point* $\boldsymbol{\mathcal{T}_p}$:

$$\mathcal{T}_p = \left(\mathcal{P}_{\mathcal{A}_{\mathcal{M}i}} \cup \{\mathcal{E}_{\mathcal{A}_{\mathcal{M}i},[t+1,t+k]}\}\right)$$

A test point $\mathcal{T}_p$ is defined as an agent prompt $\mathcal{P}_{\mathcal{A}_{\mathcal{M}i}}$ containing incremental $k$ new events $\mathcal{E}_{\mathcal{A}_{\mathcal{M}i},[t+1,t+k]}$ for testing the output of an agent $\mathcal{A}_{\mathcal{M}i}$. When expanded:

$$\mathcal{T}_p = \text{Textual}(\mathcal{R}_{\mathcal{A}_{\mathcal{M}i}}, \mathcal{C}_{\mathcal{M}_i}, \mathcal{F}_{\mathcal{M}_i}, \mathcal{SOP}_{\mathcal{A}_{\mathcal{M}i}}, \mathcal{E}_{\mathcal{A}_{\mathcal{M}i},[0,t]} \cup \mathcal{E}_{\mathcal{A}_{\mathcal{M}i},[t+1,t+k]})$$

• *Test Case $\boldsymbol{\mathcal{T}_c}$:*

$$\mathcal{T}_c = (\mathcal{T}_p, \mathcal{O}^*_{\ell\ell m})$$

A test case is the combination of a test point $\mathcal{T}_p$ and the expected reference output $\mathcal{O}^*_{\ell\ell m}$ from the LLM, where $\mathcal{O}^*_{\ell\ell m}$ consists of the expected function call to be generated and a reference reason:

$$\mathcal{O}^*_{\ell\ell m} = (\mathcal{O}^*_{reason}, \mathcal{O}^*_{fc})$$

• *Test Suite $\boldsymbol{\mathcal{T}_s}$:*

$$\mathcal{T}_s = \{\mathcal{T}_c^1, \mathcal{T}_c^2, \dots, \mathcal{T}_c^n\}$$

Test cases $\mathcal{T}_c^1$ to $\mathcal{T}_c^n$ are organized into test suite $\mathcal{T}_s$, each corresponding to a specific operational task scenario.

• *Dataset $\boldsymbol{\mathcal{D}}$:*

$$\mathcal{D} = \{\mathcal{T}_s^1, \mathcal{T}_s^2, \dots, \mathcal{T}_s^n\}$$

The complete dataset $\mathcal{D}$ consists of a collection of test suites from $\mathcal{T}_s^1$ to $\mathcal{T}_s^n$, representing various task scenarios. The dataset is used for both testing and fine-tuning the LLM for the application.

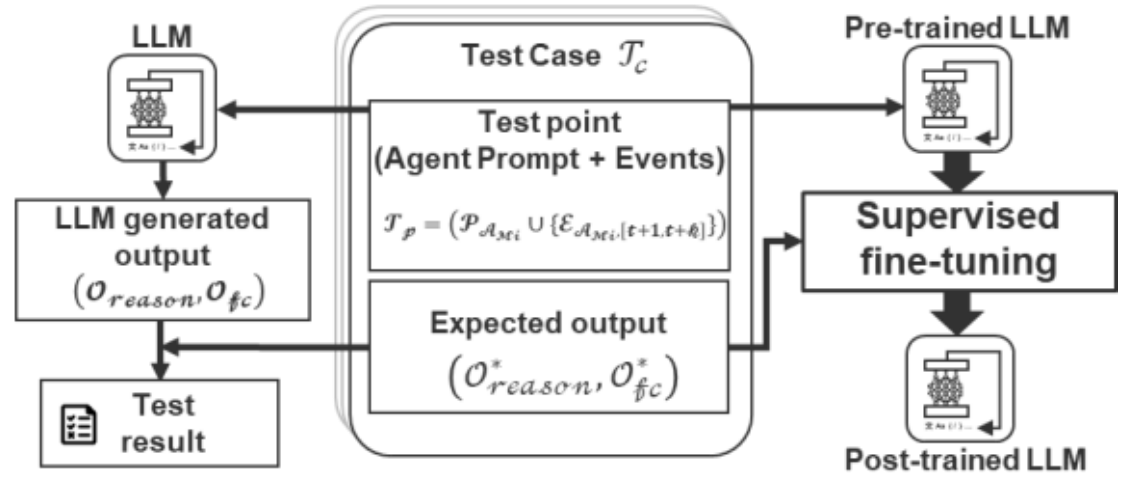

Figure 8 Two uses of the created dataset: Testing and Post-training

## V. Experiments

Using the created dataset, several open-source pre-trained models and the proprietary model GPT-4o are first evaluated. Subsequently, supervised fine-tuning is applied to further train these models on the dataset.

The objectives are twofold: firstly, to evaluate the fine-tuning enhancement of the LLM agent's performance on this specific task; and secondly, to gain insight into the cost-performance trade-offs involved.

### A. Test case creation

In contrast to normal operation where LLM agents observe events before deciding on machine operation, dataset creation involves a reverse process, as depicted in Figure 9.

In this case, the dataset is created without LLM agents, but with direct user input. With a specific task in mind, the user manually operates the command interface to interact with the automation system. As this process unfolds, the digital twin system automatically generates and records relevant events. The user finally provides a description of the intended task process. This approach captures the three essential elements for dataset: the events, the command calls, and the initial user task request. The dataset contains the knowledge necessary for successful execution of intended tasks.

Using this approach, we create various task scenarios for handling typical situations in factory operations, such as processing user orders, responding to spontaneous events, or handling abnormalities. Our initial collected dataset contains 100 test cases $\mathcal{T}_c = (\mathcal{T}_p, \mathcal{O}^*_{\ell\ell m})$, each consisting of a complete prompt and the expected LLM output. This dataset can be used to assess the performance of pre-trained LLMs in controlling industrial automation systems.

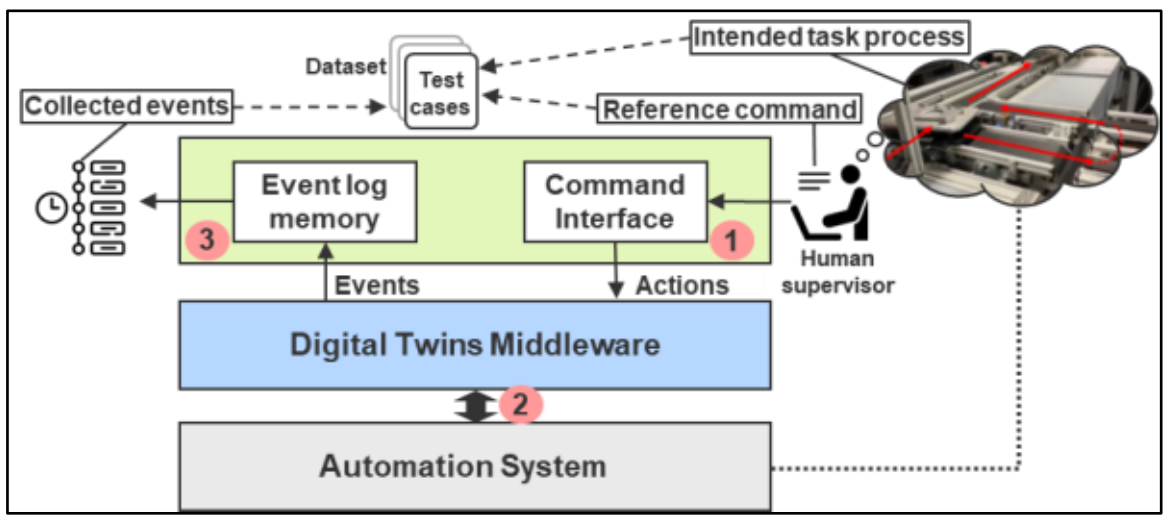

Figure 9 Dataset creation process

In our initial dataset, 68% of the control command generation involves repetitive tasks in our initially prepared dataset, where the LLM agents should follow the SOP routines. The remaining 32% consist of non-routine tasks, requiring LLM agents to respond to unprecedented events through autonomous decision-making.

### B. Metrics for evaluation

Two metrics are applied to evaluate the system's performance.

- **Correctness Rate:** Measures whether the generated command matches the reference command in the dataset.
- **Reason Plausibility:** Assessed through human evaluation, using a Likert scale from 1 to 5 to rate the plausibility of the generated reason.

These two metrics together provide a more granular metric to identify loss. In some cases, the command is incorrect while the reason may be plausible.

### C. Special requirements of automation tasks

Given the nature of industrial automation, there are two main aspects of the requirements:

- **Accuracy in Repetitive Operations:** Industrial automation tasks often require 100% accuracy in repetitive operations. This repeatability implies that some routine tasks can be anticipated. This requirement can be evaluated by assessing whether the LLM agents follow the SOP to successfully complete the tests, or whether the model improves by learning from the dataset to perform in-sample tasks.
- **Handling Unexpected Events:** The system shall be capable of responding to unexpected events not predefined in the SOP or present in the training dataset. It should demonstrate generalization by flexibly handling unforeseen situations that were not anticipated during development. The language model should use its learned knowledge to generate appropriate responses to spontaneous events—an ability typically lacking in traditional automation systems.

Based on these considerations, a comprehensive evaluation and model fine-tuning are performed.

### D. Evaluation results

Different LLMs are employed to power the agent systems. GPT-4o is selected to represent the state-of-the-art performance achievable by LLMs, while other open-source models are chosen to represent those that can be practically deployed in on-premises industrial settings. To account for the trade-off between model complexity and performance, larger models (in the 70B parameter range) are compared with smaller ones (around 7B). The models selected for the experiments include GPT-4o [2], Llama3 models [14], Qwen2 models [15] and Mistral models [16].

Table 2 Evaluation of Pre-trained and Fine-tuned LLMs

<table>
<tr><th></th><th colspan="7">Evaluation based on 100 test points***</th></tr>
<tr><th></th><th>GPT-4o</th><th>Llama-3-70B-Instruct</th><th>Llama-3-8B-Instruct</th><th>Qwen2-72B-Instruct</th><th>Qwen2-7B-Instruct</th><th>Mistral-7Bx8-Instruct-v0.2</th><th>Mistral-7B-Instruct-v0.2</th></tr>
<tr><td>Pre-trained (all)</td><td>81% | 4.7</td><td>75% | 4.3</td><td>37% | 2.8</td><td>70% | 4.0</td><td>65% | 3.7</td><td>29% | 2.4</td><td>45% | 2.9</td></tr>
<tr><td>Pre-trained (SOP)</td><td>100% | 5.0</td><td>87% | 4.5</td><td>53% | 3.1</td><td>85% | 4.5</td><td>63% | 3.6</td><td>34% | 2.4</td><td>37% | 2.5</td></tr>
<tr><td>Pre-trained (Unexpected)</td><td>41% | 4.0</td><td>50% | 3.8</td><td>3% | 2.2</td><td>38% | 3.0</td><td>69% | 4.0</td><td>19% | 2.3</td><td>63% | 3.7</td></tr>
<tr><td>SFT (all)</td><td>100% | 5.0</td><td>95% | 4.8</td><td>96% | 4.9</td><td>* 66% | 3.9</td><td>97% | 4.9</td><td>45% | 3.1</td><td>** N.A.</td></tr>
<tr><td>SFT (SOP)</td><td>100% | 5.0</td><td>94% | 4.8</td><td>99% | 4.9</td><td>* 82% | 4.4</td><td>97% | 4.9</td><td>61% | 3.6</td><td>** N.A.</td></tr>
<tr><td>SFT (Unexpected)</td><td>100% | 5.0</td><td>97% | 4.9</td><td>91% | 4.7</td><td>* 31% | 2.8</td><td>97% | 5.0</td><td>9% | 2.3</td><td>** N.A.</td></tr>
</table>

Value: (accuracy of their generated commands)% | (averaged reason plausibility 1-5)

* : we ran out of GPU capacity and used LoRA instead of full fine-tuning for the Qwen2-72B-model.

** N.A.: Our full-fine-tuning made Mistral-7B model unstable, and it generated unusable echo texts.

***: Details about the SFT, dataset and evaluation sheets on GitHub: YuchenXia/LLM4IAS

#### 1) *Evaluation of pre-trained LLM*

The evaluation starts with the original pre-trained models. In automation tasks, 1) some are routine processes where the LLM agent can follow **SOP** guidelines in agent prompts to operate the automation system, while 2) others require the agent to autonomously respond to **unexpected** events, for which reactions have not been instructed in agent prompts. The results for these two task types are presented separately in Table *2*.

Based on the evaluation results, GPT-4 generally outperforms other open-source models in interpreting agent prompts and events to generate control commands, though their performance varies significantly. Each model also exhibits a distinct “personality” in this use case—that is, different models produce different command outputs when given the same input. This leads to variations in system behavior, depending on which LLM powers the agent system.

#### 2) *Evaluation of post-trained LLM based on created dataset*

Using the collected dataset, supervised fine-tuning (SFT) is applied to evaluate how model training can enhance LLM performance for this specific automation task. This training has the potential to enable the customization of a general LLM for intelligent control of specialized automation equipment. For GPT-4o, we used OpenAI’s proprietary fine-tuning API to explore the capabilities that LLMs can achieve, even though the training methods may vary.

For two main considerations, the models are trained on the fixed dataset created during this research project: 1) the machine is designed to execute repetitive process, and 2) the purpose of the fine-tuning is limited to evaluate whether the model can effectively learn the machine’s operational knowledge. In this sense, the training process is analogous to a programming process. As indicated in Table 2, OpenAI’s model and fine-tuning services outperform other models, and the GPT-4o model quickly learns from the samples how to control the automation systems. Other models also demonstrated reasonably good performance. Interestingly, fine-tuned smaller LLMs did not necessarily underperform in this particular use case. However, our contingency LoRA [17] fine-tuning yielded poor results in our experiments and led to a decrease in model performance.

An accuracy of less than 100% means that the results are not fully reliable for directly controlling an automation system, as it could result in stops and errors during operation. However, they can be developed as an assistant system, proposing next actions for human supervisors to approve in human-machine collaboration use case scenarios.

## VI. Discussion and Implications

This section discusses the broader impact of integrating LLMs into industrial automation, drawing on key insights from the system design and experimental evaluation.

### A. *Adaptive reasoning capabilities of LLM agents*

The reasoning capability of LLMs enables agents to react flexibly to diverse event information. The results show that LLMs are able to reason adaptively when embedded in an agent software component. During the test, the models did not simply replay memorized instructions: they produced context-dependent control commands together with short natural-language reasons. These reasons reflect sound rationale ($\geqslant$ 4.7 for the best models as evaluated in Table 2), and there is a strong, monotonic relationship between the plausibility of the rationale and the correctness of the corresponding command.

This adaptive reasoning also has another effect: whenever an unexpected event (e.g., an operation failure) falls outside the Standard Operating Procedure (SOP) and is not explicitly covered by the prompt, both GPT-4o and fine-tuned open-source models still generate commands aligned with common-sense safety rationale (e.g., issuing an emergency stop or alerting a human operator). This illustrates the flexibility and knowledge generalization that conventional rule-based controllers lack.

### B. *Semantic enrichment of low-level data*

Raw sensor values or bit flips convey limited semantic meaning for an LLM agent to perform reasoning. The data-observer module is therefore designed to semantically translate changes in the information model into textual events at the appropriate level of abstraction for each agent (refer to Section II.B). This semantic enrichment is essential, because a single physical change can—and must—be interpreted differently by different agents based on their task context.

To support this requirement, the observer provides multi-view semantics for different agents: it annotates each low-level signal with textual descriptions, each tailored to the context in which a particular agent operates. A subscription mechanism then delivers only the relevant view to each agent. In this way, semantic enrichment bridges the gap between raw machine data and the natural-language interpretation capabilities of LLM agents, enabling context-aware reasoning and decision-making across the agent system.

### C. *Prompting as a new programming paradigm*

In the proposed framework, natural-language prompts serve as executable specifications that define agents’ behavior. By expressing Standard Operating Procedures (SOPs) and contextual information within the prompt, users can configure how LLM agents act—without needing to understand or modify underlying code. This lowers the barrier to entry: even users without expertise in traditional automation programming (such as PLCs or ROS) can configure agent behavior through simple text edits.

### D. Revisiting the automation pyramid

The proposed system design aligns structurally with the ISA-95/IEC-62264 automation pyramid model [18], while also introducing several key differences:

#### 1) Hierarchical layered abstraction

Manager agents operate at the MES tier, where they formulate high-level plans and delegate subtasks. Operator agents function one level below, issuing function-level commands based on equipment context. These distinct agent roles operate at different abstraction levels, closely reflecting the separation of concerns defined by the different layers in the traditional automation pyramid model [18].

#### 2) Agent-based strategic planning vs. software-logic-based repetitive execution

High-level reasoning tasks—such as production planning and exception handling—are performed by LLM agents, while repetitive and deterministic functions are carried out by PLCs, ROS nodes, or other dedicated software components. This hybrid architecture preserves the reliability and precision of low-level operations, while enabling flexible, language-driven orchestration at the upper layers of planning and control.

## VII. Conclusion

This paper presents a novel application-oriented framework that enables the use of LLMs to control industrial automation systems. From a theoretical perspective, the framework introduces formal constructs for LLM agents as autonomous software entities in automation, a structured prompting method as a new form of executable specification, and a semantics-aware data enhancement mechanism to support flexible context-based reasoning. As a contribution to practical applications, this paper provides a methodology for enabling LLM-driven end-to-end automation, in which the development is structured in two phases: 1) modularizing the system and creating interoperable interfaces to establish the physical and digital foundation for agents, and 2) applying LLM-specific prompting techniques and creating datasets for evaluation and training. The result is an end-to-end solution that allows an LLM to control automation systems, with the reconfiguration process and human-machine interactions made more intuitive through natural language. These contributions extend the existing theory and methodologies of intelligent control systems by integrating LLMs as reasoning agents, and provide a foundation for future research on intelligent system automation.

## Acknowledgment

This work was supported by Stiftung der Deutschen Wirtschaft (SDW) and the Ministry of Science, Research and the Arts of the State of Baden-Wuerttemberg within the support of the projects of the Exzellenzinitiative II.

## References

[1] Y. Fan *et al.*, “A digital-twin visualized architecture for Flexible Manufacturing System,” *J Manuf Syst*, vol. 60, pp. 176–201, Jul. 2021, doi: 10.1016/J.JMSY.2021.05.010.

[2] OpenAI *et al.*, “GPT-4 Technical Report,” Mar. 2023, [Online]. Available: https://arxiv.org/abs/2303.08774v6

[3] A. Veichtlbauer, M. Ortmayer, and T. Heistracher, “OPC UA integration for field devices,” *Proceedings - 2017 IEEE 15th International Conference on Industrial Informatics, INDIN 2017*, pp. 419–424, Nov. 2017, doi: 10.1109/INDIN.2017.8104808.

[4] T. Müller, N. Jazdi, J. P. Schmidt, and M. Weyrich, “Cyber-physical production systems: enhancement with a self-organized reconfiguration management,” *Procedia CIRP*, vol. 99, pp. 549–554, Jan. 2021, doi: 10.1016/J.PROCIR.2021.03.075.

[5] B. Ashtari Talkhestani *et al.*, “An architecture of an Intelligent Digital Twin in a Cyber-Physical Production System,” *At-Automatisierungstechnik*, vol. 67, no. 9, pp. 762–782, Sep. 2019, doi: 10.1515/AUTO-2019-0039/MACHINEREADABLECITATION/RIS.

[6] D. Dittler, P. Lierhammer, D. Braun, T. Müller, N. Jazdi, and M. Weyrich, “A Novel Model Adaption Approach for intelligent Digital Twins of Modular Production Systems,” *IEEE International Conference on Emerging Technologies and Factory Automation, ETFA*, vol. 2023-September, 2023, doi: 10.1109/ETFA54631.2023.10275384.

[7] N. Sahlab, N. Jazdi, and M. Weyrich, “An Approach for Context-Aware Cyber-Physical Automation Systems,” *IFAC-PapersOnLine*, vol. 54, no. 4, pp. 171–176, Jan. 2021, doi: 10.1016/J.IFACOL.2021.10.029.

[8] L. Wang *et al.*, “A survey on large language model based autonomous agents,” *Front Comput Sci*, vol. 18, no. 6, pp. 1–26, Dec. 2024, doi: 10.1007/S11704-024-40231-1/METRICS.

[9] H. Fan, X. Liu, J. Y. H. Fuh, W. F. Lu, and B. Li, “Embodied intelligence in manufacturing: leveraging large language models for autonomous industrial robotics,” *J Intell Manuf*, vol. 36, no. 2, pp. 1141–1157, Feb. 2024, doi: 10.1007/S10845-023-02294-Y/FIGURES/7.

[10] S. Huang *et al.*, “Instruct2Act: Mapping Multi-modality Instructions to Robotic Actions with Large Language Model,” May 2023, Accessed: Jun. 20, 2025. [Online]. Available: https://arxiv.org/pdf/2305.11176

[11] Y. Xia, M. Shenoy, N. Jazdi, and M. Weyrich, “Towards autonomous system: Flexible modular production system enhanced with large language model agents,” *IEEE International Conference on Emerging Technologies and Factory Automation, ETFA*, vol. 2023-September, 2023, doi: 10.1109/ETFA54631.2023.10275362.

[12] Y. Xia, J. Zhang, N. Jazdi, and M. Weyrich, “Incorporating Large Language Models into Production Systems for Enhanced Task Automation and Flexibility,” *Automation 2024*, Jul. 2024, doi: 10.51202/9783181024379.

[13] S. Yao *et al.*, “ReAct: Synergizing Reasoning and Acting in Language Models,” Oct. 2022, [Online]. Available: https://arxiv.org/abs/2210.03629v3

[14] A. Dubey *et al.*, “The Llama 3 Herd of Models,” Jul. 2024, [Online]. Available: https://arxiv.org/abs/2407.21783v2

[15] A. Yang *et al.*, “Qwen2 Technical Report,” Jul. 2024, [Online]. Available: https://arxiv.org/abs/2407.10671v4

[16] A. Q. Jiang *et al.*, “Mixtral of Experts,” Jan. 2024, [Online]. Available: https://arxiv.org/abs/2401.04088v1

[17] E. Hu *et al.*, “LoRA: Low-Rank Adaptation of Large Language Models,” *ICLR 2022 - 10th International Conference on Learning Representations*, 2022.

[18] International Electrotechnical Commission, “IEC 62264-1: Enterprise-control system integration,” 2.0., International Electrotechnical Commission, 2013. [Online]. Available: https://webstore.iec.ch/en/publication/6675

# LLM experiments with simulation: Large Language Model Multi-Agent System for Process Simulation Parametrization in Digital Twins

Yuchen Xia, Daniel Dittler, Nasser Jazdi, Haonan Chen, Michael Weyrich
*Institute of Industrial Automation and Software Engineering*
*University of Stuttgart*
Stuttgart Germany
{yuchen.xia; daniel.dittler; nasser.jazdi; michael.weyrich}@ias.uni-stuttgart.de

***Abstract*— This paper presents a novel design of a multi-agent system framework that applies a large language model (LLM) to automate the parametrization of process simulations in digital twins. We propose a multi-agent framework that includes four types of agents: observation, reasoning, decision and summarization. By enabling dynamic interaction between LLM agents and simulation model, the developed system can automatically explore the parametrization of the simulation and use heuristic reasoning to determine a set of parameters to control the simulation to achieve an objective. The proposed approach enhances the simulation model by infusing it with knowledge from LLM and enables autonomous search for feasible parametrization to solve a user task. Furthermore, the system has the potential to increase user-friendliness and reduce the cognitive load on human operators by assisting in complex decision-making processes. The effectiveness and functionality of the system are demonstrated through a case study, and the code and demos are available at a GitHub Repository: https://github.com/YuchenXia/LLMDrivenSimulation**

***Keywords—Digital Twins, Large Language Models, Multi-Agent System, Simulation, Intelligent Automation.***

## I. Introduction

Digital twin technology has significantly improved productivity in industries such as engineering and manufacturing by providing virtual replicas of physical systems. Digital twin system offers an infrastructure that enables **users** to monitor, simulate, predict, and control real-world processes in real-time, thereby improving decision-making and enhancing operational efficiency.[1]

Currently, users are responsible for interpreting data from digital twins, understanding complex system behaviors, and making informed decisions. Human users can only process a limited amount of information at a time, constraining the efficiency and usability of digital twin systems. Processing information requires cognitive effort and specialized knowledge, often necessitating extensive training to ensure accurate understanding and effective control of the processes that is represented in digital twins. Additionally, the availability of such skilled users can be limited, and the high knowledge barrier can lead to delays and compromised decision-making.

Large Language Models (LLMs) have demonstrated significant intelligence in understanding knowledge. Incorporating the intelligence of LLMs with digital twin technology allows for a higher degree of automation and system intelligence, presenting new opportunities to enhance digital twins.

Given the critical role users play in interacting with digital twins to monitor and predictively control processes, a pertinent question arises: Can this user role be assisted by large language models to automate the interaction with digital twin simulation, thus determining feasible parametrizations for controlling the simulated processes?

To answer this question, the key contributions of this paper include:

- **Multi-agent framework:** We present a novel LLM-agent architecture that dynamically interacts with digital twin simulation to determine feasible parameter settings for simulated physical processes.
- **Proof-of-concept:** A case study demonstrates the system's capability to automate simulation parametrization, employing heuristic knowledge reasoning from LLMs.
- **Industrial applications:** the results suggest that integrating LLMs with digital twin technologies can enhance the intelligence and user-friendliness of industrial digital twin systems, improving operational efficiency and accessibility.

## II. Background

Figure 1 selectively illustrates the key conceptual components in this work. These components include a physical entity under study, its virtual counterpart in the form of a digital twin, and an LLM multi-agent system designed for intelligent interaction.

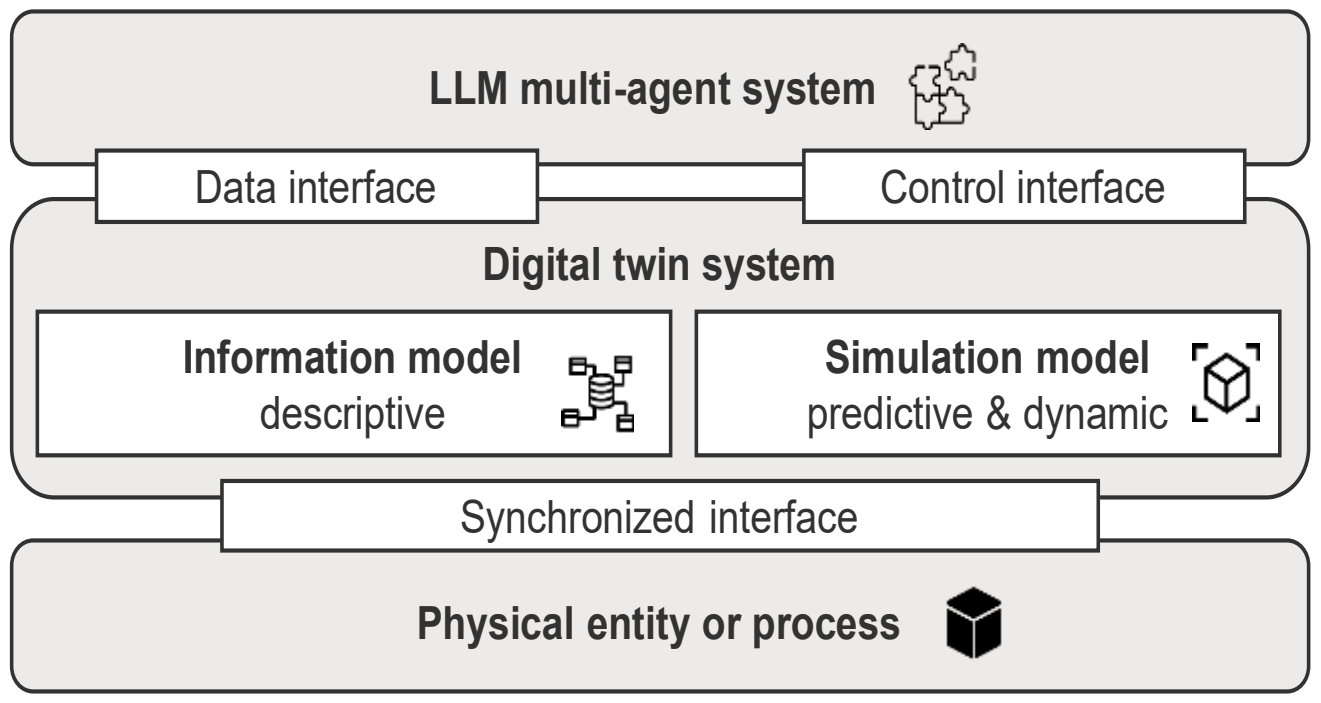

Figure 1 Key conceptual components in this work

### A. Digital twin

While definitions of digital twins differ in similar details [1][2], they converge on key aspects: A digital twin is a software system that represents a physical entity or system that mirrors real-world conditions, processes, and systems. It remains synchronized with its physical counterpart through interfaces that manage data acquisition and control physical processes (e.g., through an automation system) [3]. In the context of this paper, two primary parts of the digital twin are selectively considered:

**Information Model:** This component focuses on the structured integration and management of descriptive data. It serves as the core of data storage and retrieval within the digital twin system, facilitating access and processing of detailed information about the physical entity.[4]

**Simulation model:** In contrast to the information model, the simulation component is predictive and dynamic. It is primarily used for operational testing and scenario analysis, enabling the exploration of potential outcomes based on various inputs and conditions.[5]

The digital twin serves as a critical digital interface for the LLM system. It provides LLM agents with access to information about physical processes, thereby enabling the LLM system to develop intelligent applications based on these data.

### B. *"Divide and conquer" and LLM multi-agent system*

Due to the complexity of the task or the complex data processed from the digital twin, a single LLM often cannot produce satisfactory results. To address this, complex tasks are decomposed into smaller, more manageable components. This "divide and conquer" approach allows multiple LLM agents to collaboratively tackle these tasks, leading to a comprehensive and effective solution. This principle can be realized by designing LLM multi-agent systems.

### C. *Related works*

Several studies have explored the development of LLM-powered applications, often applying the "divide and conquer" principle to design these systems. These can be mainly organized in two level:

**Model level:** To manage complexity effectively, several frameworks have been developed. The Chain-of-Thought (CoT)[6]framework facilitates systematic, step-by-step reasoning, while the Tree-of-Thought (ToT)[7] allows LLMs to explore multiple pathways, identifying the most viable solutions. Additionally, the ReAct framework [8] generates reasoning traces and task-specific actions in an interleaved manner, effectively addressing complex tasks. These frameworks can be realized by systematic prompting to guide the reasoning processes of LLMs, enhancing their capability to solve intricate problems.

Alternative methods involve fine-tuning the LLM to modify its text generation behavior. Techniques such as Proximal Policy Optimization (PPO)[9] and Direct Policy Optimization (DPO)[10] adjust the generation policies of the LLM, enhancing its ability to produce logically connected texts and thereby improve its problem-solving capabilities.

**Agents level:** Recent studies [11],[12] have explored frameworks to design LLM agents playing different roles in task-oriented coordination within simulated social environments, emphasizing reflection and decision-making processes. Another study introduces a framework that enables LLMs to act as game players controlling real-time strategy games by analyzing complex data and managing game dynamics [13]. In our prior work, we developed a multi-agent system to autonomously plan and control automation systems via digital twin system [3]. Additionally, a comprehensive survey summarized in [14] reviews recent advancements in LLM multi-agent systems.

In the following section, we present a novel multi-agent system framework to interact with simulation models in digital twin to perform experiments to determine satisfactory parametrization of a simulated process.

## III. THEORETICAL FRAMEWORK OF A MULTI-AGENT SYSTEM

In this work, we propose a structured framework for multi-agent system with that assigns distinct roles and responsibilities to each LLM agent within the information processing pipeline. This framework includes four types of agents: observation, reasoning, decision and summarization.

**Observation agent:** This agent collects and observes real-time data from the digital twin, focusing on operational parameters, current conditions, and changes in key metrics crucial for optimal outcomes. It identifies pertinent observation data, filters out noise and irrelevant information, and distills important insights to establish a factual foundation for further analysis.

**Reasoning agent:** Incorporating data preprocessed by the observation agent, the reasoning agent interprets and analyzes the observed data and insights. It generates reasoning steps towards actionable decisions. This process can be characterized as heuristic reasoning, as it utilizes knowledge patterns that the LLM has acquired from extensive training data [Ref.].

**Decision agent:** Upon completion of the reasoning phase, the decision agent generates executable actions based on the previously generated intermediate results. These generated actions can be formulated as API calls, procedure calls or services in a software system to invoke operations.

**Summarization agent:** The summarization agent consolidates the outcomes of the observation, reasoning, and decision processes. It compiles a concise report highlighting the significant insights and results, providing an overview of system performance. This agent helps users quickly understand key information about the system's behavior.

This multi-agent system design framework resembles the scientific methodology used in empirical experiments. It provides a structured and reproducible approach for creating LLM multi-agent systems. By clearly defining roles and responsibilities in information processing, this framework aids in the systematic development, rigorous testing, and iterative improvement of systems powered by LLMs.

## IV. METHODOLOGY

### A. *System overview*

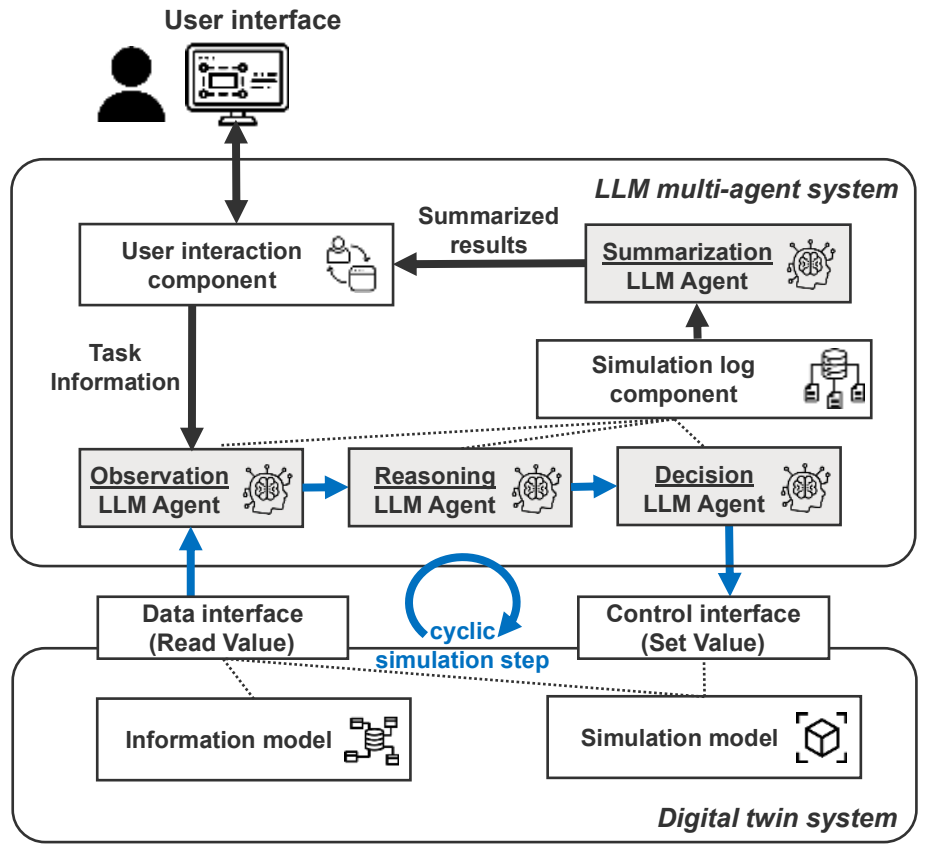

Figure 2 Architecture overview of LLM multi-agent system interacting with simulation model in a digital twin

As shown in Figure 2, this architecture comprises three layers. At the top layer, the user interface receives user requirements and the objectives of the simulation, which are set as target goals for the simulation to achieve by its conclusion. In the middle layer, an LLM multi-agent system interfaces with digital twins via data and control interfaces. At the fundamental layer, the digital twin simulation model is structured to operate sequentially. During each cyclic simulation step, data is accessed through a data interface, and simulation parameters are adjusted via a control interface.

The **observation agent** preprocesses this information to extract significant insights and distill crucial details from the extensive data retrieved, setting the stage for the next agent toward achieving targeted goals. Subsequently, the **reasoning agent** applies heuristic reasoning to analyze the situation and deduce the next viable pathways, and the **decision agent** generates actions in the form of parametrized function calls to adjust the simulation's parameters for the upcoming step. Collectively, the agents monitor the cyclic steps of the simulation and strategically control its progression.

Data from each cycle and the corresponding control parameters are recorded in a simulation log component, which captures detailed information about the LLM agents' iterative decision-making processes. The **summarization agent** compiles these logs for each simulation steps to develop a concise parametrized sequential control plan, which will be reported to the user as a feasible solution to a user task.

### B. *Information representation conversion between digital twins and LLM agents*

#### *1) Information processing of LLM agents*

In this multi-agent system, each agent functions as an information processing component. A specific prompt is crafted for each agent to direct its behavior according to the sub-task it handles within the multi-agent framework (cf. Figure 3). Both input and output information are converted into textual form. This conversion requires a translation process that transforms modeled information from the digital twin simulation model into a textual knowledge representation. This representation is then integrated into a prompt template. A concrete example of this process is demonstrated in the case study section.

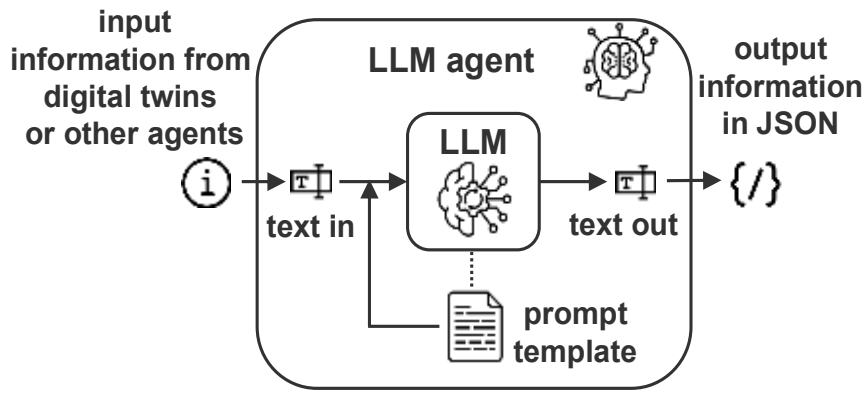

Figure 4 LLM agent as an information processor through Prompting

#### *2) JSON and function call as agent output*

The generated text is converted into JSON format to facilitate technical processing by the software. At the end of the agent information processing pipeline, the decision agent produces a parameterized function call that serves as an actionable decision. This function call is managed by the simulation model's interface, which executes the operations within the simulation model for the next simulation step.

## V. A Case Study and Implementation

For a practical demonstration of the concept, we implemented a simplified container mixing process. This simulation is akin to mixing process in a blender, where the objective is to achieve a homogeneous mixture of balls (or particles) with varying densities. The balls are added to the container in a controlled sequence, and then the container is shaken to redistribute the contents.

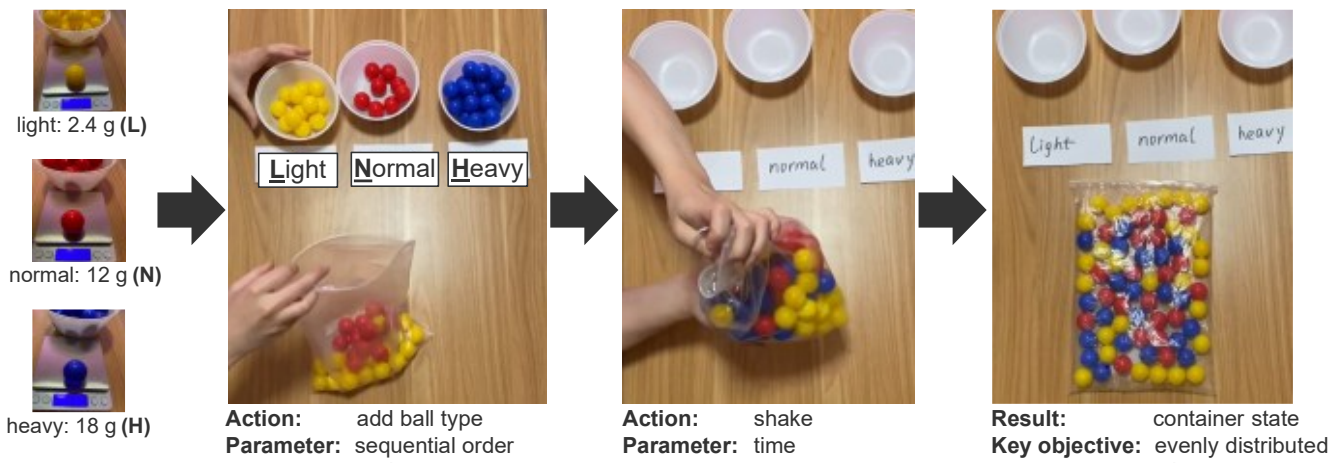

Figure 3 The simulated physical process

The case study experiments, depicted in the Figure 4, demonstrate how the order of adding the balls (light, normal, heavy) and the duration of shaking affect the mixture's homogeneity.

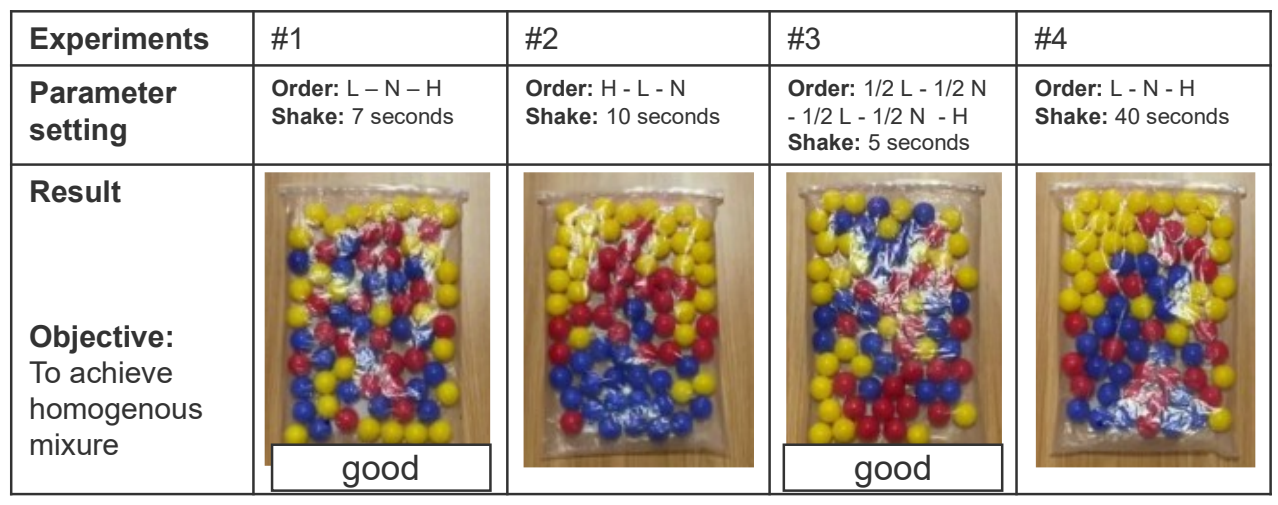

| Experiments | #1 | #2 | #3 | #4 |
|---|---|---|---|---|
| Parameter setting | **Order:** L – N – H **Shake:** 7 seconds | **Order:** H - L - N **Shake:** 10 seconds | **Order:** 1/2 L - 1/2 N - 1/2 L - 1/2 N - H **Shake:** 5 seconds | **Order:** L - N - H **Shake:** 40 seconds |
| Result<br><br>**Objective:** To achieve homogenous mixure | good | | good | |

Figure 5 The experimenting results with the physical process

We parametrized these variables to explore different combinations and their impact on the mixing outcome. Several experiments' parameter settings and results are summarized, as shown in Figure 5.

### A. *Simulation model for the case study*

In the simulation, we model a container as a 10x10 matrix where two primary actions are possible: adding different types of balls and shaking the container. With each shake of the container, there is a probability of heavier balls switching positions with nearby lighter balls beneath. The simulation begins with the user setting a goal to achieve an "evenly distributed mixture", as shown in Figure 6. The interaction with the LLM multi-agent system is facilitated through a chat box, and the simulation's progress is visualized on the user interface.

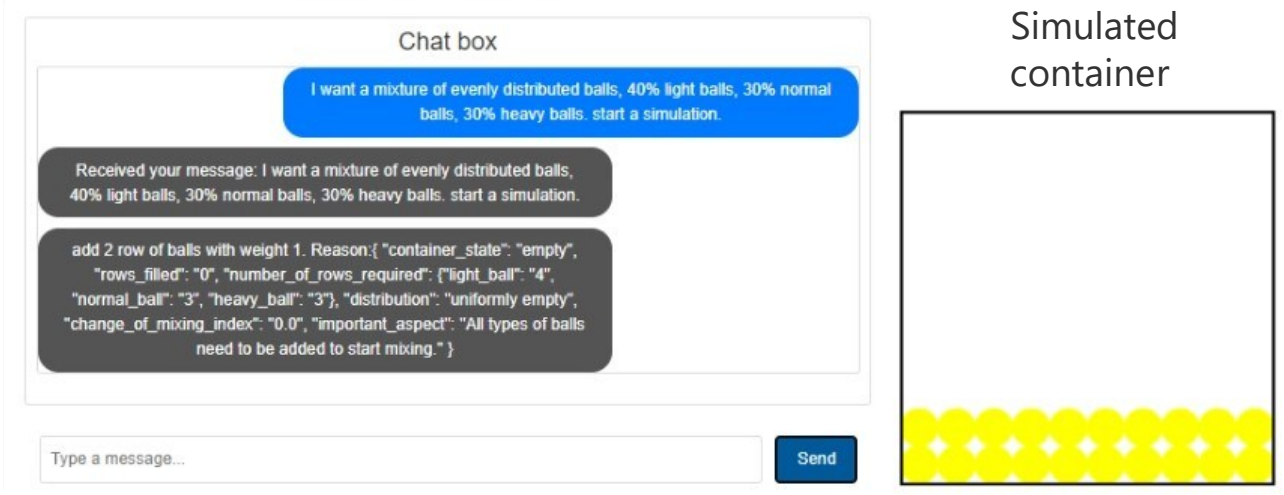

Figure 6 The user interface with chat box and visualized simulation

The LLM multi-agent system then takes over, observing, reasoning, and controlling the simulation steps to achieve the desired outcome. The final result, including a visual representation of the outcome and a summary of the control parameters used, is then presented to the user in a concise format, as illustrated in Figure 7.

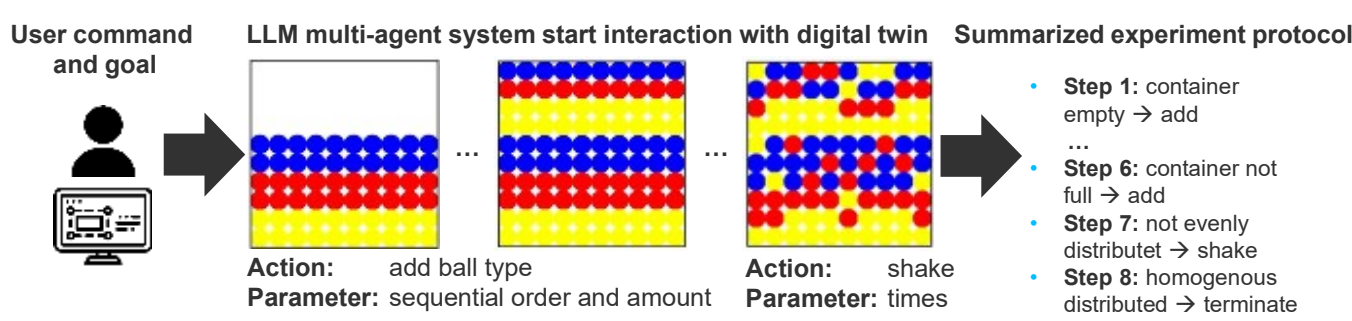

Figure 7 The LLM enhanced simulation execution process

### B. *Knowledge representation in text*

To realize this application, a key aspect is the conversion of knowledge representation: translating the simulation representation into a textual format that the LLM can interpret while preserving the knowledge. In the case study, the state of the container is represented as a matrix, where each position within the matrix is denoted by a number, as shown in Figure 8. Each number corresponds to a specific type of ball, which is explicitly defined within the agent's prompt. This textual conversion is critical for allowing the LLM to perceive and interact with the simulation data meaningfully.

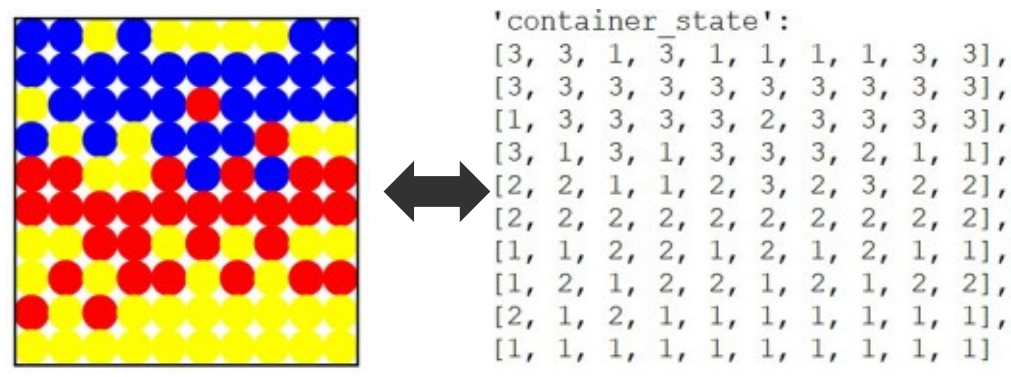

Figure 8 Conversion of simulation information into textual knowledge representation

To enhance the LLM's ability to assess the distribution within the mixture, we have integrated a quantifiable proxy metric. This metric quantitatively measures the degree of homogeneity in the distribution, and this indicator is integrated also into the prompt:

$$Degree\ of\ even\ distribution = \frac{\sum Diversity\ of\ Neighbor}{Number\ of\ balls}$$

## VI. RESULTS AND DISCUSSION

We developed this system, the code and demo video is released on the GitHub repository. The LLM operates similarly to a human experimenter within a virtual environment, interfacing with a digital twin to achieve an objective.

| **Simulation run** | #1 | #2 | #3 | #4 |
|---|---|---|---|---|
| **Parameter setting** | **Order:** 40% L – 20% H – 30% N – 10% H **Shake:** 5 times | **Order:** 20% L – 20% N – 20% H – 20% L – 10% N – 10% H **Shake:** 1 time | **Order:** 20% L – 20% N – 30% H – 20% L - 10%N **Shake:** 2 times | **Order:** 40% L – 20% H – 30% N – 10% L **Shake:** 5 times |
| **Result** | | | | |

Figure 9 The outcomes of experiments from the interaction between LLM and digital twin simulation

The system is capable of executing multiple simulation runs, allowing the LLM multi-agent system to explore a range of parameter settings and determine the most effective configurations, as demonstrated in Figure 9. This iterative process mirrors the method by which humans apply heuristic reasoning to experiment with various configurations in search of an optimal solution.

The integration of LLMs into the digital twin framework offers several advantages. Firstly, it enhances user experience by making interactions more intuitive and automating the parametrization process. Secondly, it lowers the knowledge barrier, making advanced digital twin technologies accessible to a wider range of users. This automation significantly reduces the need for manual effort in interpreting complex details and performing reasoning tasks, thereby accelerating the entire process. Moreover, utilizing a simulated environment for testing provides a risk-free setting for comprehensive scenario analysis, which minimizes potential operational risks associated with real-world testing. Finally, the combined predictive capabilities of LLMs and the simulation framework, allowing for the more efficient strategic planning and process control.

## VII. CONCLUSION AND OUTLOOK

This research demonstrates the viability of integrating Large Language Models (LLMs) into digital twin frameworks to automate simulation parametrization. By enabling intelligent multi-agent interactions within simulations, LLMs improve the efficiency, accessibility, and strategic planning capabilities of digital twins. The iterative, heuristic-based approach utilized by LLM agents not only mirrors human problem-solving techniques but also optimizes operational strategies in real-time. Moving forward, we plan to conduct further testing and evaluations and to refine the system and enhance simulation accuracy. We expect that continued advancements will expand the application scope of LLM-enhanced digital twins, potentially leading to more efficient and productive systems across various industrial applications.

## ACKNOWLEDGMENT

This work was supported by *Stiftung der Deutschen Wirtschaft (SDW)* and the Ministry of Science, Research and the Arts of the State of Baden-Wuerttemberg within the support of the projects of the *Exzellenzinitiative II*.
Special thanks to Yuye Tong for assisting with the preparation of the experiment materials.

## REFERENCES

[1] D. Dittler, D. Braun, T. Müller, V. Stegmaier, N. Jazdi, and M. Weyrich, "A procedure for the derivation of project-specific intelligent Digital Twin implementations in industrial automation," May 2022.

[2] B. Ashtari Talkhestani *et al.*, "An architecture of an Intelligent Digital Twin in a Cyber-Physical Production System," *At-Automatisierungstechnik*, Sep. 2019, doi: 10.1515/AUTO-2019-0039.

[3] Y. Xia, M. Shenoy, N. Jazdi, and M. Weyrich, "Towards autonomous system: flexible modular production system enhanced with large language model agents," in *2023 IEEE 28th ETFA*, 2023. doi: 10.1109/ETFA54631.2023.10275362.

[4] Y. Xia, Z. Xiao, N. Jazdi, and M. Weyrich, "Generation of Asset Administration Shell with Large Language Model Agents: Interoperability in Digital Twins with Semantic Node," Mar. 2024, [Online]. Available: https://arxiv.org/abs/2403.17209v1

[5] P. Häbig *et al.*, "A Modular System Architecture for an Offshore Off-grid Platform for Climate neutral Power-to-X Production in H2Mare," May 2023, [Online]. Available: https://arxiv.org/abs/2305.16285v1

[6] J. Wei *et al.*, "Chain of Thought Prompting Elicits Reasoning in Large Language Models," *NeurIPS*, 2022.

[7] S. Yao *et al.*, "Tree of Thoughts: Deliberate Problem Solving with Large Language Models," *NeurIPS*, 2023, doi: 10.48550/ARXIV.2305.10601.

[8] S. Yao *et al.*, "ReAct: Synergizing Reasoning and Acting in Language Models," Oct. 2022, [Online]. Available: https://arxiv.org/abs/2210.03629v3

[9] J. Schulman, F. Wolski, P. Dhariwal, A. Radford, and O. K. Openai, "Proximal Policy Optimization Algorithms," Jul. 2017, [Online]. Available: https://arxiv.org/abs/1707.06347v2

[10] R. Rafailov, A. Sharma, E. Mitchell, C. D. Manning, S. Ermon, and C. Finn, "Direct Preference Optimization: Your Language Model is Secretly a Reward Model," *Advances in NeurIPS*, Dec. 2023.

[11] J. C. Sung Park Joseph O *et al.*, "Generative Agents: Interactive Simulacra of Human Behavior," *UIST 2023 - Proc. of the 36th Annual ACM Symp. on User Interface Software and Technology*, Oct. 2023, doi: 10.1145/3586183.3606763.

[12] Y. Li, Y. Zhang, and L. Sun, "MetaAgents: Simulating Interactions of Human Behaviors for LLM-based Task-oriented Coordination via Collaborative Generative Agents," Oct. 2023, [Online]. Available: https://arxiv.org/abs/2310.06500v1

[13] W. Ma *et al.*, "Large Language Models Play StarCraft II: Benchmarks and A Chain of Summarization Approach," Dec. 2023, [Online]. Available: https://arxiv.org/abs/2312.11865v1

[14] T. Guo *et al.*, "Large Language Model based Multi-Agents: A Survey of Progress and Challenges," Jan. 2024, Accessed: May 24, 2024. [Online]. Available: https://arxiv.org/abs/2402.01680v2

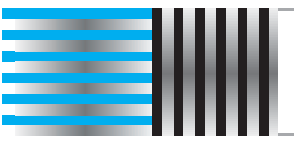

RESEARCH ARTICLE

# Generation of Asset Administration Shell With Large Language Model Agents: Toward Semantic Interoperability in Digital Twins in the Context of Industry 4.0

**YUCHEN XIA, (Member, IEEE), ZHEWEN XIAO, (Member, IEEE), NASSER JAZDI, (Senior Member, IEEE), AND MICHAEL WEYRICH, (Member, IEEE)**

Institute of Industrial Automation and Software Engineering, University of Stuttgart, 70550 Stuttgart, Germany

Corresponding authors: Yuchen Xia (yuchen.xia@ias.uni-stuttgart.de) and Michael Weyrich (michael.weyrich@ias.uni-stuttgart.de)

This work was supported in part by the Stiftung der Deutschen Wirtschaft (SDW); and in part by the Ministry of Science, Research and the Arts of the State of Baden-Wuerttemberg through the Projects of the Exzellenzinitiative II.

**ABSTRACT** This research introduces a novel approach for achieving semantic interoperability in digital twins and assisting the creation of Asset Administration Shell (AAS) as digital twin model within the context of Industry 4.0. The foundational idea of our research is that the communication based on semantics and the generation of meaningful textual data are directly linked, and we posit that these processes are equivalent if the exchanged information can be serialized in text form. Based on this, we construct a "semantic node" data structure in our research to capture the semantic essence of textual data. Then, a system powered by large language models is designed and implemented to process the "semantic node" and generate standardized digital twin models (AAS instance models in the context of Industry 4.0) from raw textual data collected from datasheets describing technical assets. Our evaluation demonstrates an effective generation rate of 62-79%, indicating a substantial proportion of the information from the source text can be translated error-free to the target digital twin instance model with the generative capability of large language models. This result has a direct application in the context of Industry 4.0, and the designed system is implemented as a data model generation tool for reducing the manual effort in creating AAS model by automatically translating unstructured textual data into a standardized AAS model. The generated AAS model can be integrated into AAS-compliant digital twin software for seamless information exchange and communication. In our evaluation, a comparative analysis of different LLMs and an in-depth ablation study of Retrieval-Augmented Generation (RAG) mechanisms provide insights into the effectiveness of LLM systems for interpreting technical concepts and translating data. Our findings emphasize LLMs' capability to automate AAS instance creation and contribute to the broader field of semantic interoperability for digital twins in industrial applications. The prototype implementation and evaluation results are presented on our GitHub Repository: https://github.com/YuchenXia/AASbyLLM.

**INDEX TERMS** Asset administration shell, large language model, semantic interoperability, digital twin, industry 4.0, generative AI, retrieval-augmented generation.

## I. INTRODUCTION

In a digital twin system, different system parts may represent entities and model data in different ways. These variations can cause interoperability issues, as data exchanged between different parts of the system may not be compatible due to mismatches in modeling and interface specifications. Consequently, communication becomes impossible without a common understanding of data. Typically, resolving this issue involves mapping data points between the various system parts. This mapping process often becomes impractically complex due to its nature of handling combinatorial complexity of O(C(n,2)), approximately $O(n^2)$, where '*n*' represents the different modeling methods.

Industry 4.0 adopts a standardization approach to address this interoperability issue. In this context, the Asset Administration Shell (AAS) [1] is a key concept designed to enable seamless interoperability within a digitalized manufacturing environment. AAS provides a standardized method for modeling information and interfaces associated with a physical asset, such as a machine, a component, or a product. If all digital twin models conform to the standardized AAS specifications, seamless communication and data exchange can be facilitated. This standardization simplifies the communication process, reducing its complexity to a linear scale, O(n), where each model and communication protocol aligns with Industry 4.0 standards [1]

Participants and users within Industry 4.0 often possess vast amounts of data. However, transforming company-specific data into the standardized Asset Administration Shell (AAS) format and manually creating AAS instance models remains a challenging and labor-intensive process. To improve the cost-benefit ratio and encourage broader adoption of the AAS technology stack, it is essential to develop new solutions for automated AAS model creation [2]. Researchers advocate in [3], [4], [5] for the development of new solutions for automated AAS creation based on available heterogeneous engineering information, and [6] poses scientific inquiries regarding how LLMs can be utilized in digital twin engineering. Automating the translation process—from existing vendor-specific information models to standardized AAS models—presents a viable and promising solution to enable interoperability in digital twins with lower costs [3], [4], [5], [6]

### A. KEY FOUNDATIONAL CONCEPTS FOR THE SOLUTION

To handle these problems, we focus on three essential conceptual building blocks: textual data serialization, semantics interpretation, and large language models.

**Textual data serialization**: digital twins' data can typically be converted into textual formats, such as code, structured data exchange files, or plain natural language text. This conversion into text makes it technically feasible to use Large Language Models (LLMs) for processing information within digital twins.

**Semantic interpretation**: the primary focus in information processing and utilization should be based on the meaning conveyed by the data, rather than the raw data itself. Systems can communicate effectively when the data communicated is clearly defined and disambiguously interpreted.

**Large language models**: they are pre-trained to acquire the knowledge patterns to process the conceptual meaning of text. This semantic interpretation capability is evidenced by studies of their internal neuron structures and behaviors [7], [8].

The combination of these three conceptual building blocks suggests that if LLMs can offer capabilities for semantic interpretation, then the information in digital twins can be seamlessly utilized on a semantic level, overcoming challenges such as raw form differences, modeling format misalignments, or vocabulary discrepancies. Reducing these barriers will enhance the interoperability of technical systems, achieving what is referred to as "semantic interoperability", where systems exchange information and operate based on the meaning of the data. Consequently, this will lead to the development of a translation mechanism that streamlines digital information flows in technical information systems.

### B. APPLY LLM TO GENERATE DIGITAL TWIN MODELS IN THE CONTEXT OF INDUSTRY 4.0

Figure 1 articulates the key concepts and the relationships introduced so far. In this paper, we aim to create a theoretical framework that employs LLMs to interpret text-based data and to translate these data according to the target system specification to achieve semantic interoperability in digital twin systems. To concretize and realize the theoretical general framework with an empirical investigation, we use the Asset Administration Shell (AAS) for a case study, which is a standard implementation of digital twins for technical assets under the German initiative "Platform Industry 4.0" [1]. AAS standards facilitate the bi-directional conversion of AAS digital twin models into serialized JSON or XML formats. Based on this foundation, LLMs can directly generate the data elements in these serialized forms and merge the generated model into a digital twin system. An LLM system is designed to process textual data collected from technical data sheets and generate AAS models that digitally represent technical components (sensor, actuator, controller, and network devices) in accordance with the standardized specifications for AAS digital twins.

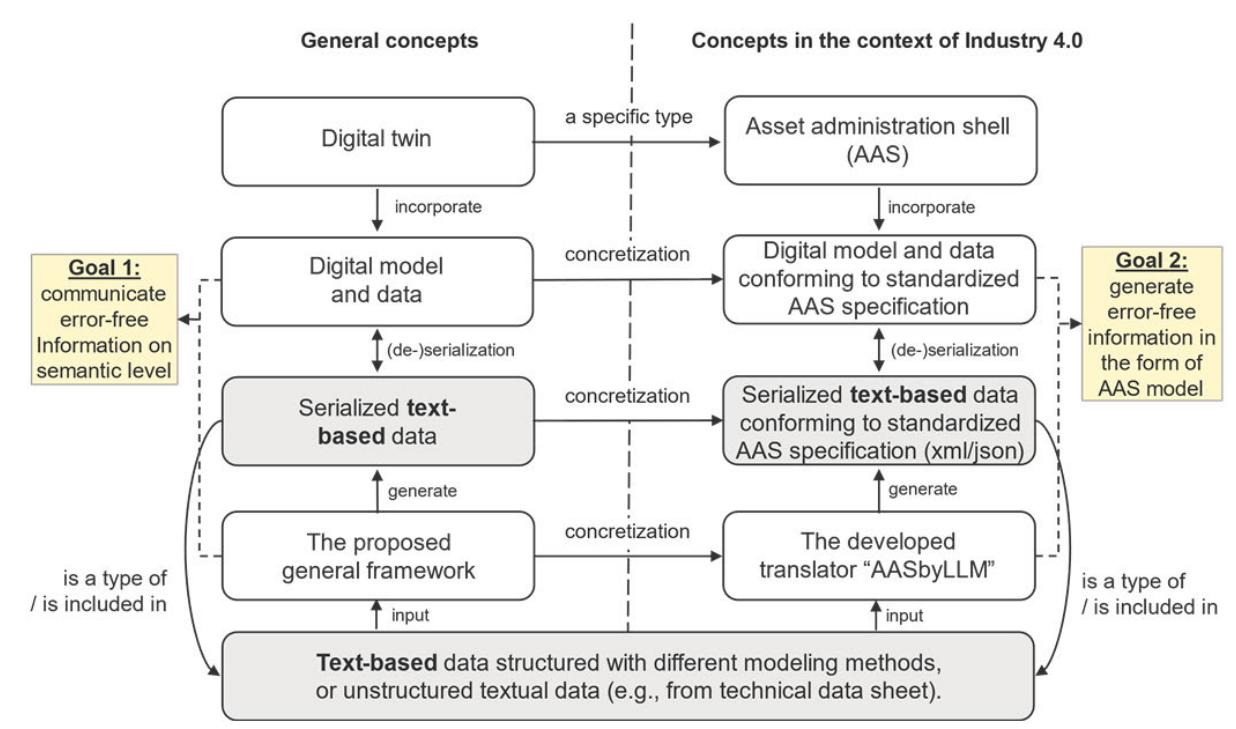

**FIGURE 1. Concept map: This paper's investigation is contextualized within the digital twin concept asset administration Shell in Industry 4.0 and maintains relevance to general digital twins.**

The foundational premise for the investigation of this paper is established on the logical equivalency of two goals:

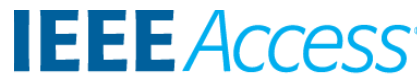

**TABLE 1.** Literature review on automated AAS model generation.

| Paper | Purpose | Transformation methods | Transformed data type / source format | Semantics in data |
|---|---|---|---|---|
| **Category 1: Using pre-defined rule-based mapping** | | | | |
| [9] | Model Transformation for AAS | Model Transformation Language based on OCL | Software model / UML | Rely on static model semantics in the context of model driven software engineering |
| [10][11] | Interoperable digital twins in IIoT systems | Detailed mapping model for conversion to AAS format | Proprietary information model / UML | Reference on ECLASS dictionary for data definition |
| [12] | AAS creation from heterogeneous Data | Python implementation for extracting and mapping engineering information | Engineering information / PDF, STP, XML, XML/AutomationML, RDF | Mentioned semantic web and semanticID in AAS with ECLASS reference |
| [13] | Mapping between AAS submodels and skill ontology CaSkMan | Rule-based mappings using RML and RDFex | CaSkMan ontology / RDF, OWL | Formal semantic in ontology |
| [14] | Semantic interoperability for AAS-based digital twins using ontologies | Ontology Modeling Language (OML) to map AAS models to OWL | Concepts in Manufacuturing Resource Capability Ontology / OWL, UML | Connecting ontology vocabularies semantics with system models |
| [15] | Interoperability between DTDL and AAS | Mapping between the Digital Twin Definition Language (DTDL) and AAS | DTDL metamodel elements / UML | Semantic annotation in JSON-LD and AAS |
| [16] | Generation of digital twins for information exchange | Formalizes the AAS meta-model in a domain-specific language and an intermediate representation | AAS model / XSD, JSON schema, RDF | Mapping based on semantic equality |
| [17] | Improve semantic discoverability | Establish a set of rules for converting RDF-based models into AAS models | RDF-based models / RDF, SHACL, OWL, SPARQL | RDF as a semantic base for AAS |
| [18] | Generating test cases to verify AAS server implementations | Transform AAS-based plant models into MaRCO Ontology instances | AAS model / UML, RDF, OWL, SPARQL | Semantic interoperability based on ontologies |
| [19] | Provide data in a standardized and semantically described manner | Mapping semantic descriptions from OPC UA information models into the AAS model | OPC UA information models / OPC UA nodes, JSON | Specified semantics in OPC UA standards |
| [20] | Model the semantics of the current operation, status and configurations of assets. | A mapping ruleset to create the semantic AAS as an RDF graph aligned with the RAMI4.0 vocabulary | AAS model / JSON, RDF, SHACL, SPARQL | Using ontological modeling and semantic technologies. |
| [21] | Generating AAS from engineering data | Collect data models into AutomationML and map attributes into the AAS format | Engineering data / AutomationML | Semantics are specified as roles and relations between objects in AutomationML |
| [22] | Interoperability in manufacturing systems for exposing information | Map XML elements to the AddressSpace of an OPC UA | AAS model elements into OPC UA node / XML | Standard semantics in OPC UA model |
| **Category 2: Applying NLP methods, specifically embedding language models** | | | | |
| [2] | Data migration into AAS | Using embedding language model to generate mappings | Technical properties / textual data | Semantic embedding with language models, ECLASS dictionary |
| [23], [24], [25] | Semantic interoperability in heterogeneous AAS | Using trained embedding language model to generate mappings | Technical properties / knowledge graph, textual data | Semantic embedding with language models, ECLASS dictionary |
| [26] | Map technical operational data for building monitoring | Using trained embedding language to automatically map heterogenous data points | Technical properties / Information model from various communication protocols | Semantic embedding with language models, ECLASS dictionary |

> ***Goal 1:*** ***achieving error-free semantic communication from a source to a target***
> *is logically equivalent to*
> ***Goal 2:*** ***ensuring error-free generation of data required by the target based on the meaning of the source data.***

This investigation serves two primary purposes: First, it demonstrates that our proposed general theory can be technically realized and empirically validated through a representative case study. Second, the software application we implemented translates universal text-based raw data into standardized AAS digital twin models, serving as a tool for automating data translation and enhancing the efficiency of AAS model creation. Finally, the generated AAS model can be integrated into AAS-compliant digital twin software (specifically, AASX Package Explorer) for seamless information exchange.[1]

[1]Prototype at: https://github.com/YuchenXia/AASbyLLM.

### C. PAPER STRUCTURE AND OUR CONTRIBUTION

This paper begins with reviewing existing approaches employed to convert diverse, heterogeneous information into AAS models. It provides a comprehensive review of their applied methods, along with an in-depth discussion of the strengths and weaknesses (c.f. **Table 1**).

Drawing insights and departing from these existing approaches, we introduce our novel theoretical framework that utilizes the large language model to analyze the semantics of textual data in technical information systems and translate them into a self-contained semantic unit called "semantic node" (Section III). We provide a narrative of development in LLM technology to justify why the LLM can be used to process the semantics of textual data (Section IV).

Diving deeper into the practical value of our proposed framework, we conducted a case study using the semantic node and LLM agent system to automate the generation of AAS instance models from raw text obtained from technical data sheets of automation components. Furthermore, we developed a translation software called "AASbyLLM" (cf. **Figure 2**), where users can use this implemented AI tool via web-browser to generate AAS instance model with their own data (Section V). The developed system is comprehensively tested and evaluated (Section VI), and the results are analytically discussed and summarized (Section VII).

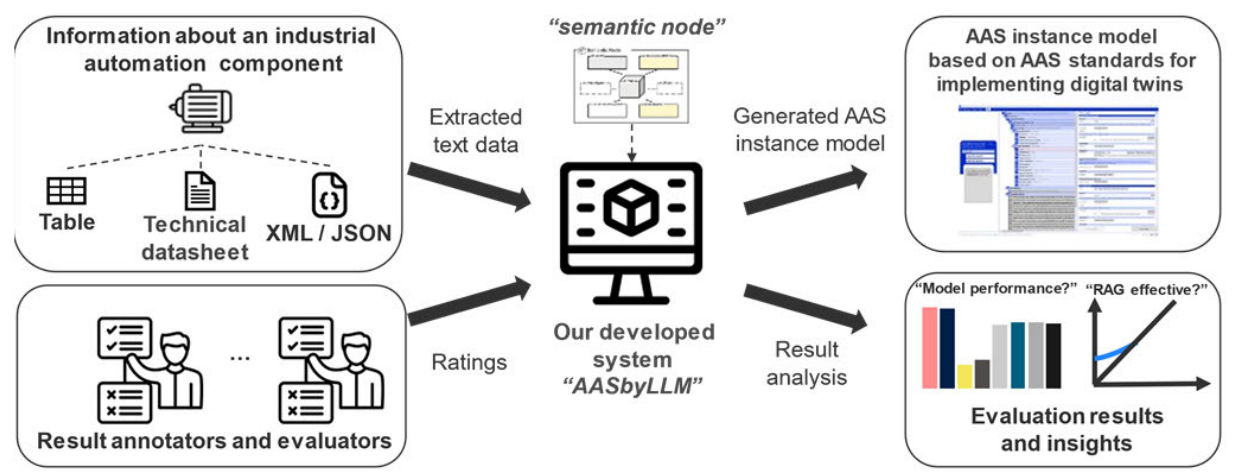

**FIGURE 2.** The presented system "AASbyLLM" in this work has two parts: an implemented prototype that delivers the proof-of-concept and an evaluation benchmark based on human feedback.

The main messages and contribution of the paper are:

- **Theory:** We conceptualize a framework to utilize the large language model to interpret the serialized textual data from digital twins to achieve semantic interoperability and construct a data structure called "semantic node" for capturing the essential semantics during the information processing. A general approach for creating LLM agents system to realize this process is designed and presented.
- **Implementation and Utility:** We show how the designed system can be implemented with an LLM agent system for processing information and generating AAS instance models. By manipulating the "semantic node", an information processing system powered by LLMs can generate informative and correct AAS instance models automatically with 62-79% effective generation rate based on human evaluation, and this indicates that at least a significant proportion of manual creation efforts can be converted into simpler validation efforts, reducing the cost of AAS instance creation. We publish our evaluation dataset and host a web-service for an end-to-end testing of the thoroughness of this approach.
- **Evaluation & Insights**: We performed model-level comparative analysis and ablation study in our evaluation. The results show that both the proprietary model from OpenAI and the open-source models can deliver satisfactory results. Our statistical analysis also revealed the conditional effectiveness and limitations of RAG mechanisms, leading us to propose the "Cheat Sheet Effect of RAG" hypothesis to elucidate these findings. This provides insight into how to design an LLM system effectively.

## II. RELATED WORKS TO AAS INSTANCE CREATION

The Asset Administration Shell (AAS) is a specific type of digital twin that acts as a structured and standardized digital representation of an asset in the context of Industry 4.0. It includes detailed submodels describing the asset's features, properties, and functions, facilitating interoperability across various applications within digitalized manufacturing systems [1].

Based on our literature review, we identified 19 papers closely related to semantic interoperability in the context of Industry 4.0 digital twin modeling and the automated creation AAS instances [2], [9], [10], [11], [12], [13], [14], [15], [16], [17], [18], [19], [20], [21], [22], they are compared and summarized in **Table 1** for better overview. After examining the broad-spectrum methods used to create AAS model and their commonalities, we distilled the essential aspects of these studies into two key categories:

### A. RULE-BASED AUTOMATED CONVERSION METHODS

The most methods focus on the automated conversion between AAS and various information sources through precisely defined and clear-cut mappings and explicitly outlined target and source models. This approach ensures structured data translation but requires significant initial setup for rule definition and managing the precise mappings between semantics between source and target models.

### B. SEMANTIC ENHANCEMENT OF AAS

This emphasizes enhancing data communication through identifiable semantic descriptions of data. It involves assigning clear semantic identifiers from standardized dictionaries or adding detailed descriptions to data elements.

These approaches connect to the concept of 'semantic interoperability', which seeks to overcome the limitations posed by varying data formats and representation styles through identifying the semantic equivalency of data.

## III. GENERAL FRAMEWORK OF THIS WORK

We define a principal proposition of "semantic communication":

> *Effective communication between two systems can be achieved provided that the data being communicated is clearly defined and disambiguously interpreted.*

As listed and summarized in **Table 1**, the most methods [9], [10], [11], [12], [13], [14], [15], [16], [17], [18], [19], [20], [21], [22] utilized the rule-based mapping to bridge the discrepancy of different modeling approach to enable interoperability. Rule-based methods adhere to this proposition and these mappings eliminate ambiguity by clearly specifying how data from one model should be converted to another. However, the drawback of this approach is indeed having these rigorous rules in place. Firstly, this mapping is fixed and constrained by a source model and a target model, making it difficult to adapt to new or evolving data models, these lead to high cost in enumerating and managing of these rules (**inflexibility**). Secondly, there are often nuanced differences between elements of two models, and a rigorous mapping may force subtle changes and lead to semantic shift. This results in information distortion, and an intuitive example is that rule-based translation often falls short in translation tasks in NLP (**distortion**).

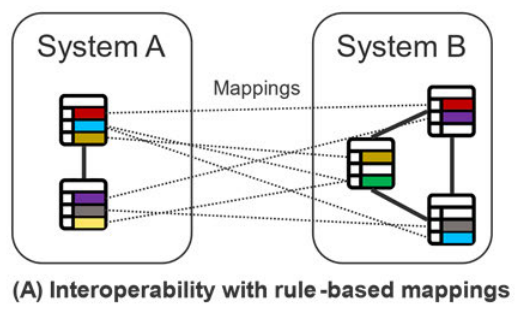

(A) Interoperability with rule-based mappings

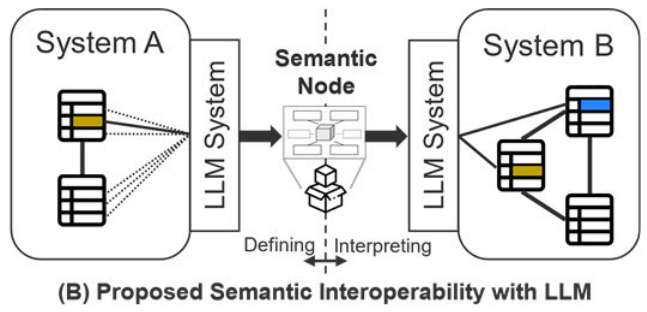

(B) Proposed Semantic Interoperability with LLM

**FIGURE 3. Interoperability between two systems with different modeling.**

The driving idea of this work is to depart from the rule-based transformation model, engage the idea of utilizing semantics, and explore a novel conceptual approach to translate information from one source technical information system to another target system. The theoretical framework of this approach is outlined in **Figure *3***, which emphasizes capturing the meaning of the data being communicated and encapsulates it into an atomic semantic unit called a ''semantic node'', as illustrated in **Figure 3 (B)**.

This implies that representation form and data format become of less concern, the disparities of diverse domain-specific languages or model specification can be overcome by semantic interpretation capability of LLM, thus achieving semantic interoperability. Following this principle, in this work, the processed data are enriched with meta-data that contribute to articulating the semantic meaning in this intermediate unit called ''semantic node''. This semantic node contains several text elements used to characterize the semantics of the communicated data: ''Name'', ''Value'', ''Conceptual definition'', ''Affordance'', ''Value type'', ''Unit'', ''Source description''. The concrete specification is detailed in **Table 2** in Section V-A. ''Semantic Node''.

## IV. BACKGROUND IN LLM TECHNOLOGY

Building on the proposed theoretical proposition of semantic communication, there remains a concrete question: How can the ''data being communicated'' be ''clearly defined'' and ''disambiguously interpreted''? To address this, we first provide a summary of the background of LLM technology and introduce the system design. We aim to conclusively answer this question by the end of this section.

### A. LARGE LANGUAGE MODELS INTERPRET SEMANTICS

In 2017, as OpenAI's scientist report in [27] their discovery of ''sentiment neuron'' – a single semantic unit in a neural network that precisely predicts customers' sentiment in review texts and only response to a specific meaning. This finding encourages further use of training objectives such as masked-language-modeling and next-token-predictions to learn high-quality concept representations from texts. Following this, the invention of transformer architecture [28], its integration into large language models, and the scaling-up of LLMs have led to further new interesting insights: in a LLM, the basic syntactic rules are learned in lower layer while high-level semantic concepts are learned in higher layer [29]; Moreover, 'poly-semantic neurons', which handle multiple unrelated concepts, enable the reuse of neurons in processing semantic through superposition, whereas 'mono-semantic neurons', representing singular meanings, enhance concept differentiation for precise detection and reasoning [8], [7]. These neurons form elemental units of LLM's processing ''circuits'' [7]. Furthermore, by investigating the correlation between neuron activation and output texts, semantic neurons can be identified [30].[2] These works provide evidence that LLMs possess an inherent structure enabling them to effectively interpret semantics.

### B. EMBEDDING LLM AND GENERATIVE LLM

The evolution of large language models has led to the emergence of two distinct types: embedding LLMs and generative LLMs. **Embedding LLMs** transform text into high-dimensional vector spaces, which can be used for indexing. Semantic relationships can be determined through mathematical operations. For example, semantic similarities between two sentences can be identified using embedding LLM, making them suitable for applications like search engines and recommendation systems. On the other hand, **generative LLMs** specialize in text production. They mostly operate on ''causal language modeling'' method, where each new token is predicted based on the preceding sequence, a process also known as ''auto-regression'' or ''next-token prediction''. This method is particularly effective for content generation, producing contextually coherent text.

### C. PROMPTING OF CAUSAL GENERATIVE LLM

A causal generative LLM operates by focusing on the important preceding tokens, predicting the next token based on probabilistic estimation. From a high-level intuition, a prompt can be viewed as an incomplete text sequence, providing an anchoring context for generating a continuation. The model's response is shaped by how these prompts trigger the LLM's learned patterns, guiding the generation process towards a specific result. This allows for a dynamic interaction with the LLM, where the content generation can be steered through careful prompt design in desired directions to elicit the learned patterns and knowledge in LLM.

### D. LLM AGENTS SYSTEM

Conventionally, *agent-oriented system design* refers to a framework used in software engineering for building complex systems that consist of multiple agents that interact with each other to perform tasks and achieve goals [31]. For LLM applications, this design methodology is especially useful for overcoming challenging tasks by applying task decomposition and modular solution strategies (''divide and conquer'').

[2]For a visual example: https://openaipublic.blob.core.windows.net/neuron-explainer/neuron-viewer/index.html.

**FIGURE 4.** From single large language model to collaborative LLM agents.

The primary advantage of LLM-agent system is its capability to simplify a complex task by breaking it down into smaller, manageable sub-tasks, each assigned to a specialized agent. A unique aspect of LLM-agents is that their behavior can be defined by the prompt in natural language. In the prompt, defining a specific role is a strategic way to guide and constrain their behavior. In our previous paper [32], we defined a unified structure for specifying prompt for LLM-agent. The method to create a LLM agents system is illustrated in **Figure 4**.

### *E. RETRIEVAL-AUGMENTED GENERATION (RAG)*

As the name indicates, Retrieval-Augmented Generation (RAG) combines the generative capabilities of LLMs with a retrieval mechanism that queries helpful information from an external knowledge base, such as a database or a vast text corpus. This approach dynamically incorporates retrieved information into the generation process (cf. **Figure 5**). As a result, RAG enhances the LLM's ability to handle tasks that demand task-specific information, especially in scenarios where up-to-date or specialized knowledge is required.

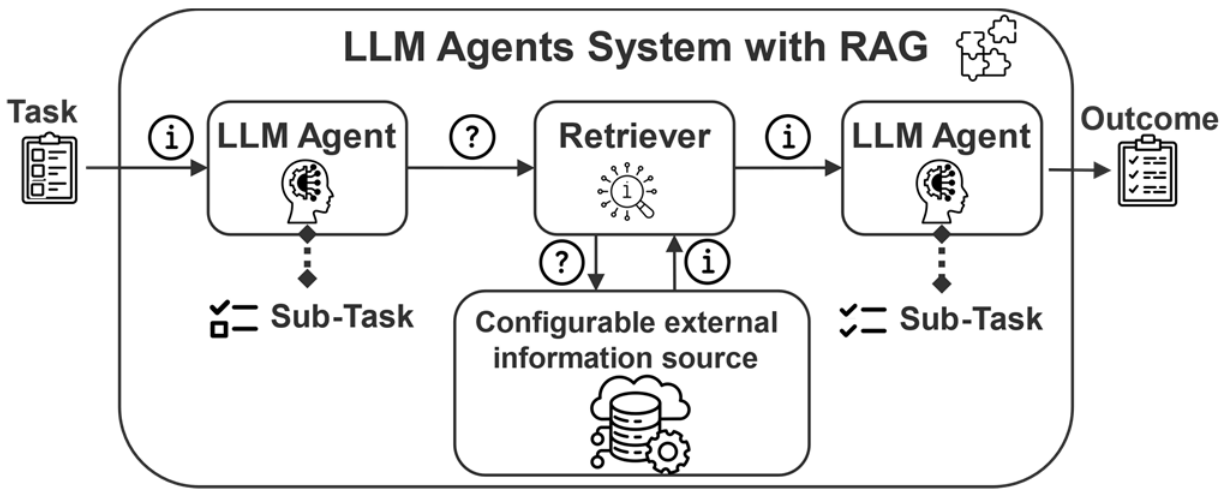

**FIGURE 5.** Integration of Retrieval-Augmented Generation (RAG) in LLM agents system.

### *F. CONCLUSION: WHY USE LLM FOR SEMANTIC PROCESSING*

Leveraging LLM is a promising approach for ensuring "data being communicated" is "clearly defined" and "disambiguously interpreted". Firstly, LLMs are trained to interpret semantics from extensive training on diverse texts and generate language that aligns with human understanding, and this capability is evidenced by the identification of specialized semantic neurons [30] and reasoning circuit [7]. Secondly, the nuanced interpretation of texts with model internal semantic neuron structures and the ability to distinguish between various meanings enable LLMs to disambiguate terms and phrases effectively. Furthermore, the application of techniques such as text embedding for capturing semantic relationships and generative next-token-prediction for text production allows for precise articulation of concepts in text. Finally, by designing LLM-agent systems and incorporating the mechanism of Retrieval-Augmented Generation (RAG), an information processing pipeline can be created for task-specific applications, for instance, the automation of AAS-instance model creation from raw textual data.

## V. DESIGN AND METHODOLOGY

To handle the semantics of a given piece of information, we propose a data structure called "semantic node", designed to encapsulate various dimensions of meaning. Building on this foundation, we develop the use case and detail the system design.

### *A. "SEMANTIC NODE"*

The proposed "semantic node" is specified in **Table 2** and visualized in **Figure 10**. The motivation for this is two folds: Firstly, text makes meaning. We want to create an atomic unit that materializes the semantic information into textual elements (c.f. Section III), wherein each text element contributes to defining an aspect of the semantics; Secondly, this data structure facilitates software implementation, and can be independently extracted, modified, evaluated, and converted through the information processing pipeline.

The proposed semantic node is an atomic unit to capture the meaning of a piece of information. This structure is processed by a system constituted of LLM-agents, which synthesizes its elements to enrich the node with specific meanings. The most semantically rich information is considered to be the name, the conceptual definition and the use of data (affordance). The value shall be extracted from the source texts to ensure the nodes' authenticity. Furthermore, the structure includes three optional fields: a source description that enriches background context but requires additional data input; and the value type and unit are not indispensable, because they contribute less to defining the meaning and can often be inferred from other elements (semantic entailment).

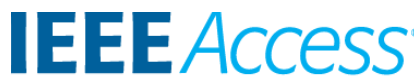

**TABLE 2. Specification of the elements in conceptualized data structure "Semantic Node."**

| Semantic node | | |
|---|---|---|
| | **Name** | A concise and specific title assigned to the semantic node. It should be concise yet descriptive enough to convey the essence of the feature at a glance. |
| | **Conceptual definition** | A definition of what the semantic node represents. |
| | **Usage of data (Affordance)** | Describes the potential usage and application of the semantic node. |
| | **Value** | The actual data that is described by the semantic node. |
| | Value type *optional* | The type of data used for representing the actual data. (by default: String) |
| | Unit *optional* | The measurement unit associated with the value. (if applicable) |
| | Source description *optional* | This is an extended explanation that situates the semantic node within the specific context of its source, making the semantic node contextually traceable. (require extra information input) |

This semantic node is a conceptualized structure that serves our theoretical proposition of semantic interoperability. In the next sections, we use this constructed semantic node to design the software application of AAS generation software powered by LLM agents.

### B. USE CASE DESCRIPTION

The designed LLM-based system serves two primary purposes: firstly, for system users, it facilitates the creation of AAS instances by automatically generating data models that describe the technical properties of automation components, thereby reducing manual effort in data migration. Secondly, for research purposes, the LLM produces semantic descriptions for technical data in text, as specified in "semantic node", as lustrated in **Figure 2** at the beginning of this paper. These generated texts are not only integral to the AAS for conveying technical information but also serve as evidence for evaluation. By examining these generated texts, we assess the LLM-system's ability to accurately interpret specialized technical concepts and gauge the overall performance of the LLM in processing and interpreting the semantics of technical data. This dual functionality underscores the system's utility in enhancing semantic interoperability in digital twin implementations and providing a measurable framework to evaluate semantic interpretation capabilities.

**Figure 6** depicts the functions of "AASbyLLM" and the interacting stakeholders in this work. The software provides a web application for users to test the thoroughness of our system and assist us (annotators and evaluators) during the evaluation process. Additionally, the validated high-quality semantic definition generated by the LLM-system can be collected by a library manager and used for enrichment of a dictionary (e.g., ECLASS).

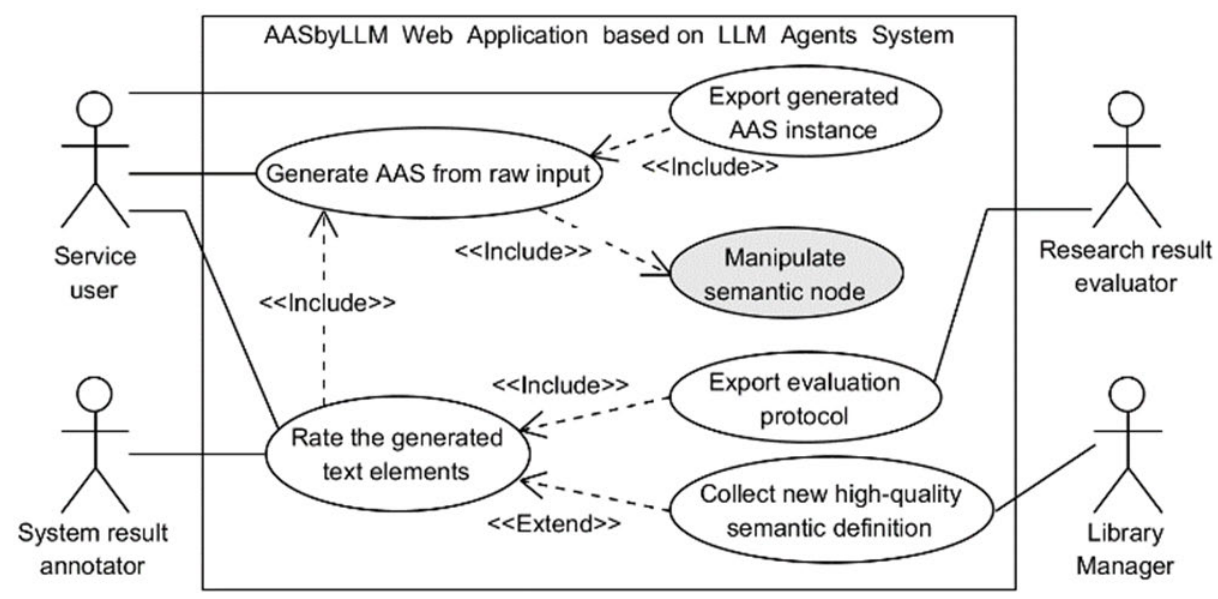

**FIGURE 6. Use case diagram of the AASbyLLM application.**

### C. SYSTEM DESIGN

The system component diagram **Figure 7** presents an overview of the components and the processing pathways involved in the handling of semantic nodes. Its primary objective is to automate the creation of a structured AAS instance model. The process begins with user interaction via a user interface, where raw data comprising technical properties is input. This data then undergoes semantic enrichment and manipulation by the LLM agents in the form of semantic node, eventually producing an AAS model instance based on AAS standard specifications in **Figure 8** [1]. The system components are introduced as follows:

**User Interface:** The user interface serves as the starting point and the user initializes the generation process by submitting the raw input textual data. This input may range from structured data such as spreadsheets or JSON-files to unstructured text such as copy-pasted paragraphs from a technical manual or text from a table of technical properties.

**Retrieval-Augmented Generation (RAG):** After initial input, the RAG sub-system group is engaged, which combines the generative LLMs with information retrieval mechanism from an external domain specific dictionary library. This RAG subsystem consists of several LLM-agents.

**- Identification and Extraction Agent:** In the RAG subsystem. An LLM-agent is designed to identify and extract the name, the value, and an initial definition for a semantic node from the input text. This LLM processes the given input text and initially creates a name, definition, and contextual description for each semantic node as output, enriching the raw data with semantic details in a data structure.

**- Semantic Search Agent:** Following identification and extraction, this system component performs a semantic search using an embedding LLM to find semantically similar entries in the ECLASS dictionary. The search mechanism is based on our previous work [2], where a vectorized embedding index, called "semantic fingerprint", is created for comparison between queried text and each ECLASS dictionary entry. The result is a list of retrieved similar definition entries from ECLASS dictionary.

**- Synthesis Agent:** This step incorporates the results from the semantic search into the generation process. An LLM-agent is prompted to generate a judgment of the relevance of the retrieved entries, accompanied by a short reason in text. The purpose of this step is two folds: firstly, semantic search is based on relationship of semantic similarity, which is a typical proxy metric for search but suboptimal for determining precise relevance, and inappropriate results shall be filtered out; Secondly, in this step, the generated judgement and reason serve as intermediate textual material for considering more nuanced relationships during the whole information analysis process by LLM. By instructing the LLM to judge and reason for each search result, the LLM generates more precise semantic node. After synthesis, a complete semantic node is created based on RAG, ready for AAS model creation. Details about this process are illustrated in our web demo under tab "LLM reasoning details", accessible per link on our GitHub Repository.

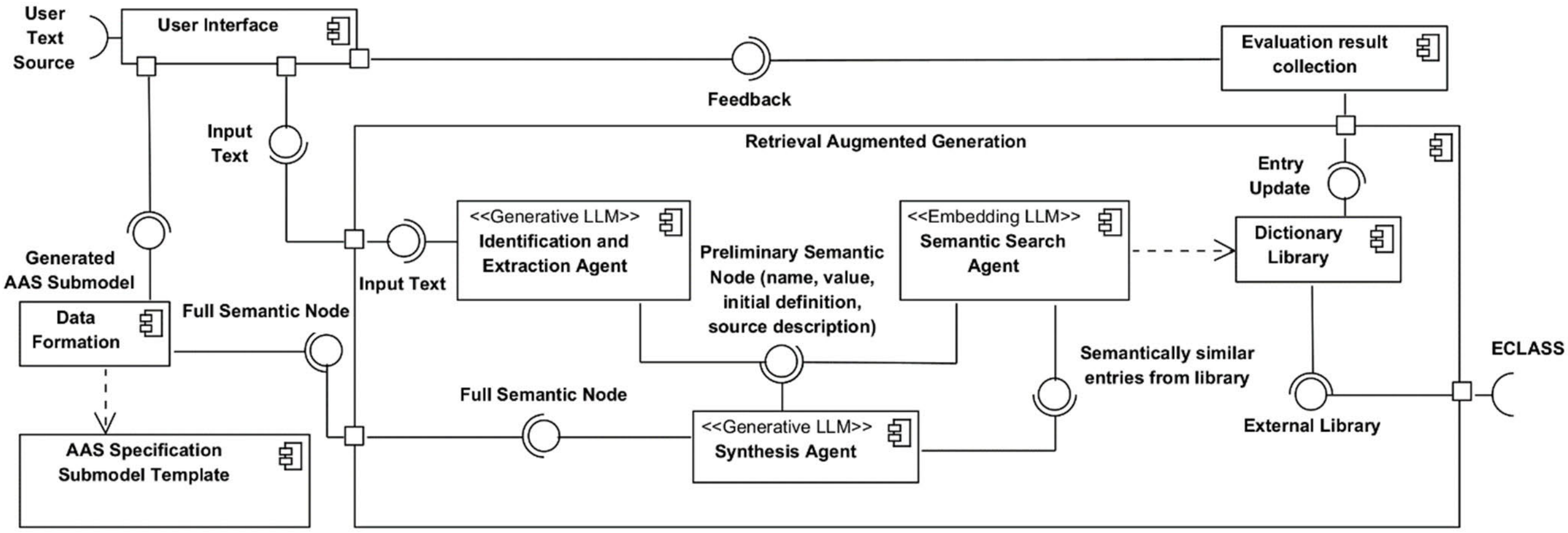

**FIGURE 7.** System components diagram of software "AASbyLLM" for AAS instance model generation with LLM agents system.

**AAS Data Formation:** The system then proceeds to form the AAS model by moving the data from the synthesized semantic nodes into an AAS JSON template. The system leverages established AAS framework to store the generated texts field from semantic node for technical properties of automation components. Specifically, it fills property name in "IdShort" and extracted value in "value", and other fields in "ConceptDescription". Furthermore, we recognize the importance of using identifiable semantics to identify the meaning of data properties and assign a unique "SemanticID" for each newly generated concept definition. This process ensures that the final AAS instance model is in the correct format and adheres to the AAS standard specification, as shown in **Figure 8**.

**User Interface Feedback:** The formed AAS model is then presented back to the user via the UI. Users can inspect the generated AAS model, assess its accuracy and quality, and download the created AAS instance model. Meanwhile, the system collects the user rating.

**Potential Dictionary Library Enhancement:** It would be infeasible for a dictionary to cover all possible meanings in the world. For this specific use case, where the dictionary entries (e.g., ECLASS dictionary library) in some cases may not be perfectly aligned with current properties, more accurate definitions of data property can be generated by the system. Once verified by human experts, these refined definitions can be used for library enrichment.

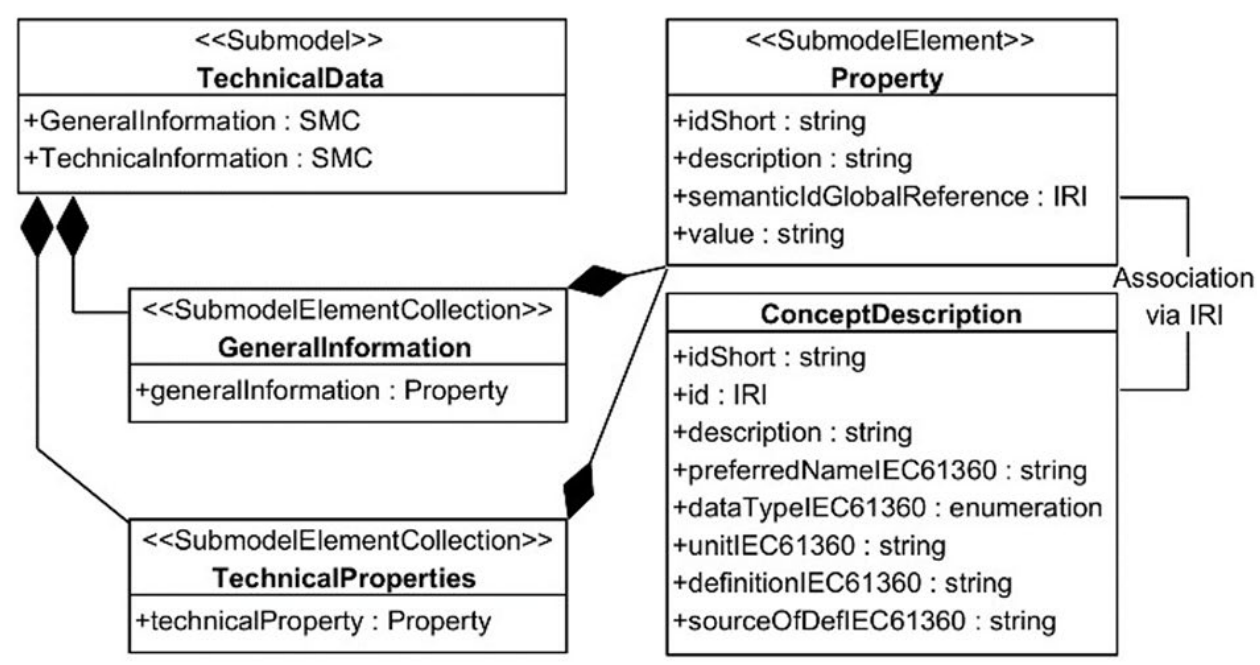

**FIGURE 8.** The AAS model elements generated by the implemented software AASbyLLM; modeling based on [1].

### *D. PROMPT DESIGN*

The LLM-agent is the elemental function component in the designed system, and it is defined and realized with prompting. We designed a structured prompt template for instructing the behavior of generative LLM-agents. This prompt template is introduced and explained in our previous work based on insights of NLP-research. The essential constitutes and key elements of this structured prompt are introduced in **Figure 9**. More concrete and detailed examples are released in the GitHub Repository.

### *E. MODEL SELECTION*

We selected four models for comparison, chosen for their similar size and comparable performance. These include the proprietary commercial model *GPT-3.5* and three open-source models deployable on local GPU servers: "*Llama_2_70B_HF*", "*Mixtral_8 × 7B_Instruct_v0.1*", and "*WizardLM_70B_v1.0*".

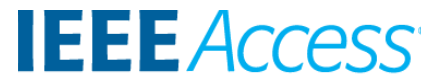

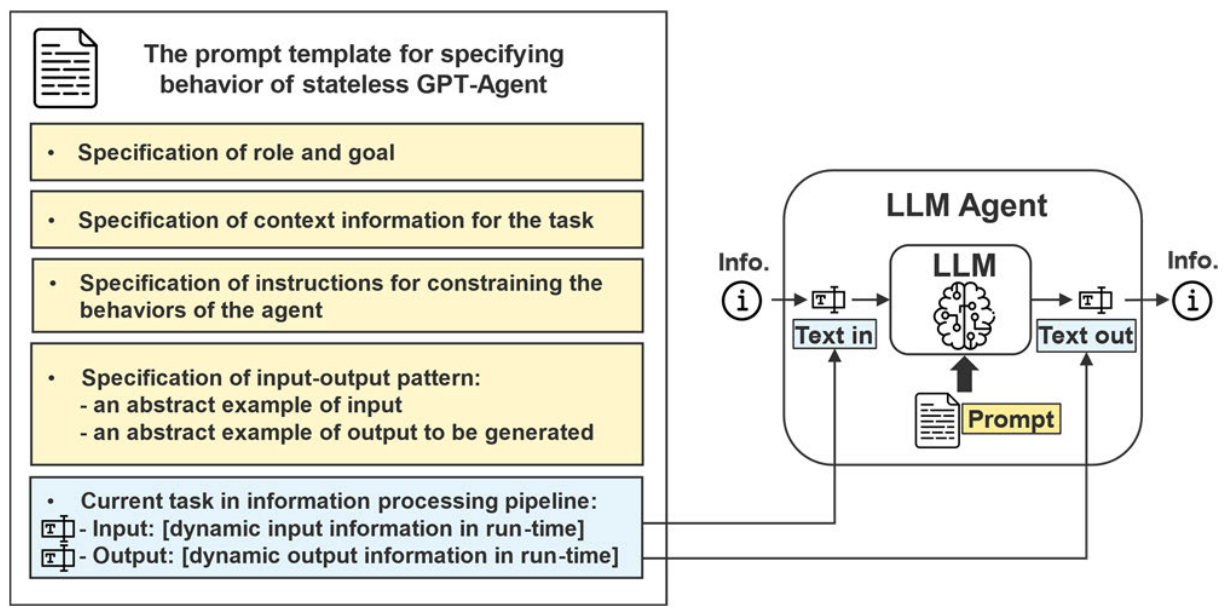

**FIGURE 9.** **The proposed prompt design for creating LLM agent.**

*"Llama_2_70B_HF"* [33] is an autoregressive language model and undergoes supervised fine-tuning (SFT) and reinforcement learning with human feedback (RLHF) post-training to align with human preferences for helpfulness. It serves as the baseline model for our comparison of open-source models.

"*Mixtral_8 × 7B_Instruct*" [34] is a model further developed from "Llama_2_7B" based on the Sparse Mixture of Experts (SMoE) architecture and fine-tuned for following instructions. It is a representative model that utilizes innovative architectural designs to improve performance, enabling it to discern complex patterns and deliver more accurate predictions.

*"WizardLM_70B_v1.0"* [35] is a model further fine-tuned from Llama. Its specialty lies in using synthetical complex instructions (called "Evol-instruct" in [35]) to train a base model solving more difficult tasks. This model exemplifies how further training on a curated dataset can enhance a model. This method is promising because it utilizes automatically generated synthetical data for scalable training of LLMs. This implies that the large language models can train themselves to become more performant.

These models are used to power the designed system. By having a comparison basis and evaluating the outcome, we can gain insights into how the model's capability and system design can have an impact on downstream task results, how effective can our theory be realized and how well the AAS-generation tool performs.

## VI. EVALUATION

Evaluating the meaning expressed by language can be a challenge due to the unlimited combination of texts in natural language, which can result in an immeasurable number of possible expressions. Automatic evaluation is not appropriate. We did a human evaluation benchmark for this work and developed a web-based software to support the evaluation under various configurations. We also formulated proxy metrics to comprehensively assess the system's effectiveness. The evaluation is categorized into three sections: 1) an end-to-end evaluation for practical usability, 2) a comparative analysis of capability of 4 different LLMs, and 3) an ablation study to investigate the importance of integrating external information through Retrieval-Augmented Generation (RAG), with or without utilizing the external ECLASS library.

The evaluation is fundamentally linked to how well humans perceive the generated texts to be **error-free** and **informative**. Though the evaluation leans on human perception, it yields quantifiable results based on statistical principle of large number, indicating the effectiveness of our methodology (pass rate, helpfulness score, element-level rating scores, and inaccuracy rate). Ultimately, the utility of the proposed Asset Administration Shell (AAS) generation system is assessed based on the extent to which it reduces human effort, highlighting the practicality of our AAS generation solution.

### A. EVALUATION SETUPS

The evaluation assesses the ability of the designed LLM-system to interpret technical concepts within an automation system. Technical properties representing the concepts were drawn from 20 automation components, achieving diverse coverage by including 5 sensors, 5 actuators, 5 controllers, and 5 connectivity components, whose technical information is collected from different source. The text results produced by the system provide the primary evidence for its assessment.

In the comparative study, 8 design variations (4 large language models × 2 mechanisms, w./w.o. RAG) of the system were examined. These variations involved four distinct models that have representative characteristics, each tested in two mechanism configurations: one with the integration of the external ECLASS library for Retrieval-Augmented Generation (RAG) and one without it. 7 graduate students with engineering backgrounds with above-average GPA were chosen as annotators to closely resemble the user knowledge profiles. The annotators' individual knowledge level and preference are confounding factors in this experiment. To mitigate potential bias, we randomly allocated the 20 groups of technical properties among 7 annotators and examinate the statistical results calculated from large sample size.

### B. EVALUATION EXECUTION WITH SOFTWARE SUPPORT

The human evaluators are first tasked with reading the technical properties from data sheet of an automation device and making binary statement whether they comprehend the meaning of these technical properties. Then, they determine whether there are errors in the generated 5 data elements ("name", "value", "conceptual definition", "affordance", "value unit"), any inaccuracy in texts or contradiction between their understanding and the generated elements is determined as an error. Then, they are tasked to rate the text quality of the two most semantic-rich elements on a scale 1 to 5: "**conceptual definition**" and "**use of data (affordance)**". Additionally, they evaluate whether the generated texts help them understand the clear meaning of the technical concept if they did not comprehend the meaning of the technical concept at the beginning ("**helpfulness**"). Finally, they determine whether the overall generated semantic node has the quality to be released without human modification and further improvement, this "**overall**" perceived quality is also on a scale 1 to 5. This evaluation process is visualized in **Figure 10**.

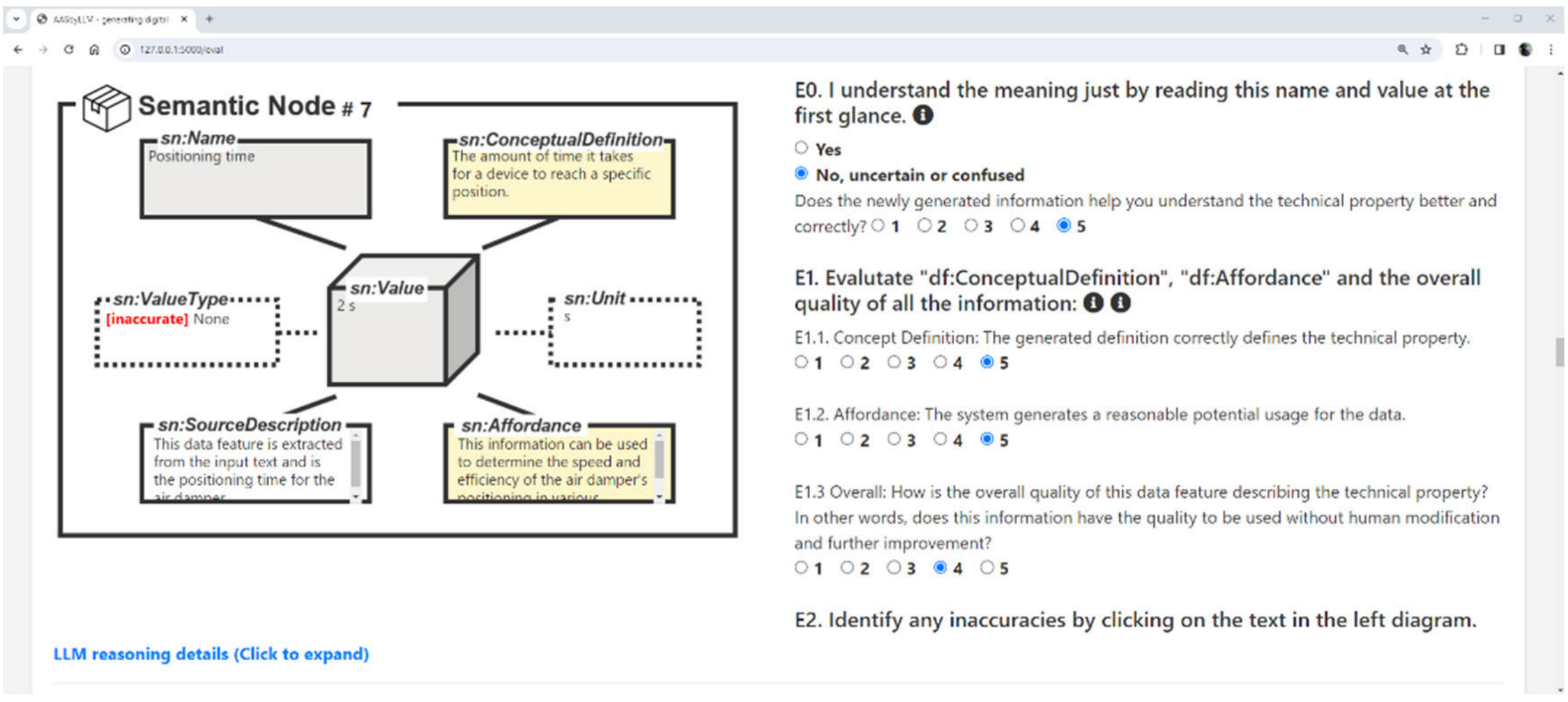

**FIGURE 10.** Screenshot of the application UI used for supporting the annotating and evaluating process.

The evaluation relies on the statistical large number to yield valid results. During the evaluation, a total of 2,259 technical property samples across all the 20 selected automation components by using 4 models and 2 design mechanisms were assessed. Derived from making 10 distinct rating decisions for each individual technical property, this comprehensive assessment resulted in a collection of 22,590 total evaluation decisions. The details of the evaluation are released on our GitHub Repository, including raw data, software web-application, evaluation raw results, calculation spreadsheets, and the analyzed results.

### *C. END-TO-END EVALUATION*

For the end-to-end evaluation of the system designed to automate the generation of Asset Administration Shell (AAS) models using Large Language Models (LLMs), two distinct metrics are proposed: **the effective generation rate**(pass rate) and **helpfulness** (ability to provide implicit knowledge).

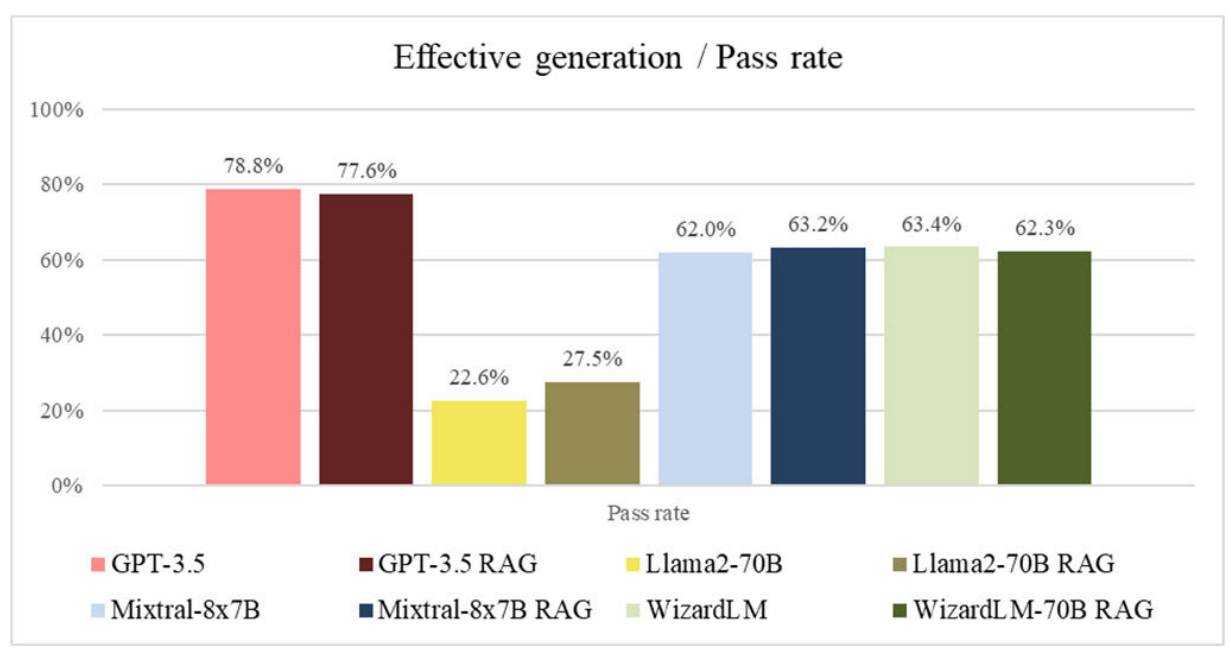

**FIGURE 11.** End-to-end pass rate of the generated instance model under different system configurations, a passed case means that the generated semantic node of a technical property is ready for release without the need for further human modifications.

#### 1) EFFECTIVE GENERATION RATE

The end-to-end evaluation aims to measure the capability of the designed LLM-system to generate semantic information, referred to as a "semantic node". This end-to-end evaluation metric is termed "pass rate" or "percentage of effective generation", which measures if the generated result is **error-free** and **informative**. For a sample to achieve the "passed" status (cf. **Figure 11** above):

$$\begin{aligned}\text{Passed (s)} = \text{NoInaccurateText (s)} \\ \wedge \text{(DefinitionRating(s)} == 5) \\ \wedge \text{(AffordanceRating(s)} == 5) \\ \wedge \text{(OverallRating(s)} == 5)\end{aligned}$$

The pass rate is a proxy metric for effectiveness of the system, indicating the readiness of generated information models for practical application without the need for further human modifications.

#### 2) HELPFULNESS SCORE

In the technical domain, descriptive texts (e.g., text in a technical datasheet) often assume that the reader possesses prior knowledge of specialized concepts, which can lead to confusion and misunderstandings. This issue of "**implicit knowledge assumption**" becomes prominent when the presented data is inherently complex or incomplete, relying on a foundation of technical knowledge of an agent (machine/human) to correctly interpret the data's meaning.

During human evaluation, the annotator first assesses whether the bare name and value of the technical property alone provide a clear sense making. If not, the system then reveals the generated texts for review. The evaluator then scores the overall helpfulness of these elements in semantic node. Based on their effectiveness in clarifying the meaning and enhancing understanding, the annotator rates on a scale from 1 to 5. Specifically, the annotator is prompt to answer the following question: "On a scale of 1 to 5, does the newly generated information help you understand the technical property better and correctly?".

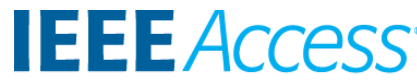

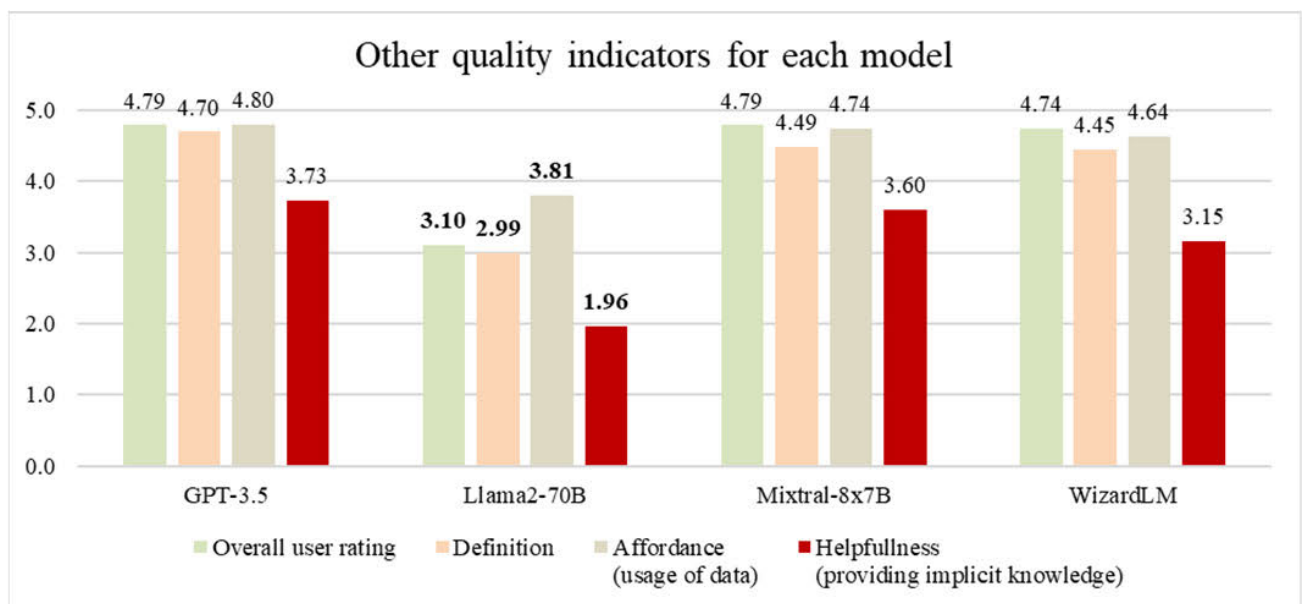

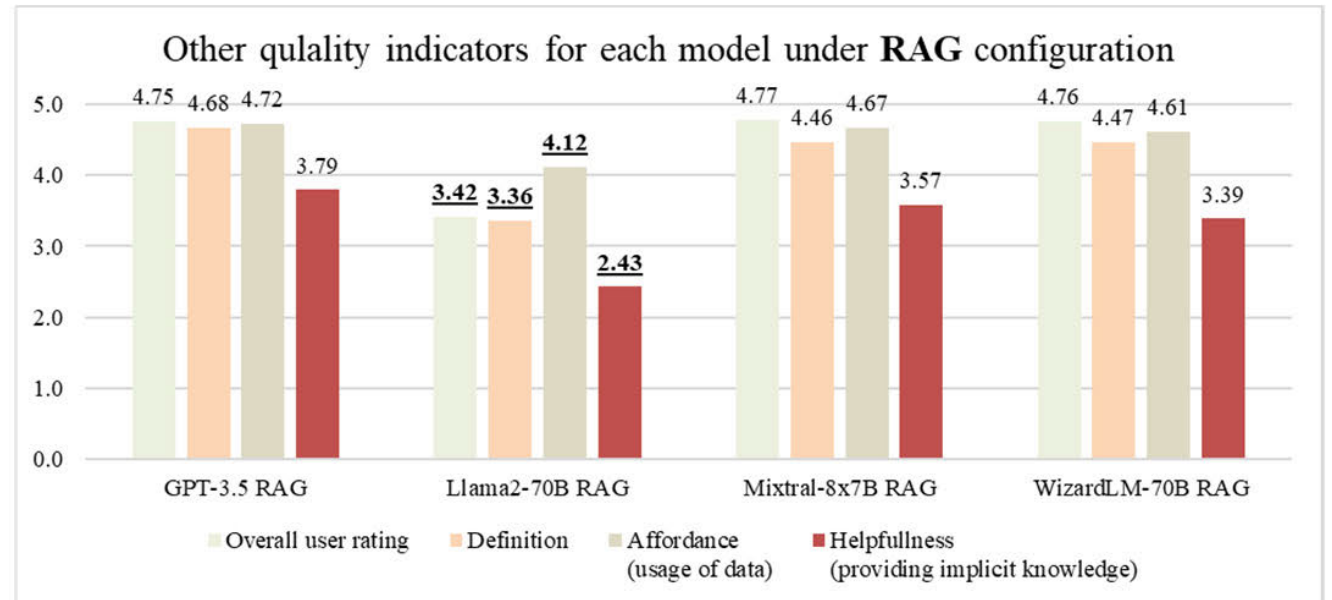

**FIGURE 12.** Comparative system performance across different configurations with and without RAG mechanism. Each group represents the performance of a system powered by a specific model, with each color indicating a proxy metric for a particular quality indicator. The comparison between the two diagrams demonstrates the effectiveness of the RAG mechanism.

The resulting so called "helpfulness score" proxy metric evaluates the system's ability to provide this implicit knowledge, i.e., to generate texts that aid human users in comprehending the clear meaning of technical properties beyond just name and value. This helpfulness proxy metric is presented in **Figure 12**, together with other detailed metrics (*overall quality, definition text quality, affordance text quality*).

$$\text{Helpfulness Score} = \frac{\sum \text{Rating of each confusing case}}{\#\,\text{confusing cases}}$$

This proxy metric reflects the knowledge capability of an LLM-powered system, specifically, the capability to determine and produce textual data aligned to specialized concepts. From a communications perspective, it indicates that even if the sender makes implicit assumption and simplify the data being communicated to merely a name and value (not loss-free), recipients can still accurately interpret the intended semantics, as long as the recipient has the knowledge to determine the semantics. Additionally, this metric highlights the LLM's ability to add informative explanation and help humans in understanding specialized technical data in scenarios of non-loss-free communication.

### 3) DETAILED VIEW ON INACCURATE GENERATION

**Figure 13** details the inaccurate generation ratio. The Llama2 based model cannot reliably generate the meaning and usage of the data, as indicated by the "definition_inaccurate" and "affordance_inaccurate" bars. While the other models rarely generate incorrect texts.

The inaccurate generation ratio in general correlates with the previous evaluation results. RAG only has significant effects for Llama-2 model.

## *D. ABLATION STUDY OF RAG MECHANISM*

Retrieval augmented generation is seen as a promising mechanism to enhance the performance of LLM-system by dynamically incorporating queried data from external knowledge base during the text generation process. To evaluate the RAG mechanism's effect, we conducted a controlled experiment comparing two configurations of our LLM system: one with RAG based on ECLASS-dictionary data definitions, and another where RAG was disabled. The impact of the Retrieval-Augmented Generation (RAG) mechanism on the overall performance of each model varies, revealing distinct patterns. The diagrams in **Figure 12** and **Figure 13** visually demonstrates this effect, comparing the results with RAG and the results without RAG. We summarize the insights as follows:

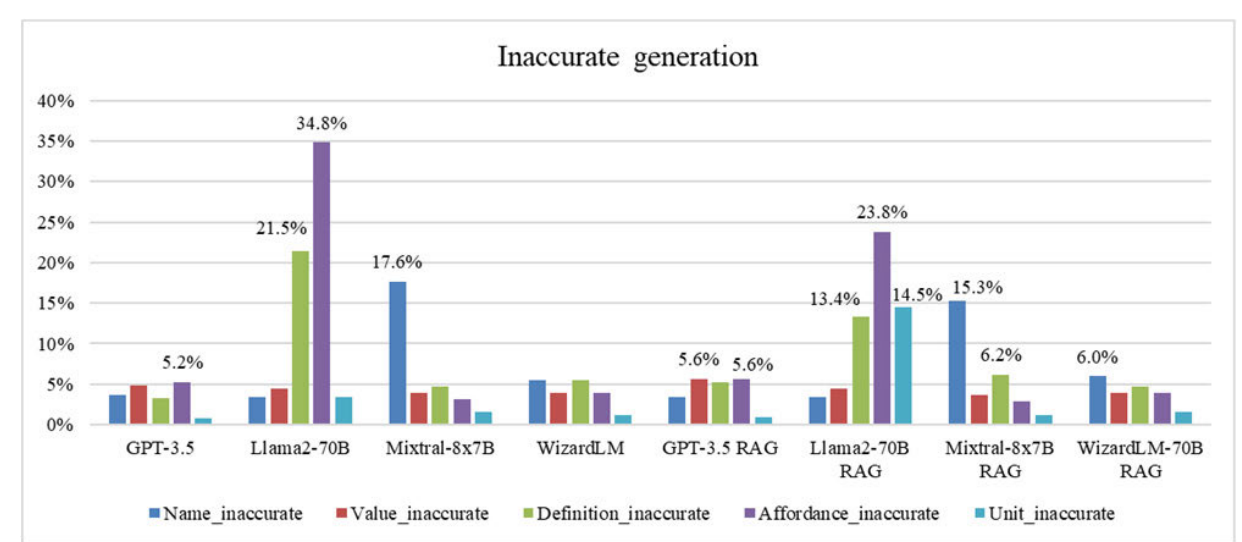

**FIGURE 13.** Comparative analysis of inaccurate generation rates by different LLMs with and without RAG.

### 1) RAG HAS SIGNIFICANT EFFECT FOR WEAKER MODEL

LLAMA-2, which initially had the lowest effective generation rate, shows significant improvement with the introduction of RAG. We calculate the statistical significance with unequal variances t-test (Welch's t-test) for each metric based on two configurations (with RAG or without RAG) with a significance threshold of 0.05. All metrics (helpfulness, overall rating, generated definition rating, affordance rating) indicate significant improvement. The t-test result is released on the GitHub Repository, together with other detailed calculation processes and results during the statistical analysis in a spreadsheet.

### 2) RAG HAS NO SIGNIFICANT EFFECT FOR STRONGER MODELS

Surprisingly, for the other three stronger models (GPT3.5, Mixtral, WizardLM), the experiment results indicate no significant effect of RAG on enhancing or degrading the qual-ity of the generated texts on all the metrics. This observation suggests that the stronger models may already possess the necessary knowledge to excel in this particular task, making additional information from the RAG mechanism redundant. An explanation is that these models have effectively learned the relevant technical concepts during their training phase, indicating (hypothetically) that a saturation point has been reached where RAG's contribution of new knowledge does not further enhance model performance.

#### 3) IMPLICATIONS FOR USING RAG

An overall intuition that can be drawn from this evaluation is that RAG can enhance the performance of weaker models, but the overall enhancement result may not surpass the performance of fine-tuned stronger models, as illustrated in **Figure 14**.

This dynamic can be metaphorically characterized as "**cheat sheet effect**" of RAG, where weaker language model can be improved with external help, while stronger model does not need to.

While RAG can guide the style and direction of the output by incorporating text from external dictionaries, the semantic similarity search component may also **introduce noise** by fetching **similar yet irrelevant** information from the external database, affecting the final output's relevance and accuracy.

Furthermore, another drawback of the RAG is its **lower efficiency**. The implemented RAG mode statistically consumes about 4.2 times the processing duration, as it needs to process more integrated texts and more numbers of model invocations to generate the final response.

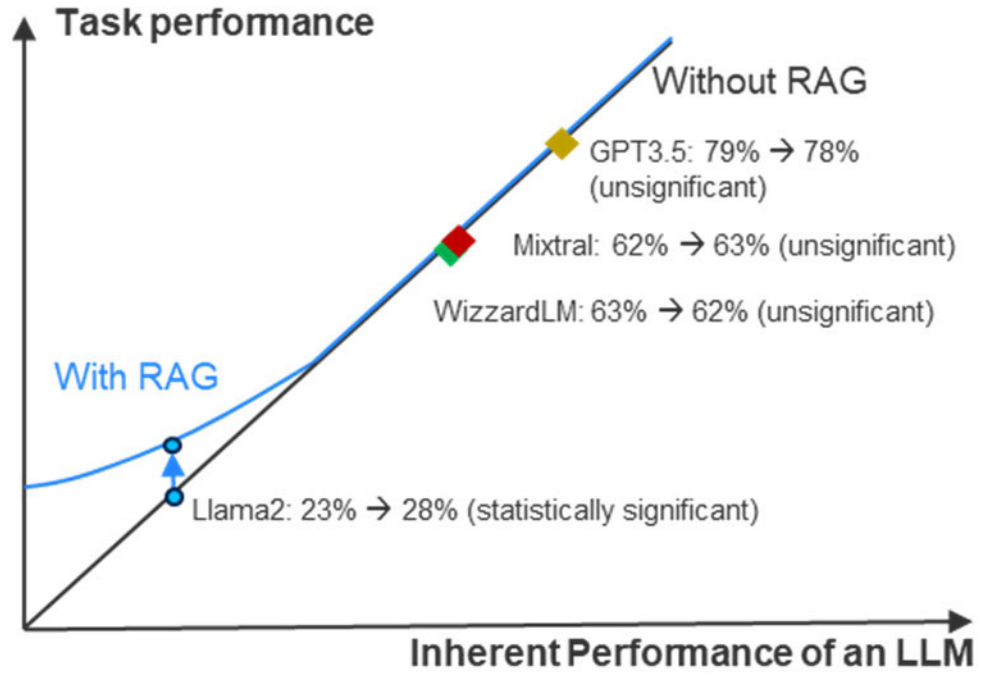

**FIGURE 14. The graphical illustration of the hypothesis how RAG affect the performance of LLM system in technical concepts understanding based on our observed results. The lines are drawn based on assumed interpolation and extrapolation. Without RAG, a LLM only uses its inherent knowledge to execute task, and task performance shall be equal to the performance of the LLM (the 1:1 line); with RAG, significant boosting effect has been observed for weaker model Llama2 in our experiment, while no significant effect has been observed for stronger models.**

From these findings and the assumption that these stronger models should have learned the conceptual meaning of technical concepts, we can formulate a hypothesis based on observed results:

> *The effectiveness of the RAG mechanism is reliant upon its ability to introduce new, previously unknown knowledge to compensate for capabilities that the LLM lacks.*

This hypothesis highlights the importance of carefully considering the effectiveness of the RAG mechanism, particularly in determining when its application is beneficial. It suggests that the decision to employ RAG should be based on an assessment of the gap between the LLM's existing knowledge basis and the required knowledge of the task at hand. Furthermore, this insight can guide the strategic development of RAG datasets, which should be tailored to fill knowledge gaps, thereby maximizing the performance of the LLM system for specific tasks.

## VII. DISCUSSION OF POTENTIAL IMPACTS

In this section, we discuss the implications and potential impact of our work, exploring how it contributes to the research community and the broader field of industrial digitalization.

### A. STREAMLINING DIGITAL TRANSFORMATION WITH LLMs

The generated AAS models can be integrated seamlessly into AAS-compliant digital twin software, facilitating seamless information exchange and communication.

The use of LLM to automate AAS-instances generation provides a technological catalysator for information exchange in manufacturing and digital twin technologies. This approach not only simplifies the creation of standardized digital representations for assets but also facilitates seamless data sharing and utilization across platforms.

The two primary goals—achieving error-free semantic communication from a source to a target, and ensuring error-free generation of data required by the target based on the source data's meaning—are demonstrated to be logically and technically equivalent based on our developed LLM translation system.

### B. SEMANTIC INTEROPERABILITY BASED ON LLM "ADAPTER"

Semantic interoperability, the ability of different systems to understand and interpret data consistently, is crucial for seamless data communication and integration across diverse technological platforms. In a broader sense, the LLMs are integrated into a data processing pipeline and act as an interpretive layer. LLMs enable a more intuitive and accurate exchange of information between disparate data sources and systems.

### C. ENHANCING STANDARDS AND OPTIMIZATION THROUGH LLMs

Based on our experience during the experiment, the static data definition in ECLASS alone often falls short in covering all the dynamic needs for accurately annotating a given technical property. LLMs emerge as a powerful solution to this challenge, offering the ability to dynamically determine semantics and generate meaningful definitions of data.

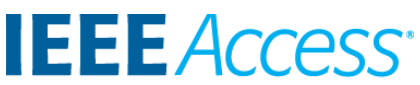

Therefore, the relationship between AAS, ECLASS dictionary, and LLMs becomes synergistic. LLMs can facilitate the continuous improvement and adaptation of ECLASS dictionary library, ensuring they remain relevant and comprehensive to cover a large number of possible technical concepts. The dynamic semantic interpretation and generation capability of LLMs is also crucial for simplifying deployment and maintaining the utility of digital twins across various industrial applications, paving the way for a more interconnected and intelligent data exchange.

### D. DATA LIMITATION

The data used to test the effectiveness of this system include descriptive data of the technical properties of automation components (sensors, actuators, controllers, and network devices) selected from datasheets. These data are selected, because it is a typical usage scenario for participants in Industry 4.0 to share technical data model of components across different companies and the information systems [1], which has practical impacts and value. The generated AAS digital twin models in this work are limited to the technical data submodel according to AAS specifications. In the future, we believe the proposed method can be extended to other data types and usage scenarios (e.g., generating other AAS submodels standardized by Industry 4.0), and we anticipate similar results as evaluated in this study.

Additionally, incorporating LLMs to process various data sources within digital twins, such as simulation models [36], [37], offers the potential to expand semantic interoperability to provide predictive and operational insights. However, a significant challenge lies in serializing different knowledge representations into a textual format suitable for LLMs.

## VIII. CONCLUSION

This paper presents a comprehensive exploration into the automated generation of Asset Administration Shells using Large Language Models, with a focus on enhancing semantic interoperability within the framework of Industry 4.0. This approach facilitates the automated generation of AAS models to reduce human effort and implementation cost in creating AAS instances. By introducing a ''semantic node'' data structure to capture the essential semantics of a piece of information and developing a novel approach that leverages LLMs for the semantic translation of technical properties in AAS model, the system converts technical data into AAS models, which can be used for error-free information exchange in digital twins. Furthermore, the semantic definitions generated by the LLM system possess the capability to dynamically annotate information, and the resulting conceptual definition can be verified by human and be used to form a dictionary collection or enrich an external dictionary such as ECLASS.

The evaluation of various LLM configurations and the impact of the Retrieval-Augmented Generation (RAG) mechanism has yielded detailed insights into the deployment of LLM systems in this context. Our research demonstrates that LLMs can substantially streamline the AAS instance creation process, achieving up to a 78% pass rate in generating error-free and informative AAS model elements. This suggests that a significant portion of the effort typically required for crafting AAS models could be shifted toward more efficient validation processes. Open-source models deliver sufficient performance for this application and can be fairly used for commercial use. However, our comparative analysis indicates that the effectiveness of these models varies, depending on their acquired knowledge during training and the strategic integration of external semantic resources (RAG based on ECLASS dictionary). Based on our observations, we propose a new hypothesis regarding the ''cheat-sheet effect of RAG'' conjecture, which we plan to explore further. This paper serves as a proof of concept for using LLMs to automate the generation of AAS instances and proves the feasibility of using LLM as information interpreter for achieving semantic interoperability in digital twin systems in the context of Industry 4.0. The research result also lays the groundwork for further collaboration with research and industrial sectors to amplify our findings' value and practicality in the context of industrial digitalization and smart manufacturing.

## REFERENCES

[1] S. Bader, E. Barnstedt, H. Bedenbender, B. Berres, M. Billmann, and M. Ristin, ''Details of the asset administration shell—Part 1—The exchange of information between partners in the value chain of Industrie 4.0 (version 3.0 RC02),'' Plattform Ind. 4.0, Berlin, Germany, Tech. Rep. 3.0RC02, 2022.

[2] Y. Xia, N. Jazdi, and M. Weyrich, ''Automated generation of asset administration shell: A transfer learning approach with neural language model and semantic fingerprints,'' in *Proc. IEEE 27th Int. Conf. Emerg. Technol. Factory Autom. (ETFA)*, Sep. 2022, pp. 1–4, doi: 10.1109/ETFA52439.2022.9921637.

[3] F. Ocker, C. Urban, B. Vogel-Heuser, and C. Diedrich, ''Leveraging the asset administration shell for agent-based production systems,'' *IFAC-PapersOnLine*, vol. 54, no. 1, pp. 837–844, Jan. 2021, doi: 10.1016/j.ifacol.2021.08.186.

[4] G. Schnauffer, D. Görzig, C. Kosel, and J. Diemer, ''Asset administration shell for the wiring harness system,'' in *Advances in Automotive Production Technology—Towards Software-Defined Manufacturing and Resilient Supply Chains*, H. D. K. Niklas and Wulle, Eds. Cham, Switzerland: Springer, 2023, pp. 324–332.

[5] H. Eichelberger and C. Niederée, ''Asset administration shells, configuration, code generation: A power trio for Industry 4.0 platforms,'' in *Proc. IEEE 28th Int. Conf. Emerg. Technol. Factory Autom. (ETFA)*, Sep. 2023, pp. 1–8, doi: 10.1109/etfa54631.2023.10275339.

[6] N. Blasek, K. Eichenmüller, B. Ernst, N. Götz, B. Nast, and K. Sandkuhl, ''Large language models in requirements engineering for digital twins,'' in *Proc. 13th Enterprise Design Eng. Working Conf.*, Jun. 2023, pp. 1–15.

[7] C. Olah, N. Cammarata, L. Schubert, G. Goh, M. Petrov, and S. Carter, ''Zoom in: An introduction to circuits,'' *Distill*, vol. 5, no. 3, Mar. 2020, Art. no. e00024.001, doi: 10.23915/distill.00024.001.

[8] W. Gurnee, N. Nanda, M. Pauly, K. Harvey, D. Troitskii, and D. Bertsimas, ''Finding neurons in a haystack: Case studies with sparse probing,'' 2023, *arXiv:2305.01610*.

[9] T. Miny, M. Thies, U. Epple, and C. Diedrich, ''Model transformation for asset administration shells,'' in *Proc. IECON 46th Annu. Conf. IEEE Ind. Electron. Soc.*, Oct. 2020, pp. 2207–2212, doi: 10.1109/IECON43393.2020.9254649.

[10] M. Platenius-Mohr, S. Malakuti, S. Grüner, J. Schmitt, and T. Goldschmidt, "File- and API-based interoperability of digital twins by model transformation: An IIoT case study using asset administration shell," *Future Gener. Comput. Syst.*, vol. 113, pp. 94–105, Dec. 2020, doi: 10.1016/j.future.2020.07.004.

[11] M. Platenius-Mohr, S. Malakuti, S. Grüner, and T. Goldschmidt, "Interoperable digital twins in IIoT systems by transformation of information models: A case study with asset administration shell," in *Proc. 9th Int. Conf. Internet Things*. New York, NY, USA: Association for Computing Machinery, Oct. 2019, doi: 10.1145/3365871.3365873.

[12] J. Zhao, B. Vogel-Heuser, F. Bi, J. Höfgen, F. Ocker, B. Vojanec, T. Markert, and A. Kraft, "A semi-automatic approach for asset administration shell creation from heterogeneous data," *IFAC-PapersOnLine*, vol. 56, no. 2, pp. 3673–3679, 2023, doi: 10.1016/j.ifacol.2023.10.1532.

[13] L. M. V. Da Silva, A. Köcher, M. S. Gill, M. Weiss, and A. Fay, "Toward a mapping of capability and skill models using asset administration shells and ontologies," in *Proc. IEEE 28th Int. Conf. Emerg. Technol. Factory Autom. (ETFA)*, Sep. 2023, pp. 1–4, doi: 10.1109/etfa54631.2023.10275459.

[14] Y. Huang, S. Dhouib, L. P. Medinacelli, and J. Malenfant, "Enabling semantic interoperability of asset administration shells through an ontology-based modeling method," in *Proc. 25th Int. Conf. Model Driven Eng. Languages Syst.* New York, NY, USA: Association for Computing Machinery, 2022, pp. 497–502, doi: 10.1145/3550356.3561606.

[15] C. Schmidt, F. Volz, L. Stojanovic, and G. Sutschet, "Increasing interoperability between digital twin standards and specifications: Transformation of DTDL to AAS," *Sensors*, vol. 23, no. 18, p. 7742, Sep. 2023, doi: 10.3390/s23187742.

[16] N. Braunisch, M. Ristin-Kaufmann, R. Lehmann, M. Wollschlaeger, and H. W. van de Venn, "Generation of digital twins for information exchange between partners in the Industrie 4.0 value chain," in *Proc. IEEE 21st Int. Conf. Ind. Informat. (INDIN)*, Jul. 2023, pp. 1–6, doi: 10.1109/indin51400.2023.10218306.

[17] S. Rongen, N. Nikolova, and M. van der Pas, "Modelling with AAS and RDF in Industry 4.0," *Comput. Ind.*, vol. 148, Jun. 2023, Art. no. 103910, doi: 10.1016/j.compind.2023.103910.

[18] Y. Huang, S. Dhouib, L. P. Medinacelli, and J. Malenfant, "Semantic interoperability of digital twins: Ontology-based capability checking in AAS modeling framework," in *Proc. IEEE 6th Int. Conf. Ind. Cyber-Physical Syst. (ICPS)*, May 2023, pp. 1–8, doi: 10.1109/ICPS58381.2023.10128003.

[19] J. Fuchs, J. Schmidt, J. Franke, K. Rehman, M. Sauer, and S. Karnouskos, "I4.0-compliant integration of assets utilizing the asset administration shell," in *Proc. 24th IEEE Int. Conf. Emerg. Technol. Factory Autom. (ETFA)*, Sep. 2019, pp. 1243–1247, doi: 10.1109/ETFA.2019.8869255.

[20] T. Moreno, T. Sobral, A. Almeida, A. L. Soares, and A. Azevedo, "Semantic asset administration shell towards a cognitive digital twin," in *Flexible Automation and Intelligent Manufacturing: Establishing Bridges for More Sustainable Manufacturing Systems*, M. A. S. Francisco, Ed. Cham, Switzerland: Springer, 2024, pp. 679–686.

[21] A. Lüder, A.-K. Behnert, F. Rinker, and S. Biffl, "Generating Industry 4.0 asset administration shells with data from engineering data logistics," in *Proc. 25th IEEE Int. Conf. Emerging Technol. Factory Automat. (ETFA)*, Sep. 2020, pp. 867–874, doi: 10.1109/ETFA46521.2020.9212149.

[22] S. Cavalieri and M. G. Salafia, "Insights into mapping solutions based on OPC UA information model applied to the Industry 4.0 asset administration shell," *Computers*, vol. 9, no. 2, p. 28, Apr. 2020, doi: 10.3390/computers9020028.

[23] J. Beermann, R. Benfer, M. Both, J. Müller, and C. Diedrich, "Comparison of different natural language processing models to achieve semantic interoperability of heterogeneous asset administration shells," in *Proc. IEEE 21st Int. Conf. Ind. Informat. (INDIN)*, Jul. 2023, pp. 1–6, doi: 10.1109/indin51400.2023.10218154.

[24] M. Both, J. Müller, and C. Diedrich, "Automated mapping of semantically heterogeneous I4.0 asset administration shells by methods of natural language processing," *At-Automatisierungstechnik*, vol. 69, no. 11, pp. 940–951, Nov. 2021, doi: 10.1515/AUTO-2021-0050.

[25] A. Cartus, M. Both, M. Nicolai, J. Müller, and C. Diedrich, "Interoperability of semantically heterogeneous digital twins through natural language processing methods," in *Proc. CLIMA*, May 2022, pp. 1–8, doi: 10.34641/clima.2022.143.

[26] M. Both, B. Kämper, A. Cartus, J. Beermann, T. Fessler, D. J. Müller, and D. C. Diedrich, "Automated monitoring applications for existing buildings through natural language processing based semantic mapping of operational data and creation of digital twins," *Energy Buildings*, vol. 300, Dec. 2023, Art. no. 113635, doi: 10.1016/j.enbuild.2023.113635.

[27] A. Radford, R. Jozefowicz, and I. Sutskever, "Learning to generate reviews and discovering sentiment," 2017, *arXiv:1704.01444*.

[28] A. Vaswani, N. Shazeer, N. Parmar, J. Uszkoreit, L. Jones, A. N. Gomez, L. Kaiser, and I. Polosukhin, "Attention is all you need," 2017, *arXiv:1706.03762*.

[29] A. Rogers, O. Kovaleva, and A. Rumshisky, "A primer in BERTology: What we know about how BERT works," *Trans. Assoc. Comput. Linguistics*, vol. 8, pp. 842–866, Dec. 2020, doi: 10.1162/tacl_a_00349.

[30] S. Bills. (2023). *Language Models Can Explain Neurons in Language Models*. Accessed: May 14, 2023. [Online]. Available: https://openaipublic.blob.core.windows.net/neuron-explainer/paperindex.html

[31] D. Dittler, P. Lierhammer, D. Braun, T. Müller, N. Jazdi, and M. Weyrich, "A novel model adaption approach for intelligent digital twins of modular production systems," in *Proc. IEEE 28th Int. Conf. Emerg. Technol. Factory Autom. (ETFA)*, Sep. 2023, pp. 1–8, doi: 10.1109/etfa54631.2023.10275384.

[32] Y. Xia, M. Shenoy, N. Jazdi, and M. Weyrich, "Towards autonomous system: Flexible modular production system enhanced with large language model agents," in *Proc. IEEE 28th Int. Conf. Emerg. Technol. Factory Automat. (ETFA)*, Apr. 2023, pp. 1–8, doi: 10.1109/ETFA54631.2023.10275362.

[33] H. Touvron, L. Martin, K. Stone, P. Albert, A. Almahairi, Y. Babaei, N. Bashlykov, S. Batra, P. Bhargava, S. Bhosale, and D. Bikel, "Llama 2: Open foundation and fine-tuned chat models," 2023, *arXiv:2307.09288*.

[34] A. Q. Jiang, A. Sablayrolles, A. Roux, A. Mensch, B. Savary, C. Bamford, D. S. Chaplot, D. D. L. Casas, E. B. Hanna, F. Bressand, and G. Lengyel, "Mixtral of experts," 2024, *arXiv:2401.04088*.

[35] C. Xu, Q. Sun, K. Zheng, X. Geng, P. Zhao, J. Feng, C. Tao, and D. Jiang, "WizardLM: Empowering large language models to follow complex instructions," 2023, *arXiv:2304.12244*.

[36] P. Liu, J. Xing, Y. Li, C. Miller, and P. Tang, "Knowledge sharing and workforce engagement using digital twins-based simulations and extended reality for process operations," in *Proc. Comput. Civil Eng.*, Jan. 2024, pp. 680–687, doi: 10.1061/9780784485231.081.

[37] Y. Xia, D. Dittler, N. Jazdi, H. Chen, and M. Weyrich, "LLM experiments with simulation: Large language model multi-agent system for process simulation parametrization in digital twins," 2024, *arXiv:2405.18092*.

# An Architecture for Integrating Large Language Models with Digital Twins and Automation Systems

Yuchen Xia, Nasser Jazdi, Michael Weyrich
Institute of Industrial Automation and Software Engineering
University of Stuttgart
Stuttgart, Germany
yuchen.xia@ias.uni-stuttgart.de; nasser.jazdi@ias.uni-stuttgart.de; michael.weyrich@ias.uni-stuttgart.de

***Abstract*— Large Language Models (LLMs) offer flexible reasoning capability but lack physical embodiment, while traditional automation systems can execute physical processes yet lack cognitive capability. This paper presents a layered architecture that bridges this gap by integrating LLMs with digital twins and physical automation systems, with reference to practical case studies as proof of concept. The proposed architecture comprises three layers: a cognitive layer powered by LLMs, a bridging layer based on digital twins, and a physical layer consisting of technical process and automation system. Within the digital twin layer, we introduce three design paradigms for structuring information to support effective LLM integration: state snapshot modeling, event message modeling, and plan sequence modeling. These paradigms are demonstrated through prototypical case studies on robotic automation control and process simulation. To address challenges such as hallucination, task complexity, and system reliability, we distill a set of practical strategies, including multi-agent system design, human validation, and test-driven development. Additionally, we propose the concept of "Return on Intelligence" as a conceptual tool for evaluating the efficacy of investments in intelligent automation. This research contributes to the theoretical foundation and the architecture design for developing intelligent, adaptive automation systems powered by LLMs.**

## I. Introduction

Traditional industrial automation systems—comprising machinery, sensors, actuators, controllers, and manufacturing execution systems—excel at performing repetitive tasks efficiently. However, as market demands shift and production requirements diversify, these rigid systems struggle to adapt. Factory customers seeking swift, cost-effective reconfigurations to maintain competitiveness often encounter knowledge barriers, since modifying and maintaining automation systems typically require specialized technical expertise.

Given these challenges, several key questions emerge:

- How can automation systems overcome their inherent rigidity to become more adaptable and flexible?
- What can provide the intelligence required to handle complexity, particularly when human expertise may be limited or unavailable?
- By what mechanisms can automation systems obtain the intelligence they need for adaptive task-solving?
- From which sources should the information and knowledge required for intelligent decision-making be obtained, and how can they be effectively utilized?

One promising answer lies in large language models that are capable of interpreting textual data and reasoning based on patterns in existing human knowledge. By integrating LLMs into automation environments, it becomes possible to enable more flexible task solving, contextual reasoning, and user interaction.

However, LLMs lack physical embodiment and cannot directly perceive or influence real-world systems. To unlock their potential in industrial automation, they must be tightly integrated with digital twins—virtual representations of physical assets and processes that act as a bridge between physical reality and digital intelligence.

This paper makes the following contributions:

- It proposes a **three-layer architecture** that connects LLMs with digital twins and physical automation systems, enabling intelligent and adaptive control in industrial environments.
- It distills **three design paradigms** for representing dynamic system information in digital twins to support effective LLMs interaction: **system snapshot modeling**, **event-log-based modeling**, **planning-based modeling**.
- The architecture and design are demonstrated with case studies in realistic use cases, and we propose the concept of "**Return on Intelligence**" as a conceptual tool for evaluating the efficacy of investments in intelligent automation.
- It outlines practical strategies for addressing common challenges associated with LLMs, such as hallucination, by applying **multi-agent design**, **human validation**, **and test-driven development**.

This paper is presented with the following structure: Section II introduces the fundamental problem and the conceptual architecture. Section III introduces the 3-layer system architecture consisting of the physical, digital twin, and cognition layers. Section IV presents the design paradigms in greater depth, focusing on how information is modeled to support LLM reasoning. Section V showcases case studies that demonstrate practical implementations of the proposed architecture. Section VI discusses key insights, design considerations, and practical recommendations based on the case studies. Finally, Section VII concludes the paper with a summary and future work.

## II. Conceptual Architecture

This section outlines the fundamental problem and presents high-level guiding ideas to form the solution.

### *A. The fundamental problem*

LLMs operate within a digital environment and lack direct grounding in the physical world. Their logical reasoning is primarily based on semantic relationships and knowledge patterns derived from textual data, making them inherently disconnected from real-world processes.

Consequently, LLM cannot directly influence physical systems without additional mechanisms or interfaces.

To address this issue, three fundamental challenges must be resolved:

1. Converting real-world information into data representation interpretable by LLM.
2. Establishing connections between the LLM outputs and actionable operations capable of influencing other systems.
3. Integrating LLM-driven reasoning to automate task as requested by users.

*B. Conceptual Architecture*

To tackle these issues, we propose a three-layer conceptual architecture that provides a bridge between digital intelligence and physical automation (Figure 1). Each layer focuses on a distinct aspect of the system:

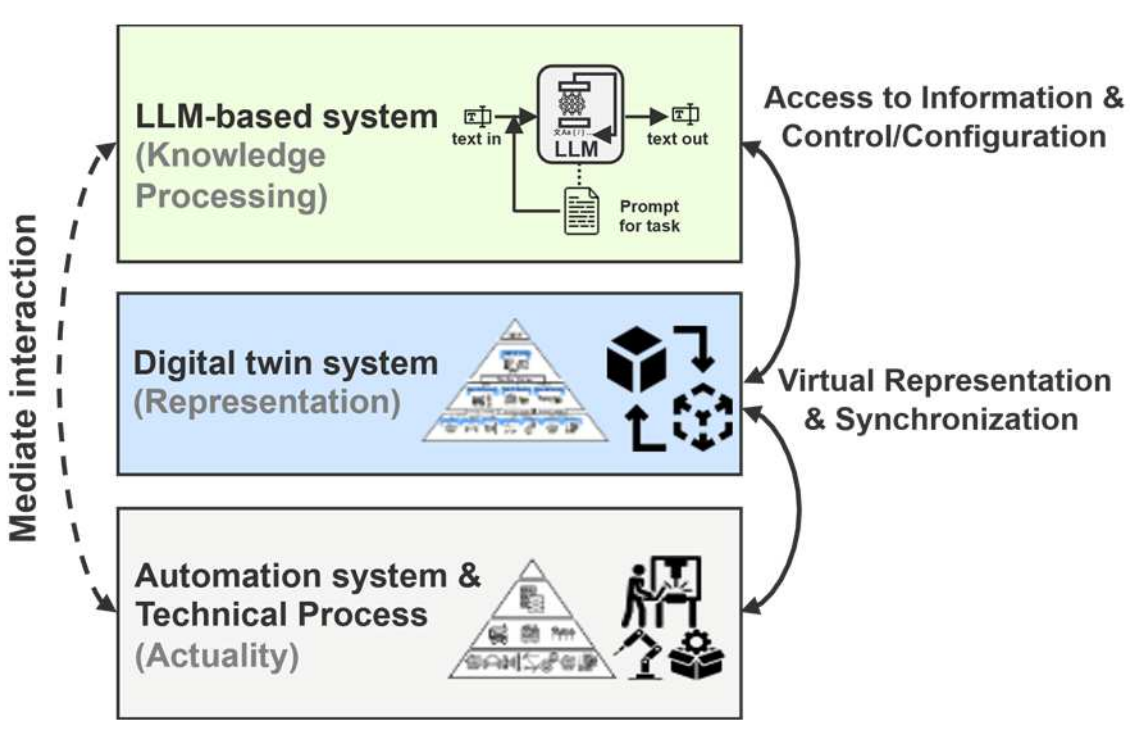

Figure 1 Three-Layer Architecture for LLM-Integrated Automation Systems

- **Automation System & Technical Process** serve as the foundational layer, responsible for monitoring and executing real-world industrial operations through machinery, sensors, actuators, and controllers.
- **Digital Twin System** acts as a virtual representation of the physical system, this middle layer maintains real-time synchronization with the automation layer. It structures representation that can be consumed by the LLM, while also receiving and mediating control or configuration commands back to the physical system.
- **LLM-based System** processes structured representation inputs provided by the digital twin. It performs reasoning, process represented knowledge, and generates outputs such as control commands, plans, or configuration settings. These outputs are then translated into executable actions through the digital twin interfaces.

Overall, the digital twin system serves as the bridge to ground the LLM to the physical reality.

## III. System Architecture

This section explains the technical necessities for system design, introduces the components required for its implementation, and illustrates them with various system diagrams with different levels of detail and focus. Figure *2* shows the system architecture overview.

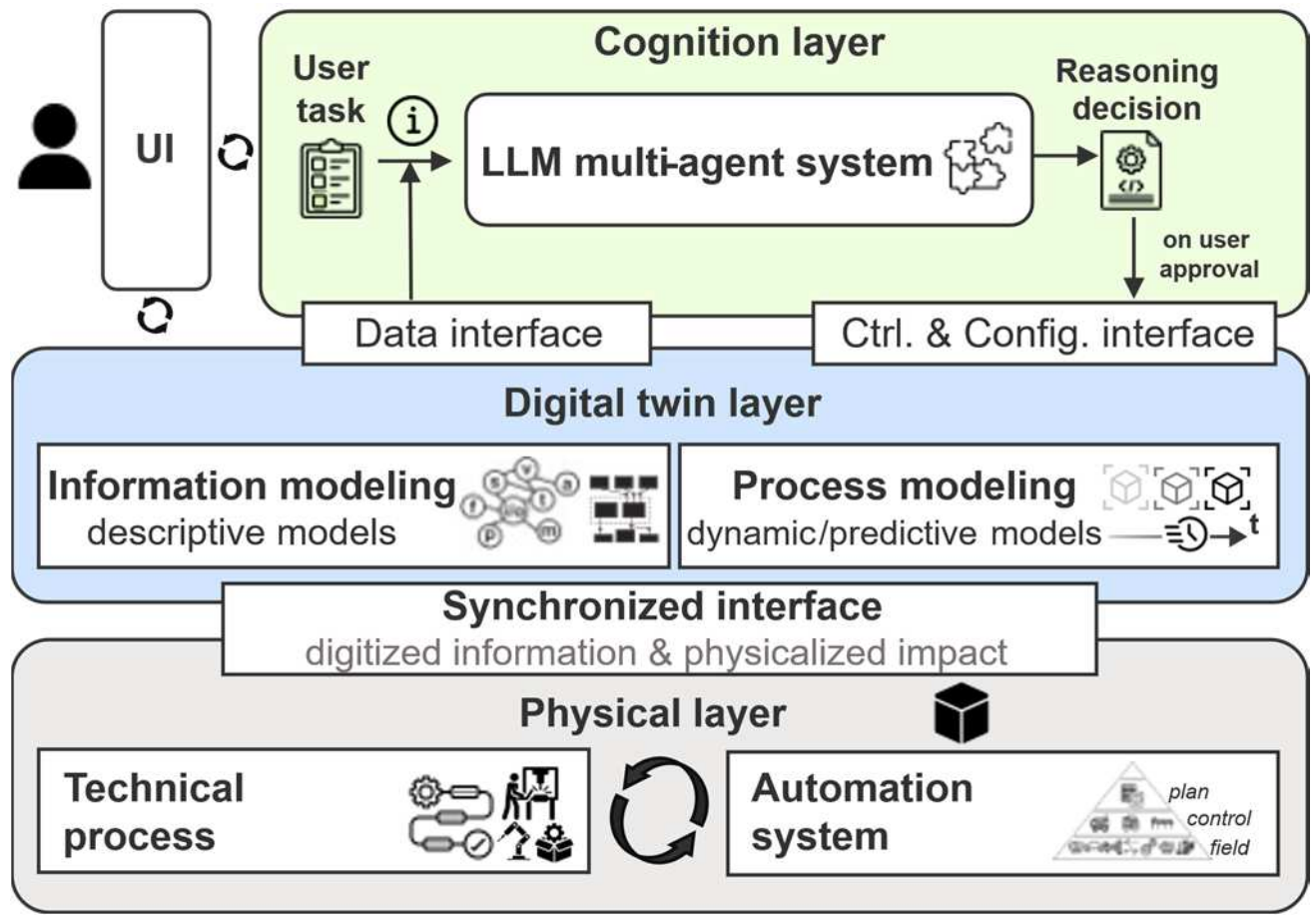

Figure 2 System Architecture Overview

*A. Physical layer:*

A **technical process** [1], which involves operations of materials and machines to achieve industrial objectives. First of all, a technical process can be executed manually, but to increase efficiency, precision, and productivity, an **automation system** is typically employed, reducing manual effort in executing and controlling the technical process.

In most industrial setups, automation follows a hierarchical structure known as the automation pyramid [1]: at the field level, sensors and actuators connect directly to machines. At the control level, Programmable Logic Controllers (PLCs) handle real-time control logic. At the planning level, a Manufacturing Execution System (MES) coordinates higher-level tasks such as production scheduling. These control and planning functions are delegated to specialized software components.

*B. Digital twin layer*

To enable advanced monitoring, control, maintenance and optimization of automated processes, the concept of digital twin is introduced [2]. A digital twin system is a software system that provides virtual model representation of physical assets, processes, or systems that enable real-time synchronization between the digital and physical domains.

To facilitate conceptual understanding and design, we classify digital twin model representations into two distinct types in the scope of this paper:

- **Information modeling** provides a static representation of entities and processes. It captures descriptive knowledge about what exists, including various factors and their relationships.
- **Process modeling**[2], in contrast, is structured around the dimension of time progression or a sequential order of execution. Once executed, these models can simulate dynamic processes and predict system behavior.

This distinction arises from how we model “time”. In information modeling, time is treated as one dimension among many dimensions in describing the system’s composition and entity relationships. By contrast, process modeling fundamentally organizes the representation around the progression of time.

[1] DIN IEC 60050-351 definition of “technical process”: the entirety of interacting operations within a system through which material, energy, or information is transformed, transported, or stored.

[2] Minor note: please note that the word “process” has different contextual meanings in “technical process” (operations within a system) and “process modeling” (way of modeling); “modeling” is an activity of creating “models”.

This distinction reflects a deeper philosophical view: our perception of reality is shaped by the experience of time moving forward—also referred to as the "arrow of time"[3]—which influences how humans reason. In [4], the author expresses skepticism about the natural existence of time and argues that it is a measure derived from how humans perceive change. These philosophical foundations are relevant when investigating how knowledge is formed and can be applied to approximate and control reality. Modeling systems based on temporal progression is not just an engineering requirement but also a natural extension of how we form and apply existing human knowledge—patterns that LLMs can also learn and utilize, as evidenced in [5], [6], [7].

Returning from our philosophical detour discussed above, we require a modeling perspective that facilitates a more transparent and analyzable understanding of how the automation system should behave over time. The core objective of (intelligent) automation is to initiate the correct control action at the appropriate time to execute operations in the technical process. To evaluate whether the system is functioning correctly as intended, its actual execution can be compared against the predefined technical process. Figure *3* introduces such a modeling perspective, illustrating the key properties and elements of the proposed system architecture.

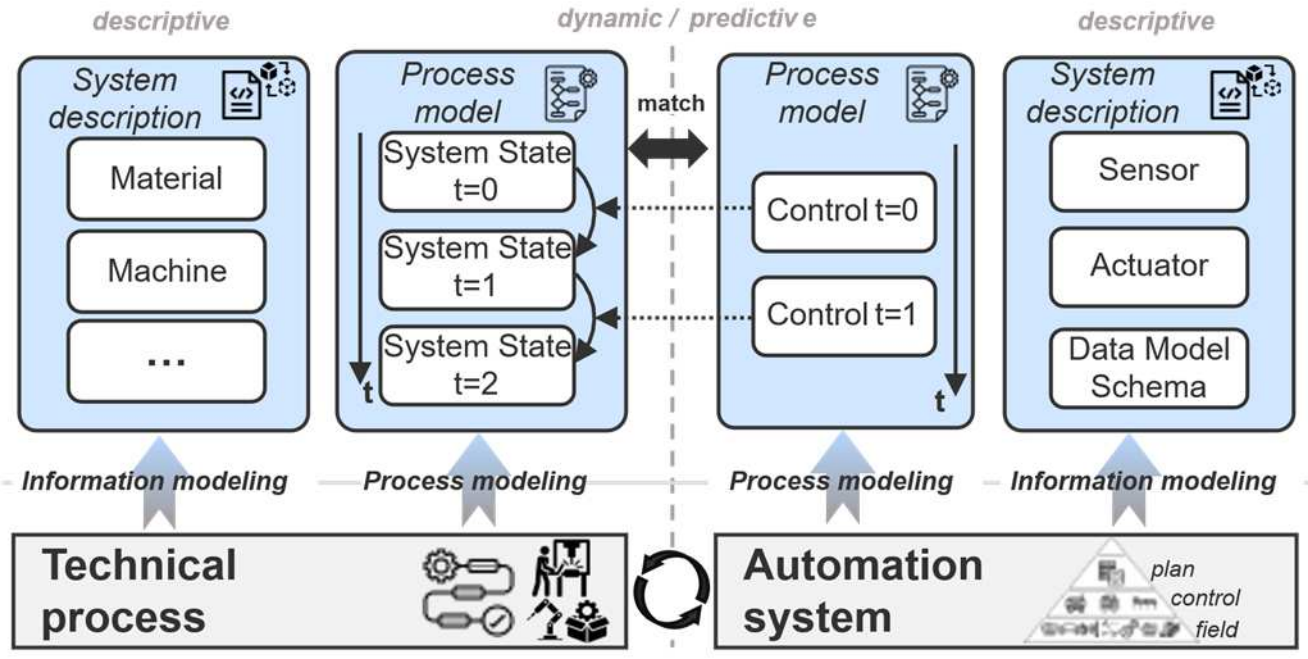

Figure 3 Representing Technical Processes and Automation Systems via Information and Process Modeling

On the *Technical Process* side (Figure *3* left), the Information Modeling provides a static description of the system, describing materials, machines and other interested entities. The process modeling structures the interested physical parameters into a sequence of system states (t=0, t=1, t=2, ...) that can represent how the process evolves over time.

On the *Automation System* side (Figure *3* right), the process modeling provides the representation of how the process is executed by applying control actions at each state. The Information Modeling captures the system's composition and functions, including sensors, actuators, and data models, which provide static knowledge about the automation system.

The central "match" between the two process models signifies that the automation system's control actions must align with (and execute) the time-based plan defined by the technical process. Meanwhile, the information modeling on each side provides the descriptions and relationships necessary for interpreting the process models.

## C. *Cognition layer: LLM-based system*

By having access to digital twins that provide detailed data and knowledge about the overall system, the LLM obtains the necessary information to perform reasoning tasks. A typical simplified integration mechanism is illustrated in Figure *4*. Within this mechanism, the LLM is implemented as a software component (referred to as the LLM agent) responsible for processing information to accomplish a specific task. It receives system information through a data interface, then processes this information based on the instruction of a structured prompt template that guides the model's behavior in generating output. The resulting output from the LLM is then parsed into structured text or executable code, which can be subsequently handled by the control or configuration interface provided by the digital twin software. This interface enables LLM's reasoning outcomes to influence and interact with other connected systems.

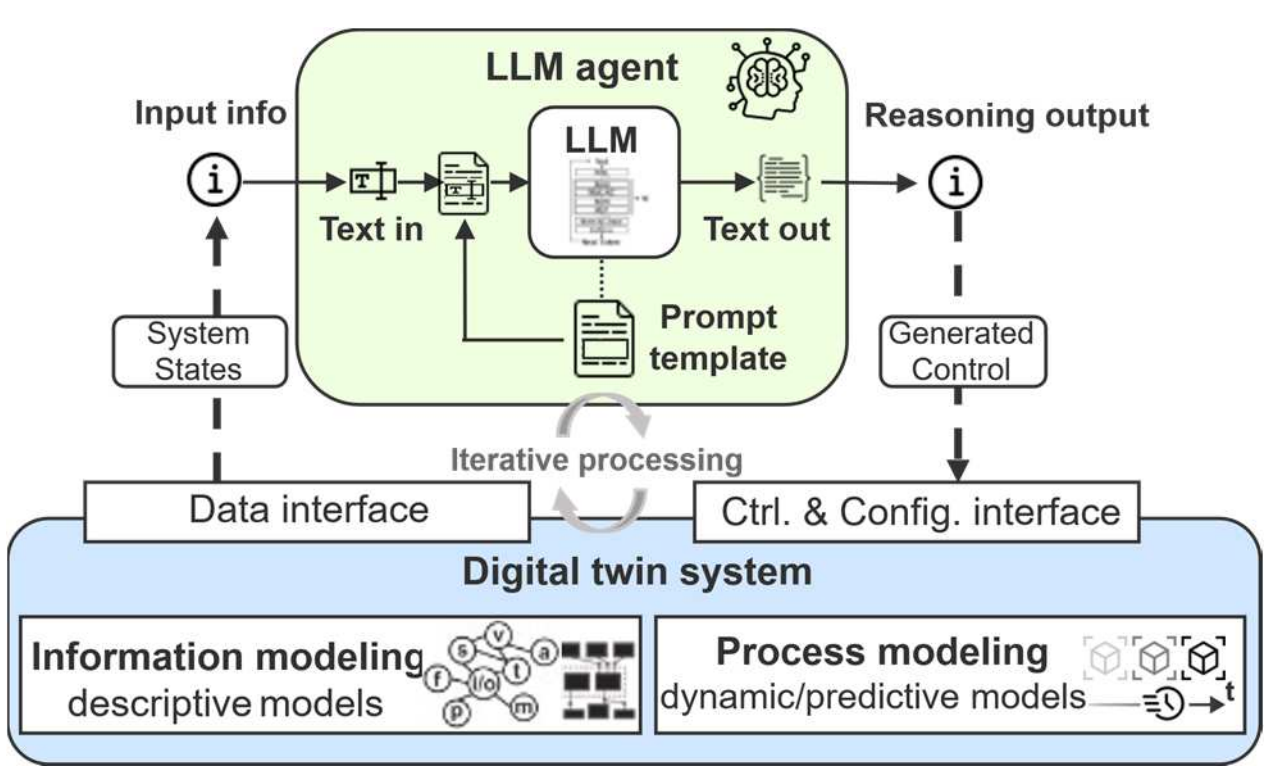

Figure 4 LLM Agent Integration with Digital Twin for Reasoning

# IV. Design Paradigms

This section provides a more detailed look at design paradigms for representing information in the digital twin layer to support LLM reasoning, as well as task-solving in the cognitive LLM layer.

## A. *Digital Twin Design: aiming for LLM integration*

Empirical investigation in [8] indicate that LLMs learn and reason with descriptive knowledge and procedural knowledge in different ways. Information provided by the information modeling (descriptive knowledge) and the process modeling (procedural knowledge) offers the contextual foundation required for LLM-based task execution. However, a key challenge lies in designing operational mechanisms that enable LLMs to effectively utilize this knowledge.

### 1) *Time-based dynamic mechanisms modeling*

Technical processes evolve continuously over time as real-world transformations unfold, with each transformation requiring specific decisions or control actions (cf. Figure *3*). From a modeling perspective, this can be represented as a series of discrete steps (t=0, t=1, t=2, …), each capturing the relevant system states information.

In practice, this can be implemented using state machines or system snapshots that record key system properties at specific moments. The LLM iteratively receives the updated system state—such as sensor readings or process status—as input. Since most LLMs operate on textual prompts, the digital twin must convert each updated system state into a structured text or code format (such as JSON). For each LLM invocation, the prompt is updated with the latest information provided by the digital twin software.

This enables the LLM to reason over dynamically updated system information and determine the next appropriate control action. The process is iterative, and further details are presented in the following section.

*2) The three design paradigms*

Building on empirical insights from our previous research and other related literature, we synthesize three primary paradigms for representing evolving system information and integrating it into LLM-based control: system snapshot modeling [9], event log-based modeling [10], [11], and planning-based modeling[12]. Each approach provides a distinct mechanism for capturing dynamic system behavior, offering different design paradigms for LLM integration, as symbolically illustrated in Figure *5*:

Figure 5 Three Design Paradigms for Structuring System Information to Support LLM Reasoning

**Type 1: System Snapshot Modeling (State Modeling)**

This mechanism relies on capturing a sequence of snapshots of the system's state at discrete time steps (t=1, t=2, t=3). Figure *5* uses colored symbols to illustrate these sequential changes in data structure. Each snapshot can use concrete modeling formats (JSON or a structured text format) to represent the state of the system at that specific moment. The LLM system reasons over these representations to derive control actions for the next state. A case study[3] illustrating this approach can be found in [9].

**Type 2: Event-Log-Based Modeling**

The information can be modeled in an event log. Planning and control rely on two key prerequisites: time and information. These can be technically captured in the form of an event. In this approach, the system state is represented as a chronological sequence of event messages. These logs contain textual information such as system activities, status updates, or triggering events recorded over time. The LLM system interprets this temporally ordered sequence to analyze context and determine appropriate next actions. A case study[4] illustrating this approach can be found in [10], [11].

**Type 3: Planning-Based Modeling**

In this approach, system states are defined in advance as part of a structured plan that outlines a sequence of predefined steps required to complete a task. The LLM system monitors the progress of the plan and compares the current execution status against the expected status. When discrepancies or failures are detected (e.g., a step fails or becomes infeasible), the LLM can propose a revised plan (replanning). A case[5] illustrating this approach can be found in [12].

**Commonality: Temporal Discretization of Information**

Although these three types involve different design patterns, they share a common constraint shaped by the operational nature of state-of-the-art LLMs (i.e., prompting and response): to be processed by the LLM, information must be discretized in time and serialized into text as part of the prompt during each reasoning cycle (cf. Figure *4*).

In Type I, each system state is captured in a snapshot model at regular time intervals and serialized into text for insertion into the LLM agent's prompt. In Type II, information is modeled as events: the system generates event entries in chronological order, the LLM agent fetches relevant information from the event log, and its outputs are also recorded as events. In Type III, the agent generates a plan covering a limited future scope. A common feature across these mechanisms is their ability to encapsulate information within discretized time slots.

*B. LLM System Design: aiming for flexible task automation*

This subsection presents a structured approach to designing LLM-based systems for flexible task automation, while addressing common challenges such as task complexity, hallucination and reliability.

*1) Task-solving as core function*

A key advantage of integrating LLMs into industrial automation is their ability to solve tasks flexibly. These tasks typically originate from user needs or from events that arise during runtime and require a response. A user interface is required to connect the LLM-agent layer with the digital twin system, enabling both real-time system monitoring and task-oriented interaction.

A typical setup uses a front-end application with a chatbot interface, allowing users to dispatch tasks, view system status, and receive results. This arrangement enables intuitive, text-based interaction with processes that may be too intricate for the user, while the LLM system performs on-demand reasoning and interprets detailed technical logic.

*2) Challenge of task complexity*

In practical scenarios, depending on the use case, tasks can be too complex for a single LLM agent to handle effectively, often leading to reduced accuracy. A solution to this challenge is task decomposition combined with a multi-agent system design (cf. Figure *6*). In this approach, the task is divided into manageable sub-tasks, which are then assigned to multiple LLM agents. Each agent specializes in a specific aspect of information processing and reasoning, thus improving overall performance and accuracy. The computational complexity of the multi-agent system shall generally scale proportionally with the complexity of the original task—that is, more difficult tasks require longer reasoning processes and more generated text by the agents.

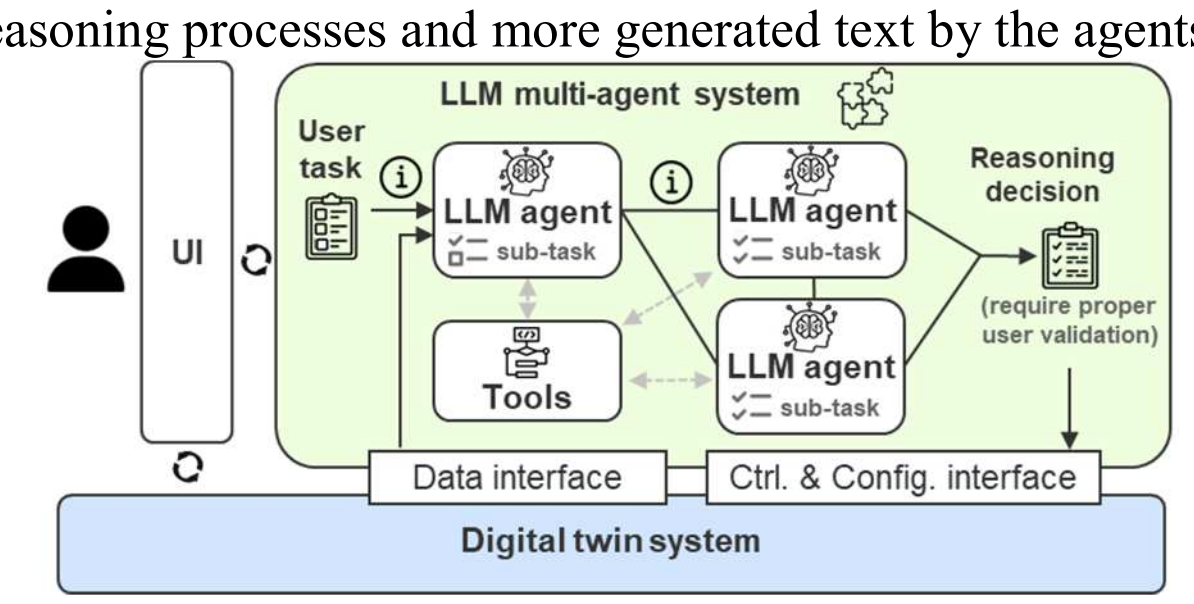

Figure 6 Multi-Agent Design and Tool Integration of LLM System

[3] State Modeling: GitHub: LLM interacts with simulation models
[4] Event-Log-Based: GitHub: LLM controlled automation based on event messages
[5] Planning-Based: GitHub: LLM agents plan flexible production tasks

It is important to note that not all subtasks need to be handled by LLM agents. Some operations, such as mathematical calculations, information retrieval, pathfinding, or executing specialized code for software functions and robotic skills, can be performed more efficiently and reliably by dedicated software components, also known as "Tools" (cf. Figure *6*). To improve overall task-solving performance, LLM agents can be designed to integrate with these tools and delegate specific functions accordingly.

*3) "Hallucination" and task reliability requirements*

Hallucination is a broad term referring to cases where an LLM generates incorrect or undesired outputs. Some studies categorize hallucinations as either model-intrinsic or model-extrinsic[13]; however, this distinction alone does not provide actionable guidance for mitigation and solution in terms of system engineering. Based on our investigation and practical experience, we identify four specific types of hallucinations, each with distinct underlying causes and corresponding countermeasures.

1. **Insufficient Model Knowledge:** The model lacks the required knowledge patterns or semantic associations for accurate reasoning. This limitation can stem from various sources, including technical limitations in the model's training process, the inherent difficulty of representing certain domain-specific knowledge in text form, or gaps in the available human knowledge captured in the training data.
   Countermeasures**:**
   - Choose a more capable base model.
   - Fine-tune the model with domain-specific data.
   - Integrate external software components for specialized tasks (e.g., RAG for information lookup [14], [15], or tool-using [16], [17], [18]).
2. **Misalign with intended task and user preference:** Models are typically trained on broad datasets and fine-tuned for varied tasks; they may not match the specific goals or style preferences of a particular context. A correct answer is not necessarily a useful one if it does not align with the user's intent. Additionally, imprecise or poorly formulated prompts can create a mismatch between the task and the expected output, preventing the LLM from performing accurate and relevant reasoning.
   Countermeasures:
   - Select and fine-tune models that align more closely with the domain and task.
   - Provide clearer and instructive prompts.
   - Iteratively improve prompts to adapt LLM agent behavior
3. **Ambiguous and general text input:** If a task is insufficiently articulated or ambiguous, the model cannot execute it with certainty or the required level of specificity. A lack of contextual clarity in the task or question description can result in unpredictable outputs, as the model must infer a direction of reasoning while generating its response.
   Countermeasures:
   - Supply additional context or supporting data.
   - Ask the user for clarifications to remove ambiguities.
4. **Complexity mismatch:** The computational complexity of the reasoning process should generally scale with the complexity of the task [19]. Short reasoning process leads to reduced result accuracy.
   Countermeasures:
   - Decompose large tasks into smaller subtasks using a multi-agent system [12].
   - Apply step-by-step reasoning (Chain-of-Thought [7], ReAct [20]).
   - Extend the reasoning process before the LLM generates a final decision [20], [21].

*4) Necessities of UI design in response to reliability issue*

While the system can streamline task-solving by reducing human effort in information retrieval, data interpretation, reasoning and content drafting, users must ultimately verify the outputs to assess plausibility and maintain accountability. Outputs with imperfect accuracy may still be useful across different use cases, depending on the specific requirements and the degree to which they reduce human workload. To meet practical reliability requirements, the LLM system shall be limited to functioning as an assistant system, with final result validation still remaining the user's responsibility. Therefore, the user must be an integral part of the system's design, acting as both the initiator and validator of the task-solving process, as illustrated in Figure *6*.

## V. Case Studies and Applications

This section presents case studies applying the proposed architecture and design paradigms across varied scenarios. The architecture is applied to each use case with different implementation details and technology stacks, prototyped in a lab setting. Each case is accompanied by demonstration videos showcasing how intelligent automated systems may look and operate.

### *A. Planning-Based Control of a Modular Automation Robotic System (Design Paradigm Type 3)*

This case study [12] demonstrates how LLMs can be applied to plan and control a modular robotic automation system. Based on a user-specified task, the system autonomously generates a corresponding production plan and executes it through the underlying automation infrastructure.

In this setup, LLM agents are designed to interpret descriptive information provided by digital twins (implemented with AAS) and control the physical system through structured service interfaces, as illustrated in Figure *7*. A key feature of this approach is the **hierarchical control service interface** within the digital twin, organized into two abstraction levels: (1) **coarse-granular skills**, representing higher-level operations at the automation module level, and (2) **fine-granular functionalities**, corresponding to low-level control actions at the component level.

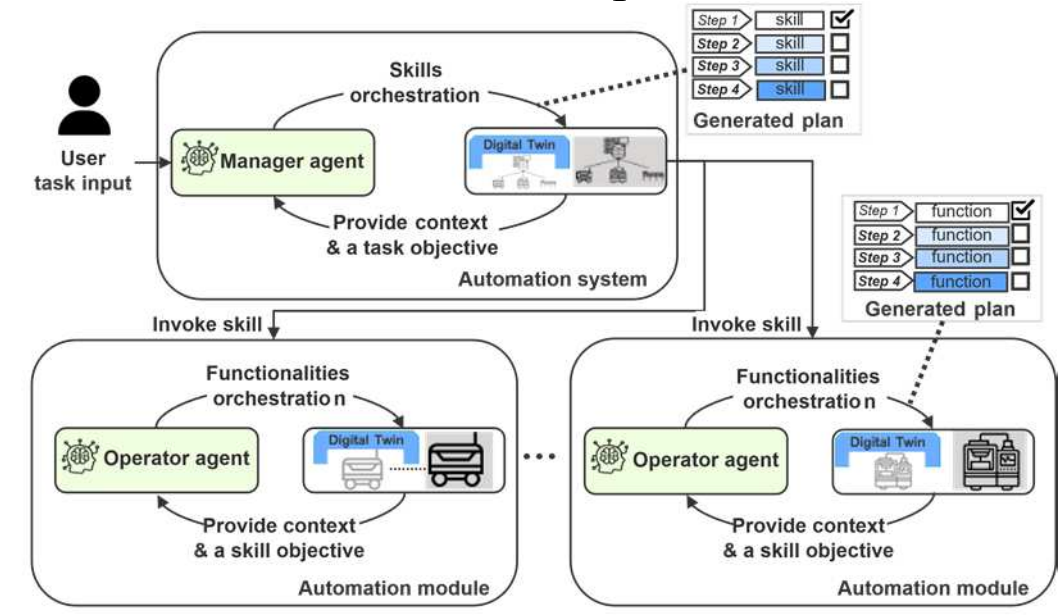

Figure 7 Task Decomposition Mechanism and the Design of Manager and Operator Agents.

The LLM agents decompose the user-defined task and orchestrate a sequential plan combining atomic skills and functions to accomplish the task. The process modeling is **planning-based**: given a user task as input, an LLM agent first generates a high-level production plan consisting of skill invocations across relevant automation modules. Each skill is then further decomposed through LLM reasoning into a detailed sequence of function calls at the component level.

The LLM agents are systematically designed using a manager-operator agent hierarchy. The **Manager Agent** is responsible for interpreting high-level user-defined tasks, decomposing them into machine skills, and distributing these skills across relevant automation modules. When a skill command reaches a specific automation module, an **Operator Agent** takes over, further decomposing the skill into a detailed sequence of executable function calls at the component level.

These plans are executed through the digital twin's service interfaces. The **system state** maintains only the **current execution status** of the plan and supports replanning when triggered by exception events or execution failures.

This approach enables dynamic adaptation and flexible task execution within a modular automation environment, supported by the intelligent reasoning capabilities of LLMs.

### B. *Event-Driven Control of Industrial Automation System with LLM (Design Paradigm Type 2)*

This case study [10], [11] presents an event-driven information modeling approach that continuously provides real-time data updates to LLM agents. This allows the LLM to interpret events within the physical automation environment and make informed decisions to control the system.

A key innovation in this approach is the semantic enhancement of raw data from field components through an event-driven information modeling mechanism. Traditional industrial automation systems often produce data at a low semantic level—such as binary signals from sensors or numeric feedback from actuators—which inherently lack sufficient semantic clarity for meaningful interpretation. To address this limitation, the approach introduces a dedicated Data Observer software component. This Data Observer continuously monitors low-level signals from field components, including sensor states, actuator signals, and controller statuses, and translates these raw data signals into semantically rich textual event descriptions through predefined semantic annotations.

These semantically enhanced events are generated dynamically during system operation and are stored within an event log memory. The event log provides a clear history of system activities, forming a basis for LLM-driven reasoning and control. LLM agents interact with this event log memory by subscribing to specific event notifications, which are generated in real-time by the digital twin middleware upon detecting changes in the automation system or the underlying technical process.

The agent system in this implementation is specifically structured around three defined roles: **Manager Agent**, **Operator Agent**, and **Summarization Agent**. The Manager Agent listens to high-level user commands or event triggers, then generates task plans based on the incoming semantically enriched events. The Operator Agent subsequently executes these plans by translating high-level task directives into executable function calls on the field automation components. The Summarization Agent continuously observes the event log memory, providing concise operational summaries to users for improved transparency and interpretability.

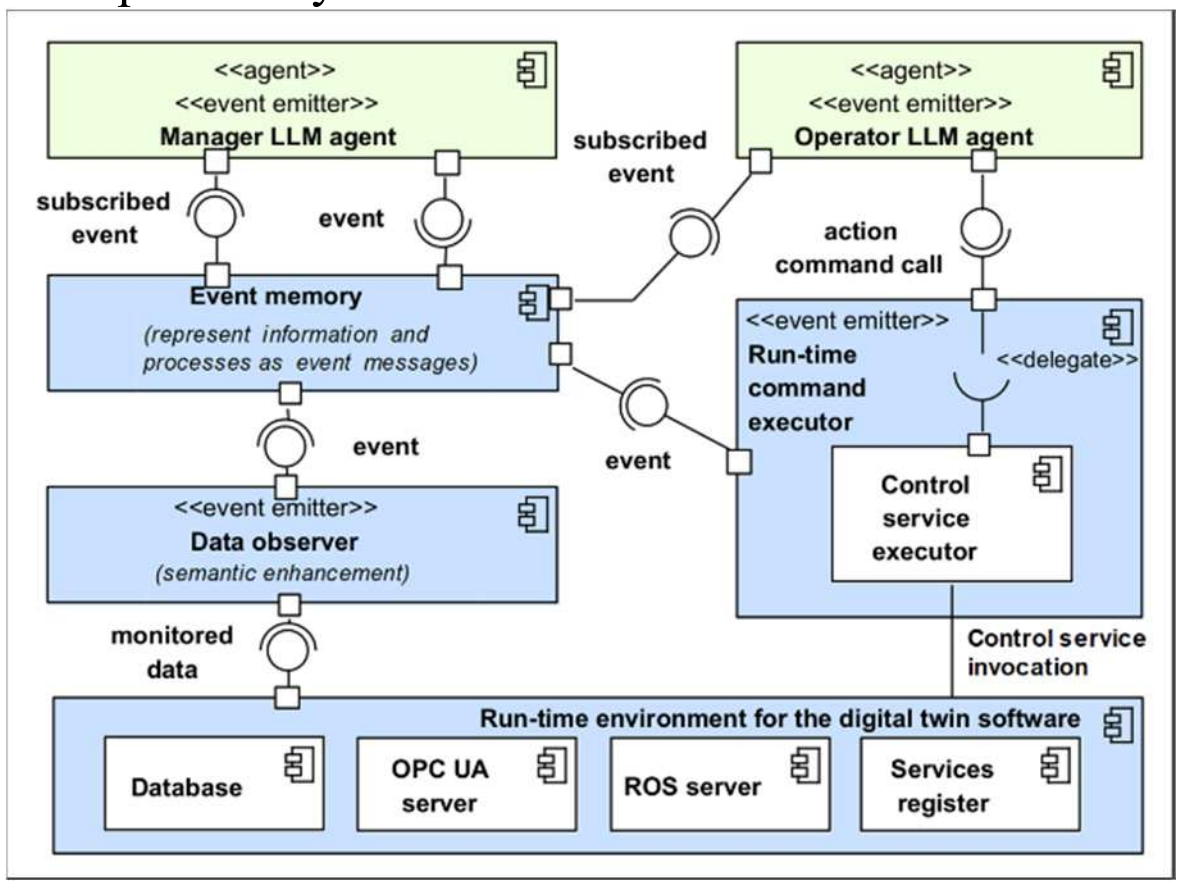

Figure 8 System Component Diagram of the Implementated System for Event-Driven Control of Industrial Automation System with LLM

The concrete implementation was realized using established industrial communication protocols and frameworks such as OPC UA and ROS, as shown in Figure *8*. Additionally, the system facilitates structured dataset creation derived from operational event logs, which further supports the supervised fine-tuning of LLM models. This allows improvement of model accuracy, tailoring the LLMs precisely to downstream industrial control tasks.

### C. *Reasoning on System State Snapshot of Simulated Process (Design Paradigm Type 1)*

This case study [9] introduces a specialized multi-agent LLM framework designed explicitly for the autonomous parametrization of digital twin simulation models. Parametrization of technical processes typically requires human knowledge and iterative experimentation. This use case directly addresses that challenge by deploying multiple specialized LLM agents to automatically explore the parameter space of digital twin simulations, identifying feasible and optimal parameter configurations for simulated technical processes.

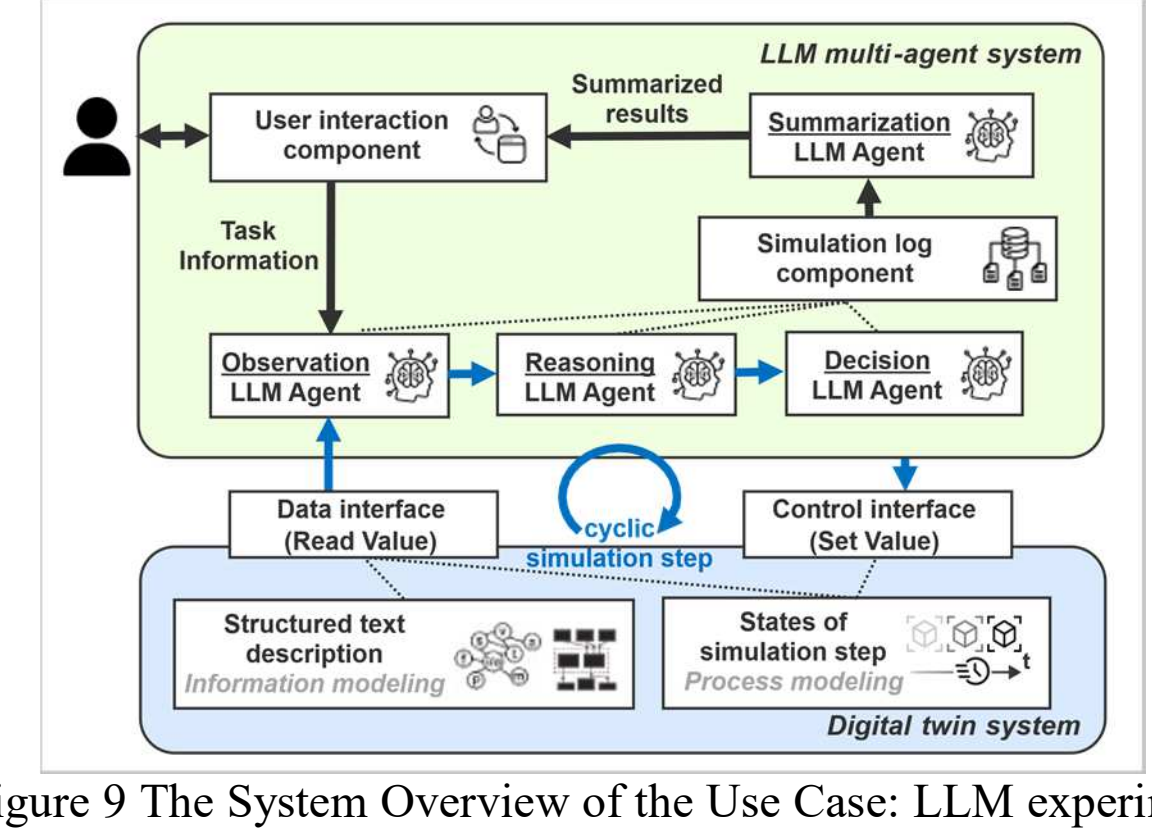

Figure 9 The System Overview of the Use Case: LLM experiments with Simulation Model to Determine Parameter Settings

In this implementation (Figure *9*), the technical process is modeled as a simulation composed of discrete execution steps, with each step associated with a corresponding system state representation. As the simulation progresses, these snapshots are generated incrementally, providing a structured context for agent-based reasoning.

The cognition layer is implemented as an LLM-based system, consisting of four distinct agent roles that operate iteratively within a structured processing pipeline:

- **Observation and Reasoning Agents** process raw data from the ongoing simulation, extract meaningful insights from state snapshots, and apply heuristic reasoning to interpret and analyze the observed states.
- **Decision Agent** generates executable control actions in the form of function calls, dynamically setting the simulation parameters to guide the simulation toward desirable outcomes.
- **Summarization Agent** consolidates agents output, executed control, and other system information, ultimately generating a concise, user-friendly summary report highlighting effective parameter configuration identified during the simulation run.

This case study demonstrates the potential of LLM-based multi-agent systems to autonomously parametrize digital twin simulations. It highlights a promising research direction for integrating LLM intelligent reasoning into process simulation and control.

## VI. Discussion

This section discusses key actionable insights from the proposed architecture and use cases, including when LLMs are suitable for industrial automation, how integration depth affects system capabilities, and practical recommendations for system development. These insights help inform future development and investment decisions.

### A. *Application goals: when to apply LLM in intelligent automation?*

The application of LLM-based automation is particularly advantageous when the following characteristics are present:

**Knowledge-intensive Tasks**: Tasks requiring semantic interpretation of textual data and perform reasoning based on existing human knowledge.

**Flexibility and On-demand:** Tasks requiring adaptability and customization, where rigid automation falls short. LLMs enable systems to respond flexibly to varying, unforeseen demands.

**Communicative Tasks:** Tasks involving text processing, interaction with user, explanation, insights analysis, recommendation, and report content generation.

### B. *The concept of "Return on Intelligence"*

To support the evaluation of LLM integration in automation systems, we introduce the concept of **"Return on Intelligence",** analogous to "Return on Investment" and illustrated in Figure *10*. It refers to the reduction in human effort, compared to conventional systems, resulting from the development of an intelligent automation solution.

For instance, LLM-enhanced systems like intelligent robots, advanced automation, or service chatbots may initially demand higher human effort during the development phase (front-loading of efforts, marked with red in Figure *10*). This increased effort can arise from the complexities involved in system design, knowledge representation modeling, and integration processes. However, this investment can be justified by reduced operational effort and greater flexibility during the system's usage phase (marked with green in Figure *10*), especially when users face cognitive challenges or knowledge barriers in operating automation systems. Furthermore, consolidating knowledge in digital twin and enabling its automated utilization through LLM agents can make the digital twin and LLM a strategic software asset.

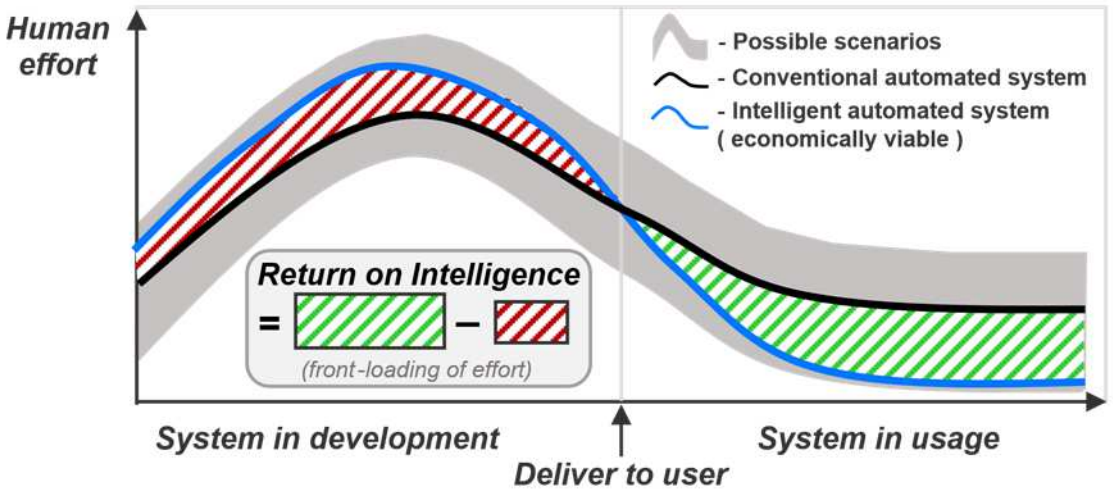

Figure 10 "Return on Intelligence" Concept Principal Illustration, The trends of the curves do not represent exact quantitative values.

For systems handling straightforward, repetitive tasks that does not require much intelligence, the LLM might yield Negative "Return on Intelligence", especially in the case that a conventional system based on simple rules and algorithmic logic can have lower development and operation costs and barely requires human intervention.

The human effort and development cost can vary depending on the task use case. For example, building a simple question-answering chatbot may require less effort than developing a system that integrates digital twins or simulation-based task-solving capabilities.

While this paper provides a design architecture and use case demonstration in a laboratory setup, realizing this "Return on Intelligence" and making LLM integration economically beneficial requires further investigation on specific use cases and product development.

### C. *Integration Levels of LLM-based Automation Systems*

To discuss how the integration of digital twins and automation systems transforms LLMs from passive knowledge processors into intelligent agents capable of real-world interaction and control, it is useful to examine the progressive levels of system integration. The table below compares three stages of LLM integration—standalone usage, combination with digital twins, and full integration with automation systems. Each stage enhances the system's capabilities in data access, modeling, and interaction, reflecting a progressive path toward more autonomous and intelligent automated systems.

Table 1 Integration Levels of LLM-powered Systems

| LLM | LLM + Digital Twins | LLM + Digital Twins + Automation System |
|---|---|---|
| **External Knowledge & Data Access** | | |
| Search engine, text file attachments, tools integration | High-fidelity models that replicate reality | Real-time data acquisition from the physical reality |
| **Modeling** | | |
| Textual data | Systematic knowledge and representation modeling (serialized as textual data) | Synchronized data for model update from physical processes |
| | Integration of simulation models (serialized as textual data) | Feedback to the LLM reasoning process and generated commands |
| **Interaction Capability** | | |
| Primarily text-based interaction | Interaction with other systems via digital twin software interfaces | Perception and actuation to interact with physical world |

### D. Test-driven development

Drawing from our projects experience, we formed a task-centric, test-driven methodology in building, integrating, and validating such systems in LLM projects:

*1) Define Tasks and Test Cases*

Begin by defining the typical tasks intended for automation. Design typical test cases based on these tasks to evaluate whether the chosen LLM has the baseline knowledge and capabilities to perform them successfully. Automated benchmarking can be first realized using these test cases, and the quality of LLM responses can be further automatically evaluated through LLM-as-a-judge methods [22].

*2) Tackle Complexity with Task Decomposition and LLM Agent Design*

If a task is too complex for a single LLM, create a structured LLM multi-agent system. Break down the task into smaller, manageable sub-tasks within a task processing pipeline. Assign specialized LLM agents to sub-tasks that match the characteristics described in Section VI.A..

*3) Integrate Domain-Specific Knowledge and Tools*

If the task requires knowledge and capability beyond the LLM's built-in capabilities, apply supporting methods such as tool usage [16], [17], [18], RAG variants [14], [15]. Some other options are also listed in IV.B.3), where "hallucination" problem is discussed.

*4) Iterative Testing and Refinement*

Use iterative testing throughout development to assess system performance, uncover issues, and refine the information processing pipeline. Concentrate the LLM agent's reasoning function on sub-tasks where it is most effective, based on task characteristics (as outlined in VI.A) and observed benchmark outcomes.

## VII. Conclusion

This paper presents a three-layer architecture that integrates LLMs, digital twins, and automation systems to enable intelligent task automation in industrial environments. Digital twins serve as the critical bridge between the physical system and cognitive reasoning, allowing LLMs to perceive and influence real-world processes. We propose three design paradigms—state snapshots, event logs, and planning sequences—to structure system information for LLM-driven control. To manage task complexity and improve reliability, we employ multi-agent design and task decomposition strategies. Realistic use cases demonstrate the practical implementation and benefits of the proposed approach. We introduce the concept of "Return on Intelligence" to assess the value of integrating LLMs into automation systems. Moving forward, future efforts could focus on connecting academic research insights with industrial needs, transforming the proposed architecture into more innovative and value-adding applications.

## Acknowledgment

This work was supported by Stiftung der Deutschen Wirtschaft (SDW) and the Ministry of Science, Research and the Arts of the State of Baden-Wuerttemberg within the support of the projects of the Exzellenzinitiative II.

## References

[1] M. Weyrich, *Industrial Automation and Information Technology*. Berlin, Heidelberg: Springer Berlin Heidelberg, 2024. doi: 10.1007/978-3-662-69243-1.

[2] B. Ashtari Talkhestani *et al.*, "An architecture of an Intelligent Digital Twin in a Cyber-Physical Production System," *At-Automatisierungstechnik*, vol. 67, no. 9, pp. 762–782, Sep. 2019, doi: 10.1515/AUTO-2019-0039/PDF.

[3] A. S. Eddington, *The nature of the physical world*, vol. 39, no. 5. Dent, 1928.

[4] J. Barbour, *The End of Time: The Next Revolution in Physics*. Weidenfeld & Nicholson, 1999.

[5] W. Gurnee and M. Tegmark, "Language Models Represent Space and Time," in *International Conference on Learning Representations*, 2023. doi: 10.48550/ARXIV.2310.02207.

[6] I. Dasgupta *et al.*, "Language models show human-like content effects on reasoning tasks," Jul. 2022, doi: 10.48550/arXiv.2207.07051.

[7] J. Wei *et al.*, "Chain-of-Thought Prompting Elicits Reasoning in Large Language Models," in *Advances in Neural Information Processing Systems*, S. Koyejo, S. Mohamed, A. Agarwal, D. Belgrave, K. Cho, and A. Oh, Eds., Curran Associates, Inc., 2022, pp. 24824–24837.

[8] L. Ruis *et al.*, "Procedural Knowledge in Pretraining Drives Reasoning in Large Language Models," in *ICLR 2025*, Nov. 2024. [Online]. Available: https://openreview.net/forum?id=1hQKHHUsMx

[9] Y. Xia, D. Dittler, N. Jazdi, H. Chen, and M. Weyrich, "LLM experiments with simulation: Large Language Model Multi-Agent System for Simulation Model Parametrization in Digital Twins," in *2024 IEEE 29th ETFA*, IEEE, Sep. 2024, pp. 1–4. doi: 10.1109/ETFA61755.2024.10710900.

[10] Y. Xia, J. Zhang, N. Jazdi, and M. Weyrich, "Incorporating Large Language Models into Production Systems for Enhanced Task Automation and Flexibility," *Automation 2024*, Jul. 2024, doi: 10.51202/9783181024379.

[11] Y. Xia, N. Jazdi, J. Zhang, C. Shah, and M. Weyrich, "Control Industrial Automation System with Large Language Models," Sep. 2024, doi: 10.48550/arXiv.2409.18009.

[12] Y. Xia, M. Shenoy, N. Jazdi, and M. Weyrich, "Towards autonomous system: Flexible modular production system enhanced with large language model agents," *IEEE ETFA*, 2023, doi: 10.1109/ETFA54631.2023.10275362.

[13] Z. Ji *et al.*, "Survey of Hallucination in Natural Language Generation," *ACM Comput Surv*, vol. 55, no. 12, Dec. 2023, [Online]. Available: https://dl.acm.org/doi/10.1145/3571730

[14] P. Lewis *et al.*, "Retrieval-Augmented Generation for Knowledge-Intensive NLP Tasks," in *Advances in Neural Information Processing Systems*, H. Larochelle, M. Ranzato, R. Hadsell, M. F. Balcan, and H. Lin, Eds., Curran Associates, Inc., 2020, pp. 9459–9474. doi: 10.48550/arXiv.2501.00309.

[15] H. Han *et al.*, "Retrieval-Augmented Generation with Graphs (GraphRAG)," Dec. 2024, doi: 10.48550/arXiv.2005.11401.

[16] B. Paranjape, S. Lundberg, S. Singh, H. Hajishirzi, L. Zettlemoyer, and M. T. Ribeiro, "ART: Automatic multi-step reasoning and tool-use for large language models," Mar. 2023.

[17] Y. Qin *et al.*, "ToolLLM Facilitating Large Language Models to Master 16000+ Real-world APIs," in *The 12th International Conference on Learning Representations*, 2024.

[18] Anthropic, "Model Context Protocol (MCP)." [Online]. Available: https://docs.anthropic.com/en/docs/agents-and-tools/mcp

[19] C. Snell, J. Lee, K. Xu, and A. Kumar, "Scaling LLM Test-Time Compute Optimally can be More Effective than Scaling Model Parameters," Aug. 2024.

[20] S. Yao *et al.*, "ReAct: Synergizing Reasoning and Acting in Language Models," in *International Conference on Learning Representations*, 2023.

[21] D. Guo *et al.*, "DeepSeek-R1: Incentivizing Reasoning Capability in LLMs via Reinforcement Learning," Jan. 2025.

[22] J. Gu *et al.*, "A Survey on LLM-as-a-Judge," Nov. 2024, [Online]. Available: https://arxiv.org/abs/2411.15594